\documentclass[twoside,12pt]{article}
\usepackage{epsfig}

\newcommand{\be}{\begin{equation}}
\newcommand{\ee}{\end{equation}}
\newcommand{\bea}{\begin{eqnarray}}
\newcommand{\eea}{\end{eqnarray}}

\usepackage{amsmath}
\usepackage{amsfonts}
\usepackage{amssymb}
\usepackage{amsxtra}
\usepackage{tikz}
\usetikzlibrary{arrows,snakes,decorations.pathmorphing}
\usepackage{supertabular}
\begin{document}
\title{ \vspace{1cm} 
Can the ``basis vectors'', describing the internal spaces of 
fermion and boson fields with the Clifford odd (for fermion) and
Clifford even (for boson) objects, explain interactions among 
fields, with gravitons included?\\
}

\author{N.S.\ Manko\v c Bor\v stnik$^{1}$,
\\
$^1$Department of Physics, University of Ljubljana\\
SI-1000 Ljubljana, Slovenia}  
 
\maketitle

\begin{abstract}
The Clifford odd and even ``basis vectors'', describing the internal
spaces of fermion and boson fields, respectively, offer in even-dimensional
spaces, like in $d=(13+1)$, the description of quarks and leptons and
antiquarks and antileptons appearing in families, as well as of all 
the corresponding gauge fields: photons, weak bosons, gluons, 
Higgs's scalars and the gravitons, which not only explain all the 
assumptions of the {\it standard model}, and  makes several predictions,  
but also explains the existence of the graviton gauge fields.
Analysing the properties of fermion and boson fields concerning how
they manifest in $d=(3+1)$, assuming space in $d=(3+1)$ flat, while
all the fields have non-zero momenta only in $d=(3+1)$, this article
illustrates that scattering of fermion and boson fields, with gravitons
included, represented by the Feynman diagrams, are determined by the
algebraic products of the corresponding ``basis vectors'' of fields,
contributing to scattering. There are two kinds of boson gauge fields
appearing in this theory, both contribute when describing scattering.
We illustrate, assuming that the internal space, which manifests in
$d=(3+1)$ origin in $d=(5+1)$, and in $d=(13+1)$, the annihilation
of an electron and positron into two photons, and the scattering of
an electron and positron into two muons.
The theory offers an elegant and promising illustration of the 
interaction among fermion and boson second quantised fields.
\end{abstract} 
%

\section{Introduction}
\label{introduction}

In a long series of works~\cite{norma92,norma93,norma95,pikanorma2005,
nh02,nd2017,2020PartIPartII,nh2021RPPNP,n2022IARD,n2023NPB,n2023MDPI,gmdn2008,
gn2009,n2014matterantimatter,JMP2013,gn2014,nm2015}
the author, with collaborators,
has found the phenomenological success with the model named the
{\it spin-charge-family} theory with the properties: The creation and
annihilation operators for fermions and bosons fields are described as tensor
products of the odd (for fermions) and the even (for bosons)
``basis vectors'' and basis in ordinary space%
~\cite{norma93,nh2021RPPNP,n2023NPB,
n2023MDPI}.

The theory explains the observed properties of fermions and
bosons and several cosmological observations, offering several predictions,
like~\footnote{ Let be pointed out that the description of internal spaces of
fermions and bosons with the group theory can elegantly be replaced by
the ``basis vectors'' for fermion and boson fields, if the ``basis vectors'' are
the superposition of the odd (for fermions) and even (for
bosons) products of $\gamma^a$'s. The ``basis vectors'' can easily be
constructed as products of nilpotents and projectors, which are eigenfunctions
of the chosen subalgebra members.
}:
{\bf \, i.\,} The existence of a fourth family to the three already observed~\cite{pikanorma2005,gmdn2008,nh2021RPPNP},
{\bf \, ii.\,}The particular symmetry of mass matrices $4 \times 4$ of quarks and
leptons (and antiquarks and antileptons)~\cite{pikanorma2005,gmdn2008,
nh2021RPPNP}, {\bf \, iii.\,} The existence of (at low energies decoupled) another
group of four families (the masses of which are determined by another
scalar fields~\cite{gn2009,JMP2013,nm2015,nh2021RPPNP}), offering the
explanation for the dark matter, {\bf \, iv.\,} The existence of scalar triplet and
antitriplet fields, offering an explanation for the matter/antimatter asymmetry in 
our universe~\cite{n2014matterantimatter}. {\bf \, v.\,} A new understanding of 
the second quantization of fermion and boson fields by describing their internal spaces by the superposition of (odd for fermions and even for bosons) products  
of $\gamma^a$'s, forming the anticommuting odd ``basis vectors'' for fermions
 and commuting even ``basis vectors'' for bosons~\footnote{
The description of internal spaces of fermions and bosons by the odd and even ``basis vectors'', respectively, which are chosen to be the eigenvectors of 
the chosen Cartan subalgebra, offers a new view on the matrix representation 
of the internal spaces of fermions and bosons.
}, 
and a unique interpretation of the internal spaces of 
fermions and bosons~\cite{nh2021RPPNP,n2023NPB,n2023MDPI}.

In Refs.~\cite{n2023NPB,n2023MDPI,nh2023dec} properties of the odd and
even ``basis vectors'' are discussed in even ($d=2(2n+1)$ or
$d=4n$) and odd (d=2n+1) dimensional spaces (all fields are assumed
to be massless, representing fields before the break of symmetries).

In even-dimensional spaces, fermion fields, described by the odd
``basis vectors'', are superpositions of odd products of operators
$\gamma^a$, chosen to be the eigenvectors of the Cartan subalgebra
members, Eq.~(\ref{cartangrasscliff}). Having $2^{\frac{d}{2}-1}$ members
(which include particles and antiparticles) appearing in $2^{\frac{d}{2}-1}$
families, ``basis vectors'' describing fermion fields anti-commute, fulfilling
together with their Hermitian conjugated partners (appearing in a separate
group) the anti-commutation relations postulated by Dirac.

In even-dimensional spaces, boson fields, described by even
``basis vectors'' --- they are superpositions of even products of operators
$\gamma^a$, chosen as well to be the eigenvectors of the Cartan
subalgebra members, Eq.~(\ref{cartangrasscliff}) --- appearing in two 
orthogonal groups, each with $2^{\frac{d}{2}-1}\times 2^{\frac{d}{2}-1}$ members, have their Hermitian conjugated partners within the same group, 
again in agreement with the Dirac postulates for the second quantization of 
boson fields~\cite{n2023NPB}.

Although there is a simple relation between both kinds of ``basis vectors''
(the multiplication of the odd ``basis vectors'' with operators 
$\gamma^a$ transforms the odd ``basis vectors'' -- appearing in two
separate groups, Hermitian conjugated to each other --- into the even
``basis vectors'', which have their Hermitian conjugated partners within the
same group, and opposite), Subsect~\ref{gammatilde}, they manifest drastic
differences~\cite{n2023NPB,n2023MDPI}.

In even dimensional space with $d= (13+1)$ the internal space of fermions,
described by the odd ``basis vectors'' and analysed concerning
the {\it standard model} groups $SO(3,1) \times SU(2)_{I}\times SU(2)_{II}
\times SU(3) \times U(1)$, includes in one family quarks and leptons and
antiquarks and antileptons with desired properties required by the
{\it standard model} before the electroweak break, as presented in
Table~\ref{Table so13+1.}~\cite{nh2021RPPNP}.
Table manifests that the $SO(7,1)$ part of the group $SO(13,1)$ are identical
for quarks and leptons and antiquarks and antileptons. Quarks and leptons
distinguish only in the $SU(3)$ and $U(1)$ part of $SO(13,1)$, and antiquarks
and antileptons distinguish in the $SU(3)$ and $U(1)$ part of $SO(13,1)$%
~\footnote{
There are two kinds of the $SU(2)$ weak charges: $SU(2)_{I}$ and
$SU(2)_{II}$. While the right-handed quarks and leptons and the left-handed
antiquarks and antileptons have the $SU(2)_{I}$ weak charge equal zero
and the $SU(2)_{II}$ weak charge equal to $+ \frac{1}{2}$ ($u_R, \nu_R,
\bar{d}_{L}, \bar{e}_{L}$) or $- \frac{1}{2}$ ($d_R, e_R, \bar{u}_{L},
\bar{\nu}_L$), have the left-handed quarks and leptons and the right-handed
antiquarks and antileptons the $SU(2)_{II}$ charge equal to zero and
the $SU(2)_{I}$ weak charge equal to $+ \frac{1}{2}$ ($u_L, \nu_L,
\bar{d}_{R}, \bar{e}_{R}$) or $- \frac{1}{2}$ ($d_L, e_L, \bar{u}_{R},
\bar{\nu}_R$)~\cite{nh2021RPPNP,n2023NPB,n2023MDPI}.
}.
Quarks and leptons and antiquarks and antileptons appear in families.

The even ``basis vectors'' carrying in addition to the ``basis 
vectors'' also the space index, either equal to $(0,1,2,3)$ (they manifest 
as photons, weak bosons of two kinds, gluons and gravitons) or to 
$(5,6,7..,14)$ (they manifest as scalar gauge fields, like higgses with
the weak and hyper charges, explaining Yukawa 
couplings~\cite{JMP2013,nh2021RPPNP}, and triplet and antitriplet scalar 
fields~\cite{n2014matterantimatter}, offering the explanation for the 
matter/antimatter asymmetry) appear elegantly as the gauge fields of 
the corresponding fermion fields
(quarks and leptons and antiquarks and antileptons)~\footnote{
There are two kinds (two groups) of the even ``basis vectors''.
They are orthogonal to each other, both having their Hermitian conjugated
partners within their group. One kind of the even ``basis
vectors'' transforms family members of each family of fermions among
themself. The second kind transforms a particular family member of one
family of fermions into the same family member of one of the families.
The internal spaces of the vector gauge fields of quarks and leptons and
antiquarks and antileptons are describable by the first kind of the
even ``basis vectors'', while the masses of quarks and leptons and
antiquarks and antileptons of the two groups of four families and the
masses of the weak gauge fields are determined by the scalar fields,
the internal space of which is described by even
``basis vectors'' of the second kind.
}.

Both, fermion and boson fields are in this paper assumed to have 
non-zero momentum only in $d=(3+1)$.

The odd-dimensional spaces, $d=(2n+1)$, offer the 
surprise~\cite{n2023NPB,n2023MDPI,n2024Bled}:
Half of ``basis vectors''
manifest properties of those in even dimensional spaces of one lower
dimension, the remaining half represents the anti-commuting ``basis vectors''
appearing in two orthogonal groups with the Hermitian conjugated partners
within the same groups, the commuting ``basis vectors'' appear in two
separate groups, Hermitian conjugated to each other, manifesting a kind of 
the Fadeev-Popov ghosts.
We indeed find out that they offer a special kind of 
the supersymmetry~\cite{n2024Bled}~\footnote{
In odd-dimensional spaces, $d=(2n+1)$, there are the anti-commuting
$2^{\frac{d-1}{2}-1}$ ``basis vectors'' appearing in $2^{\frac{d-1}{2}-1}$ families, with their Hermitian conjugated $2^{\frac{d-1}{2}-1}$ $\times$ $ 2^{\frac{d-1}{2}-1}$
partners in a separate group, and the commuting $2\times$
$ 2^{\frac{d-1}{2}-1}$ $\times$ $2^{\frac{d-1}{2}-1}$ ``basis vectors''
appearing in two orthogonal groups with their Hermitian conjugated partners
within the same group.\\
There are in addition the  anti-commuting $2^{\frac{d-1}{2}-1}$$\times$
$2^{\frac{d-1}{2}-1}$ ``basis vectors''  appearing in two
orthogonal groups with their Hermitian conjugated
partners within the same group, and the commuting $2^{\frac{d-1}{2}-1}$
``basis vectors'' appearing in $2^{\frac{d-1}{2}-1}$ families, with their
Hermitian conjugated $2^{\frac{d-1}{2}-1}$ $\times$
$ 2^{\frac{d-1}{2}-1}$ partners in a separate group, manifesting an
interesting kind of supersymmetry.
 }.
 
In Sec.~\ref{basisvectors0}, this article presents a short overview of the
definition and the properties of the odd and even
``basis vectors'', mainly following Refs.~\cite{n2023NPB,n2023MDPI,n2022epjc}.

The reader can find in App.~\ref{simplecases} demonstration of the presentations
of Sec.~\ref{basisvectors0} on two simple cases: $d=(1+1)$ and 
$d=(3+1)$.

App.~\ref{gammatilde} demonstrates that although the odd and even
``basis vectors'' in even-dimensional spaces have so different properties,
yet the algebraic multiplication of one kind of ``basis vectors'' by
$\gamma^a$ or $\tilde{\gamma}^a$ transforms one kind into another.

In Sect.~\ref{creationanihilationoperators}, the corresponding creation and
annihilation operators are described as the tensor products, $*_{T}$, of the
``basis vectors'' and the basis in ordinary space.

The main part of this article is Sect.~\ref{bosons13+1and5+1}, 
representing discussions on:

{\it a.} Internal spaces of all boson fields, with the gravitons included, using 
the even ``basis vectors'' in $d=(13 +1)$-dimensional space and 
analysing them from the point of view of their
properties in $(3+1)$-dimensional space, Sect.~\ref{bosons13+1and5+1} and
Subsects.~\ref{photonsweakbosonsgluons},~\ref{scalarfields},~\ref{gravitons},
while illustrating the behaviour of fermions and bosons on the toy model of
$d=(5+1)$ from the point of view of their properties in $(3+1)$-dimensional space, Sect.~\ref{bosons5+1},\\
{\it  a.i.} paying attention to photons, in 
Subsect.~\ref{photonsweakbosonsgluons},\\
{\it  a.ii.} on weak bosons of two kinds, in Subsect.~\ref{photonsweakbosonsgluons},\\
{\it a.iii.} on gluons, in Subsect.~\ref{photonsweakbosonsgluons},\\
{\it  a.iv.} on gravitons, in Subsect.~\ref{gravitons},\\
{\it  a.v.} on scalar fields, in Subsect.~\ref{scalarfields}.

\vspace{1mm}

{\it b.} Annihilation of fermions and anti-fermions, scattering of fermions on 
fermions,  Subsect.~\ref{feynmandiagrams}.\\

In Sect.~\ref{conclusions}, we overview new recognitions in this article
and point out open problems.\\

\section{``Basis vectors'' describing internal spaces of fermion and boson 
second quantized fields}
\label{basisvectors0}

This section presents the ``basis vectors'' describing the internal spaces of
fermions and bosons as the superposition of odd, for fermions, and even,
for bosons, products of operators $\gamma^a$'s, which are chosen to be
the eigenvectors of the Cartan subalgebra members. It is a short
overview of similar sections in several papers cited
in~\cite{nh2021RPPNP,n2023NPB,n2023MDPI}. Here, we mainly follow
Refs.~\cite{n2023NPB,n2022epjc,nh2023dec}.

We start with the Grassmann algebra, offering two kinds of Clifford algebras,
defined by two kinds of the Clifford odd operators: $\gamma^{a}$'s and
$\tilde{\gamma}^{a}$'s, as it is presented in App.~\ref{grassmannclifford},
with the properties
\begin{eqnarray}
\label{gammatildeantiher}
\{\gamma^{a}, \gamma^{b}\}_{+}&=&2 \eta^{a b}= \{\tilde{\gamma}^{a},
\tilde{\gamma}^{b}\}_{+}\,, \nonumber\\
\{\gamma^{a}, \tilde{\gamma}^{b}\}_{+}&=&0\,,\quad
(a,b)=(0,1,2,3,5,\cdots,d)\,, \nonumber\\
(\gamma^{a})^{\dagger} &=& \eta^{aa}\, \gamma^{a}\, , \quad
(\tilde{\gamma}^{a})^{\dagger} = \eta^{a a}\, \tilde{\gamma}^{a}\,.
\end{eqnarray}
Both kinds offer the description of the internal spaces of fermions with the
``basis vectors'' which are superposition of odd products of either
$\gamma^{a}$'s or $\tilde{\gamma}^{a}$'s and fulfil correspondingly,
the anti-commuting postulates of second quantized fermion fields, as well as
the description of the internal spaces of boson fields with the ``basis
vectors'' which are superposition of even products of either $\gamma^{a}$'s
or $\tilde{\gamma}^{a}$'s and fulfil correspondingly the commuting
postulates of second quantized boson fields. %

Since there are not two kinds of anti-commuting fermions, and not two
corresponding kinds of their gauge fields, the postulate of
Eq.~(\ref{tildegammareduced0}) in App.~\ref{grassmannclifford} gives the possibility that only one of the
two kinds of operators are used to describe fermions and their gauge fields,
namely $\gamma^{a}$'s.
The operators $\tilde{\gamma}^a$'s can after the {\it postulate},
Eq.~(\ref{tildegammareduced0}), be used to describe the quantum numbers
of each irreducible representation of the Lorentz group, $S^{ab}$
($=\frac{i}{4}(\gamma^a \gamma^b - \gamma^b \gamma^a)$), in the
internal space of fermions by $\tilde{S}^{ab}$ ($=\frac{i}{4}
(\tilde{\gamma}^a \tilde{\gamma}^b - \tilde{\gamma}^b
\tilde{\gamma}^a)$ --- the ``family'' quantum number ---
and the quantum numbers of each irreducible representation of the Lorentz
group ${\cal S}^{ab}$ ($= S^{ab} + \tilde{S}^{ab}$) in the internal
space of bosons~\footnote{
One can prove (or read in App.~I of Ref.~\cite{nh2021RPPNP}) that the
relations of Eq.~(\ref{gammatildeantiher}) remain valid also after the
{\it postulate}, presented in Eq.~(\ref{tildegammareduced0}).}.

All the ``basis vectors'' described by $\gamma^{a}$'s, either with the
superposition of odd products of $\gamma^{a}$'s, the odd ``basis
vectors'', or with the superposition of even products of $\gamma^{a}$'s,
the even ``basis vectors'', are chosen to be the eigenvectors of the
Cartan subalgebra members of $ S^{ab}$ (for fermions), and of
${\cal S}^{ab}$ ($= S^{ab} + \tilde{S}^{ab}$) (for bosons)
\begin{small}
\begin{eqnarray}
&&S^{03}, S^{12}, S^{56}, \cdots, S^{d-1 \;d}\,, \nonumber\\
&&\tilde{S}^{03}, \tilde{S}^{12}, \tilde{S}^{56}, \cdots,  \tilde{S}^{d-1\; d}\,, 
\nonumber\\
&&{\cal {\bf S}}^{ab} = S^{ab} +\tilde{S}^{ab}\,, 
\label{cartangrasscliff}
\end{eqnarray}
\end{small}
where $ S^{ab} +\tilde{S}^{ab}= i \, (\theta^{a} 
\frac{\partial}{\partial \theta_{b}} -  \theta^{b} 
\frac{\partial}{\partial \theta_{a}}))$\,.
%

\subsection{Relations among even 
and odd ``basis vectors''} 
\label{basisvectors}

This subsection is a short overview of similar sections of several articles,
like~\cite{n2022IARD,n2022epjc,n2023MDPI,2020PartIPartII,n2023NPB}.

After the reduction of the two Clifford sub-algebras to only one,
Eq.~(\ref{tildegammareduced0}), we need to define ``basis vectors'',
which are the superposition of odd~\cite{nh2021RPPNP} or even products of
$\gamma^{a}$'s~\cite{n2023NPB}, using the technique which makes
``basis vectors'' products of nilpotents and projectors~\cite{norma93,nh02}
which are eigenvectors of the (chosen) Cartan subalgebra members, 
Eq.~(\ref{cartangrasscliff}), of the Lorentz algebras (of $ S^{ab}$
(for fermions), and of ${\cal S}^{ab}$ ($= S^{ab} + \tilde{S}^{ab}$)
(for bosons) in the space of odd (for fermions) and even (for bosons)
products of $\gamma^{a}$'s). \\
There are in even-dimensional spaces $\frac{d}{2}$ members of the Cartan
subalgebra, Eq.~(\ref{cartangrasscliff})~\footnote{
There are $\frac{d-1}{2}$ members of the Cartan
subalgebra in odd-dimensional spaces.
}.

One finds in even dimensional spaces for any of the $\frac{d}{2}$ Cartan
subalgebra members, $S^{ab}$, 
applying on a nilpotent $\stackrel{ab}{(k)}$, which is a superposition of an
odd number of $\gamma^{a}$'s, or on a projector $\stackrel{ab}{[k]}$, which
is a superposition of an even number of $\gamma^{a}$'s the relations
\begin{small}
\begin{eqnarray}
\label{nilproj}
\stackrel{ab}{(k)}:&=&\frac{1}{2}(\gamma^a +
\frac{\eta^{aa}}{ik} \gamma^b)\,, \;\;\; (\stackrel{ab}{(k)})^2=0\, , \nonumber \\
\stackrel{ab}{[k]}:&=&
\frac{1}{2}(1+ \frac{i}{k} \gamma^a \gamma^b)\,, \;\;\;(\stackrel{ab}{[k]})^2=
\stackrel{ab}{[k]},
\end{eqnarray}
\end{small}
\begin{small}
\begin{eqnarray}
\label{signature0}
S^{ab} \,\stackrel{ab}{(k)} = \frac{k}{2} \,\stackrel{ab}{(k)}\,,\quad && \quad
\tilde{S}^{ab}\,\stackrel{ab}{(k)} = \frac{k}{2} \,\stackrel{ab}{(k)}\,,\nonumber\\
S^{ab}\,\stackrel{ab}{[k]} = \frac{k}{2} \,\stackrel{ab}{[k]}\,,\quad && \quad
\tilde{S}^{ab} \,\stackrel{ab}{[k]} = - \frac{k}{2} \,\,\stackrel{ab}{[k]}\,,
\end{eqnarray}
\end{small}
with $k^2=\eta^{aa} \eta^{bb}$~\footnote{
Let us prove one of the relations in
Eq.~(\ref{signature0}): $S^{ab}\, \stackrel{ab}{(k)}= \frac{i}{2} \gamma^a
\gamma^b \frac{1}{2} (\gamma^a +\frac{\eta^{aa}}{ik} \gamma^b)=
\frac{1}{2^2}\{ -i (\gamma^a)^2 \gamma^b + i (\gamma^b)^2 \gamma^a
\frac{\eta^{aa}}{ik}\}= \frac{1}{2} \frac{\eta^{aa}\eta^{bb}}{k}
\frac{1}{2} \{\gamma^a + \frac{k^2}{\eta^{bb} ik}\gamma^b\}$. For
$k^2 = \eta^{aa} \eta^{bb}$ the first relation follows.
},
demonstrating, together with the relations in App.~\ref{usefulrelations}, that
the eigenvalues of $S^{ab}$ on nilpotents and
projectors expressed with $\gamma^a$ differ from the eigenvalues of
$\tilde{S}^{ab}$ on nilpotents and
projectors expressed with $\gamma^a$~\footnote{
One finds that
$S^{ac}\stackrel{ab}{(k)}\stackrel{cd}{(k)} = -\frac{i}{2} \eta^{aa} \eta^{cc}
\stackrel{ab}{[-k]}\stackrel{cd}{[-k]}$, while $\tilde{S}^{ac}
\stackrel{ab}{(k)}\stackrel{cd}{(k)} = \frac{i}{2} \eta^{aa} \eta^{cc}
\stackrel{ab}{[k]}\stackrel{cd}{[k]}$. More relations can be found in
Eq.~(\ref{graficfollow1}).
}.
$\tilde{S}^{ab}$
can correspondingly be used to equip each irreducible representation of $S^{ab}$
with the ''family'' quantum number.~\footnote{
The reader can find the proof of Eq.~(\ref{signature0}) also in Ref.~\cite{nh2021RPPNP},
App.~(I).}

Taking into account Eq.~(\ref{gammatildeantiher}) one finds
\begin{small}
\begin{eqnarray}
\label{usefulrel0}
\gamma^a \stackrel{ab}{(k)}&=& \eta^{aa}\stackrel{ab}{[-k]},\; \quad
\gamma^b \stackrel{ab}{(k)}= -ik \stackrel{ab}{[-k]}, \; \quad 
\gamma^a \stackrel{ab}{[k]}= \stackrel{ab}{(-k)},\;\quad \;\;
\gamma^b \stackrel{ab}{[k]}= -ik \eta^{aa} \stackrel{ab}{(-k)}\,,\nonumber\\
\tilde{\gamma^a} \stackrel{ab}{(k)} &=& - i\eta^{aa}\stackrel{ab}{[k]},\quad
\tilde{\gamma^b} \stackrel{ab}{(k)} = - k \stackrel{ab}{[k]}, \;\qquad \,
\tilde{\gamma^a} \stackrel{ab}{[k]} = \;\;i\stackrel{ab}{(k)},\; \quad
\tilde{\gamma^b} \stackrel{ab}{[k]} = -k \eta^{aa} \stackrel{ab}{(k)}\,,
\nonumber\\ %
\stackrel{ab}{(k)}\stackrel{ab}{(-k)}& =& \eta^{aa} \stackrel{ab}{[k]}\,,\quad \;
\stackrel{ab}{(-k)}\stackrel{ab}{(k)} = \eta^{aa} \stackrel{ab}{[-k]}\,,\quad\;
\stackrel{ab}{(k)}\stackrel{ab}{[k]} =0\,,\quad \quad\,
\stackrel{ab}{(k)}\stackrel{ab}{[-k]} =
\stackrel{ab}{(k)}\,,\quad 
\nonumber\\ 
\stackrel{ab}{(-k)}\stackrel{ab}{[k]} &=& \stackrel{ab}{(-k)}\,,\quad \quad\quad
\stackrel{ab}{[k]}\stackrel{ab}{(k)}= \stackrel{ab}{(k)}\,,
\quad \quad \quad \;
\stackrel{ab}{[k]}\stackrel{ab}{(-k)} =0\,,\quad \quad\,
\stackrel{ab}{[k]}\stackrel{ab}{[-k]} =0\,,\quad
\nonumber\\
\stackrel{ab}{(k)}^{\dagger} &=& \eta^{aa}\stackrel{ab}{(-k)}\,,\quad
(\stackrel{ab}{(k)})^2 =0\,, \quad \stackrel{ab}{(k)}\stackrel{ab}{(-k)}
=\eta^{aa}\stackrel{ab}{[k]}\,,\nonumber\\
\stackrel{ab}{[k]}^{\dagger} &=& \,\stackrel{ab}{[k]}\,, \quad \quad \quad \quad
(\stackrel{ab}{[k]})^2 = \stackrel{ab}{[k]}\,,
\quad \stackrel{ab}{[k]}\stackrel{ab}{[-k]}=0\,.
\end{eqnarray}
\end{small}
More relations are presented in App.~A of Ref.~\cite{nh2023dec}. 

One expects correspondingly:\\
{\bf a.} The odd ``basis vectors'', describing fermion fields, must be
products of an odd number of nilpotents, at least one, and the rest of the projectors,
since a product of an odd number of nilpotents anti-commute with another
product of an odd number of nilpotents. \\
The even ``basis vectors'',
describing boson fields must be products of an even number of nilpotents.\\
{\bf b.} There are $2^d$ different products of $\gamma^a$'s.
Half of them are odd, and half of them are even. The odd ``basis vectors''
appear in $2^{\frac{d}{2}-1}$ irreducible representations, families, each
with $2^{\frac{d}{2}-1}$ members. There are two groups of the superposition
of  odd products of $\gamma^a$'s. Since the Hermitian conjugated
partner of a nilpotent $\stackrel{ab}{(k)}^{\dagger}$ is
$\eta^{aa}\stackrel{ab}{(-k)}$, it follows that the Hermitian conjugated
partners of the odd ``basis vectors'' with an odd number of nilpotents
must belong to a different group of $2^{\frac{d}{2}-1}$ members in
$2^{\frac{d}{2}-1}$ families~\footnote{
Neither $S^{ab}$ nor $\tilde{S}^{ab}$ can transform an odd product of
nilpotents to the corresponding Hermitian conjugated partner.
}. \\
The even ``basis vectors'' with an even number of nilpotents must have
their Hermitian conjugated partners within the same group; projectors are self
adjoint, $S^{ac}$ transforms $\stackrel{ab}{(k)} *_A \stackrel{cd}{(k')}$
into $\stackrel{ab}{[-k]} *_A \stackrel{cd}{[-k']}$, while $\tilde{S}^{ac}$
transforms $\stackrel{ab}{(k)} *_A \stackrel{cd}{(k')}$ into
$\stackrel{ab}{[k]} *_A $ $ \stackrel{cd}{[k']}$. Since the number of the
odd and the even products of $\gamma^a$'s is the same,
there must be another group of the even ``basis vectors'' with
$2^{\frac{d}{2}-1}\times 2^{\frac{d}{2}-1}$ members.

\subsection{Odd and even ``basis vectors''}
\label{cliffordoddeven}

We choose nilpotents, $\stackrel{ab}{(k)}$, and projectors, $\stackrel{ab}{[k]} $,
Eqs.~(\ref{nilproj}, \ref{usefulrel0}), which are eigenvectors of the Cartan subalgebra
members, as the ``building blocks'' of the ``basis vectors'', defining the
odd ``basis vectors'' as products of an odd number of nilpotents, and the rest of
projectors, and the even ``basis vectors'' as products of an even number of
nilpotents, and the rest of the projectors, recognizing that the properties of the
odd ``basis vectors'' essentially differ from the properties of the even
``basis vectors'', as explained in the introduction of this Sect.~\ref{basisvectors} in
points {\bf a.} and {\bf b.}.

\subsubsection{Odd ``basis vectors''  }
\label{odd}

This part overviews several papers with the same
topic~(\cite{nh2021RPPNP,n2023MDPI} and references therein).

The odd ``basis vectors'' are chosen to be products of an odd number
of nilpotents, and the rest, up to $\frac{d}{2}$, of projectors, each nilpotent
and each projector is chosen to be the ``eigenstate'' of one of the members
of the Cartan subalgebra, Eq.~(\ref{cartangrasscliff}),
correspondingly are the ``basis vectors'' eigenstates of all the members of the Lorentz
algebra: $S^{ab}$'s determine $2^{\frac{d}{2}-1}$
members of one family, $\tilde{S}^{ab}$'s transform each member of one family to
the same member of the  $2^{\frac{d}{2}-1}$ families.

Let us call the odd ``basis vectors'' $\hat{b}^{m \dagger}_{f}$ if it is the
$m^{th}$ member of the family $f$. The Hermitian conjugated partner of
$\hat{b}^{m \dagger}_{f}$ is called $\hat{b}^{m}_{f} \,(=
(\hat{b}^{m \dagger}_{f})^{\dagger}$).

Let us start in $d=2(2n+1)$ with the ``basis vector'' $\hat{b}^{1 \dagger}_{1}$
which is the product of only nilpotents, all the rest members belonging to the $f=1$
family follow by the application of $S^{01}$, $S^{03}$, $ \dots, S^{0d}, S^{15}$,
$\dots, S^{1d}, S^{5 d}\dots, S^{d-2\, d}$. They are presented on the left-hand side.
Their Hermitian conjugated partners are presented on the right-hand side.
The algebraic product mark, $*_{A}$, among nilpotents and projectors is skipped.
\begin{small}
\begin{eqnarray}
\label{allcartaneigenvec}
&& \qquad \qquad \qquad \qquad \qquad \qquad d=2(2n+1)\, ,\nonumber\\
&& \hat{b}^{1 \dagger}_{1}=\stackrel{03}{(+i)}\stackrel{12}{(+)} \stackrel{56}{(+)}
\cdots \stackrel{d-1 \, d}{(+)}\,,\qquad \qquad \qquad \quad \quad
\hat{b}^{1}_{1}=\stackrel{03}{(-i)}\stackrel{12}{(-)}\cdots \stackrel{d-1 \, d}{(-)}\,,
\nonumber\\
&&\hat{b}^{2 \dagger}_{1} = \stackrel{03}{[-i]} \stackrel{12}{[-]}
\stackrel{56}{(+)} \cdots \stackrel{d-1 \, d}{(+)}\,,\qquad \qquad \qquad \qquad\;\;
\hat{b}^{2 }_{1} = \stackrel{03}{[-i]} \stackrel{12}{[-]}
\stackrel{56}{(-)} \cdots \stackrel{d-1 \, d}{(-)}\,,\nonumber\\
&& \cdots \qquad \qquad \qquad \qquad \qquad \qquad \qquad \qquad \qquad\;
\cdots \nonumber\\
&&\hat{b}^{2^{\frac{d}{2}-1} \dagger}_{1} = \stackrel{03}{(+i)} \stackrel{12}{[-]}
\stackrel{56}{[-]} \dots \stackrel{d-3\,d-2}{[-]}\;\stackrel{d-1\,d}{[-]}\,, \qquad
\hat{b}^{2^{\frac{d}{2}-1}}_{1} = \stackrel{03}{(-i)} \stackrel{12}{[-]}
\stackrel{56}{[-]} \dots \stackrel{d-3\,d-2}{[-]}\;\stackrel{d-1\,d}{[-]}\,,
\nonumber\\
&& \cdots\,, \qquad \qquad \qquad \qquad \qquad \qquad \qquad \qquad \cdots\,.
\end{eqnarray}
\end{small}

In $d=4n$, the choice of the starting ``basis vector'' with the maximal number of nilpotents
must have one projector
\begin{small}
\begin{eqnarray}
\label{allcartaneigenvec4n}
&& \qquad \qquad \qquad \qquad\qquad \qquad d=4n\, ,\nonumber\\
&& \hat{b}^{1 \dagger}_{1}=\stackrel{03}{(+i)}\stackrel{12}{(+)}
\cdots \stackrel{d-1 \, d}{[+]}\,,\qquad \qquad \qquad \qquad\quad
\hat{b}^{1}_{1}=\stackrel{03}{(-i)}\stackrel{12}{(-)}
\cdots \stackrel{d-1 \, d}{[+]}\,
\nonumber\\
&&\hat{b}^{2 \dagger}_{1} = \stackrel{03}{[-i]} \stackrel{12}{[-]}
\stackrel{56}{(+)} \cdots \stackrel{d-1 \, d}{[+]}\,,\qquad \qquad \qquad \qquad
\hat{b}^{2 }_{1} = \stackrel{03}{[-i]} \stackrel{12}{[-]}
\stackrel{56}{(-)} \cdots \stackrel{d-1 \, d}{[+]}\,,
\nonumber\\
&& \cdots \,, \qquad \qquad \qquad \qquad\qquad \qquad \qquad \qquad \quad
\cdots\,, \nonumber\\
&&\hat{b}^{2^{\frac{d}{2}-1} \dagger}_{1} = \stackrel{03}{[-i]} \stackrel{12}{[-]}
\stackrel{56}{[-]} \dots \stackrel{d-3\,d-2}{[-]}\;\stackrel{d-1\,d}{(-)}\,, \qquad
\hat{b}^{2^{\frac{d}{2}-1} }_{1} = \stackrel{03}{[-i]} \stackrel{12}{[-]}
\stackrel{56}{[-]} \dots \stackrel{d-3\,d-2}{[-]}\;\stackrel{d-1\,d}{(+)}\,,
\nonumber\\
&& \cdots\,,\qquad \qquad \qquad \qquad\qquad \qquad \qquad \quad \quad
\cdots\,.
\end{eqnarray}
\end{small}
The Hermitian conjugated partners of the odd ``basis vectors''
$\hat{b}^{m \dagger}_{1}$, presented in Eqs.~(\ref{allcartaneigenvec},
\ref{allcartaneigenvec4n}) on the right-hand side, follow
if all nilpotents $\stackrel{ab}{(k)}$ of $\hat{b}^{m \dagger}_{1}$ are transformed
into $\eta^{aa} \stackrel{ab}{(-k)}$.

For either $d=2(2n+1)$ or for $d=4n$ all the $2^{\frac{d}{2}-1}$ families follow by
applying $\tilde{S}^{ab}$'s on all the members of the starting family~\footnote{
Or one can find the starting $ \hat{b}^{1 \dagger}_{f}$ for all families $f$ and then generate all the members $\hat{b}^{m}_{f}$ from $\hat{b}^{1\dagger}_{f}$ by the application of $S^{ab}$ on the starting member.}.

It is not difficult to see that all the ``basis vectors'' within any family, as well as the
``basis vectors'' among families, are orthogonal; that is, their algebraic product is zero.
The same is true within their Hermitian conjugated partners. Both can be proved by
the algebraic multiplication using Eqs.~(\ref{usefulrel0}).

\begin{eqnarray}
\hat{b}^{m \dagger}_f *_{A} \hat{b}^{m `\dagger }_{f `}&=& 0\,,
\quad \hat{b}^{m}_f *_{A} \hat{b}^{m `}_{f `}= 0\,, \quad \forall m,m',f,f `\,.
\label{orthogonalodd}
\end{eqnarray}

When we choose the vacuum state equal to
\begin{eqnarray}
\label{vaccliffodd}
|\psi_{oc}>= \sum_{f=1}^{2^{\frac{d}{2}-1}}\,\hat{b}^{m}_{f}{}_{*_A}
\hat{b}^{m \dagger}_{f} \,|\,1\,>\,,
\end{eqnarray}
for one of the members $m$, which can be anyone of the odd irreducible
representations $f$ it follows that the odd ``basis vectors'' obey the relations
\begin{eqnarray}
\label{almostDirac}
\hat{b}^{m}_{f} {}_{*_{A}}|\psi_{oc}>&=& 0.\, |\psi_{oc}>\,,\nonumber\\
\hat{b}^{m \dagger}_{f}{}_{*_{A}}|\psi_{oc}>&=&  |\psi^m_{f}>\,,\nonumber\\
\{\hat{b}^{m}_{f}, \hat{b}^{m'}_{f `}\}_{*_{A}+}|\psi_{oc}>&=&
 0.\,|\psi_{oc}>\,, \nonumber\\
\{\hat{b}^{m \dagger}_{f}, \hat{b}^{m' \dagger}_{f  `}\}_{*_{A}+}|\psi_{oc}>
&=& 0. \,|\psi_{oc}>\,,\nonumber\\
\{\hat{b}^{m}_{f}, \hat{b}^{m' \dagger}_{f `}\}_{*_{A}+}|\psi_{oc}>
&=& \delta^{m m'} \,\delta_{f f `}|\psi_{oc}>\,.
\end{eqnarray}\\

We conclude Sect.~\ref{even} with two simple exercises, presenting odd 
and even "basis vectors" for $d=(1+1)$ and 
$d=(3+1)$.

\subsubsection{\it Even ``basis vectors'' } 
\label{even}

This part is an overview of Subsect.~2.2.2 in Ref~\cite{n2023NPB}, proving
that the even ``basis vectors'' are in even-dimensional
spaces offering the description of the internal spaces of boson fields --- the
gauge fields of the corresponding odd ``basis vectors", offering a
new understanding of the second quantized fermion and boson
fields~\cite{n2022epjc}.

The even ``basis vectors'' must be in even-dimensional spaces
products of an even number of
nilpotents and the rest, up to $\frac{d}{2}$, of projectors; each nilpotent and
each projector is chosen to be the ``eigenstate'' of one of the members of the
Cartan subalgebra of the Lorentz algebra, ${\bf {\cal S}}^{ab}=(S^{ab} +
\tilde{S}^{ab})$, Eq.~(\ref{cartangrasscliff}). Correspondingly, the ``basis
vectors'' are the eigenstates of all the members of the Cartan subalgebra of
the Lorentz algebra.

The even ``basis vectors'' appear in two groups; each group has
$2^{\frac{d}{2}-1}\times $ $2^{\frac{d}{2}-1}$ members~\footnote{
The members of one group can not be reached by the members of another
group by either $S^{ab}$'s or $\tilde{S}^{ab}$'s or both.}.

$S^{ab}$ and $\tilde{S}^{ab}$ generate from the starting ``basis vector'' of
each group all the $2^{\frac{d}{2}-1} \times$ $2^{\frac{d}{2}-1}$ members.
Each group contains the Hermitian conjugated partner of any member;
$2^{\frac{d}{2}-1}$ members of each group are products of only (self-adjoint)
projectors.

Let the even ``basis vectors'' be denoted by
${}^{i}\hat{\cal A}^{m \dagger}_{f}$, with $i=(I,II)$ denoting one of the
two groups of the even ``basis vectors'', and $m$ and $f$ determine
membership of ``basis vectors'' in any of the two groups, $I$ or $II$.
\begin{eqnarray}
\label{allcartaneigenvecevenI}
d&=&2(2n+1)\nonumber\\
{}^I\hat{{\cal A}}^{1 \dagger}_{1}=\stackrel{03}{(+i)}\stackrel{12}{(+)}\cdots
\stackrel{d-1 \, d}{[+]}\,,\qquad &&
{}^{II}\hat{{\cal A}}^{1 \dagger}_{1}=\stackrel{03}{(-i)}\stackrel{12}{(+)}\cdots
\stackrel{d-1 \, d}{[+]}\,,\nonumber\\
{}^I\hat{{\cal A}}^{2 \dagger}_{1}=\stackrel{03}{[-i]}\stackrel{12}{[-]}
\stackrel{56}{(+)} \cdots \stackrel{d-1 \, d}{[+]}\,, \qquad &&
{}^{II}\hat{{\cal A}}^{2 \dagger}_{1}=\stackrel{03}{[+i]}\stackrel{12}{[-]}
\stackrel{56}{(+)} \cdots \stackrel{d-1 \, d}{[+]}\,,
\nonumber\\
{}^I\hat{{\cal A}}^{3 \dagger}_{1}=\stackrel{03}{(+i)} \stackrel{12}{(+)}
\stackrel{56}{(+)} \cdots \stackrel{d-3\,d-2}{[-]}\;\stackrel{d-1\,d}{(-)}\,, \qquad &&
{}^{II}\hat{{\cal A}}^{3 \dagger}_{1}=\stackrel{03}{(-i)} \stackrel{12}{(+)}
\stackrel{56}{(+)} \cdots \stackrel{d-3\,d-2}{[-]}\;\stackrel{d-1\,d}{(-)}\,, \nonumber\\
\dots \qquad && \dots \nonumber\\
d&=&4n\nonumber\\
{}^I\hat{{\cal A}}^{1 \dagger}_{1}=\stackrel{03}{(+i)}\stackrel{12}{(+)}\cdots
\stackrel{d-1 \, d}{(+)}\,,\qquad &&
{}^{II}\hat{{\cal A}}^{1 \dagger}_{1}=\stackrel{03}{(-i)}\stackrel{12}{(+)}\cdots
\stackrel{d-1 \, d}{(+)}\,,
\nonumber\\
{}^I\hat{{\cal A}}^{2 \dagger}_{1}= \stackrel{03}{[-i]}\stackrel{12}{[-i]}
\stackrel{56}{(+)} \cdots \stackrel{d-1 \, d}{(+)}\,, \qquad &&
{}^{II}\hat{{\cal A}}^{2 \dagger}_{1}= \stackrel{03}{[+i]}\stackrel{12}{[-i]}
\stackrel{56}{(+)} \cdots \stackrel{d-1 \, d}{(+)}\,, \nonumber\\
{}^I\hat{{\cal A}}^{3 \dagger}_{1}=\stackrel{03}{(+i)} \stackrel{12}{(+)}
\stackrel{56}{(+)} \cdots \stackrel{d-3\,d-2}{[-]}\;\stackrel{d-1\,d}{[-]}\,, \qquad &&
{}^{II}\hat{{\cal A}}^{3 \dagger}_{1}=\stackrel{03}{(-i)} \stackrel{12}{(+)}
\stackrel{56}{(+)} \cdots \stackrel{d-3\,d-2}{[-]}\;\stackrel{d-1\,d}{[-]}\,\nonumber\\
\dots \qquad && \dots
\end{eqnarray}
There are $2^{\frac{d}{2}-1}\times \,2^{\frac{d}{2}-1}$ even ``basis vectors'' of
the kind ${}^{I}{\hat{\cal A}}^{m \dagger}_{f}$ and there are $2^{\frac{d}{2}-1}$
$\times \, 2^{\frac{d}{2}-1}$ even ``basis vectors'' of the kind
${}^{II}{\hat{\cal A}}^{m \dagger}_{f}$.

Table~\ref{Table Clifffourplet.}, presented in App.~\ref{usefultables},
illustrates properties of the odd and even ``basis vectors'' in the
case of $d=(5+1)$. Looking at this case, it is easy to evaluate properties of
either even or odd ``basis vectors''. In this subsection, we shall discuss the
general case by carefully inspecting the properties of both kinds of ``basis vectors''.

The even ``basis vectors'' belonging to two different groups are orthogonal due
to the fact that they differ in the sign of one projector or the algebraic
product of a member of one group with a member of another group gives zero according
to the third and fourth lines of Eq.~(\ref{usefulrel0}): $\stackrel{ab}{(k)}\stackrel{ab}{[k]} =0$,
$\stackrel{ab}{[k]}\stackrel{ab}{(-k)} =0$,
$\stackrel{ab}{[k]}\stackrel{ab}{[-k]} =0$.
\begin{eqnarray}
\label{AIAIIorth}
{}^{I}{\hat{\cal A}}^{m \dagger}_{f} *_A {}^{II}{\hat{\cal A}}^{m \dagger}_{f}
&=&0={}^{II}{\hat{\cal A}}^{m \dagger}._{f} *_A
{}^{I}{\hat{\cal A}}^{m \dagger}_{f}\,.
\end{eqnarray}
The members of each of these two groups have the property.
\begin{eqnarray}
\label{ruleAAI}
{}^{i}{\hat{\cal A}}^{m \dagger}_{f} \,*_A\, {}^{i}{\hat{\cal A}}^{m' \dagger}_{f `}
\rightarrow \left \{ \begin{array} {r}
{}^{i}{\hat{\cal A}}^{m \dagger}_{f `}\,, i=(I,II) \\
{\rm or \,zero}\,.
\end{array} \right.
\end{eqnarray}
For a chosen ($m, f, f `$) there is only one $m'$ (out of $2^{\frac{d}{2}-1}$)
which gives a nonzero contribution.

Two ``basis vectors'', ${}^{i}{\hat{\cal A}}^{m \dagger}_{f}$ and
${}^{i}{\hat{\cal A}}^{m' \dagger}_{f '}$, the algebraic product, $*_{A}$, of which
gives non zero contribution, ``scatter'' into the third one
${}^{i}{\hat{\cal A}}^{m \dagger}_{f `}$, for $i=(I,II)$.

\vspace{2mm}

The algebraic application, $*_{A}$, of the even ``basis vectors''
${}^{I}{\hat{\cal A}}^{m \dagger}_{f }$ on the odd ``basis vectors''
$ \hat{b}^{m' \dagger}_{f `} $ gives
\begin{eqnarray}
\label{calIAb1234gen}
{}^{I}{\hat{\cal A}}^{m \dagger}_{f } \,*_A \, \hat{b}^{m' \dagger }_{f `}
\rightarrow \left \{ \begin{array} {r} \hat{b }^{m \dagger}_{f `}\,, \\
{\rm or \,zero}\,.
\end{array} \right.
\end{eqnarray}
Eq.~(\ref{calIAb1234gen}) demonstrates that
${}^{I}{\hat{\cal A}}^{m \dagger}_{f}$,
applying on $\hat{b}^{m' \dagger }_{f `} $, transforms the odd
``basis vector'' into another odd ``basis vector'' of the same family,
transferring to the odd ``basis vector'' integer spins or gives zero.

One finds
\begin{eqnarray}
\label{calbIA1234gen}
\hat{b}^{m \dagger }_{f } *_{A} {}^{I}{\hat{\cal A}}^{m' \dagger}_{f `} = 0\,, \quad
\forall (m, m`, f, f `)\,.
\end{eqnarray}
For ``scattering'' the even ``basis vectors''
${}^{II}{\hat{\cal A}}^{m \dagger}_{f }$ on the odd ``basis vectors''
$ \hat{b}^{m' \dagger}_{f `} $ it follows
%
\begin{eqnarray}
\label{calIIAb1234gen}
{}^{II}{\hat{\cal A}}^{m \dagger}_{f } \,*_A \, \hat{b}^{m' \dagger }_{f `}= 0\,,\;\;
\forall (m,m',f,f `)\,,
\end{eqnarray}
while we get
\begin{eqnarray}
\label{calbIIA1234gen}
\hat{b}^{m \dagger }_{f } *_{A} {}^{II}{\hat{\cal A}}^{m' \dagger}_{f `} \,
\rightarrow \left \{ \begin{array} {r} \hat{b }^{m \dagger}_{f ``}\,, \\
{\rm or \,zero}\,.
\end{array} \right.
\end{eqnarray}
Eq.~(\ref{calbIIA1234gen}) demonstrates that scattering of the odd
``basis vector'' $\hat{b}^{m \dagger }_{f }$ on
$ {}^{II}{\hat{\cal A}}^{m' \dagger}_{f `}$ transforms the odd
``basis vector'' into another odd ``basis vector''
$\hat{b }^{m \dagger}_{f ``}$ belonging to the same family member $m$
of a different family $f ``$.

\vspace{2mm}

While the odd ``basis vectors'', $\hat{b}^{m \dagger}_{f}$, offer the
description of the internal space of the second quantized anti-commuting fermion
fields, appearing in families, the even ``basis vectors'',
${}^{I,II}{\hat{\cal A}}^{m \dagger}_{f }$, offer the description of the internal
space of the second quantized commuting boson fields, having no families and
appearing in two groups. One of the two groups,
${}^{I}{\hat{\cal A}}^{m \dagger}_{f }$, transferring their integer quantum
numbers to the odd ``basis vectors'', $\hat{b}^{m \dagger}_{f}$,
changes the family members' quantum numbers, leaving the family quantum
numbers unchanged. The second group, transferring their integer quantum
numbers to the ``basis vector'', changes the family quantum
numbers leaving the family members quantum numbers unchanged.

\vspace{2mm}

{\it Both groups of even ``basis vectors'' manifest as the gauge fields
of the corresponding fermion fields: One concerning the family members
quantum numbers, the other concerning the family quantum numbers.}

 \vspace{1mm}

The reader can find in App.~\ref{simplecases} manifestation of this 
Subsect.~\ref{even} for the choice
of two simple exercises; the ``basis vectors'' for fermions and bosons for
 the case $d=(1+1)$ and $d=(3+1)$.

\subsubsection{Even ``basis vectors''  as algebraic products 
of the odd ``basis vectors'' and their Hermitian conjugated partners} 
\label{bbA}

The even ``basis vectors'',  
${}^{I}\hat{\cal A}^{m \dagger}_{f}$ and 
${}^{II}\hat{\cal A}^{m \dagger}_{f}$, can be represented as the algebraic 
products of the odd ``basis vectors'' and their Hermitian conjugated
partners: $\hat{b}^{m \dagger}_{f} $ and 
$(\hat{b}^{m' \dagger}_{f})^{\dagger}$, as presented in 
Ref.~\cite{nh2023dec}. 

{\it This means that if we know all the family members 
of all the families we can know all the even ``basis vectors''.}

\vspace{2mm}

 For ${}^{I}{\hat{\cal A}}^{m \dagger}_{f}$, all the family members 
$m$, there are $2^{\frac{d}{2}-1}$ members, for a particular family $f$ 
(out of $2^{\frac{d}{2}-1}$ families) and their Hermitian conjugated partners 
contribute
\begin{eqnarray}
\label{AIbbdagger}
{}^{I}{\hat{\cal A}}^{m \dagger}_{f}&=&\hat{b}^{m' \dagger}_{f `} *_A 
(\hat{b}^{m'' \dagger}_{f `})^{\dagger}\,.
\end{eqnarray}
Each family $f '$ of $\hat{b}^{m' \dagger}_{f `} *_A$
($\hat{b}^{m'' \dagger}_{f `})^{\dagger}$ generates the same  
$2^{\frac{d}{2}-1}\times$ $2^{\frac{d}{2}-1}$ 
even ``basis vectors'' ${}^{I}{\hat{\cal A}}^{m \dagger}_{f}$.

For ${}^{II}{\hat{\cal A}}^{m \dagger}_{f}$, all the families $f$,
$2^{\frac{d}{2}-1}$, of a particular member $m$ (out of 
$2^{\frac{d}{2}-1}$ family members) and their Hermitian conjugated 
partners contribute 
\begin{eqnarray}
\label{AIIbdaggerb}
 {}^{II}{\hat{\cal A}}^{m \dagger}_{f}&=&
(\hat{b}^{m' \dagger}_{f `})^{\dagger} *_A 
\hat{b}^{m' \dagger}_{f `'}\,. 
\end{eqnarray}
Each family member $m'$ generates 
in $(\hat{b}^{m' \dagger}_{f `})^{\dagger} *_A$ 
$\hat{b}^{m' \dagger}_{f `'}$ the same  $2^{\frac{d}{2}-1}\times$ 
$2^{\frac{d}{2}-1}$ even ``basis vectors''
${}^{II}{\hat{\cal A}}^{m \dagger}_{f}$. 

It follows that ${}^{I}{\hat{\cal A}}^{m \dagger}_{f}$, expressed by
$\hat{b}^{m' \dagger}_{f `} *_A$ ($\hat{b}^{m'' \dagger}_{f `})^{\dagger}$, 
applying on $\hat{b}^{m''' \dagger}_{f `''}$, obey Eq.~(\ref{calIAb1234gen}), 
and $\hat{b}^{m''' \dagger}_{f ``'}$ applying on 
${}^{II}\hat{\cal A}^{m \dagger}_{f}$, expressed by 
$(\hat{b}^{m' \dagger}_{f `})^{\dagger}$ $ *_A$ 
$\hat{b}^{m' \dagger}_{f ``}$, obey Eq.~(\ref{calbIIA1234gen}).\\

The reader can find the illustration of this Subsec.~\ref{bbA} in App.~\ref{simplecases}.

We shall discuss the properties of the even and odd ``basis 
vectors'' in more details for $d=(5+1)$-dimensional internal spaces in
Subsect.~\ref{cliffordoddevenbasis5+1}; 
$d=(5+1)$-dimensional internal spaces offer more possibilities than 
$d=(1+1)$ and $d=(3+1)$ and an easier understanding of 
what offer general  $d=((d-1)+1)$-dimensional internal spaces for 
even $d$.

\subsubsection{Odd and even ``basis vectors'' in  $d=(5+1)$}
\label{cliffordoddevenbasis5+1}

Subsect.~\ref{cliffordoddevenbasis5+1}, which is a short overview of
Ref.~\cite{n2023NPB} demonstrates the properties of the odd and
even ``basis vectors'' in the special case when $d=(5+1)$ to help the reader
to see the elegance of the description of the internal spaces of fermions and
bosons with the odd and even ``basis vectors''.

Table~\ref{Table Clifffourplet.} which appears in App.~\ref{usefultables}
presents the $64 \,(=2^{d=6})$ ``eigenvectors" of the Cartan subalgebra
members of the Lorentz algebra, $S^{ab}$ and ${\bf \cal{S}}^{ab}$, Eq.~(\ref{cartangrasscliff}), describing the internal space of fermions and
bosons.\\

The odd ``basis vectors'', denoted as $odd \, I \,\hat{b}^{m\dagger}_f$,
appearing in $4 \,(=2^{\frac{d=6}{2}-1})$ families, each family has $4$
members, are products of an odd number of nilpotents, either of three or one.
Their Hermitian conjugated partners appear in the separate group denoted as
$odd \, II \,\,\hat{b}^m_f$. Within each of these two groups, the members are
mutually orthogonal~\footnote{
The mutual orthogonality within the ``basis vectors'' $\hat{b}^{m\dagger}_f$,
as well as within their Hermitian conjugated partners
$(\hat{b}^{m\dagger}_f)^{\dagger}$, can be checked by using
Eq.~(\ref{usefulrel0})) taking into account that
$\stackrel{ab}{(k)}\stackrel{ab}{[k]} =0$\,,
$\stackrel{ab}{[k]}\stackrel{ab}{(-k)} =0\,,
\stackrel{ab}{[k]}\stackrel{ab}{[-k]} =0\,,$ $(\stackrel{ab}{(k)})^2 =0$;
$\hat{b}^{m\dagger} _f *_{A} \hat{b}^{m'\dagger} _{f `} =0$ for all
$(m,m', f, f `)$. Equivalently,
$\hat{b}^{m} _f *_{A} \hat{b}^{m'} _{f `} =0$ for all $(m,m', f, f `)$.
}.
The odd ``basis vectors'' and their Hermitian conjugated partners
are normalized as~\footnote{
The vacuum state $|\psi_{oc} >= \frac{1}{\sqrt{2^{\frac{d=6}{2}-1}}}$
$(\stackrel{03}{[-i]} \stackrel{12}{[-]}\stackrel{56}{[-]}+ \stackrel{03}{[-i]}
\stackrel{12}{[+]}\stackrel{56}{[+]} + \stackrel{03}{[+i]} \stackrel{12}{[-]}
\stackrel{56}{[+]}+ \stackrel{03}{[+i]} \stackrel{12}{[+]}\stackrel{56}{[-]})$
is normalized to one: $< \psi_{oc}|\psi_{oc} >=1$, Ref.~\cite{nh2021RPPNP}.
}
\begin{eqnarray}
< \psi_{oc}| \hat{b}^{m} _f *_{A} \hat{b}^{m' \dagger} _{f `} |\psi_{oc} >=
\delta^{m m'} \delta_{f f `}\,.
\label{Cliffnormalizationodd}
\end{eqnarray}

The even ``basis vectors'' are products of an even number of nilpotents of
either two or none in this case. They are presented in
Table~\ref{Table Clifffourplet.} in two groups, denoted as
$even \, {I}\,{\bf {\cal A}}^{m \dagger}_{f}$ and
$even \, II \, {\bf {\cal A}}^{m \dagger}_{f}$, each with
$16\, (=2^{\frac{d=6}{2}-1}\times 2^{\frac{d=6}{2}-1})$ members.
The two groups are mutually orthogonal, Eq.~(\ref{AIAIIorth}),
Ref.~\cite{n2022epjc}.

While the odd ``basis vectors'' have half-integer eigenvalues of the
Cartan subalgebra members, Eq.~(\ref{cartangrasscliff}), that is of $S^{03},
S^{12}, S^{56}$ (as well as of $\tilde{S}^{03}, \tilde{S}^{12},
\tilde{S}^{56}$) in this particular case of $d=(5+1)$, the even
``basis vectors'' have integer spins determined by
${\bf {\cal S}}^{03}= (S^{03}+ \tilde{S}^{03})$, ${\bf {\cal S}}^{12}=
(S^{12} +\tilde{S}^{12})$,
${\bf {\cal S}}^{56}=( S^{56}+ \tilde{S}^{56})$.

\vspace{2mm}
\begin {small}
Let us check Eq.~(\ref{calIAb1234gen}) for
${}^{I}{\hat{\cal A}}^{m=1 \dagger}_{f=4}$, presented in
Table~\ref{Table Clifffourplet.} in the first line of the fourth column of $even\,I$,
and for $\hat{b}^{m=2 \dagger}_{f=2}$, presented in
$odd \, I$ as the second member of the second column. \\
One finds:
$${}^{I}{\hat{\cal A}}^{1 \dagger}_{4} (\equiv \stackrel{03}{(+i)}
\stackrel{12}{(+)} \stackrel{56}{[+]}) *_{A} \hat{b}^{2 \dagger}_{2}
(\equiv \stackrel{03}{(-i)} \stackrel{12}{(-)} \stackrel{56}{(+)}) \rightarrow
\hat{b}^{1 \dagger}_{2} (\equiv \stackrel{03}{[+i]}
\stackrel{12}{[+]} \stackrel{56}{(+)})\,.$$
We see that ${}^{I}{\hat{\cal A}}^{1 \dagger}_{4}$ (having
${\cal S}^{03}=i, {\cal S}^{12}=1$ and ${\cal S}^{56}=0$)
transfers to $\hat{b}^{2 \dagger}_{2}$ ($S^{03}=-\frac{i}{2}$,
$S^{12}=-\frac{1}{2}, S^{56}=\frac{1}{2}$) the quantum numbers
${\cal S}^{03}=i, {\cal S}^{12}=1,{\cal S}^{56}=0$, transforming
$\hat{b}^{2 \dagger}_{2}$ to $\hat{b}^{1 \dagger}_{2}$, with
$S^{03}=\frac{i}{2}$,
$S^{12}=\frac{1}{2}, S^{56}=\frac{1}{2}$.
\end{small}

\vspace{2mm}

Calculating the eigenvalues of the Cartan subalgebra members,
Eq.~(\ref{cartangrasscliff}), before and after the algebraic multiplication, $*_A$,
assures us that ${}^{I}{\hat{\cal A}}^{m \dagger}_{f}$ carry the integer
eigenvalues of the Cartan subalgebra members, namely of
${\cal S}^{ab}$ $= (S^{ab} + \tilde{S}^{ab})$, since they transfer to
the odd ``basis vector'' integer eigenvalues of the Cartan subalgebra
members, changing the odd ``basis vector'' into another odd
``basis vector'' of the same family.

Ref.~\cite{n2023NPB}, illustrates in Fig.~1 and Fig.~2 that the
even ``basis vectors'' have the properties of the gauge fields of the
corresponding odd ``basis vectors'', by studying properties of the
$SU(3)$ $\times U(1)$ subgroups of the $SO(5,1)$ group for the
odd and even ``basis vectors'', expressing $ \tau^{3}=\frac{1}{2}
\,( -S^{1\,2} - iS^{0\,3}), \tau^{8}=\frac{1}{2\sqrt{3}}
(-i S^{0\,3} + S^{1\,2} - 2 S^{5\;6})\,,
\tau' = -\frac{1}{3}(-i S^{0\,3} + S^{1\,2} + S^{5\,6})$.

\vspace{2mm}

In Eqs.~(\ref{AIbbdagger}, \ref{AIIbdaggerb}) the formal expressions of
${}^{i}{\hat{\cal A}}^{m \dagger}_{3}, i=(I,II),$ as products of
$\hat{b}^{m \dagger}_{f}$ and $(\hat{b}^{m \dagger}_{f})^{\dagger}$
are presented for any even $d$; Tables~(\ref{transverseCliff basis5+1even I.},
\ref{S120Cliff basis5+1even I.}, \ref{transverseCliff basis5+1even II.},
\ref{S120Cliff basis5+1even II.}) in App.~\ref{usefultables}, present their
``basis vectors'' for the case $d=(5+1)$. These expressions are meant to
help the reader to follow the description of the ``basis vectors'' of photons, 
weak bosons, and gluons manifested in $d=(13+1)$ if analysed from the 
point of view of $d=(3+1)$.
In Subsect.~\ref{bosons5+1}, the even ``basis vectors'' are
illustrated as ``photons'' and ``gravitons'' in $d=(5+1)$-dimensional space, 
from the point of view of $d=(3+1)$.

The charge conjugation of fermions (in this toy model with $d=(5+1)$
they manifest as ``electrons'' and ``positrons'') and bosons (manifesting
as ``photons'') are also discussed.

%
\section{``Basis vectors'' for boson fields of $d=(13+1)$ from the point of 
view of $d=(3+1)$}
\label{bosons13+1and5+1}

In this section, we discuss:\\
{\bf i.} The internal spaces of the observed vector gauge fields --- photons,
weak bosons, gluons --- when describing them by the even ``basis vectors''
(which are a superposition of the even products of $\gamma^a$'s and are
at the same time, the eigenvectors of all the Cartan subalgebra members) in
$d=(13 +1)$-dimensional internal space from the point of view of their
properties in $(3+1)$-dimensional space,
Subsect.~\ref{photonsweakbosonsgluons}.\\
{\bf ii.} The internal space of the observed scalar gauge fields --- explaining 
the appearance of Higgs's scalars and Yukawa couplings and the appearance
of matter/antimatter asymmetry --- when describing them by even ``basis 
vectors'' in $d=(13 +1)$-dimensional space from the point of view of their 
properties in $(3+1)$-dimensional space, 
Subsect.~\ref{scalarfields}.\\
{\bf iii.} The internal space of almost not yet observed ``gravitons'', when 
describing them by the even ``basis vectors'' in $d=(13 +1)$-dimensional 
space from the point of view of their properties in $(3+1)$-dimensional 
space, Subsect.~\ref{gravitons}.\\
{\bf iv.} The internal spaces of ``photons'' and ``gravitons'' when 
describing them in a ``toy model'' by the even ``basis vectors'' in 
$d=(5+1)$-dimensional space from the point of view of their properties in
$(3+1)$-dimensional space, Subsect.~\ref{bosons5+1}.\\
{\bf v.} The scattering of ``photons'' and ``gravitons'' among themselves 
and on ``positrons'' and ``electrons'', when describing the internal spaces 
of fermions and bosons in $d=(13+1)$, and in $d=(5+1)$ in a ``toy 
model'', with the even ``basis vectors'' (for bosons) and with the odd 
``basis vectors'' (for fermions), in all cases from the point of view of
their properties in $(3+1)$-dimensional space,
Subsect.~\ref{feynmandiagrams}.

Let us start with the ``toy model'', that is, with fermions and bosons in
$d=(5+1)$-dimensional space to learn what does offer the description
of the internal spaces of fermion and bosons.

\subsection{``Basis vectors'' for boson fields of $d=(5+1)$ from the point of 
view of $d=(3+1)$}
\label{bosons5+1}

This is a short overview of the odd and even ``basis vectors'' in
$d=(5+1)$-dimensional space, having $2^{d=6}=64 $
eigenvectors of the Cartan subalgebra members, Eq.~(\ref{cartangrasscliff}),
started in Subsect.~\ref{cliffordoddevenbasis5+1}.
The ``basis vectors'' are taken from several articles (Ref.\cite{nh2021RPPNP},
and references therein). This time, the odd ``basis vectors'' are
assumed to represent ``electrons'' and ``positrons'', both appearing among
the $2^{\frac{d}{2}-1}= 4$ members of each of the $2^{\frac{d}{2}-1} = 4$
families, related by the charge conjugated operators~\footnote{
In Ref.~\cite{nhds2014}, the discrete symmetry operators for fermion fields
in $d=2(2n +1)$ with the desired properties in $d=(3+1)$ in Kaluza-Klein-like
theories are presented. The reader can also find the discrete symmetry operators
in Ref.~\cite{nh2021RPPNP}, Subsect.~3.3.5, Eq.~(72), presented as
\begin{small} 
\begin{eqnarray}
\label{CPTNlowE}
{\cal C}_{{\cal N}}  &= & \prod_{\Im \gamma^m, m=0}^{3} \gamma^m\,\, \,
\Gamma^{(3+1)} \,
K \,I_{x^6,x^8,\dots,x^{d}}  \,,\nonumber\\
{\cal T}_{{\cal N}}  &= & \prod_{\Re \gamma^m, m=1}^{3} 
\gamma^m \,\,\,\Gamma^{(3+1)}\,K \,
I_{x^0}\,I_{x^5,x^7,\dots,x^{d-1}}\,,\nonumber\\
{\cal P}^{(d-1)}_{{\cal N}}  &= & \gamma^0\,\Gamma^{(3+1)}\, 
\Gamma^{(d)}\, I_{\vec{x}_{3}}
\,,\nonumber\\
\mathbb{C}_{{ \cal N}} &=& 
\prod_{\Re \gamma^a, a=0}^{d} \gamma^a \,\,\,K
\, \prod_{\Im \gamma^m, m=0}^{3} \gamma^m \,\,\,\Gamma^{(3+1)} \,
K \,I_{x^6,x^8,\dots,x^{d}}= 
\prod_{\Re \gamma^s, s=5}^{d} \gamma^s \, \,I_{x^6,x^8,\dots,x^{d}}\,,
\nonumber\\
\mathbb{C}_{{ \cal N}}{\cal P}^{(d-1)}_{{\cal N}}  &= &\gamma^0
\prod_{\Im \gamma^a, a=5}^{d} \gamma^a \,
I_{\vec{x}_{3}} \,I_{x^6,x^8,\dots,x^{d}}
\end{eqnarray}
\end{small}
\begin{small}
Operators $I$ operate as follows in $d=2n$: 
$I_{\vec{x}_{3}} x^a = (x^0, -x^1,-x^2,-x^3,x^5, x^6,\dots, x^d)$,\\ 
$I_{x^5,x^7,\dots,x^{d-1}}$ 
$(x^0,x^1,x^2,x^3,x^5,x^6,x^7,x^8,
\dots,x^{d-1},x^d)$ $=(x^0,x^1,x^2,x^3,-x^5,x^6,-x^7,\dots,-x^{d-1},x^d)$,\\
 $I_{x^6,x^8,\dots,x^d}$ 
$(x^0,x^1,x^2,x^3,x^5,x^6,x^7,x^8,\dots,x^{d-1},x^d)$
$=(x^0,x^1,x^2,x^3,x^5,-x^6,x^7,-x^8,\dots,x^{d-1},-x^d)$. 
\end{small}

The discrete symmetry operator, transforming fermion into antifermion, is equal
to $\mathbb{C}_{{ \cal N}}{\cal P}^{(d-1)}_{{\cal N}} =\gamma^0
\prod_{\Im \gamma^a, a=5,7,..,d} \gamma^a \,
I_{\vec{x}_{3}} \,I_{x^6,x^8,\dots,x^{d}}$.
}. \\


\vspace{2mm}

{\bf a.\,\,}{\it Discrete symmetries of fermion and boson fields in $d=2(2n+1)$ 
from the point of view $d=(3+1)$}:~\footnote{
Let be pointed out that the discrete symmetry operators for fermion fields 
contain in $d=4n$ an odd number of $\gamma^{a}$'s, transforming an 
odd ``basis vector'' into the even ``basis vector''. Correspondingly, {\it 
the discrete symmetry operators for fermion fields}, 
$\mathbb{C}_{{ \cal N}}{\cal P}^{(d-1)}_{{\cal N}}$, {\it transform fermions 
into antifermions only in $d=2(2n+1)$, since only in $d=2(2n+1)$ the fermion/antifermion pairs appear within the family members of a particular family}.
}
 \vspace{2mm}

If  fermions manifest dynamics only in $d=(3+1)$ space, the charge conjugation operator transforming fermions into antifermions can be written (up to a phase)
as~\cite{nhds2014}~(\cite{nh2021RPPNP}, Subsect.~3.3.5, Eq.~(72)), Eq~(\ref{CPTNlowE}) in this paper
\begin{eqnarray}
\label{CPd}
\mathbb{C}_{{ \cal N}}{\cal P}^{(d-1)}_{{\cal N}} &=& \gamma^0
\prod_{\Im \gamma^a, a=5,7,9,..,d} \gamma^a \,
I_{\vec{x}_{3}}\,, \nonumber\\
{\rm with} \,\,\, I_{\vec{x}_{3}} \,{\rm applying\,\, on}\,x^a \,
{\rm leading \,to}
& & (x^0, -x^1,-x^2,-x^3 ,x^5,..,x^d)\,.
\end{eqnarray}
In the case of $d=(5+1)$ and when taking care of only the internal spaces of 
fermions and bosons, the discrete symmetry operator 
$\mathbb{C}_{{ \cal N}}{\cal P}^{(d-1)}_{{\cal N}}$, Eq.~(\ref{CPd}), 
simplifies to $\gamma^0  \gamma^5. $

The even ``basis vectors'' represent the corresponding gauge fields. 
For example, ``gravitons'' and ``photons'', are members of two groups, each 
group has $2^{\frac{d}{2}-1}\times 2^{\frac{d}{2}-1}= 16$ members.

While the  odd ``basis vectors'' have their Hermitian conjugated 
partners in a separate group, have the  even ``basis vectors'' their 
Hermitian conjugated partners within their group.

Not all the even ``basis vectors'' are active when observing their
behaviour in $d=(3+1)$, due to the needed breaking symmetry from 
$SO(5,1)$ to $SO(3,1)$, as illustrated in the note~\footnote{
The reader can see the influence of the broken symmetry, if assuming the 
break from $SO(6)$ to $SU(3)\times U(1)$ as discussed in Ref.~\cite{n2023NPB}:
Fig.~1, showing ``basis vectors'' of one of the four families representing 
one colour triplet (the ``quarks'' of three colours) and one colour singlet (colourless ``lepton'') and Fig.~2, showing the $16$ members of their 
boson gauge fields (one sextet with two singlets and one triplet-antitriplet 
pair with two singlets), demonstrating this influence. Before assuming the 
break of symmetry from $SO(6)$ to $SU(3)\times U(1)$ the triplet and antitriplet can transform the singlet into triplet and opposite: After the 
break, these transformations are not allowed any longer.
}.

\vspace{2mm}

It was pointed out in Subsect.~\ref{bbA}, Eqs.~(\ref{AIbbdagger}, \ref{AIIbdaggerb}), that each 
even ``basis vector'' can be written as a product of an odd ``basis vector'' 
and a Hermitian conjugated partner of the same or another odd ``basis 
vector'':\\

\;\;\;\;${}^{I}{\hat{\cal A}}^{m \dagger}_{f}=
\hat{b}^{m' \dagger}_{f `} *_A 
(\hat{b}^{m'' \dagger}_{f `})^{\dagger}$, \;\;\;\;
${}^{II}{\hat{\cal A}}^{m \dagger}_{f}=
(\hat{b}^{m' \dagger}_{f `})^{\dagger} *_A 
\hat{b}^{m' \dagger}_{f `''}$.\\

Tables~(\ref{transverseCliff basis5+1even I.},
\ref{S120Cliff basis5+1even I.}, \ref{transverseCliff basis5+1even II.},
\ref{S120Cliff basis5+1even II.}),  appearing in App.~\ref{usefultables}, 
demonstrate the ``basis vectors'', written as products of an odd ``basis 
vector'' and one of the Hermitian conjugated partners for the case 
$d=(5+1)$. 

\vspace{2mm}

Let us recognize ``electrons'' and their antiparticles ``positrons'' on  
Table~\ref{Table Clifffourplet.}, presented as 
$odd\,I\,\hat{b}^{m \dagger}_{f }$, appearing in four families, in each
family with 
the same eigenvalues of $S^{03}, S^{12}, S^{56}$, distinguishing only
in the family quantum numbers, determined by $\tilde{S}^{03}, 
\tilde{S}^{12}, \tilde{S}^{56}$.
According to Eq.~(\ref{CPTNlowE}, \ref{CPd}) the charge conjugated 
partners of the  odd ``basis vectors'' can be obtained by the 
application of  the operator $\gamma^0  \gamma^5$.

If we pay attention on the first family of ``basis vectors'', 
$\hat{b}^{m \dagger}_{f=1}$, 
we can easily find, taking into account Eq.~\ref{usefulrel0}, the boson fields
 ${}^{I}{\hat{\cal A}}^{m \dagger}_{f}$ which applying on fermion fields transform any ``basis vector'' into its charge conjugated partner
\begin{small}
\begin{eqnarray}
\label{05bf1}
\gamma^0 \gamma^5 \,*_A\, \hat{b}^{1 \dagger}_{1}\,
(\equiv \stackrel{03}{(+i)}\stackrel{12}{[+]} \stackrel{56}{[+]})\,
\rightarrow \,
\hat{b}^{3 \dagger}_{1}\, (\equiv \stackrel{03}{[-i]}\stackrel{12}{[+]} 
\stackrel{56}{(-)})\,,\quad {}^{I}{\hat{\cal A}}^{3 \dagger}_{3}
(\equiv \stackrel{03}{(-i)}\stackrel{12}{[+]} \stackrel{56}{(-)})\,
\,*_A\,\hat{b}^{1 \dagger}_{1}\,\rightarrow \hat{b}^{3 \dagger}_{1}\,,
\nonumber\\
\gamma^0 \gamma^5 \,*_A\, \hat{b}^{2 \dagger}_{1}\,
(\equiv \stackrel{03}{[-i]}\stackrel{12}{(-)} \stackrel{56}{[+]})\,
\rightarrow \,
\hat{b}^{4 \dagger}_{1}\, (\equiv \stackrel{03}{(+i)}\stackrel{12}{(-)} 
\stackrel{56}{(-)})\,,\quad {}^{I}{\hat{\cal A}}^{4 \dagger}_{4}
(\equiv \stackrel{03}{(+i)}\stackrel{12}{[-]} \stackrel{56}{(-)})\,
\,*_A\,\hat{b}^{2 \dagger}_{1}\,\rightarrow \hat{b}^{4 \dagger}_{1}\,,
\nonumber\\
\gamma^0 \gamma^5 \,*_A\, \hat{b}^{3 \dagger}_{1}\,
 (\equiv \stackrel{03}{[-i]}\stackrel{12}{[+]} \stackrel{56}{(-)})\,
\rightarrow \,
\hat{b}^{1 \dagger}_{1}\, (\equiv \stackrel{03}{(+i)}\stackrel{12}{[+]}
 \stackrel{56}{[+]})\,,\quad {}^{I}{\hat{\cal A}}^{1 \dagger}_{2}
(\equiv \stackrel{03}{(+i)}\stackrel{12}{[+]} \stackrel{56}{(+)})\,
\,*_A\,\hat{b}^{3 \dagger}_{1}\,\rightarrow \hat{b}^{1 \dagger}_{1}\,,
\nonumber\\
 \gamma^0 \gamma^5 \,*_A\, \hat{b}^{4 \dagger}_{1}\,
  (\equiv \stackrel{03}{(+i)}\stackrel{12}{(-)}\stackrel{56}{(-)})\,
\rightarrow \,\hat{b}^{2 \dagger}_{1}\,(\equiv \stackrel{03}{[-i]}
\stackrel{12}{(-)} \stackrel{56}{[+]})\,,\quad 
{}^{I}{\hat{\cal A}}^{2 \dagger}_{1}
(\equiv \stackrel{03}{(-i)}\stackrel{12}{[-]} \stackrel{56}{(+)})\,
\,*_A\,\hat{b}^{4 \dagger}_{1}\,\rightarrow \hat{b}^{2 \dagger}_{1}\,.
\end{eqnarray}
\end{small}
Since all the rest three families behave equivalently under the application
of the same ${}^{I}{\hat{\cal A}}^{m \dagger}_{f}$ 
(${}^{I}{\hat{\cal A}}^{3 \dagger}_{3}$, 
${}^{I}{\hat{\cal A}}^{4 \dagger}_{4}$, 
${}^{I}{\hat{\cal A}}^{1 \dagger}_{2}$,
${}^{I}{\hat{\cal A}}^{2 \dagger}_{1}$) as the first family,
 we can generalize to
\begin{eqnarray}
\label{05b}
&& \gamma^0 \gamma^5 \,*_A\, \hat{b}^{1 \dagger}_{f}\,\rightarrow \,
\hat{b}^{3 \dagger}_{f}\,, \quad 
{}^{I}{\hat{\cal A}}^{3 \dagger}_{3}
(\equiv \stackrel{03}{(-i)}\stackrel{12}{[+]} \stackrel{56}{(-)})\,
\,*_A\,\hat{b}^{1 \dagger}_{f}\,\rightarrow\, \hat{b}^{3 \dagger}_{f}\,,
\nonumber\\
&& \gamma^0 \gamma^5 \,*_A\, \hat{b}^{2 \dagger}_{f}\,\rightarrow \,
\hat{b}^{4 \dagger}_{f}\,, \quad {}^{I}{\hat{\cal A}}^{4 \dagger}_{4}
(\equiv \stackrel{03}{(+i)}\stackrel{12}{[-]} \stackrel{56}{(-)})\,
\,*_A\,\hat{b}^{2 \dagger}_{f}\,\rightarrow \hat{b}^{4 \dagger}_{f}\,,
\nonumber\\
&&\gamma^0 \gamma^5 \,*_A\, \hat{b}^{3 \dagger}_{f}\, \rightarrow \,
\hat{b}^{1 \dagger}_{1}\,, \quad {}^{I}{\hat{\cal A}}^{1 \dagger}_{2}
(\equiv \stackrel{03}{(+i)}\stackrel{12}{[+]} \stackrel{56}{(+)})\,
\,*_A\,\hat{b}^{3 \dagger}_{f}\,\rightarrow \hat{b}^{1 \dagger}_{f}\,,
\nonumber\\
&& \gamma^0 \gamma^5 \,*_A\, \hat{b}^{4 \dagger}_{f}\, \rightarrow \,
 \hat{b}^{2 \dagger}_{f}\,\,,\quad {}^{I}{\hat{\cal A}}^{2 \dagger}_{1}
(\equiv \stackrel{03}{(-i)}\stackrel{12}{[-]} \stackrel{56}{(+)})\,
\,*_A\,\hat{b}^{4 \dagger}_{f}\,\rightarrow \hat{b}^{2 \dagger}_{f}\,.
\end{eqnarray}
Taking into account Eq.~(\ref{calIAb1234gen}) one easily finds  the 
replacement for the discrete operator 
$\mathbb{C}_{{ \cal N}}{\cal P}^{(d-1)}_{{\cal N}} $ (which is  
$\gamma^0  \gamma^5$ in the case that $d=(5+1)$ and only the internal 
space of fermions are concerned) presented on the very right hand side of
Eqs.~(\ref{05bf1}, \ref{05b}) in terms of  
${}^{I}{\hat{\cal A}}^{m \dagger}_{f}$.


We can now ask for the relations between 
${}^{I}{\hat{\cal A}}^{m \dagger}_{f}$ generating the charge conjugated 
partners $\hat{b}^{m \dagger}_{f `}$ of $\hat{b}^{m' \dagger}_{f `}$ as 
presented on the right hand side of Eq.~(\ref{05b}), that is, in this particular 
case, between
${}^{I}{\hat{\cal A}}^{3 \dagger}_{3}$ and the charge conjugated partner
${}^{I}{\hat{\cal A}}^{1 \dagger}_{2}$,  or between
${}^{I}{\hat{\cal A}}^{4 \dagger}_{4}$ and the charge conjugated partner
${}^{I}{\hat{\cal A}}^{2 \dagger}_{1}$. 

Tables~(\ref{S120Cliff basis5+1even I.}, \ref{transverseCliff basis5+1even I.})
express ${}^{I}{\hat{\cal A}}^{m \dagger}_{f}$ in terms of the Clifford 
odd ``basis vectors'' and the ``basis vectors'' of their Hermitian conjugated 
partners as presented in~Eq.~(\ref{AIbbdagger}). Replacing the odd ``basis vectors'' and their Hermitian conjugated partners 
in~Eq.~(\ref{AIbbdagger}), ${}^{I}{\hat{\cal A}}^{m \dagger}_{f}=
\hat{b}^{m' \dagger}_{f `} *_A (\hat{b}^{m'' \dagger}_{f `})^{\dagger}$, 
by 
$\mathbb{C}_{{ \cal N}}{\cal P}^{(d-1)}_{{\cal N}} 
\hat{b}^{m' \dagger}_{f `}$
and ($\mathbb{C}_{{ \cal N}}{\cal P}^{(d-1)}_{{\cal N}}
\hat{b}^{m'' \dagger}_{f '})^{\dagger}$, one finds for the charge conjugated ``basis  vectors'' describing the internal space of bosons 
\begin{eqnarray} 
\label{05AIb} 
({\bf \mathbb{C}}_{{ \cal N}}{\bf {\cal P}}^{(d-1)}_{{\cal N}})_{b}\,\,
{}^{I}{\hat{\cal A}}^{m \dagger}_{f} =
{\bf \mathbb{C}}_{{ \cal N}}{\bf {\cal P}}^{(d-1)}_{{\cal N}}\,
\hat{b}^{m' \dagger}_{f `}\,*_{A}\,
({\bf \mathbb{C}}_{{ \cal N}}{\bf {\cal P}}^{(d-1)}_{{\cal N}}\,
\hat{b}^{m'' \dagger}_{f `})^{\dagger},
\end{eqnarray}
where index $_{b}$ in 
$({\bf \mathbb{C}}_{{ \cal N}}{\bf {\cal P}}^{(d-1)}_{{\cal N}})_{b}$
determines the operator operating on the boson ``basis vectors''.
We find for the above cases that ${\cal S}^{12}\,{}^{I}{\hat{\cal A}}^{m \dagger}_{f}=0$
$({\bf \mathbb{C}}_{{ \cal N}}{\bf {\cal P}}^{(d-1)}_{{\cal N}})_{b}\,\,
{}^{I}{\hat{\cal A}}^{3 \dagger}_{3}\,= 
{\bf \mathbb{C}}_{{ \cal N}}{\bf {\cal P}}^{(d-1)}_{{\cal N}}\,
\hat{b}^{3 \dagger}_{1}\,*_{A}\,
({\bf \mathbb{C}}_{{ \cal N}}{\bf {\cal P}}^{(d-1)}_{{\cal N}}\,
\hat{b}^{1 \dagger}_{1})^{\dagger}=
{}^{I}{\hat{\cal A}}^{1 \dagger}_{2}, \,\,\,$
$({\bf \mathbb{C}}_{{ \cal N}}{\bf {\cal P}}^{(d-1)}_{{\cal N}})_{b}\,\,
{}^{I}{\hat{\cal A}}^{1 \dagger}_{2}\,= 
{\bf \mathbb{C}}_{{ \cal N}}{\bf {\cal P}}^{(d-1)}_{{\cal N}}\,
\hat{b}^{1 \dagger}_{1}\,*_{A}\,
({\bf \mathbb{C}}_{{ \cal N}}{\bf {\cal P}}^{(d-1)}_{{\cal N}}\,
\hat{b}^{3 \dagger}_{1})^{\dagger}=
{}^{I}{\hat{\cal A}}^{3 \dagger}_{3}$. \\
Equivalently we find
$({\bf \mathbb{C}}_{{ \cal N}}{\bf {\cal P}}^{(d-1)}_{{\cal N}})_{b}
{}^{I}{\hat{\cal A}}^{4 \dagger}_{4}\,= 
{}^{I}{\hat{\cal A}}^{2 \dagger}_{1},$
$({\bf \mathbb{C}}_{{ \cal N}}{\bf {\cal P}}^{(d-1)}_{{\cal N}})_{b}
{}^{I}{\hat{\cal A}}^{2 \dagger}_{1}= 
{}^{I}{\hat{\cal A}}^{4 \dagger}_{4}\,.$~\footnote{
If ${\cal S}^{12}\,{}^{I}{\hat{\cal A}}^{m \dagger}_{f}$ is nonzero,  
as it is the case for all the members of 
Table~\ref{transverseCliff basis5+1even I.},
then $({}^{I}{\hat{\cal A}}^{m \dagger}_{f})^{\dagger}$ is not equal 
to $({\bf \mathbb{C}}_{{ \cal N}}{\bf {\cal P}}^{(d-1)}_{{\cal N}})_{b}
{}^{I}{\hat{\cal A}}^{m \dagger}_{f}$. 
}\\
\vspace{1mm}

Tables~(\ref{S120Cliff basis5+1even II.}, 
\ref{transverseCliff basis5+1even II.}) express 
${}^{II}{\hat{\cal A}}^{m \dagger}_{f}$ in terms of the odd 
``basis vectors'' and their Hermitian conjugated partners as presented 
in~Eq.~(\ref{AIIbdaggerb}).
  Discrete symmetry operators  operating on
${}^{II}{\hat{\cal A}}^{m \dagger}_{f}$ obey different relations due
to the fact that ${}^{I}{\hat{\cal A}}^{m \dagger}_{f}$ transform a
member of a family to a different member of the same family while
${}^{II}{\hat{\cal A}}^{m \dagger}_{f}$ transform a  member of
a family to the same member of another (or the same) family. 

Taking into account Eqs.~(\ref{calbIIA1234gen}, \ref{AIIbdaggerb}) one finds 
for the discrete symmetry operator 
$({\bf \tilde{\mathbb{C}}}_{{ \cal N}}{\bf  \tilde{{\cal P}}}^{(d-1)}_{{\cal N}})_{b}$, applying on ${}^{II}{\hat{\cal A}}^{m \dagger}_{f}$ the
relation
\begin{eqnarray}
\label{05AIIb}
({\bf \tilde{\mathbb{C}}}_{{ \cal N}}{\bf  \tilde{{\cal P}}}^{(d-1)}_{{\cal N}})_{b} \,{}^{II}{\hat{\cal A}}^{m \dagger}_{f}&=&
({\bf \tilde{\mathbb{C}}}_{{ \cal N}}{\bf  \tilde{{\cal P}}}^{(d-1)}_{{\cal N}}\,
\hat{b}^{m' \dagger}_{f `})^{\dagger}\,*_{A}\,
({\bf \tilde{\mathbb{C}}}_{{ \cal N}}{\bf  \tilde{{\cal P}}}^{(d-1)}_{{\cal N}}\,
\hat{b}^{m' \dagger}_{f ``})\,,
\end{eqnarray}
where index $_{b}$ in
$({\bf \tilde{\mathbb{C}}}_{{ \cal N}}{\bf \tilde{{\cal P}}}^{(d-1)}_{{\cal N}})_{b}$
determines the operator operating on the boson ``basis vectors'' of the 
second kind, while  ${\bf \tilde{\mathbb{C}}}_{{ \cal N}}{\bf  \tilde{{\cal P}}}^{(d-1)}_{{\cal N}}$\,  applying on $\hat{b}^{m' \dagger}_{f `}$ is in 
the case of $d=(5+1)$, and when only the internal space is concerned, 
proportional to
$\tilde{\gamma}^0 \tilde{\gamma}^5$.

Let us add that the same ${}^{II}{\hat{\cal A}}^{m' \dagger}_{f `}$ cause
the transformation of $\hat{b}^{m \dagger}_{f }$ to 
$\hat{b}^{m \dagger}_{f ``}$, with $m$ fixed, for all $m= (1,2,3,4)$:\\

\;\;\;\;$\hat{b}^{1 \dagger}_{1}\,*_A \,{}^{II}{\hat{\cal A}}^{1 \dagger}_{3}=
\hat{b}^{1 \dagger}_{3 }$,\,\,\,\,$\hat{b}^{2 \dagger}_{1}\,*_A \,
{}^{II}{\hat{\cal A}}^{1 \dagger}_{3}=\hat{b}^{2 \dagger}_{3 }$, 
\,\,\,\,$\hat{b}^{3 \dagger}_{1}\,*_A \,
{}^{II}{\hat{\cal A}}^{1 \dagger}_{3}=\hat{b}^{3 \dagger}_{3 },,
\,\,\,\,\hat{b}^{4 \dagger}_{1}\,*_A \,{}^{II}{\hat{\cal A}}^{1 \dagger}_{3}=
\hat{b}^{4 \dagger}_{3 }$.\\

The discrete symmetry operator 
$({\bf \tilde{\mathbb{C}}}_{{ \cal N}}{\bf \tilde{{\cal P}}}^{(d-1)}_{{\cal N}})_{b}$ operating on the second kind of boson ``basis vectors'' needs 
further study and will be discussed in a separate paper.

\vspace{2mm}

In Eqs.~(\ref{AIbbdagger}, \ref{AIIbdaggerb}) the formal expressions of
${}^{i}{\hat{\cal A}}^{m \dagger}_{3}, i=(I,II),$ as products of
$\hat{b}^{m \dagger}_{f}$ and $(\hat{b}^{m \dagger}_{f})^{\dagger}$
are presented for any even $d$; Tables~(\ref{transverseCliff basis5+1even I.},
\ref{S120Cliff basis5+1even I.}, \ref{transverseCliff basis5+1even II.},
\ref{S120Cliff basis5+1even II.}), App.~\ref{usefultables}, present their
``basis vectors'' for the case $d=(5+1)$. These expressions are meant to
 help the reader to 
easier understand the description of the ``basis vectors'' of photons, weak 
bosons, and gluons manifested in $d=(13+1)$ if analysed from the point of 
view of $d=(3+1)$.
\\

\vspace{2mm}

{\bf b.\,\,}{\it ``Electrons'', ``positrons'', ``photons'' and ``gravitons'' in 
$d=(5+1)$ as manifested from the point of view $d=(3+1)$:}

\vspace{2mm}

Let us conclude this subsection by recognizing that if the internal space of
``electron'' with spin up is represented by $\hat{b}^{1 \dagger}_{1}
(\equiv \stackrel{03}{(+i)}\stackrel{12}{[+]} \stackrel{56}{[+]})$ then
its ``positron'' is represented by $\hat{b}^{3 \dagger}_{1}
(\equiv \stackrel{03}{[-i]}\stackrel{12}{[+]} \stackrel{56}{(-)})$. Both 
have in this ``toy'' model fractional ``charge'' $S^{56}$; the ``electron's'' 
``charge'' is $\frac{1}{2}$ , the ``positron'' has the ``charge'' 
$-\frac{1}{2}$, both appear, according to Table~\ref{Table Clifffourplet.}, 
in four families.

The ``basis vectors'' of the ``electron's'' and ``positron's'' vector gauge
fields, the ``photons'', must be products of projectors since ``photons'' carry 
no charge and can correspondingly not change charges of ``electrons'' and 
``positrons''\footnote{
The ``basis vectors'' of ``photons'' must have all the quantum numbers 
${\cal S}^{ab} (=S^{ab} + \tilde{S}^{ab})$ of the Cartan subalgebra members,
Eq.~(\ref{cartangrasscliff}), equal zero (${\cal S}^{03}=0, {\cal S}^{12}=0,
{\cal S}^{56}=0$, what is the property of projectors describing bosons.
}.

There are four selfadjoint eigenvectors of the Cartan subalgebra members 
in the internal space of bosons. It then follows according to
Table~\ref{Table Clifffourplet.}
\begin{eqnarray}
\label{phelpo}
&&{}^{I}{\hat{\cal A}}^{1 \dagger}_{3 ph}
(\equiv \stackrel{03}{[+i]}\stackrel{12}{[+]} \stackrel{56}{[+]})\,
\,*_A\,\hat{b}^{1 \dagger}_{f} ({\rm for} f=1 \equiv \stackrel{03}{(+i)}\stackrel{12}{[+]} 
\stackrel{56}{[+]})\,\rightarrow \hat{b}^{1 \dagger}_{f}\,,\,\,\nonumber\\
&&{}^{I}{\hat{\cal A}}^{3 \dagger}_{2 ph}
(\equiv \stackrel{03}{[-i]}\stackrel{12}{[+]} \stackrel{56}{[-]})\,
\,*_A\,\hat{b}^{3 \dagger}_{f} ({\rm for} f=1 \equiv \stackrel{03}{[-i]}\stackrel{12}{[+]} 
\stackrel{56}{(-)})\,\rightarrow \hat{b}^{3 \dagger}_{f}\,,\,\,\nonumber\\
&&{}^{I}{\hat{\cal A}}^{2 \dagger}_{4 ph}
(\equiv \stackrel{03}{[-i]}\stackrel{12}{[-]} \stackrel{56}{[+]})\,
\,*_A\,\hat{b}^{2 \dagger}_{f} ({\rm for} f=1 \equiv \stackrel{03}{[-i])}\stackrel{12}{(-)} \stackrel{56}{[+]})\,\rightarrow \hat{b}^{2 \dagger}_{f}\,,\,\,\nonumber\\
&&{}^{I}{\hat{\cal A}}^{4 \dagger}_{1 ph}
(\equiv \stackrel{03}{[+i]}\stackrel{12}{[-]} \stackrel{56}{[-]})\,
\,*_A\,\hat{b}^{4 \dagger}_{f} ({\rm for} f=1 \equiv \stackrel{03}{(+i)}\stackrel{12}{(-)} 
\stackrel{56}{(-)})\,\rightarrow \hat{b}^{4 \dagger}_{f}\,,\,\,
\end{eqnarray}
all the rest ``scattering'' --- the application of 
${}^{I}{\hat{\cal A}}^{m \dagger}_{f ph}$ on the rest members of the same family --- give zero~\footnote{
``Scattering'' on the same member of another 
families  gives as well non zero contributions:
${}^{I}{\hat{\cal A}}^{1 \dagger}_{3 ph}\,*_A\,
\hat{b}^{1 \dagger}_{f}$ lead to $\hat{b}^{1 \dagger}_{f}$.%
}.

The ``basis vector'' of a ``photon'', applying in 
Eq.~(\ref{phelpo}) on the ``basis vector'' of a ``electron'' or ``positron''  with 
spin up or down, transforms the ``electron'' or ``positron'' back into the same
``electron'' or ``positron''~\footnote{
``Photons'' can transfer  their momentum manifesting in ordinary space to 
``electrons'' and ``positrons'', changing their momentum, but their 
``basis vectors'' do not change.
}.

In Table~\ref{S120Cliff basis5+1even I.} ``photons'', presented in 
Eq.~(\ref{phelpo}),  are marked by $\bigcirc$.\\

We expect ``gravitons'' not to have any ``charge'', $S^{56}=0$, as also
``photons'' do not have ``charge''. However, ``gravitons'' can have the spin 
and handedness (non-zero $S^{03}$ and non zero $S^{12}$) in $d=(3+1)$. 
It then follows according to 
Eq.~(\ref{calIAb1234gen}) and Table~\ref{Table Clifffourplet.} (keeping in 
mind that bosons must have an even number of nilpotents)
\begin{eqnarray}
\label{grelpo}
&&{}^{I}{\hat{\cal A}}^{3 \dagger}_{1 gr}
(\equiv \stackrel{03}{(-i)}\stackrel{12}{(+)} \stackrel{56}{[-]})\,
\,*_A\,\hat{b}^{4 \dagger}_{f}({\rm for} f=1\,
\equiv \stackrel{03}{(+i)}\stackrel{12}{(-)} 
\stackrel{56}{(-)})\,\rightarrow \hat{b}^{3 \dagger}_{f}\,,\,\,\nonumber\\
&&{}^{I}{\hat{\cal A}}^{4  \dagger}_{2 gr}
(\equiv \stackrel{03}{(+i)}\stackrel{12}{(-)} \stackrel{56}{[-]})\,
\,*_A\,\hat{b}^{3 \dagger}_{f}({\rm for} f=1\,
\equiv \stackrel{03}{[-i]}\stackrel{12}{[+]} 
\stackrel{56}{(-)})\,\rightarrow \hat{b}^{4 \dagger}_{f}\,,\,\,\nonumber\\
&&{}^{I}{\hat{\cal A}}^{2 \dagger}_{3 gr}
(\equiv \stackrel{03}{(-i)}\stackrel{12}{(-)} \stackrel{56}{[+]})\,
\,*_A\,\hat{b}^{1 \dagger}_{f}({\rm for} f=1\,
\equiv \stackrel{03}{(+i)}\stackrel{12}{[+]} 
\stackrel{56}{[+]})\,\rightarrow \hat{b}^{2 \dagger}_{f}\,,\,\,\nonumber\\
&&{}^{I}{\hat{\cal A}}^{1 \dagger}_{4 gr}
(\equiv \stackrel{03}{(+i)}\stackrel{12}{
(+)} \stackrel{56}{[+]})\,
\,*_A\,\hat{b}^{2 \dagger}_{f}({\rm for} f=1\,
\equiv \stackrel{03}{[-i]}\stackrel{12}{(-)} 
\stackrel{56}{[+]})\,\rightarrow \hat{b}^{1 \dagger}_{f}\,.\,\,
\end{eqnarray}

The same ``gravitons'' (${}^{I}{\hat{\cal A}}^{3 \dagger}_{1 gr}$,
${}^{I}{\hat{\cal A}}^{4 \dagger}_{2 gr}$,
${}^{I}{\hat{\cal A}}^{2 \dagger}_{3 gr}$,
${}^{I}{\hat{\cal A}}^{1 \dagger}_{4 gr}$) cause the equivalent
transformations of $ \hat{b}^{m \dagger}_{f}$
for any $f$. All the rest of the applications on ``electrons'' and
``positrons'' give zero.
The rest of ${}^{I}{\hat{\cal A}}^{m \dagger}_{f }$, carrying non zero
``charge'' can not represent ``gravitons''.
In Table~\ref{transverseCliff basis5+1even I.} ``gravitons'', presented in
Eq.~(\ref{grelpo}), appearing in Hermitian conjugated pairs, are marked
by $\ddagger$ (the first two in Eq.~(\ref{grelpo}) and $\odot \odot$ (the
second two in Eq.~(\ref{grelpo}).

\vspace{2mm}

There is the break of symmetries, which does not allow the rest of
${}^{I}{\hat{\cal A}}^{m \dagger}_{f}$ to be active~\footnote{
Ref.~\cite{n2023NPB}, illustrates in Fig.~1 and Fig.~2
that the even ``basis vectors'' representing the triplet and
antitriplet in Fig.~2 can not transform a colour singlet of Fig.~1 into a
member of the colour triplet after the break of symmetries.
}.\\

Let us look for the ``gravitons'' of the second kind, 
${}^{II}{\hat{\cal A}}^{m \dagger}_{f gr}$,
which transform, let say, $\hat{b}^{3 \dagger}_{3}
(\equiv \stackrel{03}{(-i)}\stackrel{12}{(+)} \stackrel{56}{(-)})$ into
$\hat{b}^{3 \dagger}_{1}(\equiv \stackrel{03}{[-i]}\stackrel{12}{[+]} 
\stackrel{56}{(-)})$ and back.

We find
\begin{eqnarray}
\label{grelpoII}
&&\hat{b}^{3 \dagger}_{3}(\equiv \stackrel{03}{(-i)}\stackrel{12}{(+)} 
\stackrel{56}{(-)})\,*_A\,{}^{II}{\hat{\cal A}}^{2 \dagger}_{3 gr}
(\equiv \stackrel{03}{(+i)}\stackrel{12}{(-)} \stackrel{56}{[+]})\,
\rightarrow\, \hat{b}^{3 \dagger}_{1}(\equiv \stackrel{03}{[-i]}\stackrel{12}{[+]} 
\stackrel{56}{(-)})\,,\nonumber\\
&& \hat{b}^{3 \dagger}_{1}(\equiv \stackrel{03}{[-i]}\stackrel{12}{[+]} 
\stackrel{56}{(-)})\,*_A\,{}^{II}{\hat{\cal A}}^{1 \dagger}_{4 gr}
(\equiv \stackrel{03}{(-i)}\stackrel{12}{(+)} \stackrel{56}{[+]})\,
\rightarrow\,\hat{b}^{3 \dagger}_{3}(\equiv \stackrel{03}{(-i)}\stackrel{12}{(+)} 
\stackrel{56}{(-)})\, .
\end{eqnarray}
We see that the ``basis vectors'' of the ``gravitons'' of the second kind transform 
the ``basis vectors'' representing a family member of a fermion into the same
family member of any other family. They do not change the spin, but the family 
quantum numbers, transferring to the odd ``basis vectors''
the family quantum numbers $\tilde{S}^{03}=+i$ and $\tilde{S}^{12}=-1$ 
on the first line of Eq.~(\ref{grelpoII}) and on the second line the family 
quantum number $\tilde{S}^{03}=-i$ and $\tilde{S}^{12}=+1$,
according to Eq.~(\ref{signature0}). 
(They are not the ``basis vectors'' of the ``gravitons'' of the kind 
${}^{I}{\hat{\cal A}}^{m \dagger}_{f gr}$.)

\vspace{2mm}

There is the break of symmetries, which chooses those ``basis
vectors'' of boson fields of both kinds, ${}^{I}{\hat{\cal A}}^{m \dagger}_{f gr}$
and ${}^{II}{\hat{\cal A}}^{m \dagger}_{f gr}$, which remain active (the break of symmetries are not studied in this contribution.)\\

Let us discuss the application of the ``basis vectors'' of ``gravitons'' on
the ``basis vectors'' of``photons'' and opposite,
``photons'' on ``gravitons'', presented in Eq.~(\ref{ruleAAI}), and
repeated below.
$${}^{I}{\hat{\cal A}}^{m \dagger}_{f} \,*_A\, 
{}^{I}{\hat{\cal A}}^{m' \dagger}_{f `}
\rightarrow  \left \{ \begin{array} {r}
 {}^{I}{\hat{\cal A}}^{m \dagger}_{f `}\,, i=(I,II) \\
{\rm or \,zero}\,.
\end{array} \right.$$

\begin{small}
\begin{eqnarray}
\label{grphgr}
{}^{I}{\hat{\cal A}}^{3 \dagger}_{1 gr}\,
(\equiv \stackrel{03}{(-i)}\stackrel{12}{(+)} \stackrel{56}{[-]})\,*_A\,
{}^{I}{\hat{\cal A}}^{4 \dagger}_{1 ph}\, (\equiv \stackrel{03}{[+i]}
\stackrel{12}{[-]} \stackrel{56}{[-]})\, \rightarrow \,
{}^{I}{\hat{\cal A}}^{3 \dagger}_{1 gr}\,,
\nonumber\\
{}^{I}{\hat{\cal A}}^{4 \dagger}_{2 gr}\,
(\equiv \stackrel{03}{(+i)}\stackrel{12}{(-)} \stackrel{56}{[-]})\,*_A\,
{}^{I}{\hat{\cal A}}^{3 \dagger}_{2 ph}\, (\equiv \stackrel{03}{[-i]}
\stackrel{12}{[+]} \stackrel{56}{[-]})\, \rightarrow \,
{}^{I}{\hat{\cal A}}^{4 \dagger}_{2 gr}\,,
\nonumber\\
{}^{I}{\hat{\cal A}}^{2 \dagger}_{3 gr}\,
(\equiv \stackrel{03}{(-i)}\stackrel{12}{(-)} \stackrel{56}{[+]})\,*_A\,
{}^{I}{\hat{\cal A}}^{1 \dagger}_{3 ph}\, (\equiv \stackrel{03}{[+i]}
\stackrel{12}{[+]} \stackrel{56}{[+]})\, \rightarrow \,
{}^{I}{\hat{\cal A}}^{2 \dagger}_{3 gr}\,,
\nonumber\\
{}^{I}{\hat{\cal A}}^{1 \dagger}_{4 gr}\,
(\equiv \stackrel{03}{(+i)}\stackrel{12}{(+)} \stackrel{56}{[+]})\,*_A\,
{}^{I}{\hat{\cal A}}^{2 \dagger}_{4 ph}\, (\equiv \stackrel{03}{[-i]}
\stackrel{12}{[-]} \stackrel{56}{[+]})\, \rightarrow \,
{}^{I}{\hat{\cal A}}^{1 \dagger}_{4 gr}\,,
\end{eqnarray}
\end{small}
all the rest applications give zero.

It remains to to see the application of the ``basis vectors'' of ``photons'' on 
the ``basis vectors'' of ``gravitons''.  It follows
\begin{small}
\begin{eqnarray}
\label{phgrgr}
{}^{I}{\hat{\cal A}}^{4 \dagger}_{1 ph}\,
(\equiv \stackrel{03}{[+i]}\stackrel{12}{[-]} \stackrel{56}{[-]})\,*_A\,
{}^{I}{\hat{\cal A}}^{4 \dagger}_{2 gr}\, (\equiv \stackrel{03}{(+i)}
\stackrel{12}{(-)} \stackrel{56}{[-]})\, \rightarrow \,
{}^{I}{\hat{\cal A}}^{4 \dagger}_{2 gr}\,,
\nonumber\\
{}^{I}{\hat{\cal A}}^{3 \dagger}_{2 ph}\,
(\equiv \stackrel{03}{[-i]}\stackrel{12}{[+]} \stackrel{56}{[-]})\,*_A\,
{}^{I}{\hat{\cal A}}^{3 \dagger}_{1 gr}\, (\equiv \stackrel{03}{(-i)}
\stackrel{12}{(+)} \stackrel{56}{[-]})\, \rightarrow \,
{}^{I}{\hat{\cal A}}^{3 \dagger}_{1 gr}\,,
\nonumber\\
{}^{I}{\hat{\cal A}}^{1 \dagger}_{3 ph}\,
(\equiv \stackrel{03}{[+i]}\stackrel{12}{[+]} \stackrel{56}{[+]})\,*_A\,
{}^{I}{\hat{\cal A}}^{1 \dagger}_{4 gr}\, (\equiv \stackrel{03}{(+i)}
\stackrel{12}{(+)} \stackrel{56}{[+]})\, \rightarrow \,
{}^{I}{\hat{\cal A}}^{1 \dagger}_{4 gr}\,,
\nonumber\\
{}^{I}{\hat{\cal A}}^{2 \dagger}_{4 ph}\,
(\equiv \stackrel{03}{[-i]}\stackrel{12}{[-]} \stackrel{56}{[+]})\,*_A\,
{}^{I}{\hat{\cal A}}^{2 \dagger}_{3 gr}\, (\equiv \stackrel{03}{(-i)}
\stackrel{12}{(-)} \stackrel{56}{[+]})\, \rightarrow \,
{}^{I}{\hat{\cal A}}^{2 \dagger}_{3 gr}\,,
\end{eqnarray}
\end{small}
all the rest applications give zero. 

Let be added that both,  ``photons'' and ``gravitons'' carry the space index 
$\alpha$ (which is in $d=(3+1)$ equal $(0,1,2,3)$).

Similar relations follow also for ${}^{II}{\hat{\cal A}}^{m \dagger}_{f}$.

Let us study ''scattering'' of ``photons'' on ``photons'', according to Eq.~(\ref{ruleAAI}).  Since 
${}^{I}{\hat{\cal A}}^{m \dagger}_{f ph}\,$ are self adjoint ``basis 
vectors'' and self adjoint ``basis vectors'' are orthogonal, it follows
\begin{small}
\begin{eqnarray}
\label{phphph}
&&{}^{I}{\hat{\cal A}}^{4 \dagger}_{1 ph}\,*_A\,
{}^{I}{\hat{\cal A}}^{4 \dagger}_{1 ph}\rightarrow \,
{}^{I}{\hat{\cal A}}^{4 \dagger}_{1 ph}\,, \quad
{}^{I}{\hat{\cal A}}^{3 \dagger}_{2 ph}\,*_A\,
{}^{I}{\hat{\cal A}}^{3 \dagger}_{2 ph}\rightarrow \,
{}^{I}{\hat{\cal A}}^{3 \dagger}_{2 ph}\,, \nonumber\\
&&{}^{I}{\hat{\cal A}}^{1 \dagger}_{3 ph}\,*_A\,
{}^{I}{\hat{\cal A}}^{1 \dagger}_{3 ph}\rightarrow \,
{}^{I}{\hat{\cal A}}^{1 \dagger}_{3 ph}\, ,\quad
{}^{I}{\hat{\cal A}}^{2 \dagger}_{4 ph}\,*_A\,
{}^{I}{\hat{\cal A}}^{2 \dagger}_{4 ph}\rightarrow \,
{}^{I}{\hat{\cal A}}^{2 \dagger}_{4 ph}\, \,
\end{eqnarray}
\end{small}
all the rest applications give zero. 

Applications of ``gravitons'' on ``gravitons'' lead to
\begin{small}
\begin{eqnarray}
\label{grgrgr}
{}^{I}{\hat{\cal A}}^{3 \dagger}_{1 gr}\,
(\equiv \stackrel{03}{(-i)}\stackrel{12}{(+)} \stackrel{56}{[-]})\,*_A\,
{}^{I}{\hat{\cal A}}^{4 \dagger}_{2 gr}\, (\equiv \stackrel{03}{(+i)}
\stackrel{12}{(-)} \stackrel{56}{[-]})\, \rightarrow \,
{}^{I}{\hat{\cal A}}^{3 \dagger}_{2 ph} \, (\equiv \stackrel{03}{[-i]}\stackrel{12}{[+]} \stackrel{56}{[-]})\,\,,
\nonumber\\
{}^{I}{\hat{\cal A}}^{4 \dagger}_{2 gr}\,
(\equiv \stackrel{03}{(+i)}\stackrel{12}{(-)} \stackrel{56}{[-]})\,*_A\,
{}^{I}{\hat{\cal A}}^{3 \dagger}_{1 gr}\, (\equiv \stackrel{03}{(-i)}
\stackrel{12}{(+)} \stackrel{56}{[-]})\, \rightarrow \,
{}^{I}{\hat{\cal A}}^{4 \dagger}_{1 ph}\,(\equiv \stackrel{03}{[+i]}\stackrel{12}{[-]} \stackrel{56}{[-]})\,,
\nonumber\\
{}^{I}{\hat{\cal A}}^{2 \dagger}_{3 gr}\,(\equiv \stackrel{03}{(-i)}\stackrel{12}{(-)} \stackrel{56}{[+]}) \,*_A\,
{}^{I}{\hat{\cal A}}^{1 \dagger}_{4 gr}\, (\equiv \stackrel{03}{(+i)}
\stackrel{12}{(+)} \stackrel{56}{[+]})\, \rightarrow \,
{}^{I}{\hat{\cal A}}^{2 \dagger}_{4 ph}\, (\equiv \stackrel{03}{[-i]}\stackrel{12}{[-]} \stackrel{56}{[+]})\,\,,
\nonumber\\
{}^{I}{\hat{\cal A}}^{1 \dagger}_{4 gr}\,
 (\equiv \stackrel{03}{(+i)} \stackrel{12}{(+)} \stackrel{56}{[+]})\,*_A\,
{}^{I}{\hat{\cal A}}^{2 \dagger}_{3 gr}\, (\equiv \stackrel{03}{(-i)}
\stackrel{12}{(-)} \stackrel{56}{[+]})\, \rightarrow \,
{}^{I}{\hat{\cal A}}^{1 \dagger}_{3 ph}(\equiv \stackrel{03}{[+i]}\stackrel{12}{[+]} \stackrel{56}{[+]})\,\,,
\end{eqnarray}
\end{small}
all the rest applications give zero.\\

One gets equivalent relations also for 
${}^{II}{\hat{\cal A}}^{m \dagger}_{f}$.
They transform the family member of the odd ``basis vector'' into 
the same family member of any other family of the odd ``basis vector''. 


\subsection{Photons, weak bosons and gluons in $(13+1)$-dimensional space 
from the point of view of $d=(3+1)$}
\label{photonsweakbosonsgluons}


In Subsect.~\ref{bosons5+1} we studied the ``basis vectors'' in $d=(5+1)$
describing internal spaces of fermions (represented by the superposition of
the odd products of $\gamma^a$'s), and the ``basis vectors'' describing
internal spaces of bosons (represented by the superposition of the even
products of $\gamma^a$'s). It was pointed out in
Subsect.~\ref{cliffordoddevenbasis5+1} that each even ``basis
vector'' can be written as a product of an odd ``basis vector'' and of 
one of the Hermitian conjugated partners of odd ``basis vectors'', 
 Eqs.~(\ref{AIbbdagger}, \ref{AIIbdaggerb}). In Subsect.~\ref{cliffordoddevenbasis5+1}, we call 
fermions ``positrons'' if they carry $S^{56}=- \frac{1}{2}$, and 
``electrons'', if they carry $S^{56}=\frac{1}{2}$. Their vector gauge 
fields, the bosons, are called ``photons'', Eq.~(\ref{phelpo}) (all the 
eigenvalues of the Cartan subalgebra members, 
Eq.~(\ref{cartangrasscliff}), of ``photons'' ``basis vectors'' are equal to
zero) and ``gravitons'', Eq.~(\ref{grelpo}), (they carry ${\cal S}^{03}=
(- i, + i)$ and ${\cal S}^{12}= (- 1, + 1)$): ``Photons'' ``scatter'' on
``electrons'' and positrons'' without changing their ``basis vectors''. They
can transfer to ``electrons'' and ``positrons'' their ordinary space
momentum (carrying the space index $\alpha$ which we do not take into
account in these considerations). ``Gravitons'' can change the spin and
handedness of the ``basis vectors'' of ``electrons'' and ``positrons'', 
and can transfer to ``electrons'' and ``positrons'' their ordinary space
momentum.

We study in Subsect.~\ref{bosons5+1} ``basis vectors'' of the boson
fields which transform ``positrons'' into ``electrons'' and back,
Eq.~(\ref{05b}).~\footnote{
These ``basis vectors'' should not be allowed after the 
break of the symmetry from $SO(5,1)$ to $SO(3,1)\times U(1)$ and so 
do not the rest of the $2^{\frac{6}{2}-1}\times 2^{\frac{6}{2}-1}$ even ``basis vectors'', which would change the spin and ``charge'' of ``basis vectors'' of ``positrons'' and ``electrons'' at the same time, if we require
that there are no ``basis vectors'' transforming ``positrons'' into 
``electrons'' or opposite.
}
In this Subsect.~\ref{bosons5+1}, we study also scattering of ``photons'' on
``gravitons'' and ``gravitons'' on ``photons'', as well as ``photons'' on
``photons'' and ``gravitons on ``gravitons''.

\vspace{2mm}

To reproduce the quantum numbers of the observed quarks and leptons and
antiquarks and antileptons, the dimension of space-time must be
$d\ge (13+1)$, as assumed by the {\it spin-charge-family}
theory~(\cite{nh2021RPPNP}, and references therein).

The ``basis vectors'' of one irreducible representation of the internal space
of fermions (of one family), are presented in Table~\ref{Table so13+1.},
analysed from the point of view $d=(3+1)$, as suggested by the {\it standard
model}. One family represents quarks and leptons and antiquarks and
antileptons (with right handed neutrinos and left handed antineutrinos included).
They carry the weak charge, $\tau^{13}$, if they are left-handed, and the
second kind of the weak charge, $\tau^{23}$, if they are right-handed;
quarks are colour triplets; antiquarks are colour antitriplets; leptons are
colour singlets and antilepons are anticolour singlets.
In the $SO(7,1)$ content of the $SO(13,1)$, the ``basis vectors'' of
quarks can not be distinguished from the ``basis vectors'' of leptons, and the
``basis vectors'' of antiquarks are not distinguishable from the ``basis vectors''
of antileptons; these can be seen when comparing the
corresponding ``basis vectors'' in Table~\ref{Table so13+1.}.
Quarks and leptons distinguish only in the $SU(3)\times U(1)$ part and
so do antiquarks and antileptons. 
In $d=(3+1)$, the internal space of quarks and leptons (and antiquarks and
antileptons) manifest, analysed from the point of view of the {\it standard
model} groups before the electroweak break, all the desired properties. 
The reader can find in the review article~\cite{nh2021RPPNP} and
Refs.~(\cite{n2023MDPI}, \cite{n2023NPB}, \cite{n2023DOI},
\cite{nh2023dec}) the achievements of the {\it spin-charge-family}
theory so far. 
In the present article, we are interested in the vector and scalar gauge
fields, in their description of the internal spaces in terms of the even
``basis vectors'' (the superposition of even products of $\gamma^a$'s), 
to better understand the meaning of the second quantisation of boson 
fields (the internal space of which is describable by the even ``basis 
vectors'').~\footnote{
We must keep in mind that when analysing the internal spaces of 
fermions and bosons from the point of view $d=(3+1)$, the breaks of
symmetries from the starting one (let say $SO(13+1)$ describing 
the internal space of fermions and bosons) to the ``observed'' ones 
before the electroweak break must be taken into account: The boson
fields, if they are gluons, can change the colour of quarks; if they are weak
bosons can change the weak charge among quarks or among leptons;
for gravitons, we expect that they can change the spins and handedness
of quarks and leptons. And, since there exist even ``basis vectors'' 
which could transform fermions into antifermions, or change the colour 
and the weak charge of fermions
at the same time, they should not be active after the breaks of
symmetries. 

An example is presented in Fig.~1 and Fig.~2, in
Ref.~\cite{n2023NPB}, pointing out that at low energies, the two 
even ``basis vectors,''
which form one triplet and one antitriplet should not transform a
``lepton'' into one of three ``quarks''.
}.

 \vspace{1mm}

Let us use the knowledge learned in Subsect.~\ref{bosons5+1}. 

 \vspace{1mm}

{\bf a.}$\;\;$
We can start with Eq.~(\ref{CPd}) and recognize in 
Table~\ref{Table so13+1.} the charge conjugated partners among the 
odd ``basis vectors'', related by $\gamma^0 \gamma^5 \gamma^7
\gamma^9 \gamma^{11} \gamma^{13}$ (paying attention  only 
to internal space of fermions).
We find in Table~\ref{Table so13+1.}, using Eq.~(\ref{usefulrel0}), that 
$u^{c1 \dagger}_{R}$, first line, and  $\bar{u}^{\bar{c1}\dagger}_{L}, 
35^{th}$ line, are 
the charge conjugated pair; also $\nu^{\dagger}_{R}, 25^{th}$ line, and  
$\bar{\nu}^{\dagger}_{L}, 59^{th}$  line, are the charge conjugated pair,
for example. 

All the members  of one irreducible representation appear in the charge 
conjugated pairs, as we learned in Subsect.~\ref{bosons5+1}.


Searching for the even ``basis vector'' which transforms 
$u^{c1 \dagger}_{R}, 1^{st}$ line, into  $\bar{u}^{\bar{c1}\dagger}_{L}, 
35^{th}$ line and opposite and the same for $\nu$ we find 
(using Eq.~(\ref{usefulrel0})~\footnote{Taking into account that
\begin{small}
\begin{eqnarray}
\label{simple}
\stackrel{ab}{(k)}\stackrel{ab}{(-k)}& =& \eta^{aa} \stackrel{ab}{[k]}\,,\quad 
\stackrel{ab}{(-k)}\stackrel{ab}{(k)} = \eta^{aa} \stackrel{ab}{[-k]}\,,\quad
\stackrel{ab}{(k)}\stackrel{ab}{[k]} =0\,,\quad 
\stackrel{ab}{(k)}\stackrel{ab}{[-k]} =
 \stackrel{ab}{(k)}\,,\quad 
 \nonumber\\
 \stackrel{ab}{(-k)}\stackrel{ab}{[k]} &=& \stackrel{ab}{(-k)}\,,\quad 
\stackrel{ab}{[k]}\stackrel{ab}{(k)}= \stackrel{ab}{(k)}\,,
\quad 
 \stackrel{ab}{[k]}\stackrel{ab}{(-k)} =0\,,\quad 
 \stackrel{ab}{[k]}\stackrel{ab}{[-k]} =0\,,\quad 
\end{eqnarray}
\end{small}
 one easily reproduced the relations of Eq.~(\ref{CPAubaru}).}  one easily reproduced the below relations) 
\begin{small}
\begin{eqnarray}
\label{CPAubaru}
&&{}^{I}{\hat{\cal A}}^{ \dagger}_{CP u^{c1}_{R}\rightarrow 
{\bar u}^{\bar c1}_{L}} 
(\equiv \stackrel{03}{(-i)}\stackrel{12}{[+]} \stackrel{56}{(-)}
 \stackrel{7 8}{(-)}\stackrel{9\,10}{(-)} \stackrel{11\,12}{(+)}
\stackrel{13\,14}{(+)})\,*_A\,u^{c1 \dagger}_{R \,1^{st}}
(\equiv \stackrel{03}{(+i)}\stackrel{12}{[+]} 
\stackrel{56}{[+]}\stackrel{78}{(+)}\stackrel{9\,10}{(+)} 
\stackrel{11\,12}{[-]} \stackrel{13\,14}{[-]})\,\rightarrow 
\nonumber\\
&&\bar{u}^{\bar{c1}\dagger}_{L\, 35^{th}}
(\equiv \stackrel{03}{[-i]}\stackrel{12}{[+]} 
\stackrel{56}{(-)}\stackrel{78}{[-]}\stackrel{9\,10}{[-]} 
\stackrel{11\,12}{(+)} \stackrel{13\,14}{(+)})
\,,\quad \quad \quad
{}^{I}{\hat{\cal A}}^{ \dagger}_{CP u^{c1}_{R}\rightarrow 
{\bar u}^{\bar c1}_{L}} = {\bar u}^{\bar{c1} \dagger}_{L\,35^{th}}\,*_A\,
(u^{c1 \dagger}_{R\, 1^{st}})^{\dagger}\,,
\nonumber\\
&&{}^{I}{\hat{\cal A}}^{ \dagger}_{CP {\bar u}^{\bar c1}_{L}\rightarrow 
u^{c1}_{R}} 
(\equiv \stackrel{03}{(+i)}\stackrel{12}{[+]} \stackrel{56}{(+)}
 \stackrel{7 8}{(+)}\stackrel{9\,10}{(+)} \stackrel{11\,12}{(-)}
\stackrel{13\,14}{(-)})\,*_A\,\bar{u}^{\bar{c1}\dagger}_{L\,35^{th}}
(\equiv \stackrel{03}{[-i]}\stackrel{12}{[+]} 
\stackrel{56}{(-)}\stackrel{78}{[-]}\stackrel{9\,10}{[-]} 
\stackrel{11\,12}{(+)} \stackrel{13\,14}{(+)})\,\rightarrow 
\nonumber\\
&&u^{c1 \dagger}_{R\, 1^{st}}
(\equiv \stackrel{03}{(+i)}\stackrel{12}{[+]} 
\stackrel{56}{[+]}\stackrel{78}{(+)}\stackrel{9\,10}{(+)} 
\stackrel{11\,12}{[-]} \stackrel{13\,14}{[-]})\,,
\quad \quad \quad
{}^{I}{\hat{\cal A}}^{ \dagger}_{CP {\bar u}^{\bar c1}_{L}\rightarrow 
u^{c1}_{R}} = u^{c1 \dagger}_{R\, 1^{st}}\,*_A\,
(\bar{u}^{\bar{c1}\dagger}_{L\, 35^{th}})^{\dagger}\,,
\nonumber\\
&&{}^{I}{\hat{\cal A}}^{ \dagger}_{CP \nu_{R}\rightarrow 
{\bar \nu}_{L}} 
(\equiv \stackrel{03}{(-i)}\stackrel{12}{[+]} \stackrel{56}{(-)}
 \stackrel{7 8}{(-)}\stackrel{9\,10}{(-)} \stackrel{11\,12}{(-)}
\stackrel{13\,14}{(-)})\,*_A\,\nu^{\dagger}_{R\, 25^{th}})
(\equiv \stackrel{03}{(+i)}\stackrel{12}{[+]} 
\stackrel{56}{[+]}\stackrel{78}{(+)}\stackrel{9\,10}{(+)} 
\stackrel{11\,12}{(+)} \stackrel{13\,14}{(+)})\,\rightarrow 
\nonumber\\
&&\bar{\nu}^{\dagger}_{L\, 59^{th}}
(\equiv \stackrel{03}{[-i]}\stackrel{12}{[+]} 
\stackrel{56}{(-)}\stackrel{78}{[-]}\stackrel{9\,10}{[-]} 
\stackrel{11\,12}{[-]} \stackrel{13\,14}{[-]})
\,,\quad \quad \quad 
{}^{I}{\hat{\cal A}}^{ \dagger}_{CP \nu_{R}\rightarrow 
{\bar \nu}_{L}} = {\bar \nu}^{ \dagger}_{L\, 59^{th}}\,*_A\,
(\nu^{\dagger}_{R\, 25^{th}})^{\dagger}\,,\nonumber\\
&&{}^{I}{\hat{\cal A}}^{ \dagger}_{CP {\bar \nu}_{L}\rightarrow 
\nu_{R}}
(\equiv \stackrel{03}{(+i)}\stackrel{12}{[+]} \stackrel{56}{(+)}
 \stackrel{7 8}{(+)}\stackrel{9\,10}{(+)} \stackrel{11\,12}{(+)}
\stackrel{13\,14}{(+)})\,*_A\,\bar{\nu}^{\dagger}_{L\, 59^{th}},
(\equiv \stackrel{03}{[-i]}\stackrel{12}{[+]} 
\stackrel{56}{(-)}\stackrel{78}{[-]}\stackrel{9\,10}{[-]} 
\stackrel{11\,12}{[-]} \stackrel{13\,14}{[-]})\,\rightarrow 
\nonumber\\
&&\nu^{\dagger}_{R\, 25^{th}}
(\equiv \stackrel{03}{(+i)}\stackrel{12}{[+]} 
\stackrel{56}{[+]}\stackrel{78}{(+)}\stackrel{9\,10}{(+)} 
\stackrel{11\,12}{(+)} \stackrel{13\,14}{(+)})\,,
\quad \quad \quad
{}^{I}{\hat{\cal A}}^{ \dagger}_{CP {\bar \nu}_{L}\rightarrow 
\nu_{R}} = \nu^{ \dagger}_{R\, 25^{th}}\,*_A\,
(\bar{\nu}^{\dagger}_{L\, 59^{th}})^{\dagger}\,,
\end{eqnarray}
\end{small}
 Eq.~(\ref{CPAubaru}),  demonstrates that any even 
``basis vector'' can be written as the algebraic product of an odd 
``basis vector''  and a Hermitian conjugated member, as demonstrated in Eq.~(\ref{AIbbdagger}).

Let us notice also in this case, as we learned in the case $d=(5+1)$,  
that the two bosonic even ``basis vectors'', causing the 
transformations of fermion odd ``basis vectors'' to their 
antifermion ``basis vectors'',   
${}^{I}{\hat{\cal A}}^{ \dagger}_{CP u^{c1}_{R}\rightarrow 
{\bar u}^{\bar c1}_{L}} $ and 
 ${}^{I}{\hat{\cal A}}^{ \dagger}_{CP {\bar u}^{\bar c1}_{L}\rightarrow 
u^{c1}_{R}} $ in the case of quarks and 
 ${}^{I}{\hat{\cal A}}^{ \dagger}_{CP \nu_{R}\rightarrow 
{\bar \nu}_{L}} $ and
  ${}^{I}{\hat{\cal A}}^{ \dagger}_{CP {\bar \nu}_{L}\rightarrow 
\nu_{R}}$  in the case of neutrinos,
are Hermitian conjugated to each other, fulfilling Eq.~(\ref{05AIb}).
 
\vspace{2mm}  
  
{\bf b.} $\;\;$  %
Let us study photons using the knowledge from Subsect.~\ref{bosons5+1}.
Photons interact at low energies with all quarks and electrons and antiquarks 
and positrons, leaving their ``basis vectors'' unchanged  and offering them
only momentum which they carry in external space. According to the observations,
they can not interact with neutrinos and antineutrinos.

 Let us look in Table~\ref{Table so13+1.} for $u^{c1 \dagger}_{R}$, first 
 line. The photon  ${}^{I}{\hat{\cal A}}^{ \dagger}_{ph\, u^{c1 \dagger}_{R}
 \rightarrow u^{c1 \dagger}_{R}}$ interacts with $u^{c1 \dagger}_{R}$ as 
 follows
\begin{small}
\begin{eqnarray}
\label{phAqeqe}
&&{}^{I}{\hat{\cal A}}^{ \dagger}_{ph \,u^{c1 \dagger}_{R}
 \rightarrow u^{c1 \dagger}_{R}}
(\equiv \stackrel{03}{[+i]}\stackrel{12}{[+]} \stackrel{56}{[+]}
 \stackrel{7 8}{[+]}\stackrel{9\,10}{[+]} \stackrel{11\,12}{[-]}
\stackrel{13\,14}{[-]})\,*_A\, u^{c1 \dagger}_{R\, 1^{st}},
(\equiv \stackrel{03}{(+i)}\stackrel{12}{[+]} 
\stackrel{56}{[+]}\stackrel{78}{(+)}\stackrel{9\,10}{(+)} 
\stackrel{11\,12}{[-]} \stackrel{13\,14}{[-]})\,\rightarrow 
\nonumber\\
&&u^{c1 \dagger}_{R\, 1^{st}},
(\equiv \stackrel{03}{(+i)}\stackrel{12}{[+]} 
\stackrel{56}{[+]}\stackrel{78}{(+)}\stackrel{9\,10}{(+)} 
\stackrel{11\,12}{[-]} \stackrel{13\,14}{[-]})\,,
 \quad \quad \quad
{}^{I}{\hat{\cal A}}^{ \dagger}_{ph\, u^{c1 \dagger}_{R}
 \rightarrow u^{c1 \dagger}_{R}} = u^{c1 \dagger}_{R\, 1^{st}},
\,*_A\, (u^{c1 \dagger}_{R\, 1^{st}})^{\dagger}\,,
\nonumber\\
&&{}^{I}{\hat{\cal A}}^{ \dagger}_{ph \,\bar{u}^{\bar{c1} \dagger}_{R}
 \rightarrow\bar{u}^{\bar{c1} \dagger}_{R}}
(\equiv \stackrel{03}{[-i]}\stackrel{12}{[+]} \stackrel{56}{[-]}
 \stackrel{7 8}{[-]}\stackrel{9\,10}{[-]} \stackrel{11\,12}{[+]}
\stackrel{13\,14}{[+]})\,*_A\, \bar{u}^{\bar{c1}\dagger}_{L\, 35^{th}},
(\equiv \stackrel{03}{[-i]}\stackrel{12}{[+]} 
\stackrel{56}{(-)}\stackrel{78}{[-]}\stackrel{9\,10}{[-]} 
\stackrel{11\,12}{(+)} \stackrel{13\,14}{(+)})\,\rightarrow 
\nonumber\\
&&\bar
{u}^{\bar{c1}\dagger}_{L\, 35^{th}}
(\equiv \stackrel{03}{[-i]}\stackrel{12}{[+]} 
\stackrel{56}{(-)}\stackrel{78}{[-]}\stackrel{9\,10}{[-]} 
\stackrel{11\,12}{(+)} \stackrel{13\,14}{(+)})\,, 
 \quad \quad \quad
 {}^{I}{\hat{\cal A}}^{ \dagger}_{ph\, \bar{u}^{\bar{c1} \dagger}_{R}
 \rightarrow\bar{u}^{\bar{c1} \dagger}_{R}}= 
 {u}^{\bar{c1}\dagger}_{L\, 35^{th}}\,*_A\,
 ({u}^{\bar{c1}\dagger}_{L\, 35^{th}})^{\dagger}\,.
\end{eqnarray}
\end{small}
Similarly, one finds photons interacting with the rest of  quarks and 
electrons and antiquarks and positrons. 

Looking at the quantum numbers of neutrinos in 
Table~\ref{Table so13+1.} we find the electromagnetic charge
of neutrinos equal zero.  In Ref.~(\cite{nh2021RPPNP}, Table~6, and in 
the references therein) there is the condensate of two right handed 
neutrinos that makes (due to taking 
care of the masses of all bosons, except the photons, gravitons, weak
bosons and gluons) the charges of neutrinos equal zero.


Since neutrinos do not carry the electric charge {\it the following 
interaction of a neutrino and a photon is not possible}
\begin{small}
\begin{eqnarray}
\label{phnu}
&&{}^{I}{\hat{\cal A}}^{ \dagger}_{ph\, \nu^{ \dagger}_{R}\rightarrow
\nu^{ \dagger}_{R}}
(\equiv \stackrel{03}{[+i]}\stackrel{12}{[+]} \stackrel{56}{[+]}
 \stackrel{7 8}{[+]}\stackrel{9\,10}{[+]} \stackrel{11\,12}{[+]}
\stackrel{13\,14}{[+]})\,*_A\,\nu^{ \dagger}_{R\, 25^{th}},
(\equiv \stackrel{03}{(+i)}\stackrel{12}{[+]} 
\stackrel{56}{[+]}\stackrel{78}{(+)}\stackrel{9\,10}{(+)} 
\stackrel{11\,12}{(+)} \stackrel{13\,14}{(+)})\,\rightarrow 
\nonumber\\
&&\nu^{ \dagger}_{R\, 25^{th}}
(\equiv \stackrel{03}{(+i)}\stackrel{12}{[+]} 
\stackrel{56}{[+]}\stackrel{78}{(+)}\stackrel{9\,10}{(+)} 
\stackrel{11\,12}{(+)} \stackrel{13\,14}{(+)})\, . 
\end{eqnarray}
\end{small}
The same is true for all the neutrinos (appearing in Table~\ref{Table so13+1.} on the lines $26,31,32$) and anti-neutrinos appearing in Table~\ref{Table so13+1.} on the lines $(59, 60,61,62)$.

The break of symmetries, which are responsible for distinguishing
among quarks and leptons ($SO(13,1)$ to $SO(7,1)\times$ 
$SU(3)\times U(1)$) and then within quarks and within leptons
($SO(7,1)$ to $SO(3,1)\times$ $SU(2)_I \times SU(2)_{II}$)
makes that the observed photons do not interact directly with neutrinos. 
The same is true for all 
the families of the neutrinos, as we learned in 
Subsect.~\ref{bosons5+1}, the same 
${}^{I}{\hat{\cal A}}^{ \dagger}_{ph \,\nu^{ \dagger}_{R}
\rightarrow \nu^{ \dagger}_{R}} $ would cause first order interactions of 
photons with neutrinos in all the families.

\vspace{2mm}


{\bf c.} $\;\;$
Let us study the properties of the weak bosons.
Table~\ref{Table so13+1.} contains one irreducible representation (one
family) of quarks and leptons and antiquarks and antileptons.
The multiplet contains the left-handed ($\Gamma^{(3+1)}=-1$) weak
$SU(2)_{I}$ charged ($\tau^{13}=\pm \frac{1}{2}$, Eq.~(\ref{so42})),
and $SU(2)_{II}$ chargeless ($\tau^{23}=0$, Eq.~(\ref{so42})),
quarks and leptons, and the right-handed ($\Gamma^{(3+1)}=1$),
weak ($SU(2)_{I}$) chargeless and $SU(2)_{II}$ charged
($\tau^{23}=\pm \frac{1}{2}$) quarks and leptons, both with the spin
$ S^{12}$ up and down ($\pm \frac{1}{2}$, respectively).
The creation operators of quarks distinguish from those of leptons only
in the $SU(3) \times U(1)$ part: Quarks are triplets of three colours
($ (\tau^{33}, \tau^{38})$ $ = [(\frac{1}{2},\frac{1}{2\sqrt{3}}),
(-\frac{1}{2},\frac{1}{2\sqrt{3}}), (0,-\frac{1}{\sqrt{3}}) $], Eq.~(\ref{so42}))
carrying the "fermion charge" ($\tau^{4}=
\frac{1}{6}$, $=-\frac{1}{3}(S^{9\,10}+ S^{11\,12}+ S^{13\,14})$.
The colourless leptons carry the "fermion charge" ($\tau^{4}=
-\frac{1}{2}$).
In the same multiplet, there are the right-handed ($\Gamma^{(3+1)}=1$
weak ($SU(2)_{I}$) charged ($\tau^{13}=\pm \frac{1}{2}$,
antiquarks and antileptons, and the right-handed
($\Gamma^{(3+1)}=1$), weak ($SU(2)_{I}$) charged and
$SU(2)_{II}$ chargeless ($\tau^{23}=\pm \frac{1}{2}$) antiquarks
and antileptons, both with the spin $ S^{12}$ up and down
($\pm \frac{1}{2}$, respectively).
Antiquarks are antitriplets carrying the ``fermion charge'' $- \frac{1}{6}$,
antileptons are ``antisinglets'' with the ``fermion charge'' $\frac{1}{2}$.

There are two kinds of weak bosons: those transforming right-handed
quarks and leptons within a $SU(2)_{II}$ doublet, and those transforming
left-handed quarks and leptons within a $SU(2)_{I}$ doublet, as well as
those transforming the right-handed antiquarks and antileptons
within a $SU(2)_{I}$ doublet, and those transforming left-handed antiquarks and
antileptons within a $SU(2)_{II}$ doublet.

Let us look for the ``basis vectors'' of weak bosons, which transform
$u^{c1 \dagger}_{R}$,
$1^{th}$ line, to the $d^{c1 \dagger}_{R}$, $3^{rd}$ line and back, in
Table~\ref{Table so13+1.}.

\begin{small}
\begin{eqnarray}
\label{w2udc1R}
&&{}^{I}{\hat{\cal A}}^{ \dagger}_{w2\, u^{c1}_{R}
\rightarrow d^{c1}_{R}}
(\equiv \stackrel{03}{[+i]}\stackrel{12}{[+]} \stackrel{56}{(-)}
 \stackrel{7 8}{(-)}\stackrel{9\,10}{[+]} \stackrel{11\,12}{[-]}
\stackrel{13\,14}{[-]})\,*_A\, u^{c1 \dagger}_{R\, 1^{st}},
(\equiv \stackrel{03}{(+i)}\stackrel{12}{[+]} 
\stackrel{56}{[+]}\stackrel{78}{(+)}\stackrel{9\,10}{(+)} 
\stackrel{11\,12}{[-]} \stackrel{13\,14}{[-]})\,\rightarrow 
\nonumber\\
&&d^{c1 \dagger}_{R\, 3^{rd}},
(\equiv \stackrel{03}{(+i)}\stackrel{12}{[+]} 
\stackrel{56}{(-)}\stackrel{78}{[-]}\stackrel{9\,10}{(+)} 
\stackrel{11\,12}{[-]} \stackrel{13\,14}{[-]})\,, 
\quad \quad \quad {}^{I}{\hat{\cal A}}^{ \dagger}_{w2\, u^{c1}_{R}
\rightarrow d^{c1}_{R}}= d^{c1 \dagger}_{R\, 3^{rd}}\, *_A\,
(u^{c1 \dagger}_{R\, 1^{st}})^{\dagger}\,,\nonumber\\
&&{}^{I}{\hat{\cal A}}^{ \dagger}_{w2\, d^{c1}_{R}
\rightarrow u^{c1}_{R}}
(\equiv \stackrel{03}{[+i]}\stackrel{12}{[+]} \stackrel{56}{(+)}
 \stackrel{7 8}{(+)}\stackrel{9\,10}{[+]} \stackrel{11\,12}{[-]}
\stackrel{13\,14}{[-]})\,*_A\, d^{c1 \dagger}_{R\,3^{rd}},
(\equiv \stackrel{03}{(+i)}\stackrel{12}{[+]} 
\stackrel{56}{(-)}\stackrel{78}{[-]}\stackrel{9\,10}{(+)} 
\stackrel{11\,12}{[-]} \stackrel{13\,14}{[-]})\,\rightarrow 
\nonumber\\
&&u^{c1 \dagger}_{R\, 1^{st}},
(\equiv \stackrel{03}{(+i)}\stackrel{12}{[+]} 
\stackrel{56}{[+]}\stackrel{78}{(+)}\stackrel{9\,10}{(+)} 
\stackrel{11\,12}{[-]} \stackrel{13\,14}{[-]})\,, 
\quad \quad \quad {}^{I}{\hat{\cal A}}^{ \dagger}_{w2\, d^{c1}_{R}
\rightarrow u^{c1}_{R}}= u^{c1 \dagger}_{R\, 1^{st}}\, *_A\,
(d^{c1 \dagger}_{R\, 3^{rd}})^{\dagger}\,.
\end{eqnarray}
\end{small}
One can notice in Eq.~(\ref{w2udc1R}) that
${}^{I}{\hat{\cal A}}^{ \dagger}_{w2\, u^{c1}_{R}\rightarrow 
d^{c1}_{R}}=  d^{c1 \dagger}_{R\, 3^{rd}}\, *_A\,
(u^{c1 \dagger}_{R\, 1^{st}})^{\dagger}$,  and 
${}^{I}{\hat{\cal A}}^{ \dagger}_{w2\, d^{c1}_{R}
\rightarrow u^{c1}_{R}}= u^{c1 \dagger}_{R\, 1^{st}}\, *_A\,
(d^{c1 \dagger}_{R\, 3^{rd}})^{\dagger}$, demonstrate 
Eq.~(\ref{AIbbdagger}),  and that 
${}^{I}{\hat{\cal A}}^{ \dagger}_{w2\, u^{c1}_{R}\rightarrow d^{c1}_{R}}$ and 
${}^{I}{\hat{\cal A}}^{ \dagger}_{w2 \,d^{c1}_{R}
\rightarrow u^{c1}_{R}}$ are Hermitian conjugated to each other, as
suggested by Eq.~(\ref{AIbbdagger}).\\
%



Looking  for the weak bosons, which transform $e^{ \dagger}_{L}$ 
 (presented in Table~\ref{Table so13+1.} in $29^{th}$ line) to  $\nu^{\dagger}_{L}$ (presented in Table~\ref{Table so13+1.} in $31^{st}$ line), and back, we find
\begin{small}
\begin{eqnarray}
\label{w1enuL}
&&{}^{I}{\hat{\cal A}}^{ \dagger}_{w1\, e_{L}
\rightarrow \nu_{L}}
(\equiv \stackrel{03}{[-i]}\stackrel{12}{[+]} \stackrel{56}{(+)}
 \stackrel{7 8}{(-)}\stackrel{9\,10}{[+]} \stackrel{11\,12}{[+]}
\stackrel{13\,14}{[+]})\,*_A\, e^{\dagger}_{L\, 29^{th}},
(\equiv \stackrel{03}{[-i]}\stackrel{12}{[+]} 
\stackrel{56}{(-)}\stackrel{78}{(+)}\stackrel{9\,10}{(+)} 
\stackrel{11\,12}{(+)} \stackrel{13\,14}{(+)})\,\rightarrow 
\nonumber\\
&&\nu^{\dagger}_{L\, 31^{st}},
(\equiv \stackrel{03}{[-i]}\stackrel{12}{[+]} 
\stackrel{56}{[+]}\stackrel{78}{[-]}\stackrel{9\,10}{(+)} 
\stackrel{11\,12}{(+)} \stackrel{13\,14}{(+)})\,, 
\quad \quad \quad {}^{I}{\hat{\cal A}}^{ \dagger}_{w1\, e_{L}
\rightarrow \nu_{L}}= \nu^{\dagger}_{L\, 31^{st}}\, *_A\,
(e^{\dagger}_{L\, 29^{th}})^{\dagger}\,,\nonumber\\
&&{}^{I}{\hat{\cal A}}^{ \dagger}_{w1 \,\nu _{L}
\rightarrow e_{L}}
(\equiv \stackrel{03}{[-i]}\stackrel{12}{[+]} \stackrel{56}{(-)}
 \stackrel{7 8}{(+)}\stackrel{9\,10}{[+]} \stackrel{11\,12}{[+]}
\stackrel{13\,14}{[+]})\,*_A\, \nu^{\dagger}_{L\,31^{st}},
(\equiv \stackrel{03}{[-i]}\stackrel{12}{[+]} 
\stackrel{56}{[+]}\stackrel{78}{[-]}\stackrel{9\,10}{(+)} 
\stackrel{11\,12}{(+)} \stackrel{13\,14}{(+)})\,\rightarrow 
\nonumber\\
&&e^{\dagger}_{L\, 29^{th}},
(\equiv \stackrel{03}{[-i]}\stackrel{12}{[+]} 
\stackrel{56}{(-)}\stackrel{78}{(+)}\stackrel{9\,10}{(+)} 
\stackrel{11\,12}{(+)} \stackrel{13\,14}{(+)})\,, 
\quad \quad \quad {}^{I}{\hat{\cal A}}^{ \dagger}_{w1\, \nu _{L}
\rightarrow e_{L}}= e^{\dagger}_{L\, 29^{th}}\, *_A\,
(\nu^{\dagger}_{L\,31^{st}})^{\dagger}\,.
\end{eqnarray}
\end{small}
We can find as well the weak bosons ``basis vectors'' which
make all other transformations, as those which have not
been observed yet, at least not at low energies.

As an example, we discuss in this part the two kinds of ``basis
vectors'' of weak bosons which transform the right-handed
quarks and leptons and left-handed quarks and leptons within
the corresponding weak doublets.
It is a {\it break of symmetries} which
does not allow transitions in which, for example, a left-handed lepton transforms to a right-handed one, or a left-handed lepton transforms to a right-handed quark.\\
\vspace{2mm}

\vspace{2mm}

{\bf d.} $\;\;$ 
Let us look  for the `basis vectors'' of gluons, transforming $u^{c1 \dagger}_{R}$, 
$1^{th}$ line, to the  $u^{c2 \dagger}_{R}$, $9^{th}$ line and back,  as presented 
in Table~\ref{Table so13+1.}.
\begin{small}
\begin{eqnarray}
\label{guc1c2R}
&&{}^{I}{\hat{\cal A}}^{ \dagger}_{gl\, u^{c1}_{R}
\rightarrow u^{c2}_{R}}
(\equiv \stackrel{03}{[+i]}\stackrel{12}{[+]} \stackrel{56}{[+]}
 \stackrel{7 8}{[+]}\stackrel{9\,10}{(-)} \stackrel{11\,12}{(+)}
\stackrel{13\,14}{[-]})\,*_A\, u^{c1 \dagger}_{R\, 1^{st}},
(\equiv \stackrel{03}{(+i)}\stackrel{12}{[+]} 
\stackrel{56}{[+]}\stackrel{78}{(+)}\stackrel{9\,10}{(+)} 
\stackrel{11\,12}{[-]} \stackrel{13\,14}{[-]})\,\rightarrow 
\nonumber\\
&&u^{c2 \dagger}_{R\, 9^{th}},
(\equiv \stackrel{03}{(+i)}\stackrel{12}{[+]} 
\stackrel{56}{[+]}\stackrel{78}{(+)}\stackrel{9\,10}{[-]} 
\stackrel{11\,12}{(+)} \stackrel{13\,14}{[-]})\,, 
\quad \quad \quad {}^{I}{\hat{\cal A}}^{ \dagger}_{gl\,u^{c1}_{R}
\rightarrow u^{c2}_{R}}= u^{c2 \dagger}_{R\, 9^{th}}\, *_A\,
(u^{c1 \dagger}_{R\, 1^{st}})^{\dagger}\,,\nonumber\\
&&{}^{I}{\hat{\cal A}}^{ \dagger}_{gl\, u^{c2}_{R}
\rightarrow u^{c1}_{R}}
(\equiv \stackrel{03}{[+i]}\stackrel{12}{[+]} \stackrel{56}{[+]}
 \stackrel{7 8}{[+]}\stackrel{9\,10}{(+)} \stackrel{11\,12}{(-)}
\stackrel{13\,14}{[-]})\,*_A\, u^{c2 \dagger}_{R\, 9^{th}}
(\equiv \stackrel{03}{(+i)}\stackrel{12}{[+]} 
\stackrel{56}{[+]}\stackrel{78}{(+)}\stackrel{9\,10}{[-]} 
\stackrel{11\,12}{(+)} \stackrel{13\,14}{[-]})\,\rightarrow 
\nonumber\\
&&u^{c1 \dagger}_{R\, 1^{st}},
(\equiv \stackrel{03}{(+i)}\stackrel{12}{[+]} 
\stackrel{56}{[+]}\stackrel{78}{(+)}\stackrel{9\,10}{(+)} 
\stackrel{11\,12}{[-]} \stackrel{13\,14}{[-]})\,, 
\quad \quad \quad {}^{I}{\hat{\cal A}}^{ \dagger}_{gl\, u^{c2}_{R}
\rightarrow u^{c1}_{R}}= u^{c1 \dagger}_{R\, 1^{st}}\, *_A\,
(u^{c2 \dagger}_{R\, 9^{th}})^{\dagger}\,.
\end{eqnarray}
\end{small}
%
The two ``basis vectors'', ${}^{I}{\hat{\cal A}}^{ \dagger}_{gl \,u^{c1}_{R}
\rightarrow u^{c2}_{R}}$ and ${}^{I}{\hat{\cal A}}^{ \dagger}_{gl\, u^{c2}_{R}
\rightarrow u^{c1}_{R}}$ are Hermitian conjugated to each other, as 
the two expressions, ${}^{I}{\hat{\cal A}}^{ \dagger}_{gl \,u^{c1}_{R}
\rightarrow u^{c2}_{R}}= u^{c2 \dagger}_{R\, 9^{th}}\, *_A\,
(u^{c1 \dagger}_{R\, 1^{st}})^{\dagger}$ and 
${}^{I}{\hat{\cal A}}^{ \dagger}_{gl \,u^{c2}_{R}
\rightarrow u^{c1}_{R}}= u^{c1 \dagger}_{R\, 1^{st}}\, *_A\,
(u^{c2 \dagger}_{R\, 9^{th}})^{\dagger}$, demonstrate
~\footnote{
Let us look, as an exercise, for ``basis vectors'' of gluons, 
${}^{I}{\hat{\cal A}}^{ \dagger}_{gl\, CPu^{c1}_{L}\rightarrow 
CPu^{c2}_{L}}$, transforming the antiparticle of $u^{c1}_{L}$ (appearing
in Table~\ref{Table so13+1.} in $7^{th}$ line), to the antiparticle of 
$u^{c2}_{L}$ (appearing in $15^{th}$ line in Table~\ref{Table so13+1.}).

Their antiparticles appear in $39^{th} line$ and $47^{th} line$, respectively, in 
Table~\ref{Table so13+1.}.
\begin{small}
\begin{eqnarray}
\label{gCPuc1CPuc2L}
&&{}^{I}{\hat{\cal A}}^{ \dagger}_{gl\, CPu^{c1}_{R}
\rightarrow CPu^{c2}_{R}}
(\equiv \stackrel{03}{[+i]}\stackrel{12}{[+]} \stackrel{56}{[-]}
 \stackrel{7 8}{[+]}\stackrel{9\,10}{(+)} \stackrel{11\,12}{(-)}
\stackrel{13\,14}{[+]})\,*_A\, \bar{u}^{\bar{c1} \dagger}_{L\, 39^{th}},
(\equiv \stackrel{03}{(+i)}\stackrel{12}{[+]} 
\stackrel{56}{(-)}\stackrel{78}{(+)}\stackrel{9\,10}{[-]} 
\stackrel{11\,12}{(+)} \stackrel{13\,14}{(+)})\,\rightarrow 
\nonumber\\
&& \bar{u}^{\bar{c2}\dagger}_{L\, 47^{th}},
(\equiv \stackrel{03}{(+i)}\stackrel{12}{[+]} 
\stackrel{56}{(-)}\stackrel{78}{(+)}\stackrel{9\,10}{(+)} 
\stackrel{11\,12}{[-]} \stackrel{13\,14}{(+)}\,, 
\quad \quad \quad {}^{I}{\hat{\cal A}}^{ \dagger}_{gl\, CPu^{c1}_{l}
\rightarrow CPu^{c2}_{l}}=  \bar{u}^{\bar{c2}\dagger}_{L\, 47^{th}}\, *_A\,
( \bar{u}^{\bar{c1} \dagger}_{L\, 39^{th}})^{\dagger}\,.
\end{eqnarray}
\end{small}
}.


%
\subsection{Scalar fields in $(13+1)$-dimensional space from the point of 
view of $d=(3+1)$}
\label{scalarfields}

There is no difference in the internal space of bosons (that is, in the 
even ``basis vectors'') if they carry concerning the ordinary space index
$\alpha=\mu =(0,1,2,3)$ or $\alpha=(5,6,7,8,...13,14)$: \\
Vectors in $d=(3+1)$ carry the ordinary space index $\mu=(0,1,2,3)$.
They represent, after the breaks of symmetry and before the electroweak
break, in $d=(3+1)$ photons ($U(1)$), weak bosons ($SU (2)_{1}$), and
gluons ($SU(3)$, embedded in ($SO(6)$)~\footnote{In
Ref.~\cite{nh2021RPPNP} (and references therein), Sect.~6, properties of
the vector and scalar gauge fields, presented with the gauge fields
$\omega_{ab \alpha}$ ($\alpha=(0,1,2,3)$ for vectors) and 
$\tilde{\omega}_{ab \alpha}$ ($\alpha=(5,6,7,8,...13,14)$ for scalars) 
in the {\it spin-charge-family} theory are discussed.%
}. \\

Scalars have in $d=(3+1)$ the space index $\alpha=(5,6,7,8,...13,14)$. 
Weak $SU (2)_{II}$ scalars, representing Higgs's scalars, are doublets 
concerning $\alpha=(7,8)$ (Ref.~\cite{nh2021RPPNP} Table 8 and 
Eq.~(110)) and are superposition of several fields with 
different ``basis vectors''~\footnote{In
Ref.~\cite{nh2021RPPNP} (and references therein), Sect.~6, properties of
the scalar gauge fields, presented with the gauge fields
$\omega_{ab \alpha}$ and $\tilde{\omega}_{ab \alpha}$ in the
{\it spin-charge-family} theory, are discussed, Eqs.~(108-112) and Table 8.%
}. \\
The {\it spin-charge-family} theory also predicts additional scalar fields,
which are triplets and antitriplets concerning the space index~\footnote{
Refs.~(\cite{n2014matterantimatter,nh2021RPPNP} (and references therein), Sect.~6, discusses triplet
and antitriplet scalar fields, responsible for transitions from antileptons
and antiquarks into quarks and leptons, Eqs.~(113-114) and Table 9, Fig.~1.
}.\\

In this article, the internal spaces of all the boson gauge fields, vectors, and
scalar ones are discussed in order to try to understand the second
quantized fermion and boson fields in this my new way. In Ref.~\cite{n2023NPB} it is
discussed how the
$\omega_{ab \alpha}$ and $\tilde{\omega}_{ab \alpha}$, used in the
{\it spin-charge-family} theory to describe the vector and scalar gauge
fields be replaced by
${\bf {}^{i}{\hat{\cal A}}^{m \dagger}_{f \alpha}} (\vec{p}), i=(I,II)$,
presented in Eqs.~(33-35). 

\vspace{2mm}

%
\subsection{``Gravitons'' in $(13+1)$-dimensional space from the point of 
view of $d=(3+1)$ }
\label{gravitons}

The assumption that the internal spaces of fermion and boson fields are
describable by the  odd and  even ``basis vectors",
respectively, leads to the conclusion that the internal space of gravitons
must also be described by even ``basis vectors'' (which are 
the superposition of even products of $\gamma^a$'s, arranged to be 
eigenvectors of all the Cartan subalgebra members and are arranged correspondingly to be an even number of nilpotents and the rest of 
projectors, Eqs.~(\ref{nilproj},\ref{signature0}).

Let us, therefore, try to describe ``gravitons'' in an equivalent way as we
do for the so far observed gauge fields that is, in terms of
${}^{i}{\hat{\cal A}}^{m \dagger}_{f},\, i=I,II$, starting with $i=I$.

We expect correspondingly that ``gravitons'' have no weak and no
colour ``charges'', as also photons do not have any charge from the
point of view of $d=(3+1)$. However, ``gravitons'' can have the spin
and handedness (non-zero ${\cal S}^{03}$ and $ {\cal S}^{12}$) in
$d=(3+1)$.
Keeping in mind that bosons must have an even number of nilpotents,
and taking into account Table~\ref{Table so13+1.} in which
$u^{c1 \dagger}_{R}$ appears in the first line of the table, we find
that the ``basis vector'' of the ``graviton''
${}^{I}{\hat{\cal A}}^{ \dagger}_{gr\, u^{c1 \dagger}_{R}
\rightarrow u^{c1 \dagger}_{R}}$ applies on $u^{c1 \dagger}_{R}$
as follows
\begin{small}
\begin{eqnarray}
\label{grAqeqe}
&&{}^{I}{\hat{\cal A}}^{ \dagger}_{gr \,u^{c1 \dagger}_{R, 1^{st}}
 \rightarrow u^{c1 \dagger}_{R, 2^{nd}}}
(\equiv \stackrel{03}{(-i)}\stackrel{12}{(-)} \stackrel{56}{[+]}
 \stackrel{7 8}{[+]}\stackrel{9\,10}{[+]} \stackrel{11\,12}{[-]}
\stackrel{13\,14}{[-]})\,*_A\, u^{c1 \dagger}_{R\, 1^{st}} 
(\equiv \stackrel{03}{(+i)}\stackrel{12}{[+]} 
\stackrel{56}{[+]}\stackrel{78}{(+)}\stackrel{9\,10}{(+)} 
\stackrel{11\,12}{[-]} \stackrel{13\,14}{[-]})\,\rightarrow 
\nonumber\\
&&u^{c1 \dagger}_{R\, 2^{nd}},
(\equiv \stackrel{03}{[-i]}\stackrel{12}{(-)} 
\stackrel{56}{[+]}\stackrel{78}{(+)}\stackrel{9\,10}{(+)} 
\stackrel{11\,12}{[-]} \stackrel{13\,14}{[-]})\,,
 \quad \quad \quad
{}^{I}{\hat{\cal A}}^{ \dagger}_{gr\, u^{c1 \dagger}_{R, 1^{st}}
 \rightarrow u^{c1 \dagger}_{R, 2^{nd}}} = u^{c1 \dagger}_{R\, 2^{nd}} 
\,*_A\, (u^{c1 \dagger}_{R\, 1^{st}})^{\dagger}\,,
\nonumber\\
&&{}^{I}{\hat{\cal A}}^{ \dagger}_{gr \,\bar{u}^{\bar{c1} \dagger}_{R}
 \rightarrow\bar{u}^{\bar{c1} \dagger}_{R}}
(\equiv \stackrel{03}{(+i)}\stackrel{12}{(-)} \stackrel{56}{[-]}
 \stackrel{7 8}{[-]}\stackrel{9\,10}{[-]} \stackrel{11\,12}{[+]}
\stackrel{13\,14}{[+]})\,*_A\, \bar{u}^{\bar{c1}\dagger}_{L\, 35^{th}} 
(\equiv \stackrel{03}{[-i]}\stackrel{12}{[+]} 
\stackrel{56}{(-)}\stackrel{78}{[-]}\stackrel{9\,10}{[-]} 
\stackrel{11\,12}{(+)} \stackrel{13\,14}{(+)})\,\rightarrow 
\nonumber\\
&&\bar{u}^{\bar{c1}\dagger}_{L\, 36^{th}}
(\equiv \stackrel{03}{(+i)}\stackrel{12}{(-)} 
\stackrel{56}{(-)}\stackrel{78}{[-]}\stackrel{9\,10}{[-]} 
\stackrel{11\,12}{(+)} \stackrel{13\,14}{(+)})\,, 
 \quad \quad \quad
 {}^{I}{\hat{\cal A}}^{\dagger}_{gr\, \bar{u}^{\bar{c1} \dagger}_{R}
 \rightarrow\bar{u}^{\bar{c1} \dagger}_{R}}= 
 {u}^{\bar{c1}\dagger}_{L\, 36^{th}}\,*_A\,
 ({u}^{\bar{c1}\dagger}_{L\, 35^{th}})^{\dagger}\,.
\end{eqnarray}
\end{small}
As all the boson gauge fields, manifesting in $d=(3+1)$ as the vector
gauge fields of the corresponding quarks and leptons and antiquarks and
antileptons, also the creation operators for ``gravitons'' must carry the
space index $\alpha$, describing the $\alpha$ component of the creation
operators for ``gravitons'' in the ordinary space,
Eq.~(\ref{wholespacebosons}). Since we pay attention to the vector
gauge fields in $d=(3+1)$ $\alpha$ must be
$\mu=(0,1,2,3)$.

The representative of a ``graviton'', its creation operator indeed,
manifesting in $d=(3+1)$ must correspondingly be of the kind
\begin{eqnarray}
\label{grmu}
{}^{I}{\bf {\hat{\cal A}}}^{\dagger}_{gr\, \mu}=
\stackrel{03}{(\pm i)} \stackrel{12}{(\pm)}
\stackrel{56}{[\pm]} \dots
\stackrel{11\,12}{[\pm]}\stackrel{13\,14}{[\pm]} {}^{I}{\cal C}_{ gr\,\mu}\,,
\end{eqnarray}
with the ``basis vectors'' of the kind as we see in Eq.~(\ref{grAqeqe}),
or rather, the superposition of
${}^{I}{\bf {\hat{\cal A}}}^{\dagger}_{gr\, \mu}$.\\

Since the only nilpotents are the first two factors, with eigenvalues of
$S^{03}$ and $S^{12}$ equal to $\pm i$ and $\pm 1$, respectively,
while all the rest factors are projectors with the corresponding aigenvalues of the cartan subalgebra members ${\cal S}^{ab}=0$,
the ``basis vectors'' of ``gravitons'' can offer to fermions on which they
apply the integer spin ${\cal S}^{12}=\pm 1$ and ${\cal S}^{03}=\pm i$.
The product of two  ``basis vectors'' of ``gravitons'', Eq.~(\ref{grmu}), 
can lead to the ``basis vector'' with only projectors.
 
The case of projectors only, we observe also in the case that
the space has $d=(5+1)$, Subsect.~\ref{bosons5+1},
Eq.~(\ref{grgrgr}): The product of two ``basis vectors'' of ``gravitons''
leads to the object being indistinguishable from the ``basis vectors'' of
photons.
``Gravitons'' ${}^{I}{\bf {\hat{\cal A}}}^{\dagger}_{gr\, \mu}$, like
photons, weak bosons and gluons offer fermions also the momentum
in ordinary space.
\vspace{1mm}

Let us discuss the case  $d=(13+1)$ looking for 
${}^{I}{\hat{\cal A}}^{ \dagger}_{gr\, u^{c1}_{L, 7^{th}}\rightarrow  
u^{c1}_{L, 8^{th}}}$,
which transforms the quark $ u^{c1 \dagger}_{L}$, presented in 
Table~\ref{Table so13+1.} on the $7^{th}$ line to the quark 
$ u^{c1 \dagger}_{L}$, presented on the $8^{th}$ line, or looking for 
${}^{I}{\hat{\cal A}}^{ \dagger}_{gr\, e_{L,29^{th}}\rightarrow  
e_{L, 30^{th}}}$, which transforms the electron $ e^{\dagger}_{L}$, 
presented in Table~\ref{Table so13+1.} on the $29^{th}$ line to the 
electron $ e^{\dagger}_{L}$, presented on the $30^{th}$ line
\begin{small}
\begin{eqnarray}
\label{grue}
{}^{I}{\hat{\cal A}}^{ \dagger}_{gr\, u^{c1}_{L, 7^{th}}\rightarrow  
u^{c1}_{L, 8^{th}}}(\equiv \stackrel{03}{(+i)} \stackrel{12}{(-)}
\stackrel{56}{[+]}\stackrel{78}{[-]}\stackrel{9\,10}{[+]}
\stackrel{11\,12}{[-]}\stackrel{13\,14}{[-]}) &*_A& 
u^{c1 \dagger}_{L,7^{th}} (\equiv \stackrel{03}{[-i]} \stackrel{12}{[+]}
\stackrel{56}{[+]}\stackrel{78}{[-]}\stackrel{9\,10}{(+)}\stackrel{11\,12}{[-]}\stackrel{13\,14}{[-]})\rightarrow
\nonumber\\
&&u^{c1 \dagger}_{L,8^{th}}
(\equiv \stackrel{03}{(+i)} \stackrel{12}{(-)}\stackrel{56}{[+]}\stackrel{78}{[-]}\stackrel{9\,10}{(+)}\stackrel{11\,12}{[-]}\stackrel{13\,14}{[-]})\,,\nonumber\\
{}^{I}{\hat{\cal A}}^{ \dagger}_{gr\, e_{L,29^{th}}\rightarrow  
e_{L, 30^{th}}} (\equiv \stackrel{03}{(+i)} \stackrel{12}{(-)}
\stackrel{56}{[-]}\stackrel{78}{[+]}\stackrel{9\,10}{[+]}
\stackrel{11\,12}{[+]}\stackrel{13\,14}{[+]}) &*_A& e^{\dagger}_{L,29^{th}}
(\equiv \stackrel{03}{[-i]} \stackrel{12}{[+]}\stackrel{56}{(-)}\stackrel{78}{(+)}\stackrel{9\,10}{(+)}\stackrel{11\,12}{(+)}\stackrel{13\,14}{(+)})\rightarrow
\nonumber\\
&&e^{\dagger}_{L,30^{th}}(\equiv \stackrel{03}{(+i)} \stackrel{12}{(-)}
\stackrel{56}{(-)}\stackrel{78}{(+)}\stackrel{9\,10}{(+)}\stackrel{11\,12}{(+)}\stackrel{13\,14}{(+)})\,.
\end{eqnarray}
\end{small}
The boson ``basis vectors'' of the first group transform fermion ``basis vectors'' to members of the same family.

The boson ``basis vectors'' of the second kind transform a member of 
a family to the same member of another family.

One finds in the case  $d=(13+1)$  for ${}^{II}{\hat{\cal A}}^{ \dagger}_{gr\, u^{c1}_{L, 7^{th}f=1}}$ that it causes
the transition of $u^{c1}_{L, 7^{th}f=1}$ appearing in  
Table~\ref{Table so13+1.} on the $7^{th}$ line  to the same member of 
another family $u^{c1}_{L, 7^{th}f\ne1}$, and similarly for $e_{L, 29^{th}f=1}$
\begin{small}
\begin{eqnarray}
\label{grueII}
&& u^{c1 \dagger}_{L,7^{th}} (\equiv \stackrel{03}{[-i]} \stackrel{12}{[+]}
\stackrel{56}{[+]}\stackrel{78}{[-]}\stackrel{9\,10}{(+)}\stackrel{11\,12}{[-]}\stackrel{13\,14}{[-]})\,*_A\,
{}^{II}{\hat{\cal A}}^{ \dagger}_{gr\, u^{c1}_{L, 7^{th}f=1}\rightarrow  
u^{c1}_{L, 7^{th}f\ne1}}(\equiv \stackrel{03}{(-i)} \stackrel{12}{(+)}
\stackrel{56}{[+]}\stackrel{78}{[-]}\stackrel{9\,10}{[-]}
\stackrel{11\,12}{[-]}\stackrel{13\,14}{[-]}) \,\rightarrow \,\nonumber\\
&&u^{c1 \dagger}_{L,7^{th}} (\equiv \stackrel{03}{(-i)} \stackrel{12}{(+)}
\stackrel{56}{[+]}\stackrel{78}{[-]}\stackrel{9\,10}{(+)}\stackrel{11\,12}{[-]}\stackrel{13\,14}{[-]}) \,,\nonumber\\
&& e_{L, 29^{th}f=1} (\equiv \stackrel{03}{[-i]} \stackrel{12}{[+]}
 \stackrel{56}{(-)}\stackrel{78}{(+)}\stackrel{9\,10}{(+)}
 \stackrel{11\,12}{(+)} \stackrel{13\,14}{(+)})\,*_A\,
 {}^{II}{\hat{\cal A}}^{ \dagger}_{gr\, e_{L,29^{th}}\rightarrow  
 e_{L, 29^{th}}} (\equiv \stackrel{03}{(-i)} \stackrel{12}{(+)}
\stackrel{56}{[+]}\stackrel{78}{[-]}\stackrel{9\,10}{[-]}
 \stackrel{11\,12}{[-]} \stackrel{13\,14}{[-]})\rightarrow \nonumber\\
&&e^{\dagger}_{L,29^{th}f\ne1}(\equiv \stackrel{03}{(-i)} \stackrel{12}{(+)}
 \stackrel{56}{(-)}\stackrel{78}{(+)}\stackrel{9\,10}{(+)}
 \stackrel{11\,12}{(+)} \stackrel{13\,14}{(+)})\,.
\end{eqnarray}
\end{small}
%


\subsection{Feynman diagrams for scattering fermions and bosons}
\label{feynmandiagrams}

In this paper, we treat massless fermion and boson fields. We do
not discuss  the breaks of symmetries, bringing masses to all
fermions and bosons except to gravitons, photons and gluons; also
the coupling constants are not discussed;  we pay attention
primarily to internal spaces of fields (in Ref.~(\cite{nh2021RPPNP}
and the references therein) the  needed breaks of symmetry are
discussed).

We assume that the scattering objects have non-zero starting
momentum (only) in ordinary $(3+1)$ space, the sum of which is
conserved.

\vspace{2mm}

Let be pointed out that since the proposed theory differs from the 
Dirac's second quantization procedure for fermions and 
bosons~\footnote{
The anti-commuting ``basis vectors'' describing fermions are 
superposition of odd products of $\gamma^{a}$ while the
commuting ``basis vectors'' describing bosons are 
superposition of odd products of $\gamma^{a}$, explaining 
correspondingly the Dirac's second quantization postulates.
} 
({\,\bf i.\,\,} the anti-commuting``basis vectors'' describing internal
spaces of fermions appear in $2^{\frac{d}{2}-1}$ families, 
{\,\bf ii.\,\,} {\it each family}, having $2^{\frac{d}{2}-1}$  members, 
{ \it includes ``basis vectors'' of fermions and anti-fermions}, 
{\,\bf iii.\,\,} the Hermitian conjugated ``basis vectors'' of fermions 
appear in a separate group, {\,\bf iv.\,\,} the commuting ``basis 
vectors'' describing internal 
spaces of boson fields appear in two orthogonal groups, having 
their Hermitian conjugated partners within each group, {\,\bf v.\,\,}
one group causes transformations among family members (the same
``basis vectors'' cause transformations among members of all the
families), {\,\bf vi.\,\,} the second group causes transformations of 
a particular member among all families (the same ``basis vectors'' 
cause transformations of any family member), {\,\bf vii.\,\,} ``basis 
vectors'' describing internal spaces of boson fields can be described
as algebraic products of ``basis vectors'' of fermions and of the 
Hermitian conjugated partners\,)   the Feynman diagrams in this
theory differ from the Feynman diagrams as presented by Feynman.
\\

We study in this Subsect.~\ref{feynmandiagrams} a few examples: 
{\bf a.\,\,\,}electron - positron annihilation, {\bf b.\,\,\,} electron - 
positron scattering, assuming first that the internal space of 
fermions and bosons concerns $d=(5+1)$ and also that the 
internal space of fermions and bosons concerns $d=(13+1)$, while 
all the fields have non zero momenta only in ordinary $(3+1)$ space.

\vspace{2mm}

{\bf a.i.\,\,\,} {\it electron - positron annihilation in the case when 
internal space of fermions and bosons is $d=(5+1)$}: In 
Table~\ref{Table Clifffourplet.} there are four families 
(four irreducible representations of $S^{ab}$), each family contains 
four ``basis vectors'' (reachable by $S^{ab}$) describing the internal 
space of fermions, one with spin up and one with spin down, and of 
antifermions, again one with spin up and one with spin down. Let 
$\hat{b}^{1 \dagger}_{1}$ be called ``electron'' with spin up (it has 
$S^{56}=\frac{1}{2}$) and let $\hat{b}^{3 \dagger}_{1}$ be 
``positron'' with spin up (it has $S^{56}= - \frac{1}{2}$). 
$\hat{b}^{1 \dagger}_{1}$ and $\hat{b}^{3 \dagger}_{1}$  are 
related by charge conjugation, $\gamma^0 \gamma^5$, 
Eqs.~(\ref{CPTNlowE}, \ref{05bf1}).~\footnote{
The boson's ``basis vector'' ${}^{I} \hat{{\cal A}}^{3 \dagger}_{3}$
does the same as $\gamma^0 \gamma^5$; it transforms $\hat{b}^{1 \dagger}_{1}$ into $\hat{b}^{3 \dagger}_{1}$:
${}^{I} \hat{{\cal A}}^{3 \dagger}_{3} 
(\equiv \stackrel{03}{(- i)} \stackrel{12}{[+]} \stackrel{56}{(-)}) 
\,*_A\,\hat{b}^{1 \dagger}_{1} (\equiv \stackrel{03}{(+ i)} 
\stackrel{12}{[+]} \stackrel{56}{[+]})\rightarrow \hat{b}^{3 \dagger}_{1}
(\equiv \stackrel{03}{[-i])} \stackrel{12}{[+]} \stackrel{56}{(- )})$, 
and ${}^{I} \hat{{\cal A}}^{1 \dagger}_{2}$ transforms 
$\hat{b}^{3 \dagger}_{1}$ into $\hat{b}^{1 \dagger}_{1}$: 
${}^{I} \hat{{\cal A}}^{1 \dagger}_{2} 
(\equiv \stackrel{03}{(+ i)} \stackrel{12}{[+]} \stackrel{56}{(+)}) 
\,*_A\,\hat{b}^{3 \dagger}_{1} (\equiv  \stackrel{03}{[-i])} 
\stackrel{12}{[+]} \stackrel{56}{(- )})\rightarrow \hat{b}^{1 \dagger}_{1}
(\equiv \stackrel{03}{(+ i)} \stackrel{12}{[+]}  \stackrel{56}{[+]})$.
}

Let us present useful relations, from  
Table~\ref{S120Cliff basis5+1even I.} (in the fifth and fourth line), 
and in Table~\ref{S120Cliff basis5+1even II.} (in the fifth line) ~\footnote{
Let me remind the reader:
Taking into account Eq.~(\ref{usefulrel0});\;
$ \stackrel{ab}{(k)}\stackrel{ab}{(-k)} = \eta^{aa} \stackrel{ab}{[k]}\,,
\,\stackrel{ab}{(-k)}\stackrel{ab}{(k)} = \eta^{aa} \stackrel{ab}{[-k]}\,,
\,\stackrel{ab}{(k)}\stackrel{ab}{[k]} =0\,,\,\stackrel{ab}{(k)}\stackrel{ab}{[-k]} =
\stackrel{ab}{(k)}\,, \,\stackrel{ab}{(-k)}\stackrel{ab}{[k]} =
\stackrel{ab}{(-k)}\,, \,\stackrel{ab}{[k]}\stackrel{ab}{(k)}= \stackrel{ab}{(k)}\,,
\,\stackrel{ab}{[k]}\stackrel{ab}{(-k)} =0\,, \,
\stackrel{ab}{[k]}\stackrel{ab}{[-k]} =0$, one easily checks the relations in
Eq.~(\ref{epscattering}).
}.
\begin{eqnarray}
\label{epscattering}
\hat{b}^{1 \dagger}_{1}\,*_A\,(\hat{b}^{1 \dagger}_{1})^{\dagger}
={}^{I} \hat{{\cal A}}^{1 \dagger}_{3} (\equiv \stackrel{03}{[+ i]} 
\stackrel{12}{[+]} \stackrel{56}{[+]})\,,\quad 
\hat{b}^{3 \dagger}_{1}\,*_A\,(\hat{b}^{3 \dagger}_{1})^{\dagger}
={}^{I} \hat{{\cal A}}^{3 \dagger}_{2} (\equiv \stackrel{03}{[- i]} 
\stackrel{12}{[+]} \stackrel{56}{[-]})\,,\nonumber\\
(\hat{b}^{1 \dagger}_{1})^{\dagger}\,*_A\,\hat{b}^{1 \dagger}_{1}
={}^{II} \hat{{\cal A}}^{1 \dagger}_{3} (\equiv \stackrel{03}{[- i]} 
\stackrel{12}{[+]} \stackrel{56}{[+]})\,,\quad 
(\hat{b}^{3 \dagger}_{1})^{\dagger}\,*_A\,\hat{b}^{3 \dagger}_{1}
={}^{II} \hat{{\cal A}}^{1 \dagger}_{3} (\equiv \stackrel{03}{[- i]} 
\stackrel{12}{[+]} \stackrel{56}{[+]})\,.
\end{eqnarray}
Let us point out the last relations in Eq.~(\ref{epscattering}),  saying that
$(\hat{b}^{1 \dagger}_{1})^{\dagger}\,*_A\,\hat{b}^{1 \dagger}_{1}=$
${}^{II} \hat{{\cal A}}^{1 \dagger}_{3}=
(\hat{b}^{3 \dagger}_{1})^{\dagger}\,*_A\,\hat{b}^{3 \dagger}_{1}$. 
\\

Let ``electron'', $\hat{b}^{1 \dagger}_{1}$, and 
``positron'', $\hat{b}^{3 \dagger}_{1}$, carry the external 
space momenta  $\vec{p}_{1}$ and $\vec{p}_{2}$, respectively. 

Scattering of an ``electron'', $e^{-}$, and ``positron'', $e^{+}$, into two 
``photons'' is studied for the case that the internal space space has $(5+1)$ dimensions while all the fields have non zero momentum in $(3+1)$; 
``Electron'' $e^{-}$,  with the internal space presented by $\hat{b}^{1 \dagger}_{1}$, and 
``positron'' $e^{+}$, presented by $\hat{b}^{3 \dagger}_{1}$, exchange 
the ``photon'' ${}^{II} \hat{{\cal A}}^{1 \dagger}_{3} 
(\equiv \stackrel{03}{[- i]} \stackrel{12}{[+]} \stackrel{56}{[+]})=
(\hat{b}^{1 \dagger}_{1})^{\dagger}\,*_A\,\hat{b}^{1 \dagger}_{1}
=(\hat{b}^{3 \dagger}_{1})^{\dagger}\,*_A\,\hat{b}^{3 \dagger}_{1}$.\\
``Electron'', $\hat{b}^{1 \dagger}_{1}$, takes from
 ${}^{II} \hat{{\cal A}}^{1 \dagger}_{3}$ the part 
$(\hat{b}^{1 \dagger}_{1})^{\dagger}$ generating the ``photon'' 
${}^{I} \hat{{\cal A}}^{1 \dagger}_{3} (\equiv \stackrel{03}{[+ i]} 
\stackrel{12}{[+]} \stackrel{56}{[+]})=$\\ $\hat{b}^{1 \dagger}_{1}
(\equiv \stackrel{03}{(+i)}\stackrel{12}{[+]} \stackrel{56}{[+]})*_A
(\hat{b}^{1 \dagger}_{1})^{\dagger}(\equiv \stackrel{03}{(-i)}
\stackrel{12}{[+]} \stackrel{56}{[+]}) $ and transfers to it  its 
momentum $\vec{p}_{1}$,
while the rest, the ``basis vector'' $\hat{b}^{1 \dagger}_{1}$, remains 
as the ``basis vector'', without momentum in ordinary space.\\
 ``Positron'', $\hat{b}^{3 \dagger}_{1}$, takes from
 ${}^{II} \hat{{\cal A}}^{1 \dagger}_{3}$ the part 
$(\hat{b}^{3 \dagger}_{1})^{\dagger}$ generating  the ``photon'' 
${}^{I} \hat{{\cal A}}^{3 \dagger}_{2} (\equiv \stackrel{03}{[- i]} 
\stackrel{12}{[+]} \stackrel{56}{[-]})=$\\$\hat{b}^{3 \dagger}_{1}
(\equiv \stackrel{03}{[-i]}\stackrel{12}{[+]} \stackrel{56}{(-)})*_A
(\hat{b}^{3 \dagger}_{1})^{\dagger} (\equiv \stackrel{03}{[-i]}
\stackrel{12}{[+]} \stackrel{56}{(+)}) $ to which it transfer its 
momentum $\vec{p}_{2}$,
while the rest, the ``basis vector'' $\hat{b}^{3 \dagger}_{1}$, remains 
as the ``basis vector'', without momentum in ordinary space. 
(It is better to say that ``electron'' and ``positron'', with momentum 
$\vec{p}_{1}$ and $\vec{p}_{2}$, respectively, exchanging a ``photon''  
${}^{II} \hat{{\cal A}}^{1 \dagger}_{3}$, transfer the momentum 
$\vec{p}_{1}+\vec{p}_{2}$ to two photons
$ {}^{I} \hat{{\cal A}}^{1 \dagger}_{3}$ and 
${}^{I} \hat{{\cal A}}^{3 \dagger}_{2}$, the two ``basis vectors'' of the
``electron'', $\hat{b}^{1 \dagger}_{1}$, and ``positron '',  
$\hat{b}^{3 \dagger}_{1}$, remain without momenta in ordinary space.) 

The ordinary presentation of positron-electron annihilation describes the
positron line with an arrow down so that the electron line on the left
hand side in Fig.~\ref{Figep2gamma5+1}
replaces ${}^{II} \hat{{\cal A}}^{1 \dagger}_{3}$,
continuing on the right-hand side with the arrow down represented a positron.
The ordinary theory has no analogous boson fields to the one with 
the internal ``basis vectors'' 
${}^{II} \hat{{\cal A}}^{1 \dagger}_{3}$, and it has not fermions and
anti-fermions appearing in the same family.


``Photons'', having the internal quantum numbers equal to zero, do carry
the external index in $d=(3+1)$. The annihilation of  ``electron'', $e^{-}$, 
and ``positron'', $e^{+}$ is presented in Fig.~\ref{Figep2gamma5+1}.\\

\vspace{2mm}

\begin{figure}
  \centering
  \begin{tikzpicture}[>=triangle 45]
\draw [->,snake]
 (4,1.5) node {$\bullet$}-- (5.5,5)
node [anchor=west]
{$
\begin{smallmatrix} \\
 \\                                  
                                  \;\;\;{\rm photon} \;
                                   {}^I\hat{{\cal A}}^{3 \dagger}_{2}\\
                                   (\equiv \stackrel{03}{[-i]}\stackrel{12}{[+]} \stackrel{56}{[-]})
                                      \end{smallmatrix} $};
\draw [snake]
(1,1.5)  node {$\bullet$} -- (4,1.5) node {$\bullet$};) 
%
\draw [->,snake]
 (1,1.5) node {$\bullet$}-- (-0.5,5)
node [anchor=east]
{$
\begin{smallmatrix} \\
\\
                                       \;\;\;\;{\rm photon}\;
                                  {}^I\hat{{\cal A}}^{1 \dagger}_{3}\\
                                   (\equiv \stackrel{03}{[+i]}\stackrel{12}{[+]} \stackrel{56}{[+]})
                                      \end{smallmatrix} $}; 
  \draw [->]                                    
(6,-2.5) node {$$}-- (4,1.5)
node [anchor=west]
{$\;\;\;\;
\begin{smallmatrix} \\
\\
\\
\\
\\
\\
\\
\\                                 
                                      \;\;\;\;{\rm positron}
                                     \;   \hat{b}^{3 \dagger}_{1}\\
                                       \;\;\;\; (\equiv \stackrel{03}{[-i]}\stackrel{12}{[+]} \stackrel{56}{(-)})
                                      \end{smallmatrix} $}; 
   \draw [->]                                    
(-0.5,-2.5) node {$$}-- (1,1.5)
node [anchor=east]
{$
\begin{smallmatrix} \\
\\
\\
\\
\\
\\
\\
\\ 
                                          {\rm elektron}\;
                                        \hat{b}^{1 \dagger}_{1}\\
                                       (\equiv \stackrel{03}{(+i)}\stackrel{12}{[+]} \stackrel{56}{[+]})\;\;\;\;\;
                                      \end{smallmatrix} $}; 
  \draw [ ] 
(3.5,2.3) node {$$}-- (3.5,2.3)  
  node [anchor=east]
{$
\begin{smallmatrix} \\
                                  {\rm photon}\;
                                  {}^{II}\hat{{\cal A}}^{1 \dagger}_{3}\\
                                   (\equiv \stackrel{03}{[-i]}\stackrel{12}{[+]} \stackrel{56}{[+]})
                                      \end{smallmatrix} $}; 
%
\end{tikzpicture} 
  Scattering of an ``electron'', $e^{-}$, and ``positron'', $e^{+}$, into two 
``photons'' is studied for the case that the internal space space has $(5+1)$ dimensions while all the fields have nonzero momentum in $(3+1)$; 
  \caption{\label{Figep2gamma5+1} 
Scattering of an ``electron'', $e^{-}$, and ``positron'', $e^{+}$, 
into two ``photons'' is studied for the case that the internal space 
has $(5+1)$ dimensions, while all the fields have non zero 
momentum  only in $(3+1)$; the internal spaces of ``electron'',
``positron'' and ``photons'' are presented 
by ``basis vectors'' from Table~\ref{Table Clifffourplet.}, $e^{-}=
\hat{b}^{1 \dagger}_{1}$, $e^{+} =\hat{b}^{3 \dagger}_{1}$, 
${}^{II} \hat{{\cal A}}^{1 \dagger}_{3}$,   ${}^{I} 
\hat{{\cal A}}^{1 \dagger}_{3}$, and 
${}^{I} \hat{{\cal A}}^{3 \dagger}_{2}$ (boson fields carry the 
space index $\mu=(0,1,2,3)$). This simple  case with
$(5+1)$-dimensional internal space, is an illustrative introduction 
into the case that the internal space has $(13 +1)$ dimensions and the electron and positron are taken 
from Table~\ref{Table so13+1.}.
``Elektron'' $e^{-}$, represented by $\hat{b}^{1 \dagger}_{1}$ and the 
momentum $\vec{p}_{1}$ in ordinary space, and ``positron'' $e^{-}$, 
presented by $\hat{b}^{3 \dagger}_{1}$ and the momentum $\vec{p}_{2}$ 
in ordinary space, exchange ${}^{II} \hat{{\cal A}}^{1 \dagger}_{3} 
(\equiv \stackrel{03}{[- i]} \stackrel{12}{[+]} \stackrel{56}{[+]})=
(\hat{b}^{1 \dagger}_{1})^{\dagger}\,*_A\,\hat{b}^{1 \dagger}_{1}
=(\hat{b}^{3 \dagger}_{1})^{\dagger}\,*_A\,\hat{b}^{3 \dagger}_{1}$.
The ``electron'' generates from  $\hat{b}^{1 \dagger}_{1}\, *_{A}\,
(\hat{b}^{1 \dagger}_{1})^{\dagger}$ the ``photon'' ${}^{I} 
\hat{{\cal A}}^{1 \dagger}_{3} (\equiv \stackrel{03}{[+ i]} 
\stackrel{12}{[+]} \stackrel{56}{[+]})$, while the ``positron'' generates 
from  $\hat{b}^{3 \dagger}_{1}\, *_{A}\, 
(\hat{b}^{3 \dagger}_{1})^{\dagger}$ the ``photon'' 
${}^{I} \hat{{\cal A}}^{3 \dagger}_{2} 
(\equiv \stackrel{03}{[- i]} \stackrel{12}{[+]} \stackrel{56}{[-]})$; both
``photons'' take away the momenta $\vec{p}_3$ and $\vec{p}_4$, 
$\vec{p}_3 + \vec{p}_4= $ $\vec{p}_1 + \vec{p}_2$, leaving the ``basis vectors'' 
$\hat{b}^{1 \dagger}_{1}$ and $\hat{b}^{3 \dagger}_{1} $ (without 
momenta in ordinary space).
}.
\end{figure}

{\bf a.ii.\,\,\,} {\it Electron - positron annihilation in the case that the 
internal space has $d=(13+1)$ dimensions}: In Table~\ref{Table so13+1.}  one family of quarks and leptons and 
antiquarks and anti-leptons are presented. The ``basis vector'' of an electron
with spin up can be found in the $29^{th}$ line carrying the charge $Q=-1$, 
we name it $e^{ - \dagger}_{L}$, the ``basis vector'' of  its positron (its anti-particle) appears 
in the $63^{rd}$ line carrying $Q= +1$ (both belong to the same family, presented in Table~\ref{Table so13+1.}),  we call it 
$e^{+ \dagger}_{R}$~\footnote{
The electron and positron are related by the charge conjugation operator, 
Eqs.~(\ref{CPTNlowE}), or by the even ``basis vector'' 
${}^{I} \hat{{\cal A}}^{ \dagger}_{CP e\rightarrow p}
 (\equiv \stackrel{03}{(+i)} 
\stackrel{12}{[+]} \stackrel{56}{(+)} \stackrel{78}{(-)} \stackrel{9\, 10}{(-)} \stackrel{11\,12}{(-)}\stackrel{13\,14}{(-)})\, *_A \,e^{-\dagger}_{L} (\equiv \stackrel{03}{[- i]} \stackrel{12}{[+]} \stackrel{56}{(-)} \stackrel{78}{(+)} 
\stackrel{9\, 10}{(+)} \stackrel{11\,12}{(+)} \stackrel{13\,14}{(+)})=
e^{+\dagger}_{R} (\equiv \stackrel{03}{(+i)} \stackrel{12}{[+]} 
\stackrel{56}{[+]} \stackrel{78}{[-]} \stackrel{9\, 10}{[-]}
 \stackrel{11\,12}{[-]} \stackrel{13\,14}{[-]})$, which transforms 
 $e^{-\dagger}_{L} $ to $e^{+\dagger}_{R}$, while 
 ${}^{I} \hat{{\cal A}}^{ \dagger}_{CP p \rightarrow e} (\equiv \stackrel{03}{(-i)} 
\stackrel{12}{[+]} \stackrel{56}{(-)} \stackrel{78}{(+)} 
\stackrel{9\, 10}{(+)} \stackrel{11\,12}{(+)}\stackrel{13\,14}{(+)}) \,*_A\,
 e^{+\dagger}_{R} (\equiv \stackrel{03}{(+i)} \stackrel{12}{[+]} 
\stackrel{56}{[+]} \stackrel{78}{[-]} \stackrel{9\, 10}{[-]} 
\stackrel{11\,12}{[-]} \stackrel{13\,14}{[-]}))=e^{-\dagger}_{L} 
(\equiv \stackrel{03}{[- i]} \stackrel{12}{[+]} \stackrel{56}{(-)} 
\stackrel{78}{(+)} \stackrel{9\, 10}{(+)} \stackrel{11\,12}{(+)} 
\stackrel{13\,14}{(+)})$,
transforms $e^{+\dagger}_{R} $ into $e^{-\dagger}_{L}$.
}. 

Let the electron carry the momentum in ordinary space $\vec{p}_1$,
while the positron carry the momentum in ordinary space $\vec{p}_2$.

We proceed equivalently as we did in the case of $d=(5+1)$: We do 
not need to know all the even ``basis vectors'', just those with which the
electron and positron interact, Eq.~(\ref{epscattering}): We need to know
photons to which the electron and positron with momentum $\vec{p}_1+
\vec{p}_2$ transfer momenta, remaining as two ``basis vectors'' of
electron and positron (without any momentum in ordinary space). We need
to know the photon exchanged by the electron and positron, as well. We
learned how to generate them from the case $d=(5+1)$: A photon
${}^{I} \hat{{\cal A}}^{ \dagger}_{phee^{\dagger}}$ to which the electron
$e^{-\dagger}_{L}$ transfers its momentum is generated by
$e^{-\dagger}_{L} \,*_{A}\,(e^{-\dagger}_{L})^{\dagger}$, a photon
${}^{I} \hat{{\cal A}}^{ \dagger}_{phpp^{\dagger}}$ to which the positron
$e^{+\dagger}_{R}$ transfers its momentum is generated by
$e^{+\dagger}_{R}\,*_{A}\,(e^{+\dagger}_{R})^{\dagger}$. The electron
and positron will exchange the photon ${}^{II} \hat{{\cal A}}^{ \dagger}_{phe^{\dagger}e}=(e^{-\dagger}_{L})^{\dagger}
\,*_{A}\,e^{-\dagger}_{L}=$ $(e^{+\dagger}_{R})^{\dagger}\,*_{A}\,
e^{+\dagger}_{R}$. Let us show how the three photons' ``basis vectors''
look like:
\begin{small}
\begin{eqnarray}
\label{13+1epscattering}
&&e^{-\dagger}_{L}(\equiv \stackrel{03}{[- i]} \stackrel{12}{[+]} 
\stackrel{56}{(-)} \stackrel{78}{(+)} \stackrel{9\, 10}{(+)} 
\stackrel{11\,12}{(+)} \stackrel{13\,14}{(+)})\,*_{A}
(e^{-\dagger}_{L})^{\dagger} (\equiv \stackrel{03}{[- i]} \stackrel{12}{[+]} 
\stackrel{56}{(+)} \stackrel{78}{(-)} \stackrel{9\, 10}{(-)} 
\stackrel{11\,12}{(-)} \stackrel{13\,14}{(-)}) \nonumber\\
&&={}^{I} \hat{{\cal A}}^{ \dagger}_{phee^{\dagger}} (\equiv \stackrel{03}{[-i]} 
\stackrel{12}{[+]} \stackrel{56}{[-]} \stackrel{78}{[+]} \stackrel{9\, 10}{[+]} \stackrel{11\,12}{[+]}\stackrel{13\,14}{[+]})\,,\nonumber\\
&&e^{+\dagger}_{R} (\equiv \stackrel{03}{(+i)} \stackrel{12}{[+]} 
\stackrel{56}{[+]} \stackrel{78}{[-]} \stackrel{9\, 10}{[-]}
 \stackrel{11\,12}{[-]} \stackrel{13\,14}{[-]})\,*_A\,(e^{+\dagger}_{R})^{\dagger} (\equiv \stackrel{03}{(-i)} \stackrel{12}{[+]} 
\stackrel{56}{[+]} \stackrel{78}{[-]} \stackrel{9\, 10}{[-]}
 \stackrel{11\,12}{[-]} \stackrel{13\,14}{[-]}) \nonumber\\
&&={}^{I} \hat{{\cal A}}^{ \dagger}_{phpp^{\dagger}} (\equiv \stackrel{03}{[+i]} 
\stackrel{12}{[+]} \stackrel{56}{[+]} \stackrel{78}{[-]} \stackrel{9\, 10}{[-]} \stackrel{11\,12}{[-]}\stackrel{13\,14}{[-]})\,,\nonumber\\
&&(e^{-\dagger}_{L})^{\dagger} (\equiv \stackrel{03}{[- i]} \stackrel{12}{[+]} 
\stackrel{56}{(+)} \stackrel{78}{(-)} \stackrel{9\, 10}{(-)} 
\stackrel{11\,12}{(-)} \stackrel{13\,14}{(-)})\,*_{A}
e^{-\dagger}_{L} (\equiv \stackrel{03}{[- i]} \stackrel{12}{[+]} 
\stackrel{56}{(-)} \stackrel{78}{(+)} \stackrel{9\, 10}{(+)} 
\stackrel{11\,12}{(+)} \stackrel{13\,14}{(+)}) \nonumber\\
&&={}^{II} \hat{{\cal A}}^{ \dagger}_{phe^{\dagger}e} (\equiv \stackrel{03}{[-i]} 
\stackrel{12}{[+]} \stackrel{56}{[+]} \stackrel{78}{([-]} \stackrel{9\, 10}{[-]} \stackrel{11\,12}{[-]}\stackrel{13\,14}{[-]})\,,\nonumber\\
&&(e^{+\dagger}_{R})^{\dagger} (\equiv \stackrel{03}{(-i)} \stackrel{12}{[+]} 
\stackrel{56}{[+]} \stackrel{78}{[-]} \stackrel{9\, 10}{[-]}
 \stackrel{11\,12}{[-]} \stackrel{13\,14}{[-]})\,*_A\,e^{+\dagger}_{R} 
 (\equiv \stackrel{03}{(+i)} \stackrel{12}{[+]} 
\stackrel{56}{[+]} \stackrel{78}{[-]} \stackrel{9\, 10}{[-]}
 \stackrel{11\,12}{[-]} \stackrel{13\,14}{[-]}) \nonumber\\
&&={}^{II} \hat{{\cal A}}^{ \dagger}_{php^{\dagger}p} (\equiv \stackrel{03}{[-i]} 
\stackrel{12}{[+]} \stackrel{56}{[+]} \stackrel{78}{[-]} \stackrel{9\, 10}{[-]} \stackrel{11\,12}{[-]}\stackrel{13\,14}{[-]})\,.
\end{eqnarray}
\end{small}

As in the case when the internal space is described in $d=(5+1)$ also 
here we read from the last relation in Eq.~(\ref{13+1epscattering}):  
$(e^{-\dagger}_{L})^{\dagger} \,*_A\,e^{-\dagger}_{L} =
{}^{II} \hat{{\cal A}}^{ \dagger}_{phe^{\dagger}e}$
$=(e^{+\dagger}_{R})^{\dagger}\,*_A\,e^{+\dagger}_{R}=
{}^{II} \hat{{\cal A}}^{ \dagger}_{php^{\dagger}p}$~\footnote{
Let me add that $(u^{c1}_{L})^{\dagger}\,*_A\,
u^{c1 }_{L}=
{}^{II} \hat{{\cal A}}^{ \dagger}_{ph u^{c1 }_{L})^{\dagger}
u^{c1}_{L}} (\equiv \stackrel{03}{[-i]} 
\stackrel{12}{[+]} \stackrel{56}{[+]} \stackrel{78}{[-]} \stackrel{9\, 10}{[-]} \stackrel{11\,12}{[-]}\stackrel{13\,14}{[-]})$, as well as for the corresponding antiparticle. Quarks and leptons exchange the same
${}^{II} \hat{{\cal A}}^{ \dagger}_{ph (u^{c1}_{L})^{\dagger}
u^{c1}_{L}}= {}^{II} \hat{{\cal A}}^{ \dagger}_{php^{\dagger}p}$.
}. \\

The annihilation of an electron and positron in the case that the internal 
space has $d=(13+1)$ dimensions, is illustrated in 
Fig.~\ref{Figep2g13+1}.

\vspace{2mm}
\begin{figure}
  \centering
  \begin{tikzpicture}[>=triangle 45]
\draw [->,snake]
 (4,1.5) node {$\bullet$}-- (5.5,5)
node [anchor=west]
{$
\begin{smallmatrix} \\
 \\                                  
                                  \;\;\;{\rm photon} \;
                                   {}^{I}\hat{{\cal A}}^{\dagger}_{phpp^{\dagger}}\\
                                   (\equiv e^{+\dagger}_{R}\,*_{A}\,(e^{+\dagger}_{R})^{\dagger})
                                      \end{smallmatrix} $};
\draw [snake]
(1,1.5)  node {$\bullet$} -- (4,1.5) node {$\bullet$};) 
%
\draw [->,snake]
 (1,1.5) node {$\bullet$}-- (-0.5,5)
node [anchor=east]
{$
\begin{smallmatrix} \\
\\
                                       \;\;\;\;{\rm photon}\;
                                  {}^{I}\hat{{\cal A}}^{\dagger}_{phee^{\dagger}}\\
                                   (\equiv e^{-\dagger}_{L}\,*_{A}\,(e^{-\dagger}_{L})^{\dagger})
                                      \end{smallmatrix} $}; 
  \draw [->]                                    
(6,-2.5) node {$$}-- (4,1.5)
node [anchor=west]
{$\;\;\;\;
\begin{smallmatrix} \\
\\
\\
\\
\\
\\
\\
\\ 
\\                                
                                      \;\;\;\;{\rm positron}\\
                                      e^{+ \dagger}_{R}
                                      \end{smallmatrix} $}; 
   \draw [->]                                    
(-0.5,-2.5) node {$$}-- (1,1.5)
node [anchor=east]
{$
\begin{smallmatrix} \\
\\
\\
\\
\\
\\
\\
\\ 
\\
                                          {\rm elektron}\,\,\,\\ 
                                          e^{- \dagger}_{L}  \,\,                                      
                                      \end{smallmatrix} $}; 
  \draw [ ] 
(3.5,2.3) node {$$}-- (3.5,2.3)  
  node [anchor=east]
{$
\begin{smallmatrix} \\
                                 \;\;\;\;\; {\rm photon}\;
                                 {}^{II}\hat{{\cal A}}^{\dagger}_{phe^{\dagger}e}
                                 \\
                                 \;\; \;\;\; (\equiv 
                       (e^{+ \dagger}_{R})^{\dagger} \,*_{A}\, e^{+\dagger}_{R})
                                      \end{smallmatrix} $}; 
%
\end{tikzpicture} 
%
  \caption{\label{Figep2g13+1} 
Annihilation of an electron, $e^{-\dagger}_{L}$, and positron, $
e^{+\dagger}_{R}$, into two photons, is studied for the case that the
internal space has $(13+1)$ dimensions; the internal spaces of 
$e^{-\dagger}_{L}$
and $e^{+\dagger}_{R}$ are taken from Table~\ref{Table so13+1.},
from the line $29^{th}$ and $63^{rd}$, respectively, and the ``photons''
are generated following the procedure for  the case that the
internal space has $(5+1)$ dimensions.
The simple $(5+1)$-dimensional case is an illustrative introduction into
the case that the internal space has $(13 +1)$ dimensions and the 
electron and
positron are taken from Table~\ref{Table so13+1.}. The ``basis vector''
of an electron carries the charge $Q=-1$ (in ordinary space the electron has
momentum $\vec{p}_{1}$), the ``basis vector'' of its anti-particle positron 
carries $Q= +1$ (in ordinary space positron has momentum 
$\vec{p}_{2}$). The ``basis vectors'' of two photons taking away the
momenta $\vec{p}_{1}$ and $\vec{p}_{2}$, named
${}^{I}\hat{{\cal A}}^{\dagger}_{phee^{\dagger}}$ and
${}^{I}\hat{{\cal A}}^{\dagger}_{phpp^{\dagger}}$, respectively,
are represented by $e^{-\dagger}_{L}\,*_{A}\,
(e^{-\dagger}_{L})^{\dagger}$ and $e^{+\dagger}_{R}\,*_{A}\,
(e^{+\dagger}_{R})^{\dagger}$, respectively, just as we learned
in the case when the internal space has $d=(5+1)$.
The ``basis vector'' of a photon, 
${}^{II}\hat{{\cal A}}^{\dagger}_{phe^{\dagger}e}=
{}^{II}\hat{{\cal A}}^{\dagger}_{php^{\dagger}p}$, exchanged by
$e^{- \dagger}_{L}$ and $e^{+ \dagger}_{R}$, is equal to
$(e^{- \dagger}_{L})^{\dagger} \,*_{A}\, e^{- \dagger}_{L} =
(e^{+ \dagger}_{R})^{\dagger} \,*_{A}\, e^{+\dagger}_{R} $
 (due to the fact that $e^{-\dagger}_{L}$ and $
e^{+\dagger}_{R}$ belong to the same family), as shown in Eq.~(\ref{13+1epscattering}). This exchange results in
transferring the momenta $\vec{p}_{1}$ and $\vec{p}_{2}$ from
$e^{- \dagger}_{L}$ and $e^{+ \dagger}_{R}$, to the two photons
${}^{I}\hat{{\cal A}}^{\dagger}_{phee^{\dagger}}$
and ${}^{I}\hat{{\cal A}}^{\dagger}_{phpp^{\dagger}}$, respectively,
leaving the ``basis vectors'' $e^{- \dagger}_{L}$ and
$e^{+ \dagger}_{R}$ without momenta in ordinary space.
}.
\end{figure}

\vspace{2mm}

 {\bf b.\,\,\,} Using the knowledge presented in this article so far,
let us study the electron - positron scattering to $\mu^{-} + \mu^{+}$.
Since we study the massless fermion and boson fields (in the
situation in which, with respect to the {\it standard model}, the fields
manifest before the electroweak phase transition, and after the phase transition caused by the neutrino condensate ~\cite{nh2021RPPNP}) 
and no coupling constants are taken into account, the theory does not 
yet at this paper
predict the observable situation; these discussions are only to 
understand the internal spaces of fields and to find out what do they 
offer in comparison with the standard theories, string theories, 
Kakuza-Klein theories ~\footnote{
The scattering of electron and positron into ``$\mu$'' and 
``$\bar{\mu}$'', let us call them $\mu^-$ 
and $\mu^+$, would be the case known as ``flavour changing 
neutral currents'' (FCNC). We shall see that such scattering can 
go only by exchanging the ``graviton'', seen in 
Table~\ref{transverseCliff basis5+1even II.} on the $3^{rd}$ line
if we do not allow all possible transformations due to the break 
of symmetries  not discussed in this paper, like 
$\hat{b}^{1 \dagger}_{1}\,*_A\,
{}^{II} \hat{{\cal A}}^{1 \dagger}_{1}\rightarrow 
\hat{b}^{1 \dagger}_{4}$, or $\hat{b}^{1 \dagger}_{1}\,*_A\,
{}^{II} \hat{{\cal A}}^{1 \dagger}_{2}\rightarrow 
\hat{b}^{1 \dagger}_{2}$.  
}. %
Let be repeated: all discussed in this article is valid only in
almost empty space, equipped by the Poincare symmetry.\\

\vspace{2mm}

{\bf b.i.\,\,\,} {\it ``Electron'' - ``positron'' scattering to  ``$\mu^- $'' and 
 ``$\mu^+ $'' in the case that the internal space has $d=(5+1)$, and
 fermions and bosons have non zero momenta in $d=(3+1)$}: 
``Electron'', $e^{-}$,  presented by $\hat{b}^{1 \dagger}_{1}$, and 
``positron'', $e^{+}$, presented by $\hat{b}^{3 \dagger}_{1}$, 
exchange the ``graviton'', ${}^{II} \hat{{\cal A}}^{1 \dagger}_{4} $,
presented in Table~\ref{transverseCliff basis5+1even II.},
\begin{eqnarray}
\label{epmumu}
&&{}^{II} \hat{{\cal A}}^{1 \dagger}_{4} 
(\equiv \stackrel{03}{(- i)} \stackrel{12}{(+)} \stackrel{56}{[+]})=
(\hat{b}^{1 \dagger}_{1})^{\dagger} (\equiv \stackrel{03}{(- i)} 
\stackrel{12}{[+]} \stackrel{56}{[+]})\,*_A\,\hat{b}^{1 \dagger}_{3}
(\equiv \stackrel{03}{[+ i]} \stackrel{12}{(+)} \stackrel{56}{[+]})
\nonumber\\
&&\qquad \qquad \qquad \qquad \quad=
(\hat{b}^{3 \dagger}_{1})^{\dagger}(\equiv \stackrel{03}{[- i]} 
\stackrel{12}{[+]} \stackrel{56}{(+)})\,*_A\,\hat{b}^{3 \dagger}_{3}
(\equiv \stackrel{03}{(- i)} 
\stackrel{12}{(+)} \stackrel{56}{(-)})\,, \nonumber\\
&&\hat{b}^{1 \dagger}_{1}\,*_A\,
{}^{II} \hat{{\cal A}}^{1 \dagger}_{4}=
\hat{b}^{1 \dagger}_3\,,\quad\quad
\hat{b}^{3 \dagger}_{1}\,*_A\,
{}^{II} \hat{{\cal A}}^{1 \dagger}_{4}=
\hat{b}^{3 \dagger}_{3}\,,
\end{eqnarray}
with  ``$\mu^-$'' (with the ``basis vector'' $\hat{b}^{1 \dagger}_{3}$)
and ``$\mu^+$'' (with the ``basis vector'' $\hat{b}^{3 \dagger}_{3}$), 
taking away momentum ($\vec{p}_1 + \vec{p}_2$) of $e^{-}$ and 
$e^{+}$.
Let us tell that the ``basis vector'' $\hat{b}^{1 \dagger}_{3}$ follows
up to a constant from the ``basis vector'' $\hat{b}^{1 \dagger}_{1}$
by applying, for example, $\tilde{S}^{01}=\frac{i}{2}\tilde{\gamma}^0 
\tilde{\gamma}^1$ on $\hat{b}^{1 \dagger}_{1}$~\footnote{
Let us demonstrate the application of $\tilde{\gamma}^0 
\tilde{\gamma}^1$ on $\hat{b}^{1 \dagger}_{1}=\frac{1}{2}
(\gamma^0-\gamma^3) \frac{1}{2}(1+i \gamma^1\gamma^2) 
\frac{1}{2}(1+i \gamma^5\gamma^2)$, using 
Eqs.~(\ref{gammatildeantiher0}, \ref{tildegammareduced0}):
\begin{eqnarray}
\label{tildeSab}
\tilde{\gamma}^0 \tilde{\gamma}^1 \frac{1}{2}
(\gamma^0-\gamma^3) \frac{1}{2}(1+i \gamma^1\gamma^2) 
\frac{1}{2}(1+i \gamma^5\gamma^6)&=&
(\frac{1}{2})^3 (-i)^2 (\gamma^0-\gamma^3) \gamma^0
(1+i \gamma^1\gamma^2) \gamma^1
(1+i \gamma^5\gamma^6)
= \nonumber\\
-\frac{1}{2}
(1+\gamma^0\gamma^3) \frac{1}{2}(\gamma^1+i\gamma^2) 
\frac{1}{2}(1+i \gamma^5\gamma^2)=-\hat{b}^{1 \dagger}_{3}\,.
\end{eqnarray}
}.
Let us point out that the ``graviton'' (with the eigenvalues of the 
Cartan subalgebra members (${\cal S}^{03}=-i,  {\cal S}^{12}=1, 
{\cal S}^{56}=0$)) leaves the family member quantum number of 
$\mu^{-}$ the same as of $e^{-}$ ($S^{03}=\frac{i}{2},
S^{12}=\frac{1}{2}, S^{56}=\frac{1}{2}$), changing the family 
quantum number of $e^{-}$ (expressed by the eigenvalues of the 
Cartan subalgebra members $\tilde{S}^{03}= \frac{i}{2},
\tilde{S}^{12}=-\frac{1}{2}, \tilde{S}^{56}=-\frac{1}{2}$) to
those of $\mu^{-}$ ($\tilde{S}^{03}=- \frac{i}{2},
\tilde{S}^{12}=\frac{1}{2}, \tilde{S}^{56}=-\frac{1}{2}$).
Equivalently, ``graviton'' changes only the family quantum 
number of $e^{+}$ ($\tilde{S}^{03}=\frac{i}{2},
\tilde{S}^{12}=-\frac{1}{2}, \tilde{S}^{56}=-\frac{1}{2}$) to 
the one of $\mu^{+}$ ($\tilde{S}^{03}=-\frac{i}{2}, 
\tilde{S}^{12}=\frac{1}{2}, \tilde{S}^{56}=-\frac{1}{2}$).

This scattering is presented in Fig.~\ref{Figepmuantimu}.\\

\vspace{2mm}
\begin{figure}
\begin{tikzpicture}[>=triangle 45]
\draw [->] 
 (4,1.5) node {$\bullet$}-- (5.5,5)
node [anchor=west]
{$
\begin{smallmatrix} \\
 \\                                  
                                  \;\;\;{\rm ``anti-muon''}^{+\dagger} \,  \hat{b}^{3 \dagger}_{3}\\\;\\
                                   (\equiv \stackrel{03}{(-i)}\stackrel{12}{(+)} \stackrel{56}{(-)})
                                      \end{smallmatrix} $};
\draw [snake]
(1,1.5)  node {$\bullet$} -- (4,1.5) node {$\bullet$};) 
%
\draw [->] 
 (1,1.5) node {$\bullet$}-- (-0.5,5)
node [anchor=east]
{$
\begin{smallmatrix} \\
\\
                                       \;\;\;\;{\rm ``muon''}\,  \hat{b}^{1 \dagger}_{3}\\\;
                                   (\equiv \stackrel{03}{[+i]}\stackrel{12}{(+)} \stackrel{56}{[+]})
                                      \end{smallmatrix} $}; 
  \draw [->]                                    
(6,-2.5) node {$$}-- (4,1.5)
node [anchor=west]
{$\;\;\;\;
\begin{smallmatrix} \\
\\
\\
\\
\\
\\
\\
\\                                 
                                      \;\;\;\;{\rm ``positron''}
                                     \;   \hat{b}^{3 \dagger}_{1}\\
                                       \;\;\;\; (\equiv \stackrel{03}{[-i]}\stackrel{12}{[+]} \stackrel{56}{(-)})
                                      \end{smallmatrix} $}; 
   \draw [->]                                    
(-0.5,-2.5) node {$$}-- (1,1.5)
node [anchor=east]
{$
\begin{smallmatrix} \\
\\
\\
\\
\\
\\
\\
\\ 
                                          {\rm ``elektron''}\;
                                        \hat{b}^{1 \dagger}_{1}\\
                                       (\equiv \stackrel{03}{(+i)}\stackrel{12}{[+]} \stackrel{56}{[+]})\;\;\;\;\;
                                      \end{smallmatrix} $}; 
  \draw [ ] 
(3.5,2.3) node {$$}-- (3.5,2.3)  
  node [anchor=east]
{$
\begin{smallmatrix} \\
                                  {\rm ``graviton''}\;
                                  {}^{II}\hat{{\cal A}}^{1 \dagger}_{4}\\
                                   (\equiv \stackrel{03}{(-i)}\stackrel{12}{(+)} \stackrel{56}{[+]})
                                      \end{smallmatrix} $}; 

\end{tikzpicture}
  \caption{\label{Figepmuantimu} 
Scattering of an ``electron'', with the internal space described by 
$\hat{b}^{1 \dagger}_{1}$ and with the momentum $\vec{p}_1$ in 
ordinary space, 
and ``positron'', with the internal space described by 
$\hat{b}^{3 \dagger}_{1}$  and with the momentum $\vec{p}_2$ in 
ordinary space 
to ``muon'', with the internal space described by 
$\hat{b}^{1 \dagger}_{3}$ and with the momentum $\vec{p}_3$ in 
ordinary space,  and ``anti-muon'', with the internal space described by 
$\hat{b}^{3 \dagger}_{3}$ and with momentum $\vec{p}_4$ in 
ordinary space, 
is presented for the case that the internal space has $(5+1)$ dimensions 
(the internal spaces of ``electron'',``positron'', ``muon'' and ``anti-muon'' are presented 
as ``basis vectors'' in Table~\ref{Table Clifffourplet.}. 
``Electron'',  represented by $\hat{b}^{1 \dagger}_{1}$  and 
with the momentum $\vec{p}_{1}$ in ordinary space, and ``positron'' 
represented by $\hat{b}^{1 \dagger}_{3}$ and with the momentum 
$\vec{p}_{2}$ in ordinary space, exchange the ``graviton''
${}^{II} \hat{{\cal A}}^{1 \dagger}_{4}$, exchanging the momenta 
to ``muon'' and ``anti-muon'', which 
take away the momenta $\vec{p}_3$ 
 $+\vec{p}_4$  $= \vec{p}_1 + \vec{p}_2$. \\
This simple $(5+1)$-dimensional internal space 
is an illustrative introduction into the case that the  internal space has 
$(13 +1)$ dimensions, while ``electrons'',``positrons'', ``muons'', 
 ``anti-muons'', gravitons and photons have non zero momenta only in  $d=(3+1)$ 
ordinary space. The ``basis vectors'' of electron and
positron are taken from Table~\ref{Table so13+1.}, while  
``muon'' and ``anti-muon'' are obtained, for example, by applying 
$\tilde{S}^{01}$ on electron and positron ``basis vectors'' 
from Table~\ref{Table so13+1.}. 
}.
\end{figure}
As it follows from Eqs.~(\ref{ruleAAI}, \ref{calIAb1234gen}, 
\ref{calbIIA1234gen}) not all possibilities are allowed; when scatter 
``electrons'' on ``positrons'', the scattering of ``electrons'' to``
anti-muons'' and ``positrons'' to ``muons'', for example, can not go.

\vspace{1mm}

{\bf b.ii.\,\,\,} {\it Electron - positron scattering to $\mu^{- \dagger}$ 
and $\mu^{+\dagger}$ in $d=(13+1)$}: We find in 
Table~\ref{Table so13+1.} the``basis vector'' of an electron with spin
up in the $29^{th}$ line carrying the charge $Q=-1$, we name it 
$e^{ - \dagger}_{L}$, and the ``basis vector'' of the positron in 
the $63^{rd}$ line carrying $Q= +1$,  we named it 
$e^{+ \dagger}_{R}$. We can generate ``basis vectors'' of
$\mu^{- \dagger}$ and $\mu^{+\dagger}$ by the application
of $\tilde{S}^{ac}$ on $e^{ - \dagger}_{L}$ and $e^{+ \dagger}_{R}$,
respectively, as demonstrated in Eq.~(\ref{tildeSab}).

We proceed as we did in the case when the internal space manifests 
the ``basis vectors'' in $d=(5+1)$. Also in the case when the internal 
space has $d=(13+1)$,  the electrons, positrons, muons, anti-muon, 
gravitons (and all the vector and scalar fields) have in ordinary space 
non zero momenta only in $d=(3+1)$. We do not need to know all the families of leptons and anti-leptons and quarks and antiquarks, and 
all the even ``basis vectors'' to demonstrate the scattering of 
families: In the lowest order of the Feynman diagrams,
just those with which the electron and 
positron scatter into, let us name those, $\mu^{-\dagger}_{L}$ and
$\mu^{+\dagger}_{R}$, respectively.

We discuss two choices for $\mu^{-\dagger}_{L}$ and 
$\mu^{+\dagger}_{R}$. In both cases we use for $e^{-\dagger}_{L}
(\equiv \stackrel{03}{[- i]} \stackrel{12}{[+]} 
\stackrel{56}{(-)} \stackrel{78}{(+)} \stackrel{9\, 10}{(+)} 
\stackrel{11\,12}{(+)} \stackrel{13\,14}{(+)})$ and 
$e^{+\dagger}_{R} (\equiv \stackrel{03}{(+i)} \stackrel{12}{[+]} 
\stackrel{56}{[+]} \stackrel{78}{[-]} \stackrel{9\, 10}{[-]}
 \stackrel{11\,12}{[-]} \stackrel{13\,14}{[-]})$,  and find 
$e^{-\dagger}_{L}$ and $e^{+ \dagger}_{R}$ belonging to two 
different families, that is the two kinds of $\mu^{-\dagger}_L$ and 
$\mu^{+\dagger}_R$, by applying either $\tilde{S}^{01}$ or 
$\tilde{S}^{57}$ on $e^{-\dagger}_{L}$ and $e^{+\dagger}_{R}$,
respectively. 

Let us start with the case, resembling the one with $d=(5+1)$, by 
choosing:\\
$\mu^{-\dagger}_L (\equiv \stackrel{03}{(-i)} \stackrel{12}{(+)} 
\stackrel{56}{(-)} \stackrel{78}{(+)} \stackrel{9\, 10}{(+)}
 \stackrel{11\,12}{(+)} \stackrel{13\,14}{(+)})=
 - \tilde{\gamma}^{0}   \tilde{\gamma}^{1}\,e^{-\dagger}_{L}$, 
 and $\mu^{+\dagger}_{R} (\equiv \stackrel{03}{[+i]} 
 \stackrel{12}{(+)} \stackrel{56}{[+]} \stackrel{78}{[-]} 
 \stackrel{9\, 10}{[-]}
 \stackrel{11\,12}{[-]} \stackrel{13\,14}{[-]})= \tilde{\gamma}^{0}  
 \tilde{\gamma}^{1}\,e^{+\dagger}_{R}$.

\begin{small}
\begin{eqnarray}
\label{13+1epmu-mu+}
&&e^{-\dagger}_{L}(\equiv \stackrel{03}{[- i]} \stackrel{12}{[+]} 
\stackrel{56}{(-)} \stackrel{78}{(+)} \stackrel{9\, 10}{(+)} 
\stackrel{11\,12}{(+)} \stackrel{13\,14}{(+)})\,*_{A}\,
{}^{II} \hat{{\cal A}}^{ \dagger}_{e-mu-} (\equiv 
\stackrel{03}{(-i)} \stackrel{12}{(+)} \stackrel{56}{[+]} 
\stackrel{78}{[-]} \stackrel{9\, 10}{[-]} \stackrel{11\,12}{[-]}
\stackrel{13\,14}{[-]})\rightarrow \nonumber\\
&&\mu^{-\dagger}_L (\equiv \stackrel{03}{(-i)} \stackrel{12}{(+)} 
\stackrel{56}{(-)} \stackrel{78}{(+)} \stackrel{9\, 10}{(+)}
 \stackrel{11\,12}{(+)} \stackrel{13\,14}{(+)})\,,\nonumber\\
&&e^{+\dagger}_{R} (\equiv \stackrel{03}{(+i)} 
\stackrel{12}{[+]} \stackrel{56}{[+]} \stackrel{78}{[-]} 
\stackrel{9\, 10}{[-]} \stackrel{11\,12}{[-]}\stackrel{13\,14}{[-]})
\,*_{A}\,{}^{II} \hat{{\cal A}}^{ \dagger}_{e+mu+} (\equiv 
\stackrel{03}{(-i)} \stackrel{12}{(+)} \stackrel{56}{[+]} 
\stackrel{78}{[-]} \stackrel{9\, 10}{[-]} \stackrel{11\,12}{[-]}
\stackrel{13\,14}{[-]})\rightarrow \nonumber\\ 
&&\mu^{+\dagger}_{R} 
(\equiv \stackrel{03}{[+i]}  \stackrel{12}{(+)} \stackrel{56}{[+]}
 \stackrel{78}{[-]}  \stackrel{9\, 10}{[-]} \stackrel{11\,12}{[-]}
  \stackrel{13\,14}{[-]})\,.
\end{eqnarray}
\end{small}
Like in the case of $d=(5+1)$, the ``graviton'',
${}^{II} \hat{{\cal A}}^{ \dagger}_{e-mu-}$
$={}^{II} \hat{{\cal A}}^{ \dagger}_{e+mu+}$,
Eq.~(\ref{13+1epmu-mu+}), has all the quantum numbers (the
eigenvalues of the Cartan subalgebra members), except
${\cal S}^{03}=-i$ and ${\cal S}^{12}=1$, equal to zero. Also in
this case the scattering of electron, $e^{-\dagger}_{L}$, and
positron, $e^{+\dagger}_{R}$, into $\mu^{-\dagger}_{L}$ and
$\mu^{+\dagger}_{R}$, would be the case known as ``flavour
changing neutral currents'' (FCNC).~\footnote{
In Ref.~\cite{nh2021RPPNP} and references therein, the masses
of twice two groups of four families are studied, and also the
possible ``flavour changing neutral currents'' (FCNC) discussed,
after the breaks of symmetries, causing the non-zero masses of
quarks and leptons and antiquarks and antileptons and the weak
boson fields. Since the gravitational coupling constant is in ordinary
conditions almost negligible, this FCNC can not be observable.
}\\

Let us make a choice  of the same family members, $e^{-\dagger}_{L}
 (\equiv \stackrel{03}{[- i]} \stackrel{12}{[+]} \stackrel{56}{(-)} 
 \stackrel{78}{(+)} \stackrel{9\, 10}{(+)} \stackrel{11\,12}{(+)} 
 \stackrel{13\,14}{(+)})$ and \\$e^{+\dagger}_{R} (\equiv 
\stackrel{03}{(+i)} \stackrel{12}{[+]} \stackrel{56}{[+]} 
\stackrel{78}{[-]} \stackrel{9\, 10}{[-]} \stackrel{11\,12}{[-]}
\stackrel{13\,14}{[-]})$, scattering this time into another family, 
obtained from the one with $e^{-\dagger}_{L}$ and 
$e^{+\dagger}_{R}$ by applying on the starting two by 
$\tilde{S}^{57}= \frac{i}{2} \tilde{\gamma^5} 
\tilde{\gamma}^{7}$:\\

$\mu^{`\,-\dagger}_L (\equiv \stackrel{03}{[- i]} \stackrel{12}{[+]} 
\stackrel{56}{[-]} \stackrel{78}{[+]} \stackrel{9\, 10}{(+)}
 \stackrel{11\,12}{(+)} \stackrel{13\,14}{(+)})=
  \tilde{\gamma}^{5}   \tilde{\gamma}^{7}\,e^{-\dagger}_{L}$, 
 and $\mu^{`\,+\dagger}_{R} (\equiv \stackrel{03}{(+i)} 
 \stackrel{12}{[+]} \stackrel{56}{(+)} \stackrel{78}{(-)} 
 \stackrel{9\, 10}{[-]}  \stackrel{11\,12}{[-]} \stackrel{13\,14}{[-]})= 
 \tilde{\gamma}^{5}   \tilde{\gamma}^{7}\,e^{+\dagger}_{R}$.
 It follows
 \begin{small}
\begin{eqnarray}
\label{13+1epmu'-mu'+}
&&e^{-\dagger}_{L}(\equiv \stackrel{03}{[- i]} \stackrel{12}{[+]} 
\stackrel{56}{(-)} \stackrel{78}{(+)} \stackrel{9\, 10}{(+)} 
\stackrel{11\,12}{(+)} \stackrel{13\,14}{(+)})\,*_{A}\,
{}^{II} \hat{{\cal A}}^{ \dagger}_{e-mu'-} (\equiv 
\stackrel{03}{[-i]} \stackrel{12}{[+]} \stackrel{56}{(+)} 
\stackrel{78}{(-)} \stackrel{9\, 10}{[-]} \stackrel{11\,12}{[-]}
\stackrel{13\,14}{[-]})\rightarrow \nonumber\\
&&\mu^{`\, -\dagger}_L (\equiv \stackrel{03}{[- i]} \stackrel{12}{[+]} 
\stackrel{56}{[-]} \stackrel{78}{[+]} \stackrel{9\, 10}{(+)}
 \stackrel{11\,12}{(+)} \stackrel{13\,14}{(+)})\,,\nonumber\\
&&e^{+\dagger}_{R} (\equiv \stackrel{03}{(+i)} 
\stackrel{12}{[+]} \stackrel{56}{[+]} \stackrel{78}{[-]} 
\stackrel{9\, 10}{[-]} \stackrel{11\,12}{[-]}\stackrel{13\,14}{[-]})
\,*_{A}\,{}^{II} \hat{{\cal A}}^{ \dagger}_{e+mu'+} (\equiv 
\stackrel{03}{[-i]} \stackrel{12}{[+]} \stackrel{56}{(+)} 
\stackrel{78}{(-)} \stackrel{9\, 10}{[-]} \stackrel{11\,12}{[-]}
\stackrel{13\,14}{[-]})\rightarrow \nonumber\\ 
&&\mu^{`\, +\dagger}_{R} (\equiv \stackrel{03}{(+i)} 
 \stackrel{12}{[+]} \stackrel{56}{(+)} \stackrel{78}{(-)} 
 \stackrel{9\, 10}{[-]}  \stackrel{11\,12}{[-]} \stackrel{13\,14}{[-]})\,.
\end{eqnarray}
\end{small}
Again, ${}^{II} \hat{{\cal A}}^{ \dagger}_{e-mu'-}=
{}^{II} \hat{{\cal A}}^{ \dagger}_{e+mu'+} $, this time the 
even ``basis vector'' has non zero only the weak charge ${\cal \tau}^{13}
= \frac{1}{2} ({\cal S}^{56}-{\cal S}^{78}) = 1$, while ${\cal \tau}^{23}
= \frac{1}{2} ({\cal S}^{56}+{\cal S}^{78})=0$.

The figure for these two cases would look like the one in
\ref{Figepmuantimu} only the electron and positron and muon and anti-muon
would be represented in $d=(13+1)$ dimensional space. We, therefore, do not present these two scatterings in figures~\footnote{
In this last case, not gravitons but weak bosons cause
the ``flavour changing neutral currents'', which the breaking of symmetry
can make small enough~(\cite{nh2021RPPNP}and references therein).
}. \\

Let us conclude this section by recognizing:\\
Describing the internal space of boson fields with the even
``basis vectors'' and the internal space of fermion fields with the 
odd ``basis vectors'', under the conditions that all the fields are 
massless and the ordinary space has Poincare symmetry, we find out 
that our method offers an interesting new understanding of the second 
quantized fermion and boson fields so far observed, with the gravity 
included, in an unique way.  All the internal spaces, described by the 
``basis vectors'', are analysed from the point of view that the ordinary 
space offers the non zero momentum only in $d=(3+1)$-dimensions, 
while all the vector gauge fields, carrying the space index 
$\mu= (0,1,2,3)$ and all the scalar gauge fields with the space index 
($\alpha=(5,6,...,14)$) have the internal spaces described by the 
``basis vectors'' with $SO(13,1)$ from the point of view as observed 
in $d=(3+1)$.\\

There is much work still to be done to relate the observed fermion
and boson fields, with gravity included.
We need to suggest to which extent, for example, relations in
Eqs.~(\ref{ruleAAI}, \ref{calIAb1234gen}, \ref{calbIIA1234gen}), not
allowing several possibilities, are responsible for breaking symmetries.

And we need to suggest
how has our universe ``decided to be active'' (to have the non-zero
momentum of fermion and boson fields) only in $d=(3+1)$.
\\

%
\section{Conclusions}
\label{conclusions}

Trying to understand what the elementary constituents of our
universe are and what are the laws of nature; physicists suggest
theories and look for predictions which need confirmation of
experiments.
What seems to be trustworthy is that the elementary
constituents are two kinds of fields: Anti-commuting fermion and
commuting boson fields, both recognized as second quantized
fields.

It is accepted in this article that the {\it elementary fermion fields
are quarks, leptons, antiquarks, and antileptons, the internal
space of which is described by the odd ``basis vectors''},
and {\it elementary boson fields are $SO(3,1)$ graviton fields,
$SU(2)_I \times SU(2)_{II}$ weak boson fields~\footnote{
So far only the $SU(2)_I$  weak boson fields have been
observed. $SU(2)_{II}$ weak field which get mass due to
neutrino condensate~(\cite{nh2021RPPNP} and references
therein) is predicted to have a much larger mass than the 
$SU(2)_I$  weak boson fields.},
 $SU(3)$ gluon fields
and  (photon) $U(1)$ fields, the internal space of which is
described by the even ``basis vectors''}, both discussed in
Subsect.~\ref{cliffordoddeven}, while recognizing that $SO(3,1)
\times$ $SU(2) \times SU(2)\times$ $SU(3)\times U(1)$ are
subgroups of the group $SO(13,1)$. It is assumed as well
that the {\it dynamics in ordinary space are non-zero only in
$d=(3+1)$ space} (that is, the momentum is non-zero only if
space concerns $x_{\mu} =(x_0,x_1,x_2,x_3)$). The
vector gauge fields (photons, weak bosons, gluons, gravitons)
carry the (additional) space index  $\alpha$$=\mu =(0,1,2,3)$,
while scalars (higgses among scalars) have the space index
$\alpha \ge 5$.

In the flat space manifesting the Poincar\' e symmetry, the
interactions among all the fields (represented by the Feynman
diagrams) are determined by the algebraic multiplication of the
``basis vectors'' as presented in Eqs.~(\ref{AIAIIorth} - %
\ref{calbIIA1234gen}), demonstrated in 
Sect.~\ref{bosons13+1and5+1}, and in 
Figs.~(\ref{Figep2gamma5+1}, \ref{Figep2g13+1},
\ref{Figepmuantimu}), for the cases $d=(13+1)$ and 
$d=(5+1)$, and in Tables~(\ref{transverseCliff basis5+1even I.},
\ref{S120Cliff basis5+1even I.}, 
\ref{transverseCliff basis5+1even II.},
\ref{S120Cliff basis5+1even II.})~\footnote{
In the case of strongly interacting fermion and boson
fields the perturbation theory is intended to be
replaced with many-body approximate models developed in
many fields in physics.}.

The description of the internal spaces of {\it all elementary fields},
with the graviton included, the fermion ones by the odd
``basis vectors'', and the boson ones by the even ``basis
vectors'', explains the postulates of the second
quantized fermion and boson fields~\cite{n2023NPB,nh2021RPPNP}.

Having an equal number of ``basis vectors'' and their Hermitian
conjugated partners of the fermion fields, and of ``basis vectors'' of
the two kinds of boson fields, this theory manifests in this sense  a 
kind of supersymmetry~\footnote{
Breaking of symmetries influences the fermion fields and the boson 
fields destroying this symmetry.
}.

Analysing the properties of fermion and boson fields from the
point of view how they manifest in $d=(3+1)$~(\cite{n2023NPB,
n2023MDPI,nh2021RPPNP,n2023DOI,n2022epjc,nh2023dec},
and the references therein),
the proposed theory discussed in this contribution promises to
be the right step to better understanding the laws of nature
in our universe~\footnote{
Since the only dimensions which do not need explanation
seem to be zero and $\infty$, we should explain why we
observe $d=(3+1)$. Recognizing that for $d\ge (13+1)$ the
internal space manifests in $d=(3+1)$ all the observed
fermion and boson fields, we start with the assumption
that the internal spaces wait for the ``push'' in ordinary
space (in the case of our universe, the ``push'' was made in
$d=(3+1)$ making active internal spaces of $d\ge (13+1)$).
To make the discussions of the internal spaces of fermion
and boson fields (manifesting in $d=(3+1)$ quarks and
leptons and antiquarks and antileptons appearing in families,
while boson fields appear in two orthogonal groups
manifesting vector and scalar gauge fields with graviton
included) transparent, the assumption is made that the
ordinary $d=(3+1)$ space is flat so that the interpretation of
the Feynman diagrams make sense.
}.
For strongly interacting fields, the approximate methods are usually
needed.

Let us shortly overview what we have learned in this proposal,
assuming the space in $d=(3+1)$ is flat, the internal spaces of
fermion and boson fields manifesting $(13+1)$ dimensions are
described by the odd ``basis vectors'' for fermions and
antifermions (both appear within the same family,
correspondingly, there is no Dirac sea in this theory), and by the
even ``basis vectors'' for bosons. The fermion
``basis vectors'' appear in $d=2(2n+1)$ and $d=4n$ dimensional 
spaces in $2^{\frac{d}{2}-1}$ families, each
family having $2^{\frac{d}{2}-1}$ members, their Hermitian
conjugated partners having as well $2^{\frac{d}{2}-1}\times
2^{\frac{d}{2}-1}$ members. The boson fields appear in two
orthogonal groups, each with $2^{\frac{d}{2}-1}\times
2^{\frac{d}{2}-1}$ members, having their Hermitian conjugated
partners within the same group~\footnote{
The starting symmetry assumed to be broken with the condensate, 
made of two right handed neutrinos coupled to spin $0$%
~(\cite{nh2021RPPNP} Table 6, and  the 
references therein), splits fermions and anti-fermions.
}.

Fermion and boson fields
manifest all the properties assumed by the {\it standard
model} before the electroweak break, with the Higgs scalars
included~\cite{nh2021RPPNP,n2023NPB} and the gravitational 
field included (provided that the beak of symmetry is caused by
the two right-handed neutrinos making the vector gauge fields 
which would carry more than one kind of charge at the 
same time heavy).

In this paper, we clarify that even
``basis vectors'' offer the explanation not only for photons,
$U(1)$, weak bosons, $SU(2)_I$ and $SU(2)_{II}$, gluons,
$SU(3)$, Subsect.~\ref{photonsweakbosonsgluons}, scalar
fields, $SU(2)$, Subsect.~\ref{scalarfields}, but also for
gravitons, $SO(3,1)$, Subsect.~\ref{gravitons}: \\

{\bf i.\,\,\,} To describe the Feynman diagrams,
both kinds of boson fields, the ``basis vectors'' of which are in 
this contribution named
${}^{I}{\hat{\cal A}}^{m \dagger}_{f}$ and
${}^{II}{\hat{\cal A}}^{m \dagger}_{f}$, are needed. The
internal spaces of fermion and boson fields, described by the
``basis vectors'' are analysed from the point of view that the
ordinary space offers the non zero momentum only in
$d=(3+1)$-dimensions, while all the vector gauge fields
carry the space index ($\mu=(0,1,2,3)$) and the scalar
gauge fields carry the space index ($\alpha=(5,6,...,14)$). 
Fermion fields (quarks and leptons, antiquarks and antileptons) 
have the internal spaces described by the odd ``basis vectors'' with
the symmetry of the subgroups ($SO(3,1), SU(2), SU(2), SU(3),
U(1)$) of the group $SO(13+1)$, as suggested by their properties in
$d=(3+1)$. Also, boson fields (the vector and the scalar ones)
have their internal spaces described by the even ``basis vectors''
with the symmetry of the subgroups ($SO(3,1), SU(2), SU(2), SU(3),
U(1)$) of the group $SO(13+1)$, as well suggested by their properties in
$d=(13+1)$.  The ``basis vectors'' of all the fields, fermion or 
boson ones, are eigenstates of the Cartan subalgebra 
members: ($S^{03}, S^{12}, S^{56},..., S^{d-3\,d-2}, 
S^{d-1\,d}$, describing the quantum numbers of the family 
members fermions) and ($\tilde{S}^{03}, \tilde{S}^{12}, 
\tilde{S}^{56} ,..., \tilde{S}^{d-3\,d-2}, \tilde{S}^{d-1\,d}$,
describing the quantum numbers of the families of fermions) 
for fermions, and (${\cal S}^{03}, {\cal S}^{12}, {\cal S}^{56}, 
..., {\cal S}^{d-3\,d-2}, {\cal S}^{d-1\,d}$, describing the 
quantum numbers of bosons), 
Eq.~(\ref{cartangrasscliff}).

All the vector and scalar ``basis vectors'', 
${}^{I}{\hat{\cal A}}^{m \dagger}_{f}$ and 
${}^{II}{\hat{\cal A}}^{m \dagger}_{f}$,
are expressible as algebraic products of the odd ``basis vectors'' and their 
Hermitian conjugated partners, as presented in 
Eqs.~(\ref{AIbbdagger}, \ref{AIIbdaggerb}) and in 
Eqs.~(\ref{CPAubaru}, \ref{phAqeqe}, \ref{w2udc1R} - \ref{grAqeqe}),
resembling the closed strings 
presentations~\cite{n2023DOI,BlumenhagenString,mil,WreyString}.


We illustrate in Subsect.~\ref{feynmandiagrams} the internal spaces of fermions and bosons (assuming that the
internal space manifesting in $d=(3+1)$ origin in $d=(5+1)$,
Subsect.~\ref{bosons5+1} and in $d=(13+1)$, Subsects.~(\ref{photonsweakbosonsgluons}, \ref{scalarfields},
\ref{gravitons}): {\bf i.a.} We illustrate
the annihilation of an electron and positron into two
photons, Figs.~(\ref{Figep2gamma5+1},
\ref{Figep2g13+1}). {\bf i.b.}  We illustrate the scattering of an
electron and positron into muon and antimuon,
Figs.~(\ref{Figepmuantimu}), 
{\bf i.c.} There are several more cases discussed in 
Sect.~\ref{bosons13+1and5+1}.

Let be pointed out that since  in $d=(3+1)$ observed 
antiparticles belong in this theory (before breaking symmetries) 
to the same  massless family and manifest as particles, in the 
Feynman diagrams antiparticles (like positrons) can not be 
described as travelling backwards in time; as presented in 
Figs.~(\ref{Figep2gamma5+1}, \ref{Figep2g13+1}), 
$e^{- \dagger}_{L}$ and $e^{+ \dagger}_{R}$ instead
exchange a photon with the ``basis vector'' 
${}^{II}\hat{{\cal A}}^{\dagger}_{phe^{\dagger}e}=
{}^{II}\hat{{\cal A}}^{\dagger}_{php^{\dagger}p}$, 
equal to
$(e^{- \dagger}_{L})^{\dagger} \,*_{A}\, e^{- \dagger}_{L} =
(e^{+ \dagger}_{R})^{\dagger} \,*_{A}\, e^{+\dagger}_{R} $.
The theory correspondingly brings a new understanding of interactions 
among particles and antiparticles.\\
The theory offers an elegant and promising illustration
of the interaction among fermion fields (with the internal space
described with the odd ``basis vectors'') and boson fields
(with the internal space described with the even``basis vectors'').

{\bf ii.\,\,\,}
We pointed out that the properties of fermion and boson fields in
even dimensional spaces with $d=4n$ distinguish from those
with $d=2(2n+1)$. The discrete symmetry operator
$\mathbb{C}_{{\cal N}}$ ${\cal P}^{(d-1)}_{{\cal N}}$,
designed concerning
$d=(3+1)$~\cite{nhds2014,nh2021RPPNP}, has in the case of
$d=4n$ an odd number of $\gamma^a$'s ($\mathbb{C}_{{\cal N}}
{\cal P}^{(8-1)}_{{\cal N}}$ part expressible with the product of
$\gamma^a$'s is in the case $n=2$, for example, equal to
$\gamma^0 \gamma^5 \gamma^7$, transforming in
$d=(7+1)$, for example,  the odd ``basis vector'' presented in
Table~\ref{Table so13+1.} in the $33^{rd}$ line, if we pay
attention only in $SO(7,1)$ part, the odd $SO(7,1)$ part into
the even $SO(7,1)$ part presented in the $3^{rd}$ line of
Table~\ref{Table so13+1.}, if we
again pay attention on $SO(7,1)$ part only).

{\bf iii.\,\,\,}
The properties of ``basis vectors'' in odd-dimensional spaces differ
essentially from the properties of those in the even-dimensional
spaces.

There are two groups of the anticommuting ``basis vectors'' and
two groups of commuting ``basis vectors'' in odd-dimensional
spaces: One group of
anticommuting ``basis vectors'' have $2^{\frac{d-1}{2}-1}$
fermions and antifermions in each of $2^{\frac{d-1}{2}-1}$
families and $2^{\frac{d-1}{2}-1}\times$
$2^{\frac{d-1}{2}-1}$ Hermitian conjugated partners. The
equivalent group of commuting ``basis vectors'' appear in two
orthogonal groups, each with $2^{\frac{d-1}{2}-1} \times$
$2^{\frac{d-1}{2}-1}$ members, manifesting properties of
commuting ``basis vectors'' in even dimensional spaces.

The rest half of the commuting twice $2^{\frac{d-1}{2}-1}$
$\times 2^{\frac{d-1}{2}-1}$ members manifest properties of
the odd ``basis vectors'' and their Hermitian conjugated
partners in the sense that they appear in $2^{\frac{d-1}{2}-1}$
families with $2^{\frac{d-1}{2}-1}$ members and have their
$2^{\frac{d-1}{2}-1}\times 2^{\frac{d-1}{2}-1}$ Hermitian
conjugated partners in a separate group.
The anticommuting part of this second group manifests
properties of the ``basis vectors'' of the two orthogonal groups,
having their Hermitian conjugated partners within their group.
This second group of commuting and anticommuting ``basis
vectors''  manifest the properties of the Fadeev-Popov
ghosts~\cite{n2023MDPI}.

{\bf iv.\,\,\,} There remain questions to be answered:\\
{\bf iv.a.\,}
Can the theory achieve renormalizability by extending the ``basis
vectors'' in $2(2n+1)$-dimensional spaces with the ''basis vectors''
of strings, or by extending the ``basis vectors'' in
$2(2n+1)$-dimensional spaces to odd-dimensional,  $2(2n+1) +1$,
spaces?\\
{\bf iv.b.\,} Do two kinds
of boson fields, ${}^{I}{\hat{\cal A}}^{m \dagger}_{f}$ and
${}^{II}{\hat{\cal A}}^{m \dagger}_{f}$, appearing in this
theory (offering the interpretation of the Feynman diagrams and
elegantly confirming the requirement of the two kinds
of fields, $\omega_{ab \alpha}$ and
$\tilde{\omega}_{ab \alpha}$, appearing in
Eq.~(\ref{wholeaction}) and used so far in the
{\it spin-charge-family} theory~(\cite{nh2021RPPNP} and
references therein)) offer the correct (true) description of the boson
fields?\\
{\bf iv.c.\,}
Does this way of describing the internal spaces of fermion and
boson fields offer easier explanation for breaking symmetries
from $SO(13,1)$ to $SO(3,1)\times U(1)\times SU(3)$?%
~\footnote{
In Ref.~(\cite{nh2021RPPNP}, and references therein) the
internal space of $SO(13,1)$ breaks to $SO(7,1)\times SU(3)
\times U(1)$ due to the condensate of the two right-handed
neutrinos with the quantum numbers, which leave $SU(3)$
gluons, $SU(2)$ weak bosons, $U(1)$ photon and also $SO(3,1)$
gravitons as well as quarks and leptons of two groups with four
family members massless, while the scalar fields at the
electroweak break give masses to all quarks and leptons and
antiquarks and antileptons and  weak bosons, leaving massless
photons and gluons and gravitons.
}
Can the breaks of symmetries explain why photons interact with
all quarks and charged leptons and antiquarks and charged
antileptons but not with neutrinos?\\
{\bf iv.c.\,} Can in this theory appear the gravitino?\\
{\bf iv.d.\,} How has our universe gotten non-zero momenta
only in $d=(3+1)$?\\
{\bf iv.e.\,} Does the description of the internal spaces of
fermion and boson fields with the ``basic vectors'' help to
understand the history of our universe better (``open the new
door'' in understanding nature)?\\
{\bf iv.f.\,} And many other problems to be solved.~\footnote{
Let us add:
We demonstrate in this paper the description of all the boson fields:
The vector gauge fields  with $\alpha = (0,1,2,3)$ and
the scalar gauge fields $\alpha =(5,6,7,...,d)$.
It turns out that all the vector gauge fields, observed in $d=(3+1)$,
have only two nilpotents all the rest are projectors; in front of
and after or in between these two nilpotents are projectors ---
photons (they are products of projectors only), gravitons (they can
have two nilpotents, $ \stackrel{03}{(\pm i)}\stackrel{12}{(\pm 1)}$,
all the rest are projectors), weak bosons have two nilpotents in
$ \stackrel{56}{(\pm 1)}\stackrel{78}{(\pm 1)}$, all the rest in front
and after these two nilpotents are projectors, gluons have two
nilpotents --- $ \stackrel{9\,10}{(\pm 1)}
\stackrel{11\,12}{(\pm 1)}\stackrel{13\,14}{[\pm 1]}$,
or $ \stackrel{9\,10}{(\pm 1)}
\stackrel{11\,12}{[\pm 1])}\stackrel{13\,14}{(\pm 1)}$
or $ \stackrel{9\,10}{[\pm ]}
\stackrel{(11\,12}{(\pm 1)}\stackrel{13\,14}{(\pm 1)}$
all the rest in front of or after these two nilpotents
are projectors.
In the case of strong fields, we can not use Feynman diagrams, not in any of
these fields, we need approximate methods in all cases.
}.

\appendix

\section{Grassmann and Clifford algebras}
\label{grassmannclifford}

This part is taken from Ref.~\cite{n2022epjc,nh2023dec}, following Refs.~~\cite{norma92,norma93,nh2021RPPNP,JMP2013}.

The internal spaces of anti-commuting or commuting second quantized fields can be
described by using either the Grassmann or the Clifford algebras.

In Grassmann $d$-dimensional space there are $d$ anti-commuting (operators)
$\theta^{a}$,
and $d$ anti-commuting operators which are derivatives with respect to $\theta^{a}$,
$\frac{\partial}{\partial \theta_{a}}$.
%
\begin{eqnarray}
\label{thetaderanti0}
\{\theta^{a}, \theta^{b}\}_{+}=0\,, \, && \,
\{\frac{\partial}{\partial \theta_{a}}, \frac{\partial}{\partial \theta_{b}}\}_{+} =0\,,
\nonumber\\
\{\theta_{a},\frac{\partial}{\partial \theta_{b}}\}_{+} &=&\delta^b_{a}\,,
\;(a,b)=(0,1,2,3,5,\cdots,d)\,.
\end{eqnarray}
Making a choice
\begin{eqnarray}
(\theta^{a})^{\dagger} &=& \eta^{a a} \frac{\partial}{\partial \theta_{a}}\,,\quad
{\rm leads \, to} \quad
(\frac{\partial}{\partial \theta_{a}})^{\dagger}= \eta^{a a} \theta^{a}\,,
\label{thetaderher0}
\end{eqnarray}
with $\eta^{a b}=diag\{1,-1,-1,\cdots,-1\}$.

$ \theta^{a}$ and $ \frac{\partial}{\partial \theta_{a}}$ are, up to the sign, Hermitian
conjugated to each other. The identity is a self-adjoint member of the algebra.
Choosing the following complex properties of $\theta^a$ in even dimensional spaces,
\begin{small}
\begin{eqnarray}
\label{complextheta}
\{\theta^a\}^* &=& (\theta^0, \theta^1, - \theta^2, \theta^3, - \theta^5,
\theta^6,...,- \theta^{d-1}, \theta^d)\,,
\end{eqnarray}
\end{small}
leads to $\;\;\, $ 
$\{\frac{\partial}{\partial \theta_{a}}\}^* = (\frac{\partial}{\partial \theta_{0}},
\frac{\partial}{\partial \theta_{1}}, - \frac{\partial}{\partial \theta_{2}},
\frac{\partial}{\partial \theta_{3}}, - \frac{\partial}{\partial \theta_{5}},
\frac{\partial}{\partial \theta_{6}},..., - \frac{\partial}{\partial \theta_{d-1}},
\frac{\partial}{\partial \theta_{d}})\,. $
%

There are in $d$-dimensional space $2^d$ superposition of products of 
$\theta^{a}$, the Hermitian conjugated partners of which are the 
corresponding superposition of products
of $\frac{\partial}{\partial \theta_{a}}$~\cite{n2019PRD}.

There exist two kinds of the Clifford algebra elements (operators), $\gamma^{a}$ and
$\tilde{\gamma}^{a}$, expressible with $\theta^{a}$'s and their conjugate momenta
$p^{\theta a}= i \,\frac{\partial}{\partial \theta_{a}}$ ~\cite{norma93},
Eqs.~(\ref{thetaderanti0}, \ref{thetaderher0}),
\begin{eqnarray}
\label{clifftheta1}
\gamma^{a} &=& (\theta^{a} + \frac{\partial}{\partial \theta_{a}})\,, \quad
\tilde{\gamma}^{a} =i \,(\theta^{a} - \frac{\partial}{\partial \theta_{a}})\,,\nonumber\\
\theta^{a} &=&\frac{1}{2} \,(\gamma^{a} - i \tilde{\gamma}^{a})\,, \quad
\frac{\partial}{\partial \theta_{a}}= \frac{1}{2} \,(\gamma^{a} + i \tilde{\gamma}^{a})\,,
\nonumber\\
\end{eqnarray}
offering together $2\cdot 2^d$ operators: $2^d$ are superposition of products of
$\gamma^{a}$ and $2^d$ of $\tilde{\gamma}^{a}$.
It is easy to prove if taking into account Eqs.~(\ref{thetaderanti0}, \ref{thetaderher0}, \ref{clifftheta1}),
that they form two anti-commuting Clifford subalgebras,
$\{\gamma^{a}, \tilde{\gamma}^{b}\}_{+} =0$, Refs.~(\cite{nh2021RPPNP} and
references therein)
\begin{eqnarray}
\label{gammatildeantiher0}
\{\gamma^{a}, \gamma^{b}\}_{+}&=&2 \eta^{a b}= \{\tilde{\gamma}^{a},
\tilde{\gamma}^{b}\}_{+}\,, \nonumber\\
\{\gamma^{a}, \tilde{\gamma}^{b}\}_{+}&=&0\,,\quad
(a,b)=(0,1,2,3,5,\cdots,d)\,, \nonumber\\
(\gamma^{a})^{\dagger} &=& \eta^{aa}\, \gamma^{a}\, , \quad
(\tilde{\gamma}^{a})^{\dagger} = \eta^{a a}\, \tilde{\gamma}^{a}\,.
\end{eqnarray}

The Grassmann algebra offers the description of the {\it anti-commuting integer spin
second quantized fields} and of the {\it commuting integer spin second quantized
fields}~\cite{2020PartIPartII,nh2021RPPNP}. The Clifford algebras, which are
superposition of odd products of either $\gamma^a$'s or $\tilde{\gamma}^a$'s offer
the description of the second quantized half integer spin fermion fields, which from
the point of the subgroups of the $SO(d-1,1)$ group manifest spins and charges of
fermions and antifermions in the fundamental representations of the group and
subgroups~\cite{nh2021RPPNP}, Table~\ref{Table so13+1.} represents one family of quarks and leptons and antiquarks and antileptons.

The superposition of even products of either $\gamma^a$'s or 
$\tilde{\gamma}^a$'s
offer the description of the commuting second quantized boson fields with integer spins
(as we can see in Refs.~\cite{n2022epjc,n2023NPB,n2023MDPI}),
manifesting from the point of the subgroups of the $SO(d-1,1)$ group, spins and charges
in the adjoint representations.

There is so far observed only one kind of anti-commuting half-integer spin 
second quantized fields. 
two kinds of fermion fields have been observed so far.

The {\it postulate}, which determines how does $\tilde{\gamma}^{a}$
operate on $\gamma^a$, reduces the two Clifford subalgebras, $\gamma^a$ and
$\tilde{\gamma}^a$, to one, to the one described by
$\gamma^a$~\cite{nh02,norma93,JMP2013}
\begin{eqnarray}
\{\tilde{\gamma}^a B &=&(-)^B\, i \, B \gamma^a\}\, |\psi_{oc}>\,,
\label{tildegammareduced0}
\end{eqnarray}
with $(-)^B = -1$, if $B$ is (a function of) odd products of $\gamma^a$'s, otherwise
$(-)^B = 1$~\cite{nh02}, the vacuum state $|\psi_{oc}>$ is defined in
Eq.~(\ref{vaccliffodd}) of Subsect.~\ref{basisvectors}.

\vspace{2mm}

After the postulate of Eq.~(\ref{tildegammareduced0}) no vector space of
$\tilde{\gamma}^{a}$'s needs to be taken into account for the description 
of the internal space of either fermions or bosons, in agreement with the 
observed properties of fermions and bosons. Also, the Grassmann algebra 
is reduced to only one of the Clifford subalgebras.

 %
\section{$\gamma^a$ and $\tilde{\gamma}^a$ applying on
 odd and even ``basis vectors''}
\label{gammatilde}

Let us notice: Although the odd ($\hat{b}^{m \dagger}_{f }$)
and the even (${}^{i}{\hat{\cal A}}^{m \dagger}_{f}\,, i=(I,II)$)
``basis vectors'' have so different properties in even dimensional spaces,
the algebraic multiplication of one kind of ``basis vectors'' by either
$\gamma^a$ or by $\tilde{\gamma}^a$ transforms one kind into
another~\footnote{
As presented in Ref.~\cite{n2023MDPI} and 
mentioned in the Introduction, the odd-dimensional spaces offer the surprise:
Half of ``basis vectors'' manifest properties of those in even-dimensional
spaces of one lower dimension, the remaining half are the anti-commuting
``basis vectors'' appearing in two orthogonal groups with the Hermitian
conjugated partners within the same groups, the commuting ``basis
vectors'' appear in two separate groups, Hermitian conjugated to each
other, manifesting a special kind of supersymmetry.
}.\\

The algebraic multiplication of any odd ``basis vector'' by
$\gamma^a$ transforms it to the corresponding even
``basis vector'' ${}^{II}{\hat{\cal A}}^{m \dagger}_{f}$.\\
The algebraic multiplication of any odd ``basis vector'' by
$\tilde{\gamma}^a$ transforms it to the corresponding even
``basis vector'' ${}^{I}{\hat{\cal A}}^{m \dagger}_{f}$.\\

The algebraic multiplication of any even ``basis vector''
${}^{II}{\hat{\cal A}}^{m \dagger}_{f}$ by $\tilde{\gamma}^a$
transforms it to the corresponding odd ``basis vector''~\footnote{
These resemble a kind of supersymmetry: The same number of the 
odd and the even ``basis vectors'', and the simple relations between
fermions and bosons. 
}.

\vspace{1mm}

\noindent
{\bf i.} While the odd "basis vectors" in even dimensional spaces appear in
$2^{\frac{d}{2}-1}$ families, each family having $2^{\frac{d}{2}-1}$
members, and have their Hermitian conjugated partners in a separate group,
again with $2^{\frac{d}{2}-1}\times 2^{\frac{d}{2}-1}$ members, appear
the even "basis vectors" in even dimensional spaces in two groups,
each with $2^{\frac{d}{2}-1}\times 2^{\frac{d}{2}-1}$ members, having
the Hermitian conjugated partners within the same group. They have no
families.\\
{\bf ii.} The odd "basis vectors" in even dimensional spaces carry the
eigenvalues of the Cartan subalgebra members, Eq.~(\ref{cartangrasscliff}),
$\pm \frac{i}{2}$ or $\pm \frac{1}{2}$.
The even "basis vectors" in even dimensional spaces carry the
eigenvalues of the Cartan subalgebra members, Eq.~(\ref{cartangrasscliff}),
$(\pm i, 0)$ or $(\pm 1,0$). \\

Correspondingly, the even ``basis vectors''
${}^{I}{\hat{\cal A}}^{m \dagger}_{1}$ applying on the odd 
``basis vectors'' transform them to another odd ``basis vectors'',
transferring to them integer values of the Cartan subalgebra eigenvalues, 
or to zero.


The properties of the even and odd ``basis vectors'' are discussed 
in details for $d=(5+1)$-dimensional internal spaces in
Subsect.~\ref{cliffordoddevenbasis5+1}.

Namely, $d=(5+1)$-dimensional 
internal spaces offer more possibilities and an easier understanding of 
what offer general  $d=((d-1)+1)$-dimensional internal spaces for 
even $d$. \\

\section {Demonstration of properties of ``basis vectors'' on 
simple cases $d=(1+1)$ and $d=(3+1)$, discussed in 
Sec.~\ref{basisvectors0}}
\label{simplecases}

\begin{small}

We present in this appendix a few simple illustrations of the properties of 
fermion and boson ``basis vectors'' in the case $d=(1+1)$ and 
$d=(3+1)$.

\vspace{2mm}

{\bf a. \,\,} {\it Let us start with the presentation of the ``basis vectors'', 
in the simplest cases $d=(1+1)$ and $d=(3+1)$.} 

The presentation is a short overview of Refs.~\cite{n2023NPB,n2023MDPI}. is a short overview of Refs.~\cite{n2023NPB,n2023MDPI}.

\vspace{1mm}

{\bf $d=(1+1)$}

\vspace{1mm}

There are $4 \, (2^{d=2})$ ``eigenvectors" of the Cartan subalgebra  
members, Eq.~(\ref{cartangrasscliff}),
$S^{01}$ and ${\bf {\cal S}}^{01}$ of the Lorentz algebra $S^{ab}$ and 
${\bf {\cal S}}^{ab}$ $= S^{01} + \tilde{S}^{01}$ ($S^{ab}= \frac{i}{4} 
\{\gamma^a ,\gamma^b\}_{-}$,  $\tilde{S}^{ab}= \frac{i}{4} 
\{\tilde{\gamma}^a ,\tilde{\gamma}^b\}_{-}$), representing one odd 
``basis vector'' $\hat{b}^{ 1 \dagger}_{1}=$ $\stackrel{01}{(+i)}$ (m=1), 
appearing in one family (f=1) and correspondingly one Hermitian conjugated 
partner $\hat{b}^{ 1}_1=$ $\stackrel{01}{(-i)}$~\footnote{
It is our choice which one, $\stackrel{01}{(+i)}$
or $\stackrel{01}{(-i)}$, we choose as the ``basis vector'' 
$ \hat{b}^{ 1 \dagger}_1$,  the remaining is  its Hermitian conjugated 
partner. The choice of the ``basis vectors''
determines the vacuum state $|\psi_{oc}>$, Eq.~(\ref{vaccliffodd}). 
For $ \hat{b}^{ 1 \dagger}_1=$ $ \stackrel{01}{(+i)}$, the vacuum state is 
$|\psi_{oc}>= \stackrel{01}{[-i]}$ 
(due to the requirement that  $\hat{b}^{ 1 \dagger}_1 |\psi_{oc}>$ is nonzero,
while $\hat{b}^{ 1}_1 |\psi_{oc}>$ is zero). $ \stackrel{01}{[-i]}$  is the Clifford 
even object.} 
and two even ``basis vector'' ${}^{I}{\bf {\cal A}}^{1 \dagger}_{1}=
\stackrel{01}{[+i]}$ and ${}^{II}{\bf {\cal A}}^{1 \dagger}_{1}=
\stackrel{01}{[-i]}$, both self adjoint.

Correspondingly we have, after using  Eqs.~(\ref{gammatildeantiher}, \ref{signature0}), 
two odd and two even eigenvectors of the Cartan subalgebra members.

\end{small}
\begin{small} 
 \begin{eqnarray}
 \label{1+1oddeven}
 && {\rm  \;Clifford \;odd}\nonumber\\
 \hat{b}^{ 1 \dagger}_{1}&=&\stackrel{01}{(+i)}\,, \quad 
 \hat{b}^{ 1 }_{1}=\stackrel{01}{(-i)}\,,\nonumber\\
 &&{\rm \;Clifford \;even} \;\nonumber\\
 {}^{I}{\bf {\cal A}}^{1 \dagger}_{1}&=&\stackrel{01}{[+i]}\,, \quad 
{}^{II}{\bf {\cal A}}^{1 \dagger}_{1}=\stackrel{01}{[-i]}\,.
 \end{eqnarray}
 The two odd ``basis vectors'' are Hermitian conjugated
 to each other, the two even ``basis vectors'' are orthogonal. 
 Taking into account Eq.~(\ref{signature0}), one finds: 
 $S^{01}\stackrel{01}{( \pm i)}=\pm \frac{i}{2}\stackrel{01}{( \pm i)}$ and 
 ${\bf {\cal S}}^{01} \stackrel{01}{[\pm i]}=0$.
 
 The reader can easily check Eqs.~(\ref{calIAb1234gen}, \ref{calbIIA1234gen}).
 \end{small}

 \begin{small}
 
 \vspace{2mm}

{\bf $d=(3+1)$}

\vspace{2mm}
 
There are $16 \, (2^{d=4})$ ``eigenvectors" of the Cartan subalgebra  
members ($S^{03}, S^{12}$) and (${\bf {\cal S}}^{03}, {\bf {\cal S}}^{12}$) of the Lorentz algebras $S^{ab}$ and ${\bf {\cal S}}^{ab}$, presented in
Eq.~(\ref{cartangrasscliff}).

Half of them are the odd ``basis vectors'', appearing in two  families 
$2^{\frac{4}{2}-1}$, $f=(1,2)$), each with two ($2^{\frac{4}{2}-1}$, $m=(1,2)$), 
members, $\hat{b}^{ m \dagger}_{f}$.   

There are  $2^{\frac{4}{2}-1}\times $
$2^{\frac{4}{2}-1} $ Hermitian conjugated partners  $\hat{b}^{ m}_{f}$ appearing in a separate group.

 Choosing the right handed  odd ``basis vectors'' as
 \end{small}
\begin{small}
\begin{eqnarray}
\label{3+1oddb}
\begin{array} {ccrr}
f=1&f=2&&\\
\tilde{S}^{03}=\frac{i}{2}, \tilde{S}^{12}=-\frac{1}{2}&
\;\;\tilde{S}^{03}=-\frac{i}{2}, \tilde{S}^{12}=\frac{1}{2}\;\;\; &S^{03}\, &S^{12}\\
\hat{b}^{ 1 \dagger}_{1}=\stackrel{03}{(+i)}\stackrel{12}{[+]}&
\hat{b}^{ 1 \dagger}_{2}=\stackrel{03}{[+i]}\stackrel{12}{(+)}&\frac{i}{2}&
\frac{1}{2}\\
\hat{b}^{ 2 \dagger}_{1}=\stackrel{03}{[-i]}\stackrel{12}{(-)}&
\hat{b}^{ 2 \dagger}_{2}=\stackrel{03}{(-i)}\stackrel{12}{[-]}&-\frac{i}{2}&
-\frac{1}{2}\,,
\end{array}
\end{eqnarray}
\end{small}
\begin{small}
we find for their Hermitian conjugated partners ($\hat{b}^{ m}_{f}=
(\hat{b}^{ m \dagger}_{f})^{\dagger}$)
\end{small}
\begin{small}
\begin{eqnarray}
\label{3+1oddHb}
\begin{array} {ccrr}
S^{03}=- \frac{i}{2}, S^{12}=\frac{1}{2}&
\;\;S^{03}=\frac{i}{2}, S^{12}=-\frac{1}{2}\;\;&\tilde{S}^{03} &\tilde{S}^{12}\\
\hat{b}^{ 1 }_{1}=\stackrel{03}{(-i)}\stackrel{12}{[+]}&
\hat{b}^{ 1 }_{2}=\stackrel{03}{[+i]}\stackrel{12}{(-)}&-\frac{i}{2}&
-\frac{1}{2}\\
\hat{b}^{ 2 }_{1}=\stackrel{03}{[-i]}\stackrel{12}{(+)}&
\hat{b}^{ 2 }_{2}=\stackrel{03}{(+i)}\stackrel{12}{[-]}&\frac{i}{2}&
\frac{1}{2}\,.
\end{array}
\end{eqnarray}
\end{small}
\begin{small}
The vacuum state on which the odd ''basis vectors'' apply is equal to:\\
$|\psi_{oc}>= \frac{1}{\sqrt{2}} (\stackrel{03}{[-i]}\stackrel{12}{[+]}
  +\stackrel{03}{[+i]}\stackrel{12}{[-]} )$~\footnote{
The case $SO(1,1)$ can be viewed as a subspace of the case $SO(3,1)$,
recognizing the ``basis vectors'' $\stackrel{03}{(+i)}\stackrel{12}{[+]}$
and $\stackrel{03}{(-i)}\stackrel{12}{[-]}$ with  $\stackrel{03}{(+i)}$ and 
$\stackrel{03}{(-i)}$, respectively, as carrying two different handedness in
$d=(1+1)$, but each of them carries also a different ``charge'' $S^{12}$. 
}.

Let us recognize that all the odd ''basis  vectors'' are orthogonal: 
$\hat{b}^{ m \dagger}_{f} *_{A} \hat{b}^{ m' \dagger}_{f '}=0, 
\forall (m,m',f,f^{`})$, and so are orthogonal among themselves their
Hermitian conjugated partners.
 
 \end{small}
 
\begin{small}
Let us present  $2^{\frac{4}{2}-1}\times 2^{\frac{4}{2}-1} $ even ''basis vectors'',
the members of the group ${}^{I}{\bf {\cal A}}^{m \dagger}_{f}$, which are Hermitian conjugated to each other or are self adjoint~\footnote{
Let be repeated that ${\bf {\cal S}}^{ab}=(S^{ab} + \tilde{S}^{ab}) $~\cite{n2022epjc}.}

\end{small}
\begin{small}
\begin{eqnarray}
\label{3+1evenAI}
\begin{array} {crrcrr}
&{\bf {\cal S}}^{03}&{\bf {\cal S}}^{12}&&{\bf {\cal S}}^{03}&{\bf {\cal S}}^{12}\\
{}^{I}{\bf {\cal A}}^{1 \dagger}_{1}= \stackrel{03}{[+i]}\stackrel{12}{[+]}&0&0&\,,
{}^{I}{\bf {\cal A}}^{1 \dagger}_{2}= \stackrel{03}{(+i)}\stackrel{12}{(+)}&i&1\\
{}^{I}{\bf {\cal A}}^{2 \dagger}_{1}= \stackrel{03}{(-i)}\stackrel{12}{(-)}&-i&-1&\,,
{}^{I}{\bf {\cal A}}^{2 \dagger}_{2}= \stackrel{03}{[-i]}\stackrel{12}{[-]}&0&0\,,
\end{array}
\end{eqnarray}
\end{small}
\begin{small}
and  $2^{\frac{4}{2}-1}\times 2^{\frac{4}{2}-1} $  even ''basis vectors'',
the members of the group ${}^{II}{\bf {\cal A}}^{m \dagger}_{f}$, $m=(1,2), f=(1,2)$,
which are again Hermitian conjugated to each other or are self adjoint
\end{small}
\begin{small}
\begin{eqnarray}
\label{3+1evenAII}
\begin{array} {crrcrr}
&{\bf {\cal S}}^{03}&{\bf {\cal S}}^{12}&&{\bf {\cal S}}^{03}&{\bf {\cal S}}^{12}\\
{}^{II}{\bf {\cal A}}^{1 \dagger}_{1}= \stackrel{03}{[+i]}\stackrel{12}{[-]}&0&0&\,,
{}^{II}{\bf {\cal A}}^{1 \dagger}_{2}= \stackrel{03}{(+i)}\stackrel{12}{(-)}&i&-1\\
{}^{II}{\bf {\cal A}}^{2 \dagger}_{1}= \stackrel{03}{(-i)}\stackrel{12}{(+)}&-i&1&\,,
{}^{II}{\bf {\cal A}}^{2 \dagger}_{2}= \stackrel{03}{[-i]}\stackrel{12}{[+]}&0&0\,.
\end{array}
\end{eqnarray}
\end{small}
\begin{small}
The even ``basis vectors'' have no families. The two groups 
$ {}^{I}{\bf {\cal A}}^{m \dagger}_{f}$ and 
${}^{II}{\bf {\cal A}}^{m \dagger}_{f}$ (they are not connected by 
${\bf {\cal S}}^{ab}$) are orthogonal.
\begin{eqnarray}
\label{AIAIIorth1}
{}^{I}{\bf {\cal A}}^{m \dagger}_{f} *_{A} {}^{II}{\bf {\cal A}}^{m' \dagger}_{f `} 
=0, \quad{\rm for \;any } \;(m, m', f, f `)\,.
\end{eqnarray}
Let us point out that in even dimensional spaces have the internal spaces
(the ``basis vectors'') of fermion
and boson fields the properties of the fermion and boson second quantized
fields~\cite{n2022epjc}.
\end{small} 
\begin{small}

\vspace{1mm}

{\bf b. \,\,} {\it Let us demonstrate  on the two cases, $d=(1+1)$ and 
$d=(3+1)$ that the even ``basis vectors'' of bosons can be presented as 
algebraic products of the odd ``basis vectors'' and one of their Hermitian 
conjugated},\\
 as demonstrated in Subsect.~\ref{even}, 
Eq.~(\ref{AIIbdaggerb}).  

The middle two rows of 
Eq.~(\ref{usefulrel0}) will help the reader to reproduce the derivations 
which follow. 

\vspace{1mm}

For $d=(1+1)$ we find: \\
${}^{I}{\hat{\cal A}}^{1\dagger}_{1} (\equiv \stackrel{01}{[+i]})=
\hat{b}^{1 \dagger}_{1} (\equiv \stackrel{01}{(+i)})
\, *_A\,\hat{b}^{1 }_{1}(\equiv \stackrel{01}{(-i)})$, and 
${}^{II}{\hat{\cal A}}^{1\dagger}_{1} (\equiv
\stackrel{01}{[-i]})=\hat{b}^{1}_{1} (\equiv \stackrel{01}{(-i)})
\, *_A\,\hat{b}^{1\dagger }_{1}(\equiv \stackrel{01}{(+i)})$,
where $\hat{b}^{m}_{f} = (\hat{b}^{m \dagger}_{f})^{\dagger}$.\\

For $d=(3+1)$ we find using first family when expressing 
${}^{I}{\hat{\cal A}}^{m\dagger}_{f}$ (the reader can prove that one 
obtains the same ${}^{I}{\hat{\cal A}}^{m\dagger}_{f}$  if using the second family), and when evaluating ${}^{II}{\hat{\cal A}}^{m\dagger}_{f}$ we  
use the first member of the two families (the reader can prove that one 
obtains the same ${}^{II}{\hat{\cal A}}^{m\dagger}_{f}$  if using the 
second second member of the two families): \\ 
${}^{I}{\hat{\cal A}}^{1\dagger}_{1} (\equiv
\stackrel{03}{[+i]}\stackrel{12}{[+]})=\hat{b}^{1 \dagger}_{1} (\equiv \stackrel{03}{(+i)}\stackrel{12}{[+]})
\, *_A\,\hat{b}^{1}_{1} (\equiv \stackrel{03}{(-i)}\stackrel{12}{[+]})$, \,
${}^{I}{\hat{\cal A}}^{2 \dagger}_{1} (\equiv
\stackrel{03}{(-i)}\stackrel{12}{(-)})=\hat{b}^{2 \dagger}_{1} (\equiv \stackrel{03}{[-i]}\stackrel{12}{(-)})
\, *_A\,\hat{b}^{1}_{1} (\equiv \stackrel{03}{(-i)}\stackrel{12}{[+]})$, \\
${}^{I}{\hat{\cal A}}^{1 \dagger}_{2} (\equiv
\stackrel{03}{(+i)}\stackrel{12}{(+)})=\hat{b}^{1 \dagger}_{1} (\equiv \stackrel{03}{(+i)}\stackrel{12}{[+]})
\, *_A\,\hat{b}^{2}_{1} (\equiv \stackrel{03}{[-i]}\stackrel{12}{(+)})$,\,
${}^{I}{\hat{\cal A}}^{2\dagger}_{2} (\equiv
\stackrel{03}{[-i]}\stackrel{12}{[-]})=\hat{b}^{2 \dagger}_{1} (\equiv \stackrel{03}{[-i]}\stackrel{12}{(-)})\, *_A\,
\hat{b}^{2}_{1} (\equiv \stackrel{03}{[-i]}\stackrel{12}{(+)})$, and\\
${}^{II}{\hat{\cal A}}^{1\dagger}_{1} (\equiv
\stackrel{03}{[+i]}\stackrel{12}{[-]})=\hat{b}^{1}_{2} (\equiv 
\stackrel{03}{[+i]}\stackrel{12}{(-)})\, *_A\,\hat{b}^{1 \dagger}_{2} (\equiv \stackrel{03}{[+i]}\stackrel{12}{(+)})$,\,
${}^{II}{\hat{\cal A}}^{2\dagger}_{1} (\equiv
\stackrel{03}{(-i)}\stackrel{12}{(+)})=\hat{b}^{1}_{1} (\equiv 
\stackrel{03}{(-i)}\stackrel{12}{[+]})\, *_A\,\hat{b}^{1 \dagger}_{2} (\equiv \stackrel{03}{[+i]}\stackrel{12}{(+)})$,\\
${}^{II}{\hat{\cal A}}^{1\dagger}_{2} (\equiv
\stackrel{03}{(+i)}\stackrel{12}{(-)})=\hat{b}^{1}_{2} (\equiv 
\stackrel{03}{[+i]}\stackrel{12}{(-)})\, *_A\,\hat{b}^{1\dagger}_{1} (\equiv \stackrel{03}{(+i)}\stackrel{12}{[+]})$,\,
${}^{II}{\hat{\cal A}}^{2\dagger}_{2} (\equiv
\stackrel{03}{[-i]}\stackrel{12}{[+]})=\hat{b}^{1}_{1} (\equiv 
\stackrel{03}{(-i)}\stackrel{12}{[+]})\, *_A\,\hat{b}^{1\dagger}_{1} (\equiv \stackrel{03}{(+i)}\stackrel{12}{[+]})$.\,\\
\end{small} 

\begin{small}

\vspace{1mm}

{\bf c. \,\,} {\it Let us demonstrate  on the two cases, $d=(1+1)$ and 
$d=(3+1)$, what does the application of $\gamma^a$ and 
$\tilde{\gamma}^a$ on the odd and even ``basis
 vectors'' cause.}\\

For $d=(1+1)$ we find, using Eq.~(\ref{usefulrel0}): \\
$\gamma^0 \,{}^{I}{\hat{\cal A}}^{1\dagger}_{1} (\equiv \stackrel{01}{[+i]})=
(\hat{b}^{1 \dagger}_{1})^{\dagger} (\equiv \stackrel{01}{(-i)})$\,,\,
$\tilde{\gamma}^0\,{}^{I}{\hat{\cal A}}^{1\dagger}_{1} (\equiv \stackrel{01}{[+i]})=i\, \hat{b}^{1 \dagger}_{1} (\equiv \stackrel{01}{(+i)})$\,,\,\\
$\gamma^0 \,\hat{b}^{1 \dagger}_{1} (\equiv \stackrel{01}{(+i)})=
{}^{II}{\hat{\cal A}}^{1\dagger}_{1} (\equiv \stackrel{01}{[-i]})$\,,\,
$\tilde{\gamma}^0\,\hat{b}^{1 \dagger}_{1} (\equiv \stackrel{01}{(+i)})=
-i \,{}^{I}{\hat{\cal A}}^{1\dagger}_{1} (\equiv \stackrel{01}{[+i]})$\,.
\\
\vspace{1mm}

For $d=(3+1)$ we find: \\
$\gamma^0 \,{}^{I}{\hat{\cal A}}^{1\dagger}_{1} (\equiv \stackrel{03}{[+i]}
\stackrel{12}{[+]})=
(\hat{b}^{1 \dagger}_{1})^{\dagger} (\equiv \stackrel{03}{(-i)}\stackrel{12}{[+]})$\,,\,
$\tilde{\gamma}^0\,{}^{I}{\hat{\cal A}}^{1\dagger}_{1} (\equiv \stackrel{03}{[+i]} \stackrel{12}{[+]})=i\, \hat{b}^{1 \dagger}_{1} (\equiv \stackrel{03}{(+i)} \stackrel{12}{[+]})$\,,\,\\
$\gamma^0 \, \hat{b}^{1 \dagger}_{1} (\equiv \stackrel{03}{(+i)}\stackrel{12}{[+]})={}^{I}{\hat{\cal A}}^{1\dagger}_{1} (\equiv \stackrel{03}{[-i]} \stackrel{12}{[+]})$\,,\,
$\tilde{\gamma}^0\, \hat{b}^{1 \dagger}_{1} (\equiv \stackrel{03}{(+i)}\stackrel{12}{[+]})=-i\,{}^{I}{\hat{\cal A}}^{1\dagger}_{1} (\equiv \stackrel{03}{[+i]} \stackrel{12}{[+]})$\,.\\
$\gamma^1 \,{}^{I}{\hat{\cal A}}^{1\dagger}_{1} (\equiv \stackrel{03}{[+i]}
\stackrel{12}{[+]})=(\hat{b}^{1 \dagger}_{2} )^{\dagger}(\equiv \stackrel{03}{[+i]} \stackrel{12}{(-)})$\,,\,
$\tilde{\gamma}^1\,{}^{I}{\hat{\cal A}}^{1\dagger}_{1} (\equiv \stackrel{03}{[+i]} \stackrel{12}{[+]})=i\, \hat{b}^{1 \dagger}_{2} (\equiv \stackrel{03}{[+i]} \stackrel{12}{(+)})$\,.\,\\
 \end{small}
 
The properties of the even and odd ``basis vectors'' are discussed 
in more details also for $d=(5+1)$-dimensional internal spaces, 
Subsect.~\ref{cliffordoddevenbasis5+1}; 
$d=(5+1)$-dimensional 
internal spaces, offering more possibilities than $d=(1+1)$ and
$d=(3+1)$, offer an easier understanding of 
what  general  $d=((d-1)+1)$-dimensional internal spaces for 
even $d$, offer. \\

\section{Creation and annihilation operators for fermions and bosons}
\label{creationanihilationoperators}

 This appendix is a short overview of the equivalent section in Ref.~\cite{n2023NPB}.
 
We learned in Subsect.~(\ref{basisvectors}) that in even
dimensional spaces ($d=2n)$ the odd and even ``basis vectors'', which 
are the superposition of the odd (for fermions) and the even (for 
bosons) products of $\gamma^a$'s, offer the
description of the internal spaces of fermion and boson
fields~\cite{n2023NPB}, respectively.

The odd ``basis vectors'', $\hat{b}^{m \dagger}_{f}$, offering
$2^{\frac{d}{2}-1}$ family members $m$ (generated by $S^{ab}$
from any member in a particular family $f$) in $2^{\frac{d}{2}-1}$ families $f$
(generated by $\tilde{S}^{ab}$ from any member), fulfil
together with their $2^{\frac{d}{2}-1}\times$ $2^{\frac{d}{2}-1}$
Hermitian conjugated partners, $\hat{b}^{m}_{f}$, the postulates 
for the second quantized fermion fields, Eq.~(\ref{almostDirac}) in 
this paper, Eq.(26) in Ref.~\cite{nh2021RPPNP}, explaining the 
second quantization postulate of Dirac.

The even ``basis vectors'', ${}^{i}{\hat{\cal A}}^{m \dagger}_{f}\,,
i =(I,II)$, appear in two orthogonal groups, each group with 
$2^{\frac{d}{2}-1}\times$ $2^{\frac{d}{2}-1}$ members 
(determined by ${\cal S}^{ab}= (S^{ab}+ 
\tilde{S}^{ab})$, each group having their Hermitian conjugated 
partners within its group) fulfil the postulates of the second quantized
boson fields manifesting as the gauge fields of fermion fields
described by the odd ``basis vectors'', $\hat{b}^{m \dagger}_{f}$, Eqs.~(\ref{AIAIIorth}-\ref{calbIIA1234gen}). The commutation 
relations of ${}^{i}{\hat{\cal A}}^{m \dagger}_{f}, i=(I,II),$
are presented in Eqs.~(\ref{AIAIIorth}, \ref{ruleAAI}).

The odd and the even ``basis vectors'' are chosen to be products of
nilpotents, $\stackrel{ab}{(k)}$ (with the odd number of nilpotents 
if describing fermions and the even number of nilpotents if 
describing bosons), and projectors, $\stackrel{ab}{[k]}$. Nilpotents 
and projectors are eigenvectors of the Cartan 
subalgebra members of the Lorentz algebra in the internal space of
$S^{ab}$ for the odd ``basis vectors'' and of ${\bf {\cal S}}^{ab}
 (=S^{ab}+ \tilde{S}^{ab}$) for the even ``basis vectors''.

\vspace{2mm}

To define the creation operators for fermion or boson fields,
besides the ``basis vectors'' defining the internal spaces of fermions 
and bosons, the basis in ordinary space in momentum or coordinate 
representation is needed. Here Ref.~\cite{nh2021RPPNP}, 
Subsect.~3.3 and App. J is overviewed. \\

Let us start by postulating the momentum part of the single-particle 
states. (The extended version is presented in 
Ref.~\cite{nh2021RPPNP} in Subsect.~3.3 and App. J.)
\begin{eqnarray}
\label{creatorp}
|\vec{p}>&=& \hat{b}^{\dagger}_{\vec{p}} \,|\,0_{p}\,>\,,\quad
<\vec{p}\,| = <\,0_{p}\,|\,\hat{b}_{\vec{p}}\,, \nonumber\\
<\vec{p}\,|\,\vec{p}'>&=&\delta(\vec{p}-\vec{p}')=
<\,0_{p}\,|\hat{b}_{\vec{p}}\; \hat{b}^{\dagger}_{\vec{p}'} |\,0_{p}\,>\,,
\nonumber\\
&&{\rm pointing \;out\;} \nonumber\\
<\,0_{p}\,| \hat{b}_{\vec{p'}}\, \hat{b}^{\dagger}_{\vec{p}}\,|\,0_{p}\, > &=&\delta(\vec{p'}-\vec{p})\,,
\end{eqnarray}
with the normalization $<\,0_{p}\, |\,0_{p}\,>=1$ (the operator 
$\hat{b}^{\dagger}_{\vec{p}}$ pushes a  single particle state with zero momentum by an amount $\vec{p}$).
While the quantized operators $\hat{\vec{p}}$ and $\hat{\vec{x}}$ commute, 
$\{\hat{p}^i\,, \hat{p}^j \}_{-}=0$ and $\{\hat{x}^k\,, \hat{x}^l \}_{-}=0$,
it follows for $\{\hat{p}^i\,, \hat{x}^j \}_{-}=i \eta^{ij}$. It then follows
\begin{small}
\begin{eqnarray}
\label{eigenvalue10}
<\vec{p}\,| \,\vec{x}>&=&<0_{\vec{p}}\,|\,\hat{b}_{\vec{p}}\;
\hat{b}^{\dagger}_{\vec{x}}
|0_{\vec{x}}\,>=(<0_{\vec{x}}\,|\,\hat{b}_{\vec{x}}\;
\hat{b}^{\dagger}_{\vec{p}} \,\,
|0_{\vec{p}}\,>)^{\dagger}\, \nonumber\\
<0_{\vec{p}}\,|\{\hat{b}^{\dagger}_{\vec{p}}\,, \,
\hat{b}^{\dagger}_{\vec{p}\,'}\}_{-}|0_{\vec{p}}\,>&=&0\,,\qquad
<0_{\vec{p}}\,|\{\hat{b}_{\vec{p}}\,, \,\hat{b}_{\vec{p}\,'}\}_{-}|0_{\vec{p}}\,>=0\,,\qquad
<0_{\vec{p}}\,|\{\hat{b}_{\vec{p}}\,, \,\hat{b}^{\dagger}_{\vec{p}\,'}\}_{-}|0_{\vec{p}}\,>=0\,,
\nonumber\\
<0_{\vec{x}}\,|\{\hat{b}^{\dagger}_{\vec{x}}\,, \,\hat{b}^{\dagger}_{\vec{x}\,'}\}_{-}|0_{\vec{x}}\,>&=&0\,,
\qquad
<0_{\vec{x}}\,|\{\hat{b}_{\vec{x}}\,, \,\hat{b}_{\vec{x}\,'}\}_{-}|0_{\vec{x}}\,>=0\,,\qquad
<0_{\vec{x}}\,|\{\hat{b}_{\vec{x}}\,, \,\hat{b}^{\dagger}_{\vec{x}\,'}\}_{-}|0_{\vec{x}}\,>=0\,,
\nonumber\\
<0_{\vec{p}}\,|\{\hat{b}_{\vec{p}}\,, \,\hat{b}^{\dagger}_{\vec{x}}\}_{-}|0_{\vec{x}}\,>&=&
e^{i \vec{p} \cdot \vec{x}} \frac{1}{\sqrt{(2 \pi)^{d-1}}}\,,\quad
<0_{\vec{x}}\,|\{\hat{b}_{\vec{x}}\,, \,\hat{b}^{\dagger}_{\vec{p}}\}_{-}|0_{\vec{p}}\,>=
e^{-i \vec{p} \cdot \vec{x}} \frac{1}{\sqrt{(2 \pi)^{d-1}}}\,.
\end{eqnarray}
\end{small}
The internal space of either fermion or boson fields has a finite number of ``basis
vectors'', $2^{\frac{d}{2}-1}\times 2^{\frac{d}{2}-1}$ for fermions (and the same
number of their Hermitian conjugated partners), and twice
$2^{\frac{d}{2}-1}\times 2^{\frac{d}{2}-1}$ for bosons; the momentum basis is
continuously infinite.\\

The creation operators for either fermions or bosons must be tensor products,
$*_{T}$, of both contributions, the ``basis vectors'' describing the internal space of
fermions or bosons and the basis in ordinary momentum or coordinate space.

The creation operators for a free massless fermion field of the energy
$p^0 =|\vec{p}|$, belonging to a family $f$ and to a superposition of
family members $m$ applying on the vacuum state
$|\psi_{oc}>\,*_{T}\, |0_{\vec{p}}>$ 
can be written as~(\cite{nh2021RPPNP}, Subsect.3.3.2, and the references therein)
\begin{small}
\begin{eqnarray}
\label{wholespacefermions}
{\bf \hat{b}}^{s \dagger}_{f} (\vec{p}) \,&=& \,
\sum_{m} c^{sm}{}_f (\vec{p}) \,\hat{b}^{\dagger}_{\vec{p}}\,*_{T}\,
\hat{b}^{m \dagger}_{f} \, \,,
\end{eqnarray}
\end{small}
where the vacuum state for fermions $|\psi_{oc}>\,*_{T}\, |0_{\vec{p}}> $
includes both spaces, the internal part, Eq.(\ref{vaccliffodd}), and the momentum
part, Eq.~(\ref{creatorp})~\footnote{
The creation operators and their Hermitian conjugated partner annihilation
operators in the coordinate representation can be written as
$\hat{\bf b}^{s \dagger}_{f }(\vec{x},x^0)=
\sum_{m} \,\hat{b}^{ m \dagger}_{f} \, *_{T}\, \int_{- \infty}^{+ \infty} \,
\frac{d^{d-1}p}{(\sqrt{2 \pi})^{d-1}} \, c^{s m }{}_{f}\;
(\vec{p}) \; \hat{b}^{\dagger}_{\vec{p}}\;
e^{-i (p^0 x^0- \varepsilon \vec{p}\cdot \vec{x})}
$
~(\cite{nh2021RPPNP}, subsect. 3.3.2., Eqs.~(55,57,64) and the references therein).}).

The creation operators $ \hat{\bf b}^{s\dagger}_{f }(\vec{p}) $ and their
Hermitian conjugated partners annihilation operators
$\hat{\bf b}^{s}_{f }(\vec{p}) $, creating and annihilating the single fermion
states, respectively, fulfil when applying the vacuum state,
$|\psi_{oc}> *_{T} |0_{\vec{p}}>$, the anti-commutation relations for the second quantized
fermions, postulated by Dirac (Ref.~\cite{nh2021RPPNP}, Subsect.~3.3.1,
Sect.~5).

\begin{small}
\begin{eqnarray}
<0_{\vec{p}}\,|
\{ \hat{\bf b}^{s' }_{f `}(\vec{p'})\,,\,
\hat{\bf b}^{s \dagger}_{f }(\vec{p}) \}_{+} \,|\psi_{oc}> |0_{\vec{p}}>&=&
\delta^{s s'} \delta_{f f'}\,\delta(\vec{p}' - \vec{p})\,\cdot |\psi_{oc}>
\,,\nonumber\\
\{ \hat{\bf b}^{s' }_{f `}(\vec{p'})\,,\,
\hat{\bf b}^{s}_{f }(\vec{p}) \}_{+} \,|\psi_{oc}> |0_{\vec{p}}>&=&0\, \cdot \,
|\psi_{oc}> |0_{\vec{p}}>
\,,\nonumber\\
\{ \hat{\bf b}^{s' \dagger}_{f '}(\vec{p'})\,,\,
\hat{\bf b}^{s \dagger}_{f }(\vec{p}) \}_{+}\, |\psi_{oc}> |0_{\vec{p}}>&=&0\, \cdot
\,|\psi_{oc}> |0_{\vec{p}}>
\,,\nonumber\\
\hat{\bf b}^{s \dagger}_{f }(\vec{p}) \,|\psi_{oc}> |0_{\vec{p}}>&=&
|\psi^{s}_{f}(\vec{p})>\,,\nonumber\\
\hat{\bf b}^{s}_{f }(\vec{p}) \, |\psi_{oc}> |0_{\vec{p}}>&=&0\, \cdot\,
\,|\psi_{oc}> |0_{\vec{p}}>\,, \nonumber\\
|p^0| &=&|\vec{p}|\,.
\label{Weylpp'comrel}
\end{eqnarray}
\end{small}

The creation operators for boson gauge fields must carry the space index
$\alpha$, describing the $\alpha$ component of the boson field in the
ordinary space~\footnote{
In the {\it spin-charge-family} theory, the vector gauge fields of quarks and
leptons and antiquarks and antileptons have the space index $\alpha=
(0,1,2,3)$, while the Higgs's scalars origin
in the boson gauge fields with the space index $\alpha=(7,8)$,
Refs.~(\cite{nh2021RPPNP},
Sect.~6.2, and the references therein; \cite{n2023NPB}, Eq.~(35)).}.
We, therefore, add the space index $\alpha$ as follows
\begin{eqnarray}
\label{wholespacebosons}
{\bf {}^{i}{\hat{\cal A}}^{m \dagger}_{f \alpha}} (\vec{p}) \,&=&
{}^{i}{\hat{\cal C}}^{ m}{}_{f \alpha} (\vec{p})\,*_{T}\,
{}^{i}{\hat{\cal A}}^{m \dagger}_{f} \, \,, i=(I,II)\,,
\end{eqnarray}
with ${}^{i}{\hat{\cal C}}^{ m}{}_{f \alpha} (\vec{p})=
{}^{i}{\cal C}^{ m}{}_{f \alpha}\,\hat{b}^{\dagger}_{\vec{p}}$, with
$\hat{b}^{\dagger}_{\vec{p}}$ defined in Eqs.~(\ref{creatorp}, \ref{eigenvalue10}).
We treat free massless bosons of momentum $\vec{p}$ and energy $p^0=|\vec{p}|$
and of particular ``basis vectors'' ${}^{i}{\hat{\cal A}}^{m \dagger}_{f}$'s which are
eigenvectors of all the Cartan subalgebra members~\footnote{
In the general case, the energy eigenstates of bosons are in a superposition of
${\bf {}^{i}{\hat{\cal A}}^{m \dagger}_{f}}$, for either $i=I$ or $i=II$.
}.
%

One example, in which the superposition of the Cartan subalgebra eigenstates manifest
the $SU(3)\times U(1)$ subgroups of the group $SO(5,1)$, is presented in Fig.~2 in Ref.~\cite{n2023NPB}, as the gauge field of any of the four families, presented in
Fig.~1 of Ref.~\cite{n2023NPB} in the case that $d=(5+1)$.

The creation operators for bosons operate on the vacuum state
$|\psi_{oc_{ev}}>\,*_{T}\, |0_{\vec{p}}> $ with the internal space part just a constant,
$|\psi_{oc_{ev}}>=$ $|\,1>$, and for a starting single boson state with zero momentum
from which one obtains the other single boson states with the same ``basis vector'' by
the operators $\hat{b}^{\dagger}_{\vec{p}}$ which push the momentum by an amount
$\vec{p}$, making also ${}^{i}{\cal C}^{ m}{}_{f \alpha}$ depending on $\vec{p}$:
${}^{i}{\hat {\cal C}}^{ m}{}_{f \alpha}(\vec{p})$~\footnote{.
%
%
%
\begin{small}
For the creation operators for boson fields in a coordinate
representation one finds using Eqs.~(\ref{creatorp}, \ref{eigenvalue10})
\begin{eqnarray}
{\bf {}^{i}{\hat{\cal A}}^{m \dagger}_{f \alpha}}
(\vec{x}, x^0)& =& {}^{i}{\hat{\cal A}}^{m \dagger}_{f} \,*_{T}\,
\int_{- \infty}^{+ \infty} \,
\frac{d^{d-1}p}{(\sqrt{2 \pi})^{d-1}} \,{}^{i}{\cal C}^{ m}{}_{f \alpha}\,
\hat{b}^{\dagger}_{\vec{p}}\,
e^{-i (p^0 x^0- \varepsilon \vec{p}\cdot \vec{x})}|_{p^0=|\vec{p}|}\,,i=(I,II)\,.
\label{Weylbosonx}
\end{eqnarray}
\end{small}
%
It is obvious that the even ``basis vectors'', determining the internal space
of bosons in Eq.~(\ref{Weylbosonx}), are the vector gauge fields of the fermion
fields, the creation operator of which is presented in Eq.~(\ref{wholespacefermions}).
}.

In all the papers (discussing the ability that the {\it spin-charge-family} theory, 
reviewed in Ref.~\cite{nh2021RPPNP}, is offering the explanation for all the 
assumptions of the {\it standard model} before the electroweak break, and that 
this theory is offering new recognitions and predictions) written before this new understanding of the internal space of boson gauge fields, presented in the 
Refs.~(\cite {n2022IARD,n2023NPB} and the references therein), the two kinds of 
the vector gauge fields interacting with the fermion fields, Eq.~(\ref{wholeaction}), 
$\omega_{ab\alpha}$ and $\tilde{\omega}_{ab\alpha}$ were assumed. All the 
derivations and calculations were done with these two boson fields presented in
the simple starting action in Eq.~(\ref{wholeaction}). 
\begin{eqnarray}
{\cal A}\,  &=& \int \; d^dx \; E\;\frac{1}{2}\, (\bar{\psi} \, \gamma^a p_{0a} \psi) 
+ h.c. +
\nonumber\\  
               & & \int \; d^dx \; E\; (\alpha \,R + \tilde{\alpha} \, \tilde{R})\,,
\nonumber\\
           p_{0\alpha} &=&  p_{\alpha}  - \frac{1}{2}  S^{ab} \omega_{ab \alpha} - 
                    \frac{1}{2}  \tilde{S}^{ab}   \tilde{\omega}_{ab \alpha} \,,
                    \nonumber\\  
           p_{0a } &=& f^{\alpha}{}_a p_{0\alpha} + \frac{1}{2E}\, \{ p_{\alpha},
E f^{\alpha}{}_a\}_- \,,\nonumber\\                  
R &=&  \frac{1}{2} \, \{ f^{\alpha [ a} f^{\beta b ]} \;(\omega_{a b \alpha, \beta} 
- \omega_{c a \alpha}\,\omega^{c}{}_{b \beta}) \} + h.c. \,, \nonumber \\
\tilde{R}  &=&  \frac{1}{2} \, \{ f^{\alpha [ a} f^{\beta b ]} 
\;(\tilde{\omega}_{a b \alpha,\beta} - \tilde{\omega}_{c a \alpha} \,
\tilde{\omega}^{c}{}_{b \beta})\} + h.c.\,.               
\label{wholeaction}
\end{eqnarray}
Here~\footnote{$f^{\alpha}{}_{a}$ are inverted vielbeins to 
$e^{a}{}_{\alpha}$ with the properties $e^a{}_{\alpha} f^{\alpha}{\!}_b = 
\delta^a{\!}_b,\; e^a{\!}_{\alpha} f^{\beta}{\!}_a = \delta^{\beta}_{\alpha} $, 
$ E = \det(e^a{\!}_{\alpha}) $.
Latin indices  
$a,b,..,m,n,..,s,t,..$ denote a tangent space (a flat index),
while Greek indices $\alpha, \beta,..,\mu, \nu,.. \sigma,\tau, ..$ denote an Einstein 
index (a curved index). Letters  from the beginning of both the alphabets
indicate a general index ($a,b,c,..$   and $\alpha, \beta, \gamma,.. $ ), 
from the middle of both the alphabets   
the observed dimensions $0,1,2,3$ ($m,n,..$ and $\mu,\nu,..$), indexes from 
the bottom of the alphabets
indicate the compactified dimensions ($s,t,..$ and $\sigma,\tau,..$). 
We assume the signature $\eta^{ab} =
diag\{1,-1,-1,\cdots,-1\}$.} 
$f^{\alpha [a} f^{\beta b]}= f^{\alpha a} f^{\beta b} - f^{\alpha b} f^{\beta a}$.

The vielbeins, $f^a_{\alpha}$, and the two kinds of the spin connection fields,
$\omega_{ab \alpha}$ (the gauge fields of $S^{ab}$) and $\tilde{\omega}_{ab \alpha}$
(the gauge fields of $\tilde{S}^{ab}$), manifest in $d=(3+1)$ as the known vector
gauge fields and the scalar gauge fields taking care of masses of quarks and leptons and antiquarks and antileptons and of the weak boson fields~\cite{nd2017}~\footnote{
Since the multiplication with either $\gamma^a$'s or $\tilde{\gamma}^a$'s changes
the odd ``basis vectors'' into the even objects, 
and even ``basis vectors'' commute, the action for fermions can not include 
an odd numbers of $\gamma^a$'s or $\tilde{\gamma}^a$'s, what the simple starting action of Eq.~(\ref{wholeaction}) does not. In the starting action 
$\gamma^a$'s and $\tilde{\gamma}^a$'s appear as $\gamma^0 
\gamma^a \hat{p}_{0a}$ or as $\gamma^0 \gamma^c \, 
S^{ab}\omega_{abc}$ and as$\gamma^0 \gamma^c \,\tilde{S}^{ab}\tilde{\omega}_{abc} $.}.

According to two groups of the even ``basis vectors", it follows that 
one of the groups, presented in Eq.~(\ref{Weylbosonx}) as
${\bf {}^{I}{\hat{\cal A}}^{m \dagger}_{f \alpha}}(\vec{x}, x^0)$ must 
be related to $S^{ab} \omega_{ab \alpha}$ (this one takes care of 
interaction among family members of fermion fields, which is the same 
for any of families $f$), the second group, presented in 
Eq.~(\ref{Weylbosonx}) as 
${\bf {}^{II}{\hat{\cal A}}^{m \dagger}_{f \alpha}}(\vec{x}, x^0)$ 
must be related to ${\tilde {S}}^{ab} {\tilde{\omega}}_{ab \alpha}$ 
(this one takes care of interaction
among families of a particular family member).

Correspondingly it is expected that the covariant derivative $ p_{0\alpha} = p_{\alpha}
- \frac{1}{2} S^{ab} \omega_{ab \alpha} - \frac{1}{2} \tilde{S}^{ab}
\tilde{\omega}_{ab \alpha}$, presented in Eq.~(\ref{wholeaction}), is 
replaced by
\begin{eqnarray}
p_{0\alpha} &=& p_{\alpha} - 
\sum_{m f} {}^{I}{ \hat {\cal A}}^{m \dagger}_{f}
\,\, {}^{I}{\cal C}^{m}_{f \alpha} -
\sum_{m f} {}^{II}{\hat {{\cal A}}}^{m \dagger}_{f}\,\,
{}^{II}{\cal C}^{m}_{f \alpha}\,,
\label{covderAIAII}
\end{eqnarray}
%
since as we have seen in Eqs.~(\ref{calIAb1234gen}, 
\ref{calbIIA1234gen}), $ {}^{I}{ \hat {\cal A}}^{m \dagger}_{f}$ 
transform the family members of a family $f$ among themselves, 
while $ {}^{II}{ \hat {\cal A}}^{m \dagger}_{f}$ transform a 
particular family member of one family into the same family member 
of any other family.

The simple starting action in even-dimensional spaces must include both 
boson gauge fields (the internal space of which is described
by the even ``basis vectors''), ${}^{I}{ \hat {\cal A}}^{m \dagger}_{f}$ 
and $ {}^{II}{ \hat {\cal A}}^{m \dagger}_{f}$, replacing the so far 
assumed $S^{ab} \omega_{ab \alpha}$ and 
$ \tilde{S}^{ab}\tilde{\omega}_{ab \alpha}$.

\vspace{2mm}

To understand what new the ``basis vectors'' description of the internal 
spaces of fermion and boson fields, Eqs.~(\ref{wholespacebosons}, \ref{Weylbosonx}, \ref{wholespacefermions}), bring to our understanding 
of the second quantized fermion and boson fields in comparison with what 
we have learned from the {\it spin-charge-family} theory so far while 
using the action presented in Eq.~\ref{wholeaction}, we need to relate 
$\sum_{ab} S^{ab}  \omega_{ab \alpha}$ and 
$ \sum_{m f} {}^{I}{\hat{\cal A}}^{m \dagger}_{f} \,
{}^{I}{\cal C}^{m}{}_{f\alpha}$, recognizing that 
${}^{I}{\hat{\cal A}}^{m \dagger}_{f} \,
{}^{I}{\cal C}^{m}{}_{f\alpha}$ are eigenstates of the Cartan 
subalgebra members, while $\omega_{ab \alpha}$ are not. And, 
equivalently, we need to relate 
$\sum_{ab} {\cal \tilde{S}}^{ab}  \tilde{\omega}_{ab \alpha}$ and
$ \sum_{m f} {}^{II}{\hat{\cal A}}^{m \dagger}_{f}\, 
{}^{II}{\cal C}^{m}{}_{f\alpha}$.

The gravity fields, the vielbeins and the two kinds of spin connection fields,
$f^{a}{}_{\alpha}$, $\omega_{ab \alpha}$, $\tilde{\omega}_{ab \alpha}$,
respectively, are in the {\it spin-charge-family} theory
(unifying spins, charges and families of fermions and offering not only the
explanation for all the assumptions of the {\it standard model} but also for
the increasing number of phenomena observed so far) the only boson fields in
$d=(13+1)$, observed in $d=(3+1)$: We must all the boson fields,
gravity, gluons, weak bosons, photons, with the Higgs's scalars
included~\cite{nd2017} express by the two kinds of the even
``basis vectors''.

We, therefore, need to relate:
\begin{eqnarray}
\label{relationomegaAmf0}
{\bf \cal S}^{cd}\, \sum_{ab} S^{ab}\, \omega_{ab \alpha}) &{\rm and}&
{\bf \cal S}^{cd}\, ({}^{I}{\hat{\cal A}}^{m \dagger}_{f}\, {\cal C}^{m f}_{\alpha}\,,\,\, \forall \,(m,f)\,, \forall \,{\cal S}^{cd}, \nonumber\\
 \sum_{ab} S^{ab}\, \omega_{ab \alpha}\, 
{ \hat b}^{m\dagger}_{f } &{\rm and}&
 {}^{I}{\hat{\cal A}}^{m' \dagger}_{f '} \,
{\cal C}^{m' f '}_{\alpha} 
 {\hat b}^{m\dagger}_{f }\,,\,\,
\forall (f ,m\,,f', m')\,, 
\end{eqnarray}
end equivalently for ${}^{II}{\hat{\cal A}}^{m' \dagger}_{f '} $~\footnote{
For the case that internal space is described by $d=(5+1)$ and the gravitons
are concerned, the reader can see the footnotes (35,36) in 
Ref.~\cite{n2024NPBarxiv240709482}
.}.
Let be repeated that ${}^{I}{\hat{\cal A}}^{m \dagger}_{f } $ are 
the eigenvectors of the Cartan subalgebra members, Eq.~(\ref{cartangrasscliff}).

Let us conclude this section by pointing out that either the odd ``basis
vectors'', $\hat{b}^{m \dagger}_{f}$, or the even ``basis vectors'',
${}^{i}{\hat{\cal A}}^{m \dagger}_{f}, i=(I,II) $, have each in any even $d$,
$2^{\frac{d}{2}-1}$ $\times \,2^{\frac{d}{2}-1}$ members, while
$\omega_{ab \alpha}$ as well as $\tilde{\omega}_{ab \alpha}$ have each for
a particular $\alpha$ $\frac{d}{2}(d-1)$ members.

Let be pointed out that the description of the internal space of bosons with the
even ``basis vectors'' supports (confirms) the existence of two kinds 
of boson fields,
suggested by the {\it spin-charge-family} theory while including in the action
presented in Eq.~(\ref{wholeaction}) $\omega_{ab \alpha}$ and
$\tilde{\omega}_{ab \alpha}$.

\section{Some useful relations}
\label{usefulrelations}

 In this appendix some useful relations, needed in this paper in 
 Sects.~(\ref{basisvectors0}, \ref{bosons13+1and5+1}, \ref{13+1representation}) are presented.

One can find if taking into account Eq.~(\ref{gammatildeantiher0})
\begin{small}
\begin{eqnarray}
\label{graficfollow1}
S^{ac}\stackrel{ab}{(k)}\stackrel{cd}{(k)} &=& -\frac{i}{2} \eta^{aa} \eta^{cc} 
\stackrel{ab}{[-k]}\stackrel{cd}{[-k]}\,, \quad 
S^{ac}\stackrel{ab}{[k]}\stackrel{cd}{[k]} = 
\frac{i}{2} \stackrel{ab}{(-k)}\stackrel{cd}{(-k)}\,,\nonumber\\
S^{ac}\stackrel{ab}{(k)}\stackrel{cd}{[k]} &=& -\frac{i}{2} \eta^{aa}  
\stackrel{ab}{[-k]}\stackrel{cd}{(-k)}\,, \quad
S^{ac}\stackrel{ab}{[k]}\stackrel{cd}{(k)} = \frac{i}{2} \eta^{cc}  
\stackrel{ab}{(-k)}\stackrel{cd}{[-k]}\,, \nonumber\\
\tilde{S}^{ac} \stackrel{ab}{(k)}\stackrel{cd}{(k)} &=& \frac{i}{2} \eta^{aa} \eta^{cc} 
\stackrel{ab}{[k]}\stackrel{cd}{[k]}\,, \quad 
\tilde{S}^{ac}\stackrel{ab}{[k]}\stackrel{cd}{[k]} = 
-\frac{i}{2} \stackrel{ab}{(k)}\stackrel{cd}{(k)}\,,\nonumber\\
\tilde{S}^{ac}\stackrel{ab}{(k)}\stackrel{cd}{[k]} &=& -\frac{i}{2} \eta^{aa}  
\stackrel{ab}{[k]}\stackrel{cd}{(k)}\,, \quad
\tilde{S}^{ac}\stackrel{ab}{[k]}\stackrel{cd}{(k)} = \frac{i}{2} \eta^{cc}  
\stackrel{ab}{(k)}\stackrel{cd}{[k]}\,. 
\end{eqnarray}
\end{small}

The reader can calculate all the quantum numbers  of Table~\ref{Table so13+1.},
 App.~\ref{13+1representation}
if taking into account
the generators of the two $SU(2)$  ($\subset SO(3,1)$ $\subset SO(7,1) \subset SO(13,1)$) groups, describing  spins of fermions and  the corresponding family quantum numbers
\begin{eqnarray}
\label{so1+3}
&&\vec{N}_{\pm}(= \vec{N}_{(L,R)}): = \,\frac{1}{2} (S^{23}\pm i S^{01},
S^{31}\pm i S^{02}, S^{12}\pm i S^{03} )\,,\nonumber\\
&&\vec{\tilde{N}}_{\pm}(=\vec{\tilde{N}}_{(L,R)}): =
 \,\frac{1}{2} (\tilde{S}^{23}\pm i \tilde{S}^{01}\,,
\tilde{S}^{31}\pm i \tilde{S}^{02}, \tilde{S}^{12}\pm i \tilde{S}^{03} )\,,
\end{eqnarray}
the generators of the two $SU(2)$ ($SU(2)$ $\subset SO(4)$ $\subset SO(7,1) 
\subset SO(13,1)$) groups, describing  the weak charge, $\vec{\tau}^{1}$, and
the second kind of the weak charge, $\vec{\tau}^{2}$,  of fermions and 
the corresponding family quantum numbers
%
 \begin{eqnarray}
 \label{so42}
 \vec{\tau}^{1}:&=&\frac{1}{2} (S^{58}-  S^{67}, \,S^{57} + S^{68}, \,S^{56}-  S^{78} )\,,
 \quad
 \vec{\tau}^{2}:= \frac{1}{2} (S^{58}+  S^{67}, \,S^{57} - S^{68}, \,S^{56}+  S^{78} )\,,
 \nonumber\\
 \vec{\tilde{\tau}}^{1}:&=&\frac{1}{2} (\tilde{S}^{58}-  \tilde{S}^{67}, \,\tilde{S}^{57} + 
 \tilde{S}^{68}, \,\tilde{S}^{56}-  \tilde{S}^{78} )\,, \quad 
 \vec{\tilde{\tau}}^{2}:=\frac{1}{2} (\tilde{S}^{58}+  \tilde{S}^{67}, \,\tilde{S}^{57} - 
 \tilde{S}^{68}, \,\tilde{S}^{56}+  \tilde{S}^{78} ),\,\,\;\;
 \end{eqnarray}
and the generators of $SU(3)$ and $U(1)$ subgroups of $SO(6)$ $\subset SO(13,1)$, describing  the colour charge and the ''fermion'' charge  of fermions as well as the corresponding 
family quantum number $\tilde{\tau}^4$
%
 \begin{eqnarray}
 \label{so64}
 \vec{\tau}^{3}: = &&\frac{1}{2} \,\{  S^{9\;12} - S^{10\;11} \,,
  S^{9\;11} + S^{10\;12} ,\, S^{9\;10} - S^{11\;12}\, ,  
  S^{9\;14} -  S^{10\;13} ,\,  \nonumber\\
  && S^{9\;13} + S^{10\;14} \,,  S^{11\;14} -  S^{12\;13}\,, 
  S^{11\;13} +  S^{12\;14} ,\,  \frac{1}{\sqrt{3}} ( S^{9\;10} + S^{11\;12} - 
 2 S^{13\;14})\}\,,\nonumber\\
 \tau^{4}: = &&-\frac{1}{3}(S^{9\;10} + S^{11\;12} + S^{13\;14})\,,\;\;\nonumber\\
 \tilde{\tau}^{4}: = &&-\frac{1}{3}(\tilde{S}^{9\;10} + \tilde{S}^{11\;12} + \tilde{S}^{13\;14})\,.
 \end{eqnarray}
The (chosen) Cartan subalgebra operators, determining the commuting operators in the 
above equations,
is presented in Eq.~(\ref{cartangrasscliff}). 

The  hypercharge $Y$ and the electromagnetic charge $Q$ and the corresponding family
 quantum numbers then follows as
 \begin{eqnarray}
 \label{YQY'Q'andtilde}
 Y:= \tau^{4} + \tau^{23}\,,\;\; Q: =  \tau^{13} + Y\,,\;\; 
 Y':= -\tau^{4}\tan^2\vartheta_2 + \tau^{23}\,, 
 \;\; Q':= -Y \tan^2\vartheta_1 + \tau^{13} \,,&&\nonumber\\
  \tilde{Y}:= \tilde{\tau}^{4} + \tilde{\tau}^{23}\,,\,\;\tilde{Q}:= 
  \tilde{Y} + \tilde{\tau}^{13}\,,\;\;
   \tilde{Y'}:= -\tilde{\tau}^{4} 
  \tan^2 \vartheta_2 + \tilde{\tau}^{23}\,,\;
  \;\; \tilde{Q'}= -\tilde{Y} \tan^2 \vartheta_1 
  + \tilde{\tau}^{13}\,. &&\,
  \end{eqnarray}
 %
 \begin{small}
Below are some of the above expressions written in terms of  nilpotents and projectors
 \begin{eqnarray}
\label{plusminus}
 N^{\pm}_{+}         &=& N^{1}_{+} \pm i \,N^{2}_{+} = 
 - \stackrel{03}{(\mp i)} \stackrel{12}{(\pm )}\,, \quad N^{\pm}_{-}= 
 N^{1}_{-} \pm  i\,N^{2}_{-} = 
   \stackrel{03}{(\pm i)} \stackrel{12}{(\pm )}\,,
\nonumber\\
 \tilde{N}^{\pm}_{+} &=& - \stackrel{03}{\tilde{(\mp i)}} 
 \stackrel{12}{\tilde{(\pm )}}\,, \quad 
 \tilde{N}^{\pm}_{-}= 
   \stackrel{03} {\tilde{(\pm i)}} \stackrel{12} {\tilde{(\pm )}}\,,\nonumber\\ 
 \tau^{1\pm}         &=& (\mp)\, \stackrel{56}{(\pm )} \stackrel{78}{(\mp )} \,, \quad   
 \tau^{2\mp}=            (\mp)\, \stackrel{56}{(\mp )} \stackrel{78}{(\mp )} \,,\nonumber\\ 
 \tilde{\tau}^{1\pm} &=& (\mp)\, \stackrel{56}{\tilde{(\pm )}} 
 \stackrel{78}{\tilde{(\mp )}}\,,\quad   
 \tilde{\tau}^{2\mp}= (\mp)\, \stackrel{56}{\tilde{(\mp )}} \stackrel{78}{\tilde{(\mp )}}\,.
 \end{eqnarray}
\end{small}

 For fermions, the operator of handedness $\Gamma^d$ is determined as follows:
  \begin{small}
\begin{eqnarray}
\label{Gamma}
 \Gamma^{(d)}= \prod_a (\sqrt{\eta^{aa}} \gamma^a)  \cdot \left \{ \begin{array}{l l}
 (i)^{\frac{d}{2}} \,, &\rm{ for\, d \,even}\,,\\
 (i)^{\frac{d-1}{2}}\,,&\rm{for \, d \,odd}\,.
  \end{array} \right.
 \end{eqnarray}
 \end{small}

\section{Useful tables}
\label{usefultables}

In this appendix, the even and odd ``basis vectors''
are presented for the choice $d=(5+1)$, needed in particular in 
Subsects.~(\ref{basisvectors0}, \ref{bosons5+1} ).

Table~\ref{Table Clifffourplet.}
presents $2^{d=6}=64$ ``eigenvectors" of the Cartan subalgebra,
Eq.~(\ref{cartangrasscliff}), members of the odd and even
``basis vectors'' which are the superposition of odd
(${\hat b}^{m \dagger}_f$, $({\hat b}^{m \dagger}_f)^{\dagger})$
and even (${}^{I}{\cal A}^{m}_f$, ${}^{II}{\cal A}^{m}_f$) products
of $\gamma^{a}$'s, needed in Sects.~(\ref{basisvectors0},
\ref{bosons13+1and5+1}). Table~\ref{Table Clifffourplet.} is presented in
several papers~(\cite{n2023NPB,nh2021RPPNP}, and references therein).

Tables~(\ref{transverseCliff basis5+1even I.},
\ref{S120Cliff basis5+1even I.}, \ref{transverseCliff basis5+1even II.},
\ref{S120Cliff basis5+1even II.}), needed in Sects.~(\ref{basisvectors0},
\ref{bosons13+1and5+1}), present even ``basis vectors'' as
algebraic products of ${\hat b}^{m \dagger}_f$ and
$({\hat b}^{m \dagger}_f)^{\dagger})$. They are
taken from Ref.~\cite{nh2023dec}.

%

\begin{table*}
\begin{small}
\caption{\label{Table Clifffourplet.}  This table, taken from~\cite{n2023NPB}, represents $2^d=64$ ``eigenvectors" of the Cartan subalgebra
members of the Clifford odd and even ``basis vectors'' which are the superposition of odd 
and even products of $\gamma^{a}$'s  in $d=(5+1)$-dimensional space,
divided into four groups. The first group, $odd \,I$, is chosen to represent ``basis vectors", named  ${\hat b}^{m \dagger}_f$,
appearing in $2^{\frac{d}{2}-1}=4$ 
``families" ($f=1,2,3,4$), each ''family'' having $2^{\frac{d}{2}-1}=4$
``family'' members ($m=1,2,3,4$).
The second group, $odd\,II$, contains Hermitian conjugated partners of the first
group for each ``family'' separately, ${\hat b}^{m}_f=$
$({\hat b}^{m \dagger}_f)^{\dagger}$. Either $odd \,I$ or $odd \,II$ are products
of an odd number of nilpotents (one or three) and projectors (two or none).
The ``family" quantum numbers of ${\hat b}^{m \dagger}_f$, that is the eigenvalues of
$(\tilde{S}^{03}, \tilde{S}^{12},\tilde{S}^{56})$, appear for the first {\it odd I }
group above each ``family", the quantum
numbers of the ``family'' members $(S^{03}, S^{12}, S^{56})$ are 
written in the last three columns. 
For the Hermitian conjugated partners of {\it odd I}, presented in the group {\it odd II},
the quantum numbers $(S^{03}, S^{12}, S^{56})$ are presented above each group of the
Hermitian conjugated partners, the last three columns 
tell eigenvalues of $(\tilde{S}^{03}, \tilde{S}^{12},\tilde{S}^{56})$.
The two groups with the even number of $\gamma^a$'s, {\it even \,I} and {\it even \,II},
each group has their Hermitian conjugated partners within its group,
have the quantum numbers $f$, that is the eigenvalues of
$(\tilde{S}^{03}, \tilde{S}^{12},\tilde{S}^{56})$, written above column of
four members, the quantum numbers of the members, $(S^{03}, S^{12}, S^{56})$, are
written in the last three columns. To find the quantum numbers of $({\cal {\bf S}}^{03},
{\cal {\bf S}}^{12}, {\cal {\bf S}}^{56})$ one has to take into account that
${\cal {\bf S}}^{ab}$ $= S^{ab} + \tilde{S}^{ab} $.
 \vspace{2mm}}
 \end{small}
\begin{tiny}
\begin{center}
  \begin{tabular}{|c|c|c|c|c|c|r|r|r|}
\hline
$ $&$$&$ $&$ $&$ $&&$$&$$&$$\\
$''basis\, vectors'' $&$m$&$ f=1$&$ f=2 $&$ f=3 $&
$ f=4 $&$$&$$&$$\\ 
$(\tilde{S}^{03}, \tilde{S}^{12}, \tilde{S}^{56})$&$\rightarrow$&$(\frac{i}{2},- \frac{1}{2},-\frac{1}{2})$&$(-\frac{i}{2},-\frac{1}{2},\frac{1}{2})$&
$(-\frac{i}{2},\frac{1}{2},-\frac{1}{2})$&$(\frac{i}{2},\frac{1}{2},\frac{1}{2})$&$S^{03}$
 &$S^{12}$&$S^{56}$\\ 
\hline
$ $&$$&$ $&$ $&$ $&&$$&$$&$$\\ 
$odd \,I\; {\hat b}^{m \dagger}_f$&$1$& 
$\stackrel{03}{(+i)}\stackrel{12}{[+]}\stackrel{56}{[+]}$&
                        $\stackrel{03}{[+i]}\stackrel{12}{[+]}\stackrel{56}{(+)}$ & 
                        $\stackrel{03}{[+i]}\stackrel{12}{(+)}\stackrel{56}{[+]}$ &  
                        $\stackrel{03}{(+i)}\stackrel{12}{(+)}\stackrel{56}{(+)}$ &
                        $\frac{i}{2}$&$\frac{1}{2}$&$\frac{1}{2}$\\ 
$$&$2$&    $[-i](-)[+] $ & $(-i)(-)(+) $ & $(-i)[-][+] $ & $[-i][-](+) $ &$-\frac{i}{2}$&
$-\frac{1}{2}$&$\frac{1}{2}$\\ 
$$&$3$&    $[-i] [+](-)$ & $(-i)[+][-] $ & $(-i)(+)(-) $ & $[-i](+)[-] $&$-\frac{i}{2}$&
$\frac{1}{2}$&$-\frac{1}{2}$\\ 
$$&$4$&    $(+i)(-)(-)$ & $[+i](-)[-] $ & $[+i][-](-) $ & $(+i)[-][-]$&$\frac{i}{2}$&
$-\frac{1}{2}$&$-\frac{1}{2}$\\ 
\hline
$ $&$$&$ $&$ $&$ $&&$$&$$&$$\\ 
$(S^{03}, S^{12}, S^{56})$&$\rightarrow$&$(-\frac{i}{2}, \frac{1}{2},\frac{1}{2})$&
$(\frac{i}{2},\frac{1}{2},-\frac{1}{2})$&
$(\frac{i}{2},- \frac{1}{2},\frac{1}{2})$&$(-\frac{i}{2},-\frac{1}{2},-\frac{1}{2})$&
$\tilde{S}^{03}$
&$\tilde{S}^{12}$&$\tilde{S}^{56}$\\ 
&&
$\stackrel{03}{\;\,}\;\;\,\stackrel{12}{\;\,}\;\;\,\stackrel{56}{\;\,}$&
$\stackrel{03}{\;\,}\;\;\,\stackrel{12}{\;\,}\;\;\,\stackrel{56}{\;\,}$&
$\stackrel{03}{\;\,}\;\;\,\stackrel{12}{\;\,}\;\;\,\stackrel{56}{\;\,}$&
$\stackrel{03}{\;\,}\;\;\,\stackrel{12}{\;\,}\;\;\,\stackrel{56}{\;\,}$&
&&\\
\hline
$ $&$$&$ $&$ $&$ $&&$$&$$&$$\\ 
$odd\,II\; {\hat b}^{m}_f$&$1$ &$(-i)[+][+]$ & $[+i][+](-)$ & $[+i](-)[+]$ & $(-i)(-)(-)$&
$-\frac{i}{2}$&$-\frac{1}{2}$&$-\frac{1}{2}$\\ 
$$&$2$&$[-i](+)[+]$ & $(+i)(+)(-)$ & $(+i)[-][+]$ & $[-i][-](-)$&
$\frac{i}{2}$&$\frac{1}{2}$&$-\frac{1}{2}$\\ 
$$&$3$&$[-i][+](+)$ & $(+i)[+][-]$ & $(+i)(-)(+)$ & $[-i](-)[-]$&
$\frac{i}{2}$&$-\frac{1}{2}$&$\frac{1}{2}$\\ 
$$&$4$&$(-i)(+)(+)$ & $[+i](+)[-]$ & $[+i][-](+)$ & $(-i)[-][-]$&
$-\frac{i}{2}$&$\frac{1}{2}$&$\frac{1}{2}$\\ 
\hline
&&&&&&&&\\ 
\hline
$ $&$$&$ $&$ $&$ $&&$$&$$&$$\\ 
$(\tilde{S}^{03}, \tilde{S}^{12}, \tilde{S}^{56})$&$\rightarrow$&
$(-\frac{i}{2},\frac{1}{2},\frac{1}{2})$&$(\frac{i}{2},-\frac{1}{2},\frac{1}{2})$&
$(-\frac{i}{2},-\frac{1}{2},-\frac{1}{2})$&$(\frac{i}{2},\frac{1}{2},-\frac{1}{2})$&
$S^{03}$&$S^{12}$&$S^{56}$\\ 
&& 
$\stackrel{03}{\;\,}\;\;\,\stackrel{12}{\;\,}\;\;\,\stackrel{56}{\;\,}$&
$\stackrel{03}{\;\,}\;\;\,\stackrel{12}{\;\,}\;\;\,\stackrel{56}{\;\,}$&

$\stackrel{03}{\;\,}\;\;\,\stackrel{12}{\;\,}\;\;\,\stackrel{56}{\;\,}$&
$\stackrel{03}{\;\,}\;\;\,\stackrel{12}{\;\,}\;\;\,\stackrel{56}{\;\,}$&
&&\\ 
\hline
$ $&$$&$ $&$ $&$ $&&$$&$$&$$\\ 
$even\,I \; {}^{I}{\cal A}^{m}_f$&$1$&$[+i](+)(+) $ & $(+i)[+](+) $ & $[+i][+][+] $ & $(+i)(+)[+] $ &$\frac{i}{2}$&
$\frac{1}{2}$&$\frac{1}{2}$\\ 
$$&$2$&$(-i)[-](+) $ & $[-i](-)(+) $ & $(-i)(-)[+] $ & $[-i][-][+] $ &$-\frac{i}{2}$&
$-\frac{1}{2}$&$\frac{1}{2}$\\ 
$$&$3$&$(-i)(+)[-] $ & $[-i][+][-] $ & $(-i)[+](-) $ & $[-i](+)(-) $&$-\frac{i}{2}$&
$\frac{1}{2}$&$-\frac{1}{2}$\\ 
$$&$4$&$[+i][-][-] $ & $(+i)(-)[-] $ & $[+i](-)(-) $ & $(+i)[-](-) $&$\frac{i}{2}$&
$-\frac{1}{2}$&$-\frac{1}{2}$\\ 
\hline
$ $&$$&$ $&$ $&$ $&&$$&$$&$$\\ 
$(\tilde{S}^{03}, \tilde{S}^{12}, \tilde{S}^{56})$&$\rightarrow$&
$(\frac{i}{2},\frac{1}{2},\frac{1}{2})$&$(-\frac{i}{2},-\frac{1}{2},\frac{1}{2})$&
$(\frac{i}{2},-\frac{1}{2},-\frac{1}{2})$&$(-\frac{i}{2},\frac{1}{2},-\frac{1}{2})$&
$S^{03}$&$S^{12}$&$S^{56}$\\ 
&& 
$\stackrel{03}{\;\,}\;\;\,\stackrel{12}{\;\,}\;\;\,\stackrel{56}{\;\,}$&
$\stackrel{03}{\;\,}\;\;\,\stackrel{12}{\;\,}\;\;\,\stackrel{56}{\;\,}$&
$\stackrel{03}{\;\,}\;\;\,\stackrel{12}{\;\,}\;\;\,\stackrel{56}{\;\,}$&
$\stackrel{03}{\;\,}\;\;\,\stackrel{12}{\;\,}\;\;\,\stackrel{56}{\;\,}$&
&&\\ 
\hline
$ $&$$&$ $&$ $&$ $&&$$&$$&$$\\ 
$even\,II \; {}^{II}{\cal A}^{m}_f$&$1$& $[-i](+)(+) $ & $(-i)[+](+) $ & $[-i][+][+] $ & 
$(-i)(+)[+] $ &$-\frac{i}{2}$&
$\frac{1}{2}$&$\frac{1}{2}$\\ 
$$&$2$&    $(+i)[-](+) $ & $[+i](-)(+) $ & $(+i)(-)[+] $ & $[+i][-][+] $ &$\frac{i}{2}$&
$-\frac{1}{2}$&$\frac{1}{2}$ \\ 
$$&$3$&    $(+i)(+)[-] $ & $[+i][+][-] $ & $(+i)[+](-) $ & $[+i](+)(-) $&$\frac{i}{2}$&
$\frac{1}{2}$&$-\frac{1}{2}$\\ 
$$&$4$&    $[-i][-][-] $ & $(-i)(-)[-] $ & $[-i](-)(-) $ & $(-i)[-](-) $&$-\frac{i}{2}$&
$-\frac{1}{2}$&$-\frac{1}{2}$\\ 
\hline
 \end{tabular}
\end{center}
\end{tiny}
\end{table*}
%
\begin{table}
\begin{tiny}
\caption{
The even ``basis vectors'' ${}^{I}{\hat{\cal A}}^{m \dagger}_{f}$,
belonging to transverse momentum in internal space, ${\cal S}^{12}=$
$1$, the first half of ${}^{I}{\hat{\cal A}}^{m \dagger}_{f}$, and
${\cal S}^{12}= -1$, the
second half of ${}^{I}{\hat{\cal A}}^{m \dagger}_{f}$, for $d=(5+1)$, are
presented as algebraic products of the $f=1$ family ``basis vectors''
$\hat{b}^{m' \dagger}_{1}$ and their Hermitian conjugated partners
($\hat{b}^{m'' \dagger}_{1})^{\dagger}$: $\hat{b}^{m' \dagger}_{1} *_{A}$
($\hat{b}^{m'' \dagger}_{1})^{\dagger}$. Two
${}^{I}{\hat{\cal A}}^{m \dagger}_{f}$ which are the Hermitian conjugated
partners are marked with the same symbol ($\star \star$, $\ddagger$, $\otimes$,
$\odot \odot$).
The even ``basis vectors'' ${}^{I}{\hat{\cal A}}^{m \dagger}_{f}$
are products of one projector and two nilpotents, the odd ``basis
vectors'' and their Hermitian conjugated partners are products of one nilpotent
and two projectors or of three nilpotents. The even and odd
objects are eigenvectors of all the corresponding Cartan subalgebra members,
Eq.~(\ref{cartangrasscliff}). There are $\frac{1}{2} \times 2^{\frac{6}{2}-1}
\times 2^{\frac{6}{2}-1}$ algebraic products of $\hat{b}^{m' \dagger}_{1} *_{A}$
($\hat{b}^{m'' \dagger}_{1})^{\dagger}$ with ${\cal S}^{12}$ equal
to $ 1$ or $-1$ . The rest $8$ of $16$ members
present ${}^{I}{\hat{\cal A}}^{m \dagger}_{f}$ with ${\cal S}^{12}=0$.
The members $\hat{b}^{m' \dagger}_{f}$ together with their Hermitian
conjugated partners of each of the four families, $f=(1,2,3,4)$, offer the
same ${}^{I}{\hat{\cal A}}^{m \dagger}_{f}$ with ${\cal S}^{12}=\pm1$
as the ones presented in this table.
(And equivalently for ${\cal S}^{12}=0$.) Table is taken from
Ref.~\cite{nh2023dec}.
\vspace{3mm}}
\label{transverseCliff basis5+1even I.} 
 %
 \begin{center}
 \begin{tabular}{ |c| c| c c|}
 \hline
 $$&$$&$$&$$\\
${\cal S}^{12} $&$symbol$&${}^{I}\hat{\cal A}^{m \dagger}_f=$
&$\hat{b}^{m' \dagger}_{f `} *_A (\hat{b}^{m'' \dagger}_{f `})^{\dagger}$\\
\\
\hline
%
$$&$$&$$&$$\\
$1$&$\star \star$&${}^{I}\hat{\cal A}^{1 \dagger}_1=$&
$\hat{b}^{1 \dagger}_{1} *_{A} (\hat{b}^{4 \dagger}_{1})^{\dagger}$\\
$$&$$&$$&$$\\
$ $&$$ &$\stackrel{03}{[+i]}\,\stackrel{12}{(+)} \stackrel{56}{(+)}$&
$\stackrel{03}{(+i)}\,\stackrel{12}{[+]} \stackrel{56}{[+]} *_{A} 
\stackrel{03}{(-i)}\,\stackrel{12}{(+)} \stackrel{56}{(+)}$\\
\hline
$$&$$&$$&$$\\
$1$&$\ddagger$&${}^{I}\hat{\cal A}^{3 \dagger}_1=$&
$\hat{b}^{3 \dagger}_{1} *_{A} (\hat{b}^{4 \dagger}_{1})^{\dagger}$\\
$$&$$&$$&$$\\
$ $&$$ &$\stackrel{03}{(-i)}\,\stackrel{12}{(+)} \stackrel{56}{[-]}$&
$\stackrel{03}{[-i]}\,\stackrel{12}{[+]} \stackrel{56}{(-)} *_{A} 
\stackrel{03}{(-i)}\,\stackrel{12}{(+)} \stackrel{56}{(+)}$\\
\hline
$$&$$&$$&$$\\
$1$&$\odot \odot$&${}^{I}\hat{\cal A}^{1 \dagger}_4=$&
$\hat{b}^{1 \dagger}_{1} *_{A} (\hat{b}^{2 \dagger}_{1})^{\dagger}$\\
$$&$$&$$&$$\\
$ $&$$ &$\stackrel{03}{(+i)}\,\stackrel{12}{(+)} \stackrel{56}{[+]}$&
$\stackrel{03}{(+i)}\,\stackrel{12}{[+]} \stackrel{56}{[+]} *_{A} 
\stackrel{03}{[-i]}\,\stackrel{12}{(+)} \stackrel{56}{[+]}$\\
\hline
$$&$$&$$&$$\\
$1$&$\otimes$&${}^{I}\hat{\cal A}^{3 \dagger}_4=$&
$\hat{b}^{3 \dagger}_{1} *_{A} (\hat{b}^{2 \dagger}_{1})^{\dagger}$\\
$$&$$&$$&$$\\
$ $&$$ &$\stackrel{03}{[-i]}\,\stackrel{12}{(+)} \stackrel{56}{(-)}$&
$\stackrel{03}{[-i]}\,\stackrel{12}{[+]} \stackrel{56}{(-)} *_{A} 
\stackrel{03}{[-i]}\,\stackrel{12}{(+)} \stackrel{56}{[+]}$\\
\hline
\hline
$$&$$&$$&$$\\
$-1$&$\otimes$&${}^{I}\hat{\cal A}^{2 \dagger}_2=$&
$\hat{b}^{2 \dagger}_{1} *_{A} (\hat{b}^{3 \dagger}_{1})^{\dagger}$\\
$$&$$&$$&$$\\
$ $&$$ &$\stackrel{03}{[-i]}\,\stackrel{12}{(-)} \stackrel{56}{(+)}$&
$\stackrel{03}{[-i]}\,\stackrel{12}{(-)} \stackrel{56}{[+]} *_{A} 
\stackrel{03}{[-i]}\,\stackrel{12}{[+]} \stackrel{56}{(+)}$\\
\hline
$$&$$&$$&$$\\
$-1$&$\ddagger$&${}^{I}\hat{\cal A}^{4 \dagger}_2=$&
$\hat{b}^{4 \dagger}_{1} *_{A} (\hat{b}^{3 \dagger}_{1})^{\dagger}$\\
$$&$$&$$&$$\\
$ $&$$ &$\stackrel{03}{(+i)}\,\stackrel{12}{(-)} \stackrel{56}{[-]}$&
$\stackrel{03}{(+i)}\,\stackrel{12}{(-)} \stackrel{56}{(-)} *_{A} 
\stackrel{03}{[-i]}\,\stackrel{12}{[+]} \stackrel{56}{(+)}$\\
\hline
$$&$$&$$&$$\\
$-1$&$\odot \odot$&${}^{I}\hat{\cal A}^{2 \dagger}_3=$&
$\hat{b}^{2 \dagger}_{1} *_{A} (\hat{b}^{1 \dagger}_{1})^{\dagger}$\\
$$&$$&$$&$$\\
$ $&$$ &$\stackrel{03}{(-i)}\,\stackrel{12}{(-)} \stackrel{56}{[+]}$&
$\stackrel{03}{[-i]}\,\stackrel{12}{(-)} \stackrel{56}{[+]} *_{A} 
\stackrel{03}{(-i)}\,\stackrel{12}{[+]} \stackrel{56}{[+]}$\\
\hline
$$&$$&$$&$$\\
$-1$&$\star \star$&${}^{I}\hat{\cal A}^{4 \dagger}_3=$&
$\hat{b}^{4 \dagger}_{1} *_{A} (\hat{b}^{1 \dagger}_{1})^{\dagger}$\\
$$&$$&$$&$$\\
$ $&$$ &$\stackrel{03}{[+i]}\,\stackrel{12}{(-)} \stackrel{56}{(-)}$&
$\stackrel{03}{(+i)}\,\stackrel{12}{(-)} \stackrel{56}{(-)} *_{A} 
\stackrel{03}{(-i)}\,\stackrel{12}{[+]} \stackrel{56}{[+]}$\\
\hline
 \end{tabular}
 \end{center}
\end{tiny}
\end{table}
%
\begin{table}
\begin{tiny}
\caption{
The even ``basis vectors'' ${}^{I}{\hat{\cal A}}^{m \dagger}_{f}$,
belonging to zero momentum in internal space, ${\cal S}^{12}=$ $0$,
for $d=(5+1)$, are presented as algebraic products of the $f=1$ family
``basis vectors'' $\hat{b}^{m' \dagger}_{1}$ and their Hermitian conjugated
partners ($\hat{b}^{m'' \dagger}_{1})^{\dagger}$:
$\hat{b}^{m' \dagger}_{1} *_{A}$ ($\hat{b}^{m'' \dagger}_{1})^{\dagger}$.
The two ${}^{I}{\hat{\cal A}}^{m \dagger}_{f}$ which are Hermitian
conjugated partners, are marked with the same symbol (either $\bigtriangleup$ or $\bullet$). The symbol $\bigcirc$ presents selfadjoint members.
The even ``basis vectors'' ${}^{I}{\hat{\cal A}}^{m \dagger}_{f}$
are products of one projector and two nilpotents or three projectors (they are
self-adjoint), the odd ``basis
vectors'' and their Hermitian conjugated partners are products of one nilpotent
and two projectors or of three nilpotents. The even and odd
objects are eigenvectors of all the corresponding Cartan subalgebra members,
Eq.~(\ref{cartangrasscliff}). There are $\frac{1}{2} \times 2^{\frac{6}{2}-1}
\times 2^{\frac{6}{2}-1}$ algebraic products of $\hat{b}^{m' \dagger}_{1}
*_{A}$ ($\hat{b}^{m'' \dagger}_{1})^{\dagger}$. The rest $8$ of $16$
members have ${}^{I}{\hat{\cal A}}^{m \dagger}_{f}$ with
${\cal S}^{12}=+ 1$ (four) and with ${\cal S}^{12}=- 1$ (four), present
in Table~\ref{transverseCliff basis5+1even I.}.
The members $\hat{b}^{m' \dagger}_{f}$ together with their Hermitian
conjugated partners of each of the four families, $f=(1,2,3,4)$, offer the
same ${}^{I}{\hat{\cal A}}^{m \dagger}_{f}$ with ${\cal S}^{12}=0$
as the ones presented in this table. The table is taken from
Ref.~\cite{nh2023dec}.
\vspace{3mm}}
\label{S120Cliff basis5+1even I.} 
 %
 \begin{center}
 \begin{tabular}{ |c| c| c c|}
 \hline
 $$&$$&$$&$$\\
${\cal S}^{12} $&$symbol$&${}^{I}\hat{\cal A}^{m \dagger}_f=$
&$\hat{b}^{m' \dagger}_{f `} *_A (\hat{b}^{m'' \dagger}_{f `})^{\dagger}$\\
\\
\hline
%
$$&$$&$$&$$\\
$0$&$\bigtriangleup$&${}^{I}\hat{\cal A}^{2 \dagger}_1=$&
$\hat{b}^{2 \dagger}_{1} *_{A} (\hat{b}^{4 \dagger}_{1})^{\dagger}$\\
$$&$$&$$&$$\\
$ $&$$ &$\stackrel{03}{(-i)}\,\stackrel{12}{[-]} \stackrel{56}{(+)}$&
$\stackrel{03}{[-i]}\,\stackrel{12}{(-)} \stackrel{56}{[+]} *_{A} 
\stackrel{03}{(-i)}\,\stackrel{12}{(+)} \stackrel{56}{(+)}$\\
\hline
$$&$$&$$&$$\\
$0$&$\bigcirc$&${}^{I}\hat{\cal A}^{4 \dagger}_1=$&
$\hat{b}^{4 \dagger}_{1} *_{A} (\hat{b}^{4 \dagger}_{1})^{\dagger}$\\
$$&$$&$$&$$\\
$ $&$$ &$\stackrel{03}{[+i]}\,\stackrel{12}{[-]} \stackrel{56}{[-]}$&
$\stackrel{03}{(+i)}\,\stackrel{12}{(-)} \stackrel{56}{(-)} *_{A} 
\stackrel{03}{(-i)}\,\stackrel{12}{(+)} \stackrel{56}{(+)}$\\
\hline
$$&$$&$$&$$\\
$0$&$\bullet$&${}^{I}\hat{\cal A}^{1 \dagger}_2=$&
$\hat{b}^{1 \dagger}_{1} *_{A} (\hat{b}^{3 \dagger}_{1})^{\dagger}$\\
$$&$$&$$&$$\\
$ $&$$ &$\stackrel{03}{(+i)}\,\stackrel{12}{[+]} \stackrel{56}{(+)}$&
$\stackrel{03}{(+i)}\,\stackrel{12}{[+]} \stackrel{56}{[+]} *_{A} 
\stackrel{03}{[-i]}\,\stackrel{12}{[+]} \stackrel{56}{(+)}$\\
\hline
$$&$$&$$&$$\\
$0$&$\bigcirc$&${}^{I}\hat{\cal A}^{3 \dagger}_2=$&
$\hat{b}^{3 \dagger}_{1} *_{A} (\hat{b}^{3 \dagger}_{1})^{\dagger}$\\
$$&$$&$$&$$\\
$ $&$$ &$\stackrel{03}{[-i]}\,\stackrel{12}{[+]} \stackrel{56}{[-]}$&
$\stackrel{03}{[-i]}\,\stackrel{12}{[+]} \stackrel{56}{(-)} *_{A} 
\stackrel{03}{[-i]}\,\stackrel{12}{[+]} \stackrel{56}{(+)}$\\
\hline
\hline
$$&$$&$$&$$\\
$0$&$\bigcirc$&${}^{I}\hat{\cal A}^{1 \dagger}_3=$&
$\hat{b}^{1 \dagger}_{1} *_{A} (\hat{b}^{1 \dagger}_{1})^{\dagger}$\\
$$&$$&$$&$$\\
$ $&$$ &$\stackrel{03}{[+i]}\,\stackrel{12}{[+]} \stackrel{56}{[+]}$&
$\stackrel{03}{(+i)}\,\stackrel{12}{[+]} \stackrel{56}{[+]} *_{A} 
\stackrel{03}{(-i)}\,\stackrel{12}{[+]} \stackrel{56}{[+]}$\\
\hline
$$&$$&$$&$$\\
$0$&$\bullet$&${}^{I}\hat{\cal A}^{3 \dagger}_3=$&
$\hat{b}^{3 \dagger}_{1} *_{A} (\hat{b}^{1 \dagger}_{1})^{\dagger}$\\
$$&$$&$$&$$\\
$ $&$$ &$\stackrel{03}{(-i)}\,\stackrel{12}{[+]} \stackrel{56}{(-)}$&
$\stackrel{03}{[-i]}\,\stackrel{12}{[+]} \stackrel{56}{(-)} *_{A} 
\stackrel{03}{(-i)}\,\stackrel{12}{[+]} \stackrel{56}{[+]}$\\
\hline
$$&$$&$$&$$\\
$0$&$\bigcirc$&${}^{I}\hat{\cal A}^{2 \dagger}_4=$&
$\hat{b}^{2 \dagger}_{1} *_{A} (\hat{b}^{2 \dagger}_{1})^{\dagger}$\\
$$&$$&$$&$$\\
$ $&$$ &$\stackrel{03}{[-i]}\,\stackrel{12}{[-]} \stackrel{56}{[+]}$&
$\stackrel{03}{[-i]}\,\stackrel{12}{(-)} \stackrel{56}{[+]} *_{A} 
\stackrel{03}{[-i]}\,\stackrel{12}{(+)} \stackrel{56}{[+]}$\\
\hline
$$&$$&$$&$$\\
$0$&$\bigtriangleup$&${}^{I}\hat{\cal A}^{4 \dagger}_4=$&
$\hat{b}^{4 \dagger}_{1} *_{A} (\hat{b}^{2 \dagger}_{1})^{\dagger}$\\
$$&$$&$$&$$\\
$ $&$$ &$\stackrel{03}{(+i)}\,\stackrel{12}{[-]} \stackrel{56}{(-)}$&
$\stackrel{03}{(+i)}\,\stackrel{12}{(-)} \stackrel{56}{(-)} *_{A} 
\stackrel{03}{[-i]}\,\stackrel{12}{(+)} \stackrel{56}{[+]}$\\
\hline
 \end{tabular}
 \end{center}
\end{tiny}
\end{table}
\begin{table}
\begin{tiny}
\caption{
The even ``basis vectors'' ${}^{II}{\hat{\cal A}}^{m \dagger}_{f}$,
belonging to transverse momentum in internal space, ${\cal S}^{12}=$ $1$,
the first half ${}^{II}{\hat{\cal A}}^{m \dagger}_{f}$, and
${\cal S}^{12}=-1$, the
second half ${}^{II}{\hat{\cal A}}^{m \dagger}_{f}$, for $d=(5+1)$, are
presented as algebraic products of the first, $m=1$, member of ``basis
vectors'' $\hat{b}^{m'=1 \dagger}_{f '}$ and the Hermitian conjugated
partners ($\hat{b}^{m' =1 \dagger}_{f ''})^{\dagger}$. Two
${}^{II}{\hat{\cal A}}^{m \dagger}_{f}$ which are the Hermitian
conjugated partners are marked with the same symbol.
The even ``basis vectors'' ${}^{II}{\hat{\cal A}}^{m \dagger}_{f}$
are products of one projector and two nilpotents, the odd ``basis
vectors'' and the Hermitian conjugated partners are products of one nilpotent
and two projectors or of three nilpotents. Even and odd
objects are eigenvectors of the corresponding Cartan subalgebra members,
Eq.~(\ref{cartangrasscliff}). There are $2^{\frac{6}{2}-1}\times
2^{\frac{6}{2}-1}$ algebraic products of
$(\hat{b}^{m' \dagger}_{f `})^{\dagger}$ and
$\hat{b}^{m' \dagger}_{f ``}$, $f `$ and $f ``$ run over all
four families. The rest of the $16$ members present
${}^{II}{\hat{\cal A}}^{m \dagger}_{f}$ with ${\cal S}^{12}=0$.
The members $(\hat{b}^{m' \dagger}_{f `})^{\dagger}$ together with
$\hat{b}^{m' \dagger}_{f `'}$, $m'=(1,2,3,4)$, offer the
same ${}^{II}{\hat{\cal A}}^{m \dagger}_{f}$ with
${\cal S}^{12}=\pm1$ as the ones presented in this table.
(And equivalently for ${\cal S}^{12}=0$.) The table is taken from
Ref.~\cite{nh2023dec}.
\vspace{3mm}}
\label{transverseCliff basis5+1even II.} 
 %
 \begin{center}
 \begin{tabular}{ |c| c| c c|}
 \hline
 $$&$$&$$&$$\\
${\cal S}^{12} $&$symbol$&${}^{II}\hat{\cal A}^{m \dagger}_f=$
&$(\hat{b}^{1 \dagger}_{f `})^{\dagger} *_A \hat{b}^{1 \dagger}_{f ``}$\\
\\
\hline
%
$$&$$&$$&$$\\
$1$&$\star \star$&${}^{II}\hat{\cal A}^{1 \dagger}_1=$&
$(\hat{b}^{1 \dagger}_{1})^{\dagger} *_{A} \hat{b}^{1 \dagger}_{4}$\\
$$&$$&$$&$$\\
$ $&$$ &$\stackrel{03}{[-i]}\,\stackrel{12}{(+)} \stackrel{56}{(+)}$&
$\stackrel{03}{(-i)}\,\stackrel{12}{[+]} \stackrel{56}{[+]} *_{A} 
\stackrel{03}{(+i)}\,\stackrel{12}{(+)} \stackrel{56}{(+)}$\\
\hline
$$&$$&$$&$$\\
$1$&$\odot \odot$&${}^{II}\hat{\cal A}^{3 \dagger}_1=$&
$(\hat{b}^{1 \dagger}_{2})^{\dagger} *_{A} \hat{b}^{1\dagger}_{4}$\\
$$&$$&$$&$$\\
$ $&$$ &$\stackrel{03}{(+i)}\,\stackrel{12}{(+)} \stackrel{56}{[-]}$&
$\stackrel{03}{[+i]}\,\stackrel{12}{[+]} \stackrel{56}{(-)} *_{A} 
\stackrel{03}{(+i)}\,\stackrel{12}{(+)} \stackrel{56}{(+)}$\\
\hline
$$&$$&$$&$$\\
$1$&$\ddagger$&${}^{II}\hat{\cal A}^{1 \dagger}_4=$&
$(\hat{b}^{1 \dagger}_{1})^{\dagger} *_{A} \hat{b}^{1 \dagger}_{3}$\\
$$&$$&$$&$$\\
$ $&$$ &$\stackrel{03}{(-i)}\,\stackrel{12}{(+)} \stackrel{56}{[+]}$&
$\stackrel{03}{(-i)}\,\stackrel{12}{[+]} \stackrel{56}{[+]} *_{A} 
\stackrel{03}{[+i]}\,\stackrel{12}{(+)} \stackrel{56}{[+]}$\\
\hline
$$&$$&$$&$$\\
$1$&$\otimes$&${}^{II}\hat{\cal A}^{3 \dagger}_4=$&
$(\hat{b}^{1 \dagger}_{2})^{\dagger} *_{A} \hat{b}^{1 \dagger}_{3}$\\
$$&$$&$$&$$\\
$ $&$$ &$\stackrel{03}{[+i]}\,\stackrel{12}{(+)} \stackrel{56}{(-)}$&
$\stackrel{03}{[+i]}\,\stackrel{12}{[+]} \stackrel{56}{(-)} *_{A} 
\stackrel{03}{[+i]}\,\stackrel{12}{(+)} \stackrel{56}{[+]}$\\
\hline
\hline
$$&$$&$$&$$\\
$-1$&$\otimes$&${}^{II}\hat{\cal A}^{2 \dagger}_2=$&
$(\hat{b}^{1 \dagger}_{3})^{\dagger} *_{A} \hat{b}^{1 \dagger}_{2}$\\
$$&$$&$$&$$\\
$ $&$$ &$\stackrel{03}{[+i]}\,\stackrel{12}{(-)} \stackrel{56}{(+)}$&
$\stackrel{03}{[+i]}\,\stackrel{12}{(-)} \stackrel{56}{[+]} *_{A} 
\stackrel{03}{[+i]}\,\stackrel{12}{[+]} \stackrel{56}{(+)}$\\
\hline
$$&$$&$$&$$\\
$-1$&$\otimes \otimes$&${}^{II}\hat{\cal A}^{4 \dagger}_2=$&
$(\hat{b}^{1 \dagger}_{4})^{\dagger} *_{A} \hat{b}^{1 \dagger}_{2}$\\
$$&$$&$$&$$\\
$ $&$$ &$\stackrel{03}{(-i)}\,\stackrel{12}{(-)} \stackrel{56}{[-]}$&
$\stackrel{03}{(-i)}\,\stackrel{12}{(-)} \stackrel{56}{(-)} *_{A} 
\stackrel{03}{[+i]}\,\stackrel{12}{[+]} \stackrel{56}{(+)}$\\
\hline
$$&$$&$$&$$\\
$-1$&$\ddagger$&${}^{II} \hat{\cal A}^{2 \dagger}_3=$&
$(\hat{b}^{1 \dagger}_{3})^{\dagger} *_{A} \hat{b}^{1 \dagger}_{1}$\\
$$&$$&$$&$$\\
$ $&$$ &$\stackrel{03}{(+i)}\,\stackrel{12}{(-)} \stackrel{56}{[+]}$&
$\stackrel{03}{[+i]}\,\stackrel{12}{(-)} \stackrel{56}{[+]} *_{A} 
\stackrel{03}{(+i)}\,\stackrel{12}{[+]} \stackrel{56}{[+]}$\\
\hline
$$&$$&$$&$$\\
$-1$&$\star \star$&${}^{II}\hat{\cal A}^{4 \dagger}_3=$&
$(\hat{b}^{1 \dagger}_{4})^{\dagger} *_{A} \hat{b}^{1 \dagger}_{1}$\\
$$&$$&$$&$$\\
$ $&$$ &$\stackrel{03}{[-i]}\,\stackrel{12}{(-)} \stackrel{56}{(-)}$&
$\stackrel{03}{(-i)}\,\stackrel{12}{(-)} \stackrel{56}{(-)} *_{A} 
\stackrel{03}{(+i)}\,\stackrel{12}{[+]} \stackrel{56}{[+]}$\\
\hline
 \end{tabular}
 \end{center}
\end{tiny}
\end{table}
%

%
\begin{table}
\begin{tiny}
\caption{
The even ``basis vectors'' ${}^{II}{\hat{\cal A}}^{m \dagger}_{f}$,
belonging to 
${\cal S}^{12}=0$ in internal space,  for $d=(5+1)$, are 
presented as algebraic products of the first, $m=1$, member of ``basis 
vectors'' $\hat{b}^{m'=1 \dagger}_{f '}$ and the Hermitian conjugated 
partners ($\hat{b}^{m' =1 \dagger}_{f ''})^{\dagger}$. The Hermitian 
conjugated partners  of two ${}^{II}{\hat{\cal A}}^{m \dagger}_{f}$ are 
marked with the same symbol (either $\bigtriangleup$ or $\bullet$). The
symbol $\bigcirc$ presents selfadjoint members.
The even ``basis vectors''  ${}^{II}{\hat{\cal A}}^{m \dagger}_{f}$
are the products of one projector and two nilpotents, or of three projectors
(they are self adjoint), the odd ``basis 
vectors'' and the Hermitian conjugated partners are products of one nilpotent 
and two projectors or of three nilpotents. Even and odd
objects are eigenvectors of all the corresponding Cartan subalgebra members, 
Eq.~(\ref{cartangrasscliff}).  There are $\frac{1}{2}\times 2^{\frac{6}{2}-1}
\times 2^{\frac{6}{2}-1}$ algebraic products of $
(\hat{b}^{m'\dagger}_{f `})^{\dagger} *_{A}$  
$\hat{b}^{m' \dagger}_{f ``}$, $f `$ and $f ``$ run over all
four families. The rest of $16$ members present
${}^{II}{\hat{\cal A}}^{m \dagger}_{f}$ with ${\cal S}^{12}=\pm 1$.
The members $(\hat{b}^{m' \dagger}_{f `})^{\dagger}$ together with
$\hat{b}^{m' \dagger}_{f ``}$ $m'=(1,2,3,4)$, offer the same 
 ${}^{II}{\hat{\cal A}}^{m \dagger}_{f}$, all with ${\cal S}^{12}=0$.
 Table is taken from ~\cite{nh2023dec}.
\vspace{3mm}}
\label{S120Cliff basis5+1even II.} 
 %
 \begin{center}
 \begin{tabular}{ |c| c| c c|}
 \hline
 $$&$$&$$&$$\\
${\cal S}^{12} $&$symbol$&${}^{II}\hat{\cal A}^{m \dagger}_f=$
&$(\hat{b}^{1 \dagger}_{f `})^{\dagger} *_A \hat{b}^{1 \dagger}_{f ``}$\\
\\
\hline
%
$$&$$&$$&$$\\
$0$&$\bigtriangleup$&${}^{II}\hat{\cal A}^{2 \dagger}_1=$&
$(\hat{b}^{1 \dagger}_{3})^{\dagger} *_{A} \hat{b}^{1 \dagger}_{4}$\\
$$&$$&$$&$$\\
$ $&$$ &$\stackrel{03}{(+i)}\,\stackrel{12}{[-]} \stackrel{56}{(+)}$&
$\stackrel{03}{[+i]}\,\stackrel{12}{(-)} \stackrel{56}{[+]} *_{A} 
\stackrel{03}{(+i)}\,\stackrel{12}{(+)} \stackrel{56}{(+)}$\\
\hline
$$&$$&$$&$$\\
$0$&$\bigcirc$&${}^{II}\hat{\cal A}^{4 \dagger}_1=$&
$(\hat{b}^{1 \dagger}_{4})^{\dagger} *_{A} \hat{b}^{1\dagger}_{4}$\\
$$&$$&$$&$$\\
$ $&$$ &$\stackrel{03}{[-i]}\,\stackrel{12}{[-]} \stackrel{56}{[-]}$&
$\stackrel{03}{(-i)}\,\stackrel{12}{(-)} \stackrel{56}{(-)} *_{A} 
\stackrel{03}{(+i)}\,\stackrel{12}{(+)} \stackrel{56}{(+)}$\\
\hline
$$&$$&$$&$$\\
$0$&$\bullet$&${}^{II}\hat{\cal A}^{1 \dagger}_2=$&
$(\hat{b}^{1 \dagger}_{1})^{\dagger} *_{A} \hat{b}^{1 \dagger}_{2}$\\
$$&$$&$$&$$\\
$ $&$$ &$\stackrel{03}{(-i)}\,\stackrel{12}{[+]} \stackrel{56}{(+)}$&
$\stackrel{03}{(-i)}\,\stackrel{12}{[+]} \stackrel{56}{[+]} *_{A} 
\stackrel{03}{[+i]}\,\stackrel{12}{[+]} \stackrel{56}{(+)}$\\
\hline
$$&$$&$$&$$\\
$0$&$\bigcirc$&${}^{II}\hat{\cal A}^{3 \dagger}_2=$&
$(\hat{b}^{1 \dagger}_{2})^{\dagger} *_{A} \hat{b}^{1 \dagger}_{2}$\\
$$&$$&$$&$$\\
$ $&$$ &$\stackrel{03}{[+i]}\,\stackrel{12}{[+]} \stackrel{56}{[-]}$&
$\stackrel{03}{[+i]}\,\stackrel{12}{[+]} \stackrel{56}{(-)} *_{A} 
\stackrel{03}{[+i]}\,\stackrel{12}{[+]} \stackrel{56}{(+)}$\\
\hline
\hline
$$&$$&$$&$$\\
$0$&$\bigcirc$&${}^{II}\hat{\cal A}^{1 \dagger}_3=$&
$(\hat{b}^{1 \dagger}_{1})^{\dagger} *_{A} \hat{b}^{1 \dagger}_{1}$\\
$$&$$&$$&$$\\
$ $&$$ &$\stackrel{03}{[-i]}\,\stackrel{12}{[+]} \stackrel{56}{[+]}$&
$\stackrel{03}{(-i)}\,\stackrel{12}{([+]} \stackrel{56}{[+]} *_{A} 
\stackrel{03}{(+i)}\,\stackrel{12}{[+]} \stackrel{56}{[+]}$\\
\hline
$$&$$&$$&$$\\
$0$&$\bullet$&${}^{II}\hat{\cal A}^{3 \dagger}_3=$&
$(\hat{b}^{1 \dagger}_{2})^{\dagger} *_{A} \hat{b}^{1 \dagger}_{1}$\\
$$&$$&$$&$$\\
$ $&$$ &$\stackrel{03}{(+i)}\,\stackrel{12}{[+]} \stackrel{56}{(-)}$&
$\stackrel{03}{[+i]}\,\stackrel{12}{[+]} \stackrel{56}{(-)} *_{A} 
\stackrel{03}{(+i)}\,\stackrel{12}{[+]} \stackrel{56}{[+]}$\\
\hline
$$&$$&$$&$$\\
$0$&$\bigcirc$&${}^{II} \hat{\cal A}^{2 \dagger}_4=$&
$(\hat{b}^{1 \dagger}_{3})^{\dagger} *_{A} \hat{b}^{1 \dagger}_{3}$\\
$$&$$&$$&$$\\
$ $&$$ &$\stackrel{03}{[+i]}\,\stackrel{12}{[-]} \stackrel{56}{[+]}$&
$\stackrel{03}{[+i]}\,\stackrel{12}{(-)} \stackrel{56}{[+]} *_{A} 
\stackrel{03}{[+i]}\,\stackrel{12}{(+)} \stackrel{56}{[+]}$\\
\hline
$$&$$&$$&$$\\
$0$&$\bigtriangleup$&${}^{II}\hat{\cal A}^{4 \dagger}_4=$&
$(\hat{b}^{1 \dagger}_{4})^{\dagger} *_{A} \hat{b}^{1 \dagger}_{3}$\\
$$&$$&$$&$$\\
$ $&$$ &$\stackrel{03}{(-i)}\,\stackrel{12}{[-]} \stackrel{56}{(-)}$&
$\stackrel{03}{(-i)}\,\stackrel{12}{(-)} \stackrel{56}{(-)} *_{A} 
\stackrel{03}{[+i]}\,\stackrel{12}{(+)} \stackrel{56}{[+]}$\\
\hline
 \end{tabular}
 \end{center}
\end{tiny}
\end{table}

\section{One family representation of odd ``basis vectors'' in $d=(13+1)$
}
\label{13+1representation}  

This appendix, is following App.~D of Ref.~\cite{n2023MDPI}, with a short comment
on the corresponding gauge vector and scalar fields and fermion and boson
representations in $d=(13+1)$-dimensional internal space included.

In even dimensional space $d=(13 +1)$~(\cite{n2022epjc}, App.~A), one irreducible
representation of the odd ``basis vectors'', analysed from the point of view of
the subgroups $SO(3,1)\times SO(4) $ (included in $SO(7,1)$) and $SO(7,1)\times
SO(6)$ (included in $SO(13,1)$, while $SO(6)$ breaks into $SU(3)\times U(1)$),
contains the odd ``basis vectors'' describing internal spaces of quarks and
leptons and antiquarks, and antileptons with the quantum numbers assumed by the
{\it standard model} before the electroweak break. Since $SO(4)$ contains two
$SU(2)$ groups, $SU(2)_I$ and $SU(2)_{II}$, with the hypercharge of the 
{\it standard model} $Y=\tau^{23} + \tau^4$, one irreducible representation 
includes the right-handed neutrinos and the left-handed antineutrinos, which are 
not in the {\it standard model} scheme.

%
The even ``basis vectors'', analysed to the same subgroups,
offer the description of the internal spaces of the corresponding vector and scalar
gauge fields, appearing in the {\it standard model} before the electroweak
break~\cite{n2021SQ,n2022epjc}; as explained in Sect.~\ref{bosons5+1}.

This contribution manifests that the even ``basis vectors'' not only 
offer the description of the internal spaces of the corresponding vector and scalar
gauge fields, appearing in the {\it standard model}, but also describe the 
graviton gauge fields.

For an overview of the properties of the vector and scalar gauge fields in the
{\it spin-charge-family} theory, the reader is invited to see Refs.~%
(\cite{nh2021RPPNP,nd2017} and the references therein). The vector gauge
fields, expressed as the superposition of spin connections and vielbeins, carrying 
the space
index $m=(0,1,2,3)$, manifest properties of the observed boson fields. The scalar
gauge fields, causing the electroweak break, carry the space index $s=(7,8)$ and
determine the symmetry of mass matrices of quarks and leptons.
~\cite{
gmdn2008,gn2014}).

In this Table~\ref{Table so13+1.}, one can check the quantum numbers of the
odd ``basis vectors'' representing quarks and leptons {\it and antiquarks and
antileptons} if taking into account that all the nilpotents and projectors are eigenvectors
of one of the Cartan subalgebra members, ($S^{03}, S^{12}, S^{56}, \dots, $
$S^{13\,14}$), with the eigenvalues $\pm \frac{i}{2}$ for $\stackrel{ab}{(\pm i)}$
and $\stackrel{ab}{[\pm i]}$, and with the eigenvalues $\pm \frac{1}{2}$ for
$\stackrel{ab}{(\pm 1)}$ and $\stackrel{ab}{[\pm 1]}$.

Taking into account that the third component of the {\it standard model} weak 
charge, $\tau^{13}=\frac{1}{2} (S^{56}-S^{78})$, of the third component of
the second $SU(2)$ charge not appearing in the {\it standard model}, 
$\tau^{23}=\frac{1}{2} (S^{56}+ S^{78})$, of the colour charge [$\tau^{33}=
\frac{1}{2} (S^{9\, 10}-S^{11\,12})$ and $\tau^{38}=
\frac{1}{2\sqrt{3}} (S^{9\, 10}+S^{11\,12} - 2 S^{13\,14})$], of the ``fermion
charge'' $\tau^4=-\frac{1}{3} (S^{9\, 10}+S^{11\,12} + S^{13\,14})$, of the hyper
charge $Y=\tau^{23} + \tau^4$, and electromagnetic charge $Q=Y + \tau^{13}$,
one reproduces all the quantum numbers of quarks, leptons, and {\it antiquarks, and
antileptons}. One notices that the $SO(7,1)$ part is the same for quarks and leptons
and the same for antiquarks and antileptons. Quarks distinguish from leptons only in
the colour and ``fermion'' quantum numbers and antiquarks distinguish from 
antileptons only in the anti-colour and ``anti-fermion'' quantum numbers.\\


In odd dimensional internal space, $d=(14+1)$, the eigenstates of handedness are the superposition
of one irreducible representation of $SO(13,1)$, presented in Table~\ref{Table so13+1.},
and the one obtained if on each ``basis vector'' appearing in $SO(13,1)$ the operator
$S^{0 \, (14+1)}$ applies, 
Ref.~\cite{n2023MDPI}.

Let me point out that in addition to the electroweak break of the {\it standard
model} the break at $\ge 10^{16}$ GeV is needed~(\cite{nh2021RPPNP}, and references
therein).
The condensate of the two right-handed neutrinos causes this break
(Ref.~\cite{nh2021RPPNP}, Table 6); it  interacts with all the scalar and vector gauge
fields, except the weak, $U(1), SU(3)$ and the gravitational field in $d=(3+1)$, leaving
these gauge fields massless up to the electroweak break, when the scalar fields, leaving
massless only the electromagnetic, colour and gravitational fields, cause masses of 
fermions and weak bosons.

The theory predicts  at low energies two groups of four families: To the lower 
group of four families, the three so far observed contribute. The theory predicts the symmetry of both groups to be $SU(2)\times SU(2) \times U(1)$, Ref.~(\cite{nh2021RPPNP}, Sect. 7.3), what 
enable to calculate mixing matrices of quarks and leptons for the accurately enough
measured $3\times 3$ sub-matrix of the $4\times 4$ unitary matrix. No sterile neutrinos
are needed, and no symmetry of the mass matrices must be guessed~\cite{gn2013}.

The stable of the upper four families predicted by the {\it spin-charge-family} theory
is a candidate for the dark matter, as discussed in Refs.~\cite{gn2009,nh2021RPPNP}. 
In the literature, there are several works suggesting candidates for the dark matter
and also for matter/antimatter asymmetry. 

%

\bottomcaption{\label{Table so13+1.}%
\begin{small}
The left-handed ($\Gamma^{(13,1)} = -1$, Eq.~(\ref{Gamma})) irreducible representation
of one family of spinors --- the product of the odd number of nilpotents and of projectors,
which are eigenvectors of the Cartan subalgebra of the $SO(13,1)$ 
group~\cite{n2014matterantimatter,nh02}, manifesting the subgroup $SO(7,1)$ of the
colour charged quarks and antiquarks and the colourless leptons and antileptons ---
is presented.
It contains the left-handed ($\Gamma^{(3,1)}=-1$) weak ($SU(2)_{I}$) charged
($\tau^{13}=\pm \frac{1}{2}$), 
and $SU(2)_{II}$ chargeless ($\tau^{23}=0$) 
quarks and leptons, and the right-handed
($\Gamma^{(3,1)}=1$) weak ($SU(2)_{I}$) chargeless and $SU(2)_{II}$ charged
($\tau^{23}=\pm \frac{1}{2}$) quarks and leptons, both with the spin $ S^{12}$ up
and down ($\pm \frac{1}{2}$, respectively).
Quarks distinguish from leptons only in the $SU(3) \times U(1)$ part: Quarks are triplets
of three colours ($c^i$ $= (\tau^{33}, \tau^{38})$ $ = [(\frac{1}{2},\frac{1}{2\sqrt{3}}),
(-\frac{1}{2},\frac{1}{2\sqrt{3}}), (0,-\frac{1}{\sqrt{3}}) $, 
carrying the "fermion charge" ($\tau^{4}=\frac{1}{6}$). 
The colourless leptons carry the "fermion charge" ($\tau^{4}=-\frac{1}{2}$).
The same multiplet contains also the left handed weak ($SU(2)_{I}$) chargeless and
$SU(2)_{II}$ charged antiquarks and antileptons and the right handed weak
($SU(2)_{I}$) charged and $SU(2)_{II}$ chargeless antiquarks and antileptons.
Antiquarks distinguish from antileptons again only in the $SU(3) \times U(1)$ part:
Antiquarks are anti-triplets carrying the "fermion charge" ($\tau^{4}=-\frac{1}{6}$).
The anti-colourless antileptons carry the "fermion charge" ($\tau^{4}=\frac{1}{2}$).
$Y=(\tau^{23} + \tau^{4})$ is the hyper charge, the electromagnetic charge
is $Q=(\tau^{13} + Y$).
%
\end{small}
}

\tablehead{\hline
i&$$&$|^a\psi_i>$&$\Gamma^{(3,1)}$&$ S^{12}$&
$\tau^{13}$&$\tau^{23}$&$\tau^{33}$&$\tau^{38}$&$\tau^{4}$&$Y$&$Q$\\
\hline
&& ${\rm (Anti)octet},\,\Gamma^{(7,1)} = (-1)\,1\,, \,\Gamma^{(6)} = (1)\,-1$&&&&&&&&& \\
&& ${\rm of \;(anti) quarks \;and \;(anti)leptons}$&&&&&&&&&\\
\hline\hline}
\tabletail{\hline \multicolumn{12}{r}{\emph{Continued on next page}}\\}
\tablelasttail{\hline}
\begin{tiny}
\begin{supertabular}{|r|c||c||c|c||c|c||c|c|c||r|r|}
1&$ u_{R}^{c1}$&$ \stackrel{03}{(+i)}\,\stackrel{12}{[+]}|
\stackrel{56}{[+]}\,\stackrel{78}{(+)}
||\stackrel{9 \;10}{(+)}\;\;\stackrel{11\;12}{[-]}\;\;\stackrel{13\;14}{[-]} $ &1&$\frac{1}{2}$&0&
$\frac{1}{2}$&$\frac{1}{2}$&$\frac{1}{2\,\sqrt{3}}$&$\frac{1}{6}$&$\frac{2}{3}$&$\frac{2}{3}$\\
\hline
2&$u_{R}^{c1}$&$\stackrel{03}{[-i]}\,\stackrel{12}{(-)}|\stackrel{56}{[+]}\,\stackrel{78}{(+)}
||\stackrel{9 \;10}{(+)}\;\;\stackrel{11\;12}{[-]}\;\;\stackrel{13\;14}{[-]}$&1&$-\frac{1}{2}$&0&
$\frac{1}{2}$&$\frac{1}{2}$&$\frac{1}{2\,\sqrt{3}}$&$\frac{1}{6}$&$\frac{2}{3}$&$\frac{2}{3}$\\
\hline
3&$d_{R}^{c1}$&$\stackrel{03}{(+i)}\,\stackrel{12}{[+]}|\stackrel{56}{(-)}\,\stackrel{78}{[-]}
||\stackrel{9 \;10}{(+)}\;\;\stackrel{11\;12}{[-]}\;\;\stackrel{13\;14}{[-]}$&1&$\frac{1}{2}$&0&
$-\frac{1}{2}$&$\frac{1}{2}$&$\frac{1}{2\,\sqrt{3}}$&$\frac{1}{6}$&$-\frac{1}{3}$&$-\frac{1}{3}$\\
\hline
4&$ d_{R}^{c1} $&$\stackrel{03}{[-i]}\,\stackrel{12}{(-)}|
\stackrel{56}{(-)}\,\stackrel{78}{[-]}
||\stackrel{9 \;10}{(+)}\;\;\stackrel{11\;12}{[-]}\;\;\stackrel{13\;14}{[-]} $&1&$-\frac{1}{2}$&0&
$-\frac{1}{2}$&$\frac{1}{2}$&$\frac{1}{2\,\sqrt{3}}$&$\frac{1}{6}$&$-\frac{1}{3}$&$-\frac{1}{3}$\\
\hline
5&$d_{L}^{c1}$&$\stackrel{03}{[-i]}\,\stackrel{12}{[+]}|\stackrel{56}{(-)}\,\stackrel{78}{(+)}
||\stackrel{9 \;10}{(+)}\;\;\stackrel{11\;12}{[-]}\;\;\stackrel{13\;14}{[-]}$&-1&$\frac{1}{2}$&
$-\frac{1}{2}$&0&$\frac{1}{2}$&$\frac{1}{2\,\sqrt{3}}$&$\frac{1}{6}$&$\frac{1}{6}$&$-\frac{1}{3}$\\
\hline
6&$d_{L}^{c1} $&$ - \stackrel{03}{(+i)}\,\stackrel{12}{(-)}|\stackrel{56}{(-)}\,\stackrel{78}{(+)}
||\stackrel{9 \;10}{(+)}\;\;\stackrel{11\;12}{[-]}\;\;\stackrel{13\;14}{[-]} $&-1&$-\frac{1}{2}$&
$-\frac{1}{2}$&0&$\frac{1}{2}$&$\frac{1}{2\,\sqrt{3}}$&$\frac{1}{6}$&$\frac{1}{6}$&$-\frac{1}{3}$\\
\hline
7&$ u_{L}^{c1}$&$ - \stackrel{03}{[-i]}\,\stackrel{12}{[+]}|\stackrel{56}{[+]}\,\stackrel{78}{[-]}
||\stackrel{9 \;10}{(+)}\;\;\stackrel{11\;12}{[-]}\;\;\stackrel{13\;14}{[-]}$ &-1&$\frac{1}{2}$&
$\frac{1}{2}$&0 &$\frac{1}{2}$&$\frac{1}{2\,\sqrt{3}}$&$\frac{1}{6}$&$\frac{1}{6}$&$\frac{2}{3}$\\
\hline
8&$u_{L}^{c1}$&$\stackrel{03}{(+i)}\,\stackrel{12}{(-)}|\stackrel{56}{[+]}\,\stackrel{78}{[-]}
||\stackrel{9 \;10}{(+)}\;\;\stackrel{11\;12}{[-]}\;\;\stackrel{13\;14}{[-]}$&-1&$-\frac{1}{2}$&
$\frac{1}{2}$&0&$\frac{1}{2}$&$\frac{1}{2\,\sqrt{3}}$&$\frac{1}{6}$&$\frac{1}{6}$&$\frac{2}{3}$\\
\hline\hline
\shrinkheight{0.25\textheight}
9&$ u_{R}^{c2}$&$ \stackrel{03}{(+i)}\,\stackrel{12}{[+]}|
\stackrel{56}{[+]}\,\stackrel{78}{(+)}
||\stackrel{9 \;10}{[-]}\;\;\stackrel{11\;12}{(+)}\;\;\stackrel{13\;14}{[-]} $ &1&$\frac{1}{2}$&0&
$\frac{1}{2}$&$-\frac{1}{2}$&$\frac{1}{2\,\sqrt{3}}$&$\frac{1}{6}$&$\frac{2}{3}$&$\frac{2}{3}$\\
\hline
10&$u_{R}^{c2}$&$\stackrel{03}{[-i]}\,\stackrel{12}{(-)}|\stackrel{56}{[+]}\,\stackrel{78}{(+)}
||\stackrel{9 \;10}{[-]}\;\;\stackrel{11\;12}{(+)}\;\;\stackrel{13\;14}{[-]}$&1&$-\frac{1}{2}$&0&
$\frac{1}{2}$&$-\frac{1}{2}$&$\frac{1}{2\,\sqrt{3}}$&$\frac{1}{6}$&$\frac{2}{3}$&$\frac{2}{3}$\\
\hline
11&$d_{R}^{c2}$&$\stackrel{03}{(+i)}\,\stackrel{12}{[+]}|\stackrel{56}{(-)}\,\stackrel{78}{[-]}
||\stackrel{9 \;10}{[-]}\;\;\stackrel{11\;12}{(+)}\;\;\stackrel{13\;14}{[-]}$
&1&$\frac{1}{2}$&0&
$-\frac{1}{2}$&$ - \frac{1}{2}$&$\frac{1}{2\,\sqrt{3}}$&$\frac{1}{6}$&$-\frac{1}{3}$&$-\frac{1}{3}$\\
\hline
12&$ d_{R}^{c2} $&$\stackrel{03}{[-i]}\,\stackrel{12}{(-)}|
\stackrel{56}{(-)}\,\stackrel{78}{[-]}
||\stackrel{9 \;10}{[-]}\;\;\stackrel{11\;12}{(+)}\;\;\stackrel{13\;14}{[-]} $
&1&$-\frac{1}{2}$&0&
$-\frac{1}{2}$&$-\frac{1}{2}$&$\frac{1}{2\,\sqrt{3}}$&$\frac{1}{6}$&$-\frac{1}{3}$&$-\frac{1}{3}$\\
\hline
13&$d_{L}^{c2}$&$\stackrel{03}{[-i]}\,\stackrel{12}{[+]}|\stackrel{56}{(-)}\,\stackrel{78}{(+)}
||\stackrel{9 \;10}{[-]}\;\;\stackrel{11\;12}{(+)}\;\;\stackrel{13\;14}{[-]}$
&-1&$\frac{1}{2}$&
$-\frac{1}{2}$&0&$-\frac{1}{2}$&$\frac{1}{2\,\sqrt{3}}$&$\frac{1}{6}$&$\frac{1}{6}$&$-\frac{1}{3}$\\
\hline
14&$d_{L}^{c2} $&$ - \stackrel{03}{(+i)}\,\stackrel{12}{(-)}|\stackrel{56}{(-)}\,\stackrel{78}{(+)}
||\stackrel{9 \;10}{[-]}\;\;\stackrel{11\;12}{(+)}\;\;\stackrel{13\;14}{[-]} $&-1&$-\frac{1}{2}$&
$-\frac{1}{2}$&0&$-\frac{1}{2}$&$\frac{1}{2\,\sqrt{3}}$&$\frac{1}{6}$&$\frac{1}{6}$&$-\frac{1}{3}$\\
\hline
15&$ u_{L}^{c2}$&$ - \stackrel{03}{[-i]}\,\stackrel{12}{[+]}|\stackrel{56}{[+]}\,\stackrel{78}{[-]}
||\stackrel{9 \;10}{[-]}\;\;\stackrel{11\;12}{(+)}\;\;\stackrel{13\;14}{[-]}$ &-1&$\frac{1}{2}$&
$\frac{1}{2}$&0 &$-\frac{1}{2}$&$\frac{1}{2\,\sqrt{3}}$&$\frac{1}{6}$&$\frac{1}{6}$&$\frac{2}{3}$\\
\hline
16&$u_{L}^{c2}$&$\stackrel{03}{(+i)}\,\stackrel{12}{(-)}|\stackrel{56}{[+]}\,\stackrel{78}{[-]}
||\stackrel{9 \;10}{[-]}\;\;\stackrel{11\;12}{(+)}\;\;\stackrel{13\;14}{[-]}$&-1&$-\frac{1}{2}$&
$\frac{1}{2}$&0&$-\frac{1}{2}$&$\frac{1}{2\,\sqrt{3}}$&$\frac{1}{6}$&$\frac{1}{6}$&$\frac{2}{3}$\\
\hline\hline
17&$ u_{R}^{c3}$&$ \stackrel{03}{(+i)}\,\stackrel{12}{[+]}|
\stackrel{56}{[+]}\,\stackrel{78}{(+)}
||\stackrel{9 \;10}{[-]}\;\;\stackrel{11\;12}{[-]}\;\;\stackrel{13\;14}{(+)} $ &1&$\frac{1}{2}$&0&
$\frac{1}{2}$&$0$&$-\frac{1}{\sqrt{3}}$&$\frac{1}{6}$&$\frac{2}{3}$&$\frac{2}{3}$\\
\hline
18&$u_{R}^{c3}$&$\stackrel{03}{[-i]}\,\stackrel{12}{(-)}|\stackrel{56}{[+]}\,\stackrel{78}{(+)}
||\stackrel{9 \;10}{[-]}\;\;\stackrel{11\;12}{[-]}\;\;\stackrel{13\;14}{(+)}$&1&$-\frac{1}{2}$&0&
$\frac{1}{2}$&$0$&$-\frac{1}{\sqrt{3}}$&$\frac{1}{6}$&$\frac{2}{3}$&$\frac{2}{3}$\\
\hline
19&$d_{R}^{c3}$&$\stackrel{03}{(+i)}\,\stackrel{12}{[+]}|\stackrel{56}{(-)}\,\stackrel{78}{[-]}
||\stackrel{9 \;10}{[-]}\;\;\stackrel{11\;12}{[-]}\;\;\stackrel{13\;14}{(+)}$&1&$\frac{1}{2}$&0&
$-\frac{1}{2}$&$0$&$-\frac{1}{\sqrt{3}}$&$\frac{1}{6}$&$-\frac{1}{3}$&$-\frac{1}{3}$\\
\hline
20&$ d_{R}^{c3} $&$\stackrel{03}{[-i]}\,\stackrel{12}{(-)}|
\stackrel{56}{(-)}\,\stackrel{78}{[-]}
||\stackrel{9 \;10}{[-]}\;\;\stackrel{11\;12}{[-]}\;\;\stackrel{13\;14}{(+)} $&1&$-\frac{1}{2}$&0&
$-\frac{1}{2}$&$0$&$-\frac{1}{\sqrt{3}}$&$\frac{1}{6}$&$-\frac{1}{3}$&$-\frac{1}{3}$\\
\hline
21&$d_{L}^{c3}$&$\stackrel{03}{[-i]}\,\stackrel{12}{[+]}|\stackrel{56}{(-)}\,\stackrel{78}{(+)}
||\stackrel{9 \;10}{[-]}\;\;\stackrel{11\;12}{[-]}\;\;\stackrel{13\;14}{(+)}$&-1&$\frac{1}{2}$&
$-\frac{1}{2}$&0&$0$&$-\frac{1}{\sqrt{3}}$&$\frac{1}{6}$&$\frac{1}{6}$&$-\frac{1}{3}$\\
\hline
22&$d_{L}^{c3} $&$ - \stackrel{03}{(+i)}\,\stackrel{12}{(-)}|\stackrel{56}{(-)}\,\stackrel{78}{(+)}
||\stackrel{9 \;10}{[-]}\;\;\stackrel{11\;12}{[-]}\;\;\stackrel{13\;14}{(+)} $&-1&$-\frac{1}{2}$&
$-\frac{1}{2}$&0&$0$&$-\frac{1}{\sqrt{3}}$&$\frac{1}{6}$&$\frac{1}{6}$&$-\frac{1}{3}$\\
\hline
23&$ u_{L}^{c3}$&$ - \stackrel{03}{[-i]}\,\stackrel{12}{[+]}|\stackrel{56}{[+]}\,\stackrel{78}{[-]}
||\stackrel{9 \;10}{[-]}\;\;\stackrel{11\;12}{[-]}\;\;\stackrel{13\;14}{(+)}$ &-1&$\frac{1}{2}$&
$\frac{1}{2}$&0 &$0$&$-\frac{1}{\sqrt{3}}$&$\frac{1}{6}$&$\frac{1}{6}$&$\frac{2}{3}$\\
\hline
24&$u_{L}^{c3}$&$\stackrel{03}{(+i)}\,\stackrel{12}{(-)}|\stackrel{56}{[+]}\,\stackrel{78}{[-]}
||\stackrel{9 \;10}{[-]}\;\;\stackrel{11\;12}{[-]}\;\;\stackrel{13\;14}{(+)}$&-1&$-\frac{1}{2}$&
$\frac{1}{2}$&0&$0$&$-\frac{1}{\sqrt{3}}$&$\frac{1}{6}$&$\frac{1}{6}$&$\frac{2}{3}$\\
\hline\hline
25&$ \nu_{R}$&$ \stackrel{03}{(+i)}\,\stackrel{12}{[+]}|
\stackrel{56}{[+]}\,\stackrel{78}{(+)}
||\stackrel{9 \;10}{(+)}\;\;\stackrel{11\;12}{(+)}\;\;\stackrel{13\;14}{(+)} $ &1&$\frac{1}{2}$&0&
$\frac{1}{2}$&$0$&$0$&$-\frac{1}{2}$&$0$&$0$\\
\hline
26&$\nu_{R}$&$\stackrel{03}{[-i]}\,\stackrel{12}{(-)}|\stackrel{56}{[+]}\,\stackrel{78}{(+)}
||\stackrel{9 \;10}{(+)}\;\;\stackrel{11\;12}{(+)}\;\;\stackrel{13\;14}{(+)}$&1&$-\frac{1}{2}$&0&
$\frac{1}{2}$ &$0$&$0$&$-\frac{1}{2}$&$0$&$0$\\
\hline
27&$e_{R}$&$\stackrel{03}{(+i)}\,\stackrel{12}{[+]}|\stackrel{56}{(-)}\,\stackrel{78}{[-]}
||\stackrel{9 \;10}{(+)}\;\;\stackrel{11\;12}{(+)}\;\;\stackrel{13\;14}{(+)}$&1&$\frac{1}{2}$&0&
$-\frac{1}{2}$&$0$&$0$&$-\frac{1}{2}$&$-1$&$-1$\\
\hline
28&$ e_{R} $&$\stackrel{03}{[-i]}\,\stackrel{12}{(-)}|
\stackrel{56}{(-)}\,\stackrel{78}{[-]}
||\stackrel{9 \;10}{(+)}\;\;\stackrel{11\;12}{(+)}\;\;\stackrel{13\;14}{(+)} $&1&$-\frac{1}{2}$&0&
$-\frac{1}{2}$&$0$&$0$&$-\frac{1}{2}$&$-1$&$-1$\\
\hline
29&$e_{L}$&$\stackrel{03}{[-i]}\,\stackrel{12}{[+]}|\stackrel{56}{(-)}\,\stackrel{78}{(+)}
||\stackrel{9 \;10}{(+)}\;\;\stackrel{11\;12}{(+)}\;\;\stackrel{13\;14}{(+)}$&-1&$\frac{1}{2}$&
$-\frac{1}{2}$&0&$0$&$0$&$-\frac{1}{2}$&$-\frac{1}{2}$&$-1$\\
\hline
30&$e_{L} $&$ - \stackrel{03}{(+i)}\,\stackrel{12}{(-)}|\stackrel{56}{(-)}\,\stackrel{78}{(+)}
||\stackrel{9 \;10}{(+)}\;\;\stackrel{11\;12}{(+)}\;\;\stackrel{13\;14}{(+)} $&-1&$-\frac{1}{2}$&
$-\frac{1}{2}$&0&$0$&$0$&$-\frac{1}{2}$&$-\frac{1}{2}$&$-1$\\
\hline
31&$ \nu_{L}$&$ - \stackrel{03}{[-i]}\,\stackrel{12}{[+]}|\stackrel{56}{[+]}\,\stackrel{78}{[-]}
||\stackrel{9 \;10}{(+)}\;\;\stackrel{11\;12}{(+)}\;\;\stackrel{13\;14}{(+)}$ &-1&$\frac{1}{2}$&
$\frac{1}{2}$&0 &$0$&$0$&$-\frac{1}{2}$&$-\frac{1}{2}$&$0$\\
\hline
32&$\nu_{L}$&$\stackrel{03}{(+i)}\,\stackrel{12}{(-)}|\stackrel{56}{[+]}\,\stackrel{78}{[-]}
||\stackrel{9 \;10}{(+)}\;\;\stackrel{11\;12}{(+)}\;\;\stackrel{13\;14}{(+)}$&-1&$-\frac{1}{2}$&
$\frac{1}{2}$&0&$0$&$0$&$-\frac{1}{2}$&$-\frac{1}{2}$&$0$\\
\hline\hline
33&$ \bar{d}_{L}^{\bar{c1}}$&$ \stackrel{03}{[-i]}\,\stackrel{12}{[+]}|
\stackrel{56}{[+]}\,\stackrel{78}{(+)}
||\stackrel{9 \;10}{[-]}\;\;\stackrel{11\;12}{(+)}\;\;\stackrel{13\;14}{(+)} $ &-1&$\frac{1}{2}$&0&
$\frac{1}{2}$&$-\frac{1}{2}$&$-\frac{1}{2\,\sqrt{3}}$&$-\frac{1}{6}$&$\frac{1}{3}$&$\frac{1}{3}$\\
\hline
34&$\bar{d}_{L}^{\bar{c1}}$&$\stackrel{03}{(+i)}\,\stackrel{12}{(-)}|\stackrel{56}{[+]}\,\stackrel{78}{(+)}
||\stackrel{9 \;10}{[-]}\;\;\stackrel{11\;12}{(+)}\;\;\stackrel{13\;14}{(+)}$&-1&$-\frac{1}{2}$&0&
$\frac{1}{2}$&$-\frac{1}{2}$&$-\frac{1}{2\,\sqrt{3}}$&$-\frac{1}{6}$&$\frac{1}{3}$&$\frac{1}{3}$\\
\hline
35&$\bar{u}_{L}^{\bar{c1}}$&$ - \stackrel{03}{[-i]}\,\stackrel{12}{[+]}|\stackrel{56}{(-)}\,\stackrel{78}{[-]}
||\stackrel{9 \;10}{[-]}\;\;\stackrel{11\;12}{(+)}\;\;\stackrel{13\;14}{(+)}$&-1&$\frac{1}{2}$&0&
$-\frac{1}{2}$&$-\frac{1}{2}$&$-\frac{1}{2\,\sqrt{3}}$&$-\frac{1}{6}$&$-\frac{2}{3}$&$-\frac{2}{3}$\\
\hline
36&$ \bar{u}_{L}^{\bar{c1}} $&$ - \stackrel{03}{(+i)}\,\stackrel{12}{(-)}|
\stackrel{56}{(-)}\,\stackrel{78}{[-]}
||\stackrel{9 \;10}{[-]}\;\;\stackrel{11\;12}{(+)}\;\;\stackrel{13\;14}{(+)} $&-1&$-\frac{1}{2}$&0&
$-\frac{1}{2}$&$-\frac{1}{2}$&$-\frac{1}{2\,\sqrt{3}}$&$-\frac{1}{6}$&$-\frac{2}{3}$&$-\frac{2}{3}$\\
\hline
37&$\bar{d}_{R}^{\bar{c1}}$&$\stackrel{03}{(+i)}\,\stackrel{12}{[+]}|\stackrel{56}{[+]}\,\stackrel{78}{[-]}
||\stackrel{9 \;10}{[-]}\;\;\stackrel{11\;12}{(+)}\;\;\stackrel{13\;14}{(+)}$&1&$\frac{1}{2}$&
$\frac{1}{2}$&0&$-\frac{1}{2}$&$-\frac{1}{2\,\sqrt{3}}$&$-\frac{1}{6}$&$-\frac{1}{6}$&$\frac{1}{3}$\\
\hline
38&$\bar{d}_{R}^{\bar{c1}} $&$ - \stackrel{03}{[-i]}\,\stackrel{12}{(-)}|\stackrel{56}{[+]}\,\stackrel{78}{[-]}
||\stackrel{9 \;10}{[-]}\;\;\stackrel{11\;12}{(+)}\;\;\stackrel{13\;14}{(+)} $&1&$-\frac{1}{2}$&
$\frac{1}{2}$&0&$-\frac{1}{2}$&$-\frac{1}{2\,\sqrt{3}}$&$-\frac{1}{6}$&$-\frac{1}{6}$&$\frac{1}{3}$\\
\hline
39&$ \bar{u}_{R}^{\bar{c1}}$&$\stackrel{03}{(+i)}\,\stackrel{12}{[+]}|\stackrel{56}{(-)}\,\stackrel{78}{(+)}
||\stackrel{9 \;10}{[-]}\;\;\stackrel{11\;12}{(+)}\;\;\stackrel{13\;14}{(+)}$ &1&$\frac{1}{2}$&
$-\frac{1}{2}$&0 &$-\frac{1}{2}$&$-\frac{1}{2\,\sqrt{3}}$&$-\frac{1}{6}$&$-\frac{1}{6}$&$-\frac{2}{3}$\\
\hline
40&$\bar{u}_{R}^{\bar{c1}}$&$\stackrel{03}{[-i]}\,\stackrel{12}{(-)}|\stackrel{56}{(-)}\,\stackrel{78}{(+)}
||\stackrel{9 \;10}{[-]}\;\;\stackrel{11\;12}{(+)}\;\;\stackrel{13\;14}{(+)}$
&1&$-\frac{1}{2}$&
$-\frac{1}{2}$&0&$-\frac{1}{2}$&$-\frac{1}{2\,\sqrt{3}}$&$-\frac{1}{6}$&$-\frac{1}{6}$&$-\frac{2}{3}$\\
\hline\hline
41&$ \bar{d}_{L}^{\bar{c2}}$&$ \stackrel{03}{[-i]}\,\stackrel{12}{[+]}|
\stackrel{56}{[+]}\,\stackrel{78}{(+)}
||\stackrel{9 \;10}{(+)}\;\;\stackrel{11\;12}{[-]}\;\;\stackrel{13\;14}{(+)} $
&-1&$\frac{1}{2}$&0&
$\frac{1}{2}$&$\frac{1}{2}$&$-\frac{1}{2\,\sqrt{3}}$&$-\frac{1}{6}$&$\frac{1}{3}$&$\frac{1}{3}$\\
\hline
42&$\bar{d}_{L}^{\bar{c2}}$&$\stackrel{03}{(+i)}\,\stackrel{12}{(-)}|\stackrel{56}{[+]}\,\stackrel{78}{(+)}
||\stackrel{9 \;10}{(+)}\;\;\stackrel{11\;12}{[-]}\;\;\stackrel{13\;14}{(+)}$
&-1&$-\frac{1}{2}$&0&
$\frac{1}{2}$&$\frac{1}{2}$&$-\frac{1}{2\,\sqrt{3}}$&$-\frac{1}{6}$&$\frac{1}{3}$&$\frac{1}{3}$\\
\hline
43&$\bar{u}_{L}^{\bar{c2}}$&$ - \stackrel{03}{[-i]}\,\stackrel{12}{[+]}|\stackrel{56}{(-)}\,\stackrel{78}{[-]}
||\stackrel{9 \;10}{(+)}\;\;\stackrel{11\;12}{[-]}\;\;\stackrel{13\;14}{(+)}$
&-1&$\frac{1}{2}$&0&
$-\frac{1}{2}$&$\frac{1}{2}$&$-\frac{1}{2\,\sqrt{3}}$&$-\frac{1}{6}$&$-\frac{2}{3}$&$-\frac{2}{3}$\\
\hline
44&$ \bar{u}_{L}^{\bar{c2}} $&$ - \stackrel{03}{(+i)}\,\stackrel{12}{(-)}|
\stackrel{56}{(-)}\,\stackrel{78}{[-]}
||\stackrel{9 \;10}{(+)}\;\;\stackrel{11\;12}{[-]}\;\;\stackrel{13\;14}{(+)} $
&-1&$-\frac{1}{2}$&0&
$-\frac{1}{2}$&$\frac{1}{2}$&$-\frac{1}{2\,\sqrt{3}}$&$-\frac{1}{6}$&$-\frac{2}{3}$&$-\frac{2}{3}$\\
\hline
45&$\bar{d}_{R}^{\bar{c2}}$&$\stackrel{03}{(+i)}\,\stackrel{12}{[+]}|\stackrel{56}{[+]}\,\stackrel{78}{[-]}
||\stackrel{9 \;10}{(+)}\;\;\stackrel{11\;12}{[-]}\;\;\stackrel{13\;14}{(+)}$
&1&$\frac{1}{2}$&
$\frac{1}{2}$&0&$\frac{1}{2}$&$-\frac{1}{2\,\sqrt{3}}$&$-\frac{1}{6}$&$-\frac{1}{6}$&$\frac{1}{3}$\\
\hline
46&$\bar{d}_{R}^{\bar{c2}} $&$ - \stackrel{03}{[-i]}\,\stackrel{12}{(-)}|\stackrel{56}{[+]}\,\stackrel{78}{[-]}
||\stackrel{9 \;10}{(+)}\;\;\stackrel{11\;12}{[-]}\;\;\stackrel{13\;14}{(+)} $
&1&$-\frac{1}{2}$&
$\frac{1}{2}$&0&$\frac{1}{2}$&$-\frac{1}{2\,\sqrt{3}}$&$-\frac{1}{6}$&$-\frac{1}{6}$&$\frac{1}{3}$\\
\hline
47&$ \bar{u}_{R}^{\bar{c2}}$&$\stackrel{03}{(+i)}\,\stackrel{12}{[+]}|\stackrel{56}{(-)}\,\stackrel{78}{(+)}
||\stackrel{9 \;10}{(+)}\;\;\stackrel{11\;12}{[-]}\;\;\stackrel{13\;14}{(+)}$
 &1&$\frac{1}{2}$&
$-\frac{1}{2}$&0 &$\frac{1}{2}$&$-\frac{1}{2\,\sqrt{3}}$&$-\frac{1}{6}$&$-\frac{1}{6}$&$-\frac{2}{3}$\\
\hline
48&$\bar{u}_{R}^{\bar{c2}}$&$\stackrel{03}{[-i]}\,\stackrel{12}{(-)}|\stackrel{56}{(-)}\,\stackrel{78}{(+)}
||\stackrel{9 \;10}{(+)}\;\;\stackrel{11\;12}{[-]}\;\;\stackrel{13\;14}{(+)}$
&1&$-\frac{1}{2}$&
$-\frac{1}{2}$&0&$\frac{1}{2}$&$-\frac{1}{2\,\sqrt{3}}$&$-\frac{1}{6}$&$-\frac{1}{6}$&$-\frac{2}{3}$\\
\hline\hline
49&$ \bar{d}_{L}^{\bar{c3}}$&$ \stackrel{03}{[-i]}\,\stackrel{12}{[+]}|
\stackrel{56}{[+]}\,\stackrel{78}{(+)}
||\stackrel{9 \;10}{(+)}\;\;\stackrel{11\;12}{(+)}\;\;\stackrel{13\;14}{[-]} $ &-1&$\frac{1}{2}$&0&
$\frac{1}{2}$&$0$&$\frac{1}{\sqrt{3}}$&$-\frac{1}{6}$&$\frac{1}{3}$&$\frac{1}{3}$\\
\hline
50&$\bar{d}_{L}^{\bar{c3}}$&$\stackrel{03}{(+i)}\,\stackrel{12}{(-)}|\stackrel{56}{[+]}\,\stackrel{78}{(+)}
||\stackrel{9 \;10}{(+)}\;\;\stackrel{11\;12}{(+)}\;\;\stackrel{13\;14}{[-]} $&-1&$-\frac{1}{2}$&0&
$\frac{1}{2}$&$0$&$\frac{1}{\sqrt{3}}$&$-\frac{1}{6}$&$\frac{1}{3}$&$\frac{1}{3}$\\
\hline
51&$\bar{u}_{L}^{\bar{c3}}$&$ - \stackrel{03}{[-i]}\,\stackrel{12}{[+]}|\stackrel{56}{(-)}\,\stackrel{78}{[-]}
||\stackrel{9 \;10}{(+)}\;\;\stackrel{11\;12}{(+)}\;\;\stackrel{13\;14}{[-]} $&-1&$\frac{1}{2}$&0&
$-\frac{1}{2}$&$0$&$\frac{1}{\sqrt{3}}$&$-\frac{1}{6}$&$-\frac{2}{3}$&$-\frac{2}{3}$\\
\hline
52&$ \bar{u}_{L}^{\bar{c3}} $&$ - \stackrel{03}{(+i)}\,\stackrel{12}{(-)}|
\stackrel{56}{(-)}\,\stackrel{78}{[-]}
||\stackrel{9 \;10}{(+)}\;\;\stackrel{11\;12}{(+)}\;\;\stackrel{13\;14}{[-]}  $&-1&$-\frac{1}{2}$&0&
$-\frac{1}{2}$&$0$&$\frac{1}{\sqrt{3}}$&$-\frac{1}{6}$&$-\frac{2}{3}$&$-\frac{2}{3}$\\
\hline
53&$\bar{d}_{R}^{\bar{c3}}$&$\stackrel{03}{(+i)}\,\stackrel{12}{[+]}|\stackrel{56}{[+]}\,\stackrel{78}{[-]}
||\stackrel{9 \;10}{(+)}\;\;\stackrel{11\;12}{(+)}\;\;\stackrel{13\;14}{[-]} $&1&$\frac{1}{2}$&
$\frac{1}{2}$&0&$0$&$\frac{1}{\sqrt{3}}$&$-\frac{1}{6}$&$-\frac{1}{6}$&$\frac{1}{3}$\\
\hline
54&$\bar{d}_{R}^{\bar{c3}} $&$ - \stackrel{03}{[-i]}\,\stackrel{12}{(-)}|\stackrel{56}{[+]}\,\stackrel{78}{[-]}
||\stackrel{9 \;10}{(+)}\;\;\stackrel{11\;12}{(+)}\;\;\stackrel{13\;14}{[-]} $&1&$-\frac{1}{2}$&
$\frac{1}{2}$&0&$0$&$\frac{1}{\sqrt{3}}$&$-\frac{1}{6}$&$-\frac{1}{6}$&$\frac{1}{3}$\\
\hline
55&$ \bar{u}_{R}^{\bar{c3}}$&$\stackrel{03}{(+i)}\,\stackrel{12}{[+]}|\stackrel{56}{(-)}\,\stackrel{78}{(+)}
||\stackrel{9 \;10}{(+)}\;\;\stackrel{11\;12}{(+)}\;\;\stackrel{13\;14}{[-]} $ &1&$\frac{1}{2}$&
$-\frac{1}{2}$&0 &$0$&$\frac{1}{\sqrt{3}}$&$-\frac{1}{6}$&$-\frac{1}{6}$&$-\frac{2}{3}$\\
\hline
56&$\bar{u}_{R}^{\bar{c3}}$&$\stackrel{03}{[-i]}\,\stackrel{12}{(-)}|\stackrel{56}{(-)}\,\stackrel{78}{(+)}
||\stackrel{9 \;10}{(+)}\;\;\stackrel{11\;12}{(+)}\;\;\stackrel{13\;14}{[-]} $&1&$-\frac{1}{2}$&
$-\frac{1}{2}$&0&$0$&$\frac{1}{\sqrt{3}}$&$-\frac{1}{6}$&$-\frac{1}{6}$&$-\frac{2}{3}$\\
\hline\hline
57&$ \bar{e}_{L}$&$ \stackrel{03}{[-i]}\,\stackrel{12}{[+]}|
\stackrel{56}{[+]}\,\stackrel{78}{(+)}
||\stackrel{9 \;10}{[-]}\;\;\stackrel{11\;12}{[-]}\;\;\stackrel{13\;14}{[-]} $ &-1&$\frac{1}{2}$&0&
$\frac{1}{2}$&$0$&$0$&$\frac{1}{2}$&$1$&$1$\\
\hline
58&$\bar{e}_{L}$&$\stackrel{03}{(+i)}\,\stackrel{12}{(-)}|\stackrel{56}{[+]}\,\stackrel{78}{(+)}
||\stackrel{9 \;10}{[-]}\;\;\stackrel{11\;12}{[-]}\;\;\stackrel{13\;14}{[-]}$&-1&$-\frac{1}{2}$&0&
$\frac{1}{2}$ &$0$&$0$&$\frac{1}{2}$&$1$&$1$\\
\hline
59&$\bar{\nu}_{L}$&$ - \stackrel{03}{[-i]}\,\stackrel{12}{[+]}|\stackrel{56}{(-)}\,\stackrel{78}{[-]}
||\stackrel{9 \;10}{[-]}\;\;\stackrel{11\;12}{[-]}\;\;\stackrel{13\;14}{[-]}$&-1&$\frac{1}{2}$&0&
$-\frac{1}{2}$&$0$&$0$&$\frac{1}{2}$&$0$&$0$\\
\hline
60&$ \bar{\nu}_{L} $&$ - \stackrel{03}{(+i)}\,\stackrel{12}{(-)}|
\stackrel{56}{(-)}\,\stackrel{78}{[-]}
||\stackrel{9 \;10}{[-]}\;\;\stackrel{11\;12}{[-]}\;\;\stackrel{13\;14}{[-]} $&-1&$-\frac{1}{2}$&0&
$-\frac{1}{2}$&$0$&$0$&$\frac{1}{2}$&$0$&$0$\\
\hline
61&$\bar{\nu}_{R}$&$\stackrel{03}{(+i)}\,\stackrel{12}{[+]}|\stackrel{56}{(-)}\,\stackrel{78}{(+)}
||\stackrel{9 \;10}{[-]}\;\;\stackrel{11\;12}{[-]}\;\;\stackrel{13\;14}{[-]}$&1&$\frac{1}{2}$&
$-\frac{1}{2}$&0&$0$&$0$&$\frac{1}{2}$&$\frac{1}{2}$&$0$\\
\hline
62&$\bar{\nu}_{R} $&$ - \stackrel{03}{[-i]}\,\stackrel{12}{(-)}|\stackrel{56}{(-)}\,\stackrel{78}{(+)}
||\stackrel{9 \;10}{[-]}\;\;\stackrel{11\;12}{[-]}\;\;\stackrel{13\;14}{[-]} $&1&$-\frac{1}{2}$&
$-\frac{1}{2}$&0&$0$&$0$&$\frac{1}{2}$&$\frac{1}{2}$&$0$\\
\hline
63&$ \bar{e}_{R}$&$\stackrel{03}{(+i)}\,\stackrel{12}{[+]}|\stackrel{56}{[+]}\,\stackrel{78}{[-]}
||\stackrel{9 \;10}{[-]}\;\;\stackrel{11\;12}{[-]}\;\;\stackrel{13\;14}{[-]}$ &1&$\frac{1}{2}$&
$\frac{1}{2}$&0 &$0$&$0$&$\frac{1}{2}$&$\frac{1}{2}$&$1$\\
\hline
64&$\bar{e}_{R}$&$\stackrel{03}{[-i]}\,\stackrel{12}{(-)}|\stackrel{56}{[+]}\,\stackrel{78}{[-]}
||\stackrel{9 \;10}{[-]}\;\;\stackrel{11\;12}{[-]}\;\;\stackrel{13\;14}{[-]}$&1&$-\frac{1}{2}$&
$\frac{1}{2}$&0&$0$&$0$&$\frac{1}{2}$&$\frac{1}{2}$&$1$\\
\hline
\end{supertabular}
\end{tiny}

\vspace {3mm}
 
\section{Acknowledgement}
 The author thanks Department of Physics, FMF, University of Ljubljana, 
Society of  Mathematicians, Physicists and Astronomers of Slovenia for supporting  
the research on the  {\it spin-charge-family} theory, and
 Matja\v z Breskvar of  Beyond Semiconductor for donations, in particular for 
sponsoring the annual workshops entitled  "What comes beyond the standard
 models" at Bled, in which the ideas and realizations, presented in this paper,
 were discussed. The author thanks Holger Beck Nielsen and Milutin 
 Blagojevi\' c   for fruitful discussions.



\begin{thebibliography}{}
\bibitem{n2024NPBarxiv240709482}  N. S. Manko\v c Bor\v stnik,
 "Can the ``basis vectors'', describing the internal spaces of fermion and boson fields with the Clifford odd 
(for fermion) and Clifford even (for boson) objects, explain interactions among fields, with gravitons 
included?" NPB-D-24-00256  [arxiv: 2407.09482].
\end{thebibliography}
\end{document}